\documentclass[12pt]{article}
\usepackage{customstyle}
\usepackage{macros}

\title{\Large Chiseling: Powerful and Valid Subgroup Selection via Interactive Machine Learning}
\author[1]{Nathan Cheng}
\author[2]{Asher Spector}
\author[1]{Lucas Janson}
\date{}
\affil[1]{Department of Statistics, Harvard University}
\affil[2]{Department of Statistics, Stanford University}

\begin{document}
\maketitle

\begin{abstract}
In regression and causal inference, \textit{controlled subgroup selection} aims to identify, with inferential guarantees, a subgroup (defined as a subset of the covariate space) on which the average response or treatment effect is above a given threshold. E.g., in a clinical trial, it may be of interest to find a subgroup with a positive average treatment effect. However, existing methods either lack inferential guarantees, heavily restrict the search for the subgroup, or sacrifice efficiency by naive data splitting. We propose a novel framework called \textit{chiseling} that allows the analyst to \textit{interactively} refine and test a candidate subgroup by iteratively shrinking it. The sole restriction is that the shrinkage direction only depends on the points outside the current subgroup, but otherwise the analyst may leverage any prior information or machine learning algorithm. Despite this flexibility, chiseling controls the probability that the discovered subgroup is null (e.g., has a non-positive average treatment effect) under minimal assumptions: for example, in randomized experiments, this inferential validity guarantee holds under only bounded moment conditions. When applied to a variety of simulated datasets and a real survey experiment, chiseling identifies substantially better subgroups than existing methods with inferential guarantees.
\end{abstract}

\FloatBarrier

\section{Introduction}

\subsection{Motivation}

When a drug is tested in a clinical trial, the primary question of interest is typically: ``Will people benefit on average?" This leads to inference on the drug's population average treatment effect. But a strictly more informative question is: ``\emph{Who} will benefit from the drug?" It could be that a drug does not benefit the population on average but provides substantial average benefit to a subgroup. The task of finding such a subgroup is called \emph{subgroup selection} and has a long history of study, but methods for subgroup selection face the fundamental challenge of using the same data for both identifying a subgroup and performing statistical inference on that subgroup. Thus, the analyst either faces a difficult tradeoff between discovery and validation, or else imposes structural assumptions (such as a parametric model) to address the efficiency loss. However, such approaches typically rule out the use of complex but powerful modern machine learning algorithms whose theoretical properties are poorly understood.

The statistical problem of subgroup selection---which can also be framed as the discovery and evaluation of a treatment policy---has causal applications beyond clinical trials: Under what conditions does a crop benefit from a fertilizer? Who will be swayed by a political ad? Who will benefit from a proposed economic policy? Further applications arise in non-causal regression: Who is likely to get cancer? Who is likely to purchase a new product? What proteins are likely to have high binding affinity to a target molecule? This paper's proposed method will be equally equipped to answer such questions.

We emphasize the critical importance of not just identifying a subgroup but also providing statistical assurance of that subgroup's merit. There are myriad methods that can be used to \emph{propose} a subgroup, but especially when benefiting subgroups are small or nonexistent in the population, these methods can be unreliable and prone to identifying invalid subgroups (i.e. subgroups where the average benefit is zero or negative). Confidence in the validity of the subgroup is crucial when substantial cost (e.g. in terms of time, money, or reputation) is associated with the identification of invalid subgroups.

\subsection{Our contribution}

In this paper, we describe a new methodological framework for \emph{controlled subgroup selection} in causal inference and regression. Controlled subgroup selection, formally defined in Section~\ref{section:problem-statement}, is the task of identifying a data-dependent subgroup (a region of covariate space) endowed with a high probability guarantee that it is valid (i.e. that its average response is positive). In causal inference, the identification of such a subgroup immediately yields a \emph{treatment policy} (the policy that treats all units in the subgroup) with positive benefit over the policy that treats no one. Our key insight is the ability to improve a subgroup by shrinking it exclusively based on the data outside that subgroup, leaving \emph{all} the data inside the subgroup untouched and usable for inference. By deploying this insight incrementally and interactively, we allow the analyst to leverage any machine learning algorithm or domain knowledge to craft the best subgroup they can while ensuring high power to test the efficacy of the learned subgroup. We call this approach \emph{chiseling} and explain the analogy at the end of Section~\ref{section:thought-experiment}.

Our approach does not limit the complexity of the discovered subgroup, and the analyst may leverage state-of-the-art, black-box machine learning (ML) methods to flexibly search the covariate space without fear of violating the statistical guarantee. Our method produces valid subgroups under very weak assumptions. In the case of regression with binary outcomes, our method is finite-sample valid and exact under only the assumption that the data points are independent and identically distributed (i.i.d.). For causal inference, in randomized experiments, we guarantee asymptotic validity under only mild, dimension-free moment assumptions (in addition to standard causal identification assumptions). We show in extensive simulations and in a re-analysis of a real randomized experiment that our method produces consistently, and sometimes dramatically, better subgroups than existing methods with comparable validity guarantees. All code to reproduce numerical results is available at \url{https://github.com/ncheng4/chiseling_public/}.

\subsection{Related work}
\label{section:related-work}

Because subgroup selection is relevant across many disciplines, the literature is extensive. We focus in the main text on giving a bird's-eye view of works that provide similar frequentist inferential guarantees to ours \emph{without} naive data splitting (as data splitting is general and can be applied to any subgroup selection procedure), that are comparably flexible, and are valid under comparably weak assumptions. We defer to Appendix~\ref{appendix:extended-lit-review} discussion of works that focus on the design of selection criteria without in-sample evaluation \parencite[e.g.][]{Manski2004, Kitagawa2018, Lipkovich2011, Athey2016, spiess2023findingsubgroupssignificanttreatment}, that rely on a parametric model for validity \parencite[e.g.][]{Ma2017, Wan2024}, that utilize a Bayesian approach \parencite[e.g.][]{Dixon1991, Berger2014, Schnell2016, Hill2011}, and that leverage ideas in conformal inference \parencite[e.g.][]{Medarametla2021,Jin2023}. We also defer to Appendix~\ref{appendix:extended-lit-review} discussion of works in sequential hypothesis testing \parencite[e.g.][]{Wald1945,Lan1983,Lai2014}, with which we share some high level similarities, but review works in interactive hypothesis testing, which is more closely related to our method, in this subsection.

\textbf{Methods that select and evaluate subgroups from a structured class.} Some works take as given a prespecified, structured class of subgroups such as a finite set or low-complexity policy class (e.g. with finite Vapnik--Chervonenkis dimension). Examples of the latter include parametric classes, stochastic mixtures of a finite set of policies, and the set of subgroups that arise from continuously thresholding an \textit{a priori} fixed prognostic marker (which may be a machine learning score trained out of sample). Inference is then accomplished by either simultaneous inferences over all candidate subgroups \parencite{Bonetti2004, Song2004, li2023statisticalperformanceguaranteesubgroup, chernozhukov2025policylearningconfidence} or inference on a data-dependent estimand such as the one corresponding to the largest estimated effect \parencite{Guo2020, Andrews2023}. Central to these approaches is the \textit{a priori} fixed class of subgroups (usually with some low dimensional structure), thus limiting the use of machine learning methods unless those models are trained out of sample (i.e. via data splitting).

\textbf{Conditional average treatment effect (CATE) estimation.} There is a line of work on efficient estimation of the CATE \parencite{Knzel2019, Nie2020, Kennedy2023}. Some works produce confidence bands for the CATE either via efficient estimation \parencite{Wager2018, ritzwoller2024simultaneousinferencelocalstructural, Armstrong2015} or by leveraging other structural assumptions such as smoothness or isotonicity \parencite{Reeve2023, muller2023isotonicsubgroupselection}. CATE inference may be implausible absent these assumptions, which are especially restrictive in high dimensions. A compromise is to consider inference on a \emph{calibrated projection} of the CATE, either onto a  small subset of variables \parencite{Fan2020, zimmert2019nonparametricestimationcausalheterogeneity, Semenova2020} or a parametric family \parencite{Cai2010, Cai2011}. However, the complexity of the projection class can alter the finite-sample behavior of these approaches, and thus in these works there is intrinsically a tradeoff between flexibility and validity.

\textbf{Repeated data splitting.} Data splitting \parencite{Cox1975} is a flexible and essentially assumption-free solution to controlled subgroup selection. However, \textcite{fithian2017optimalinferencemodelselection} observe that naively applying data splitting is inadmissible in many situations because it discards more information than is used by selection. Likewise, we empirically improve over naive data splitting under similarly weak assumptions. \textcite{Chernozhukov2018} leverage \emph{repeated} data splitting to reduce dependence on a single split, but this yields inference for an \emph{aggregated} estimand that does not correspond to the effect for any one subgroup. Cross-validation, though we have not seen it applied to subgroup selection, can be used to validate a selected subgroup under stability assumptions on the selection algorithm \parencite{bayle2020, austern2020asymptoticscrossvalidation}. \textcite{jia2024crammethodefficientsimultaneous} split the data into folds to \emph{sequentially} learn and validate an improving sequence of policies, but they require stability conditions similar to those of cross-validation. Stability can be difficult to guarantee for complex algorithms, high-dimensional data, and data with weak signals. For demonstrations of the inferential invalidity of cross-validation in simple settings, see \textcite{Bates2023, bayle2025relativeinstabilitymodelcomparison}. Because our method does not require stability assumptions, it is more broadly valid.

\textbf{Interactive hypothesis testing.} Interactive hypothesis testing was introduced in \textcite{Lei2018} and extended in \textcite{Lei2020, chao2021adaptgmmpowerfulrobustcovariateassisted} for interactive control of the false discovery rate. It has been extended to tests of the global null \parencite{Duan2020Global}, family-wise error rate control \parencite{Duan2020FWER}, and tests of distributional equality \parencite{Duan2022}. At a high level, these interactive procedures all rely on a construction that iteratively reveals some information to the analyst while keeping the remaining information hidden. Notably, \textcite{Duan2024}, working in a causal setting, describe an interactive procedure that produces a subset of units of a given data set while controlling the expected proportion of units for whom the treatment was not effective. However, the paper concerns inferences on the actual experimented-upon units and thus does not yield a usable treatment policy, while our work concerns guarantees for \emph{distributional properties} of subregions of the covariate space such as average effects. Broadly, our work draws inspiration from the interactive literature while deviating in the important regard that previous works concern inferences on properties of a finite sample such as the proportion of nulls or proportion of benefiting individuals in the sample. We are motivated by the spirit of interactive procedures but require significantly different technical tools to solve our problem.

\section{Problem statement: controlled subgroup selection}
\label{section:problem-statement}

Although we are primarily motivated by problems in causal inference, we start with a regression formulation. We observe $n$ i.i.d. data points $(X_1, Y_1),...,(X_n, Y_n) \in \mathcal{X} \times \mathcal{Y}$. Let $\D:=(X_i,Y_i)_{i=1}^n$ and let $(X,Y)$ denote a generic sample from the distribution of $(X_i,Y_i)$ which is independent of $\D$. We seek to identify a region $\reg \subseteq \mathcal{X}$ such that $\Ec{Y}{X \in \reg}$ is strictly larger than some user-specified cutoff $\cutoff \in \mathbb{R}$.

In the context of causal inference, let $Y'(1), Y'(0) \in \mathbb{R}$ denote a unit's potential outcomes for treatment and control, respectively, and let $X \in \mathcal{X}$ denote that unit's covariates. We seek to find a subset of the covariate space where the average treatment effect for that subgroup is above $\cutoff$. In a randomized controlled trial (RCT) with treatment indicator $W$, under standard causal assumptions, we can inverse probability weight (IPW) each unit's potential outcomes to produce a regression problem of the above form. Letting
\begin{equation}
\label{eq:ipw-transform}
Y = \frac{W}{\Ec{W}{X}}Y'(1) - \frac{1-W}{\Ec{1-W}{X}}Y'(0),
\end{equation}
then $Y$ is observable and $\Ec{Y}{X}$ equals the conditional average treatment effect (CATE) $\Ec{Y'(1)-Y'(0)}{X}$. In particular,
\begin{equation*}
\Ec{Y}{X\in \reg} = \Ec{Y'(1)-Y'(0)}{X\in\reg}.
\end{equation*}
Hence, the regression formulation we will consider throughout this paper in some sense subsumes the causal one. We will occasionally use causal language to build intuition, in which case it should be understood that the regression has been converted from an underlying causal problem via Equation~\eqref{eq:ipw-transform}. Note that this problem statement and our method can also use the more efficient augmented IPW (AIPW) transformation, but since this does not preserve the i.i.d. structure of the data, it requires additional care; see Appendix~\ref{appendix:proofs-aipw} for details.

For a fixed subgroup $\reg$, define the \emph{subgroup mean}
\begin{equation*}
\mean(\reg) := \Ec{Y}{X \in \reg},
\end{equation*}
let $\mean(\emptyset)$ be undefined, and let the statement ``$\mean(\emptyset) \leq \cutoff$" evaluate as false. This is also the \emph{subgroup average treatment effect} when $Y$ comes from the IPW transformation. By slight abuse of notation, we will let $\mean(x) := \mean(\set{x}) = \Ec{Y}{X = x}$ for $x \in \mathcal{X}$. Given a nominal level $\alpha \in [0,1]$, the precise inferential validity guarantee we seek to provide is that our data-dependent subgroup $\reg$ satisfies
\begin{equation}
\label{eq:type1error}
\Prb\paren{\mean(\reg)\leq \cutoff} \leq \alpha.
\end{equation}
To expand on this criteria: when $\reg'$ is fixed, a classical subgroup analysis reports $\reg = \reg'$ if $H_0 : \mean(\reg') \leq \cutoff$ is rejected and $\reg = \emptyset$ otherwise. Then Equation~\eqref{eq:type1error} exactly reduces to describing a Type I error guarantee for $H_0$. By analogy, we generically refer to Equation~\eqref{eq:type1error} as Type I error control, even when $\reg$ is data-dependent. When $\cutoff = 0$, this guarantees that we erroneously report a subgroup whose average effect is nonpositive no more than $\alpha$ of the time (and where a vacuous discovery $\reg = \emptyset$ is not considered an error). We call this task \emph{controlled subgroup selection}, since we seek to select a subgroup from the data while controlling the Type I error.

Subject to the Type I error constraint, we measure the quality of a subgroup selection procedure by how well it maximizes the \emph{expected utility} of the reported subgroup for some chosen notion of utility. Abstractly, if $\util(\reg)$ measures the utility of $\reg$, then we seek to maximize the expected utility,
\begin{equation}
\label{eq:utility}
\E[\util(\reg)],
\end{equation}
subject to Equation~\eqref{eq:type1error}. Different notions of utility may be more or less appropriate depending on the inferential goals and the task at hand. One natural choice is
\begin{equation}
\label{eq:policy-utility}
\begin{aligned}
\util(\reg) := \E[(Y - \cutoff) \indic\set{X \in \reg}] = (\mean(\reg) - \cutoff) \Vol(\reg)
\end{aligned}
\end{equation}
where $\Vol(\reg) := \Prb(X \in \reg)$. This is maximized by the upper level set of the conditional mean function: $\reg^* := \set{x \in \mathcal{X} : \E[Y \mid X = x] > \cutoff}$. In the policy learning literature $\cutoff$ is often taken to be $0$, in which case treating units in $\reg^*$ is an optimal policy (Appendix~\ref{appendix:policy-max-equivalence}). By construction, ``$\mean(\reg^*) \leq \cutoff$" is false. Using Equation~\eqref{eq:policy-utility} as the utility rewards regions that are more similar to $\reg^*$, which is precisely the subgroup of units whose predictable (i.e. $X$-conditional) means exceed $\cutoff$.

Other natural utility functions include $\mathcal{U}(\reg) := \indic\set{\reg \neq \emptyset}$ and $\mathcal{U}(\reg) := \Vol(\reg)$. The expectation of the former is \emph{power}, and the expectation of the latter is \emph{expected probability mass}. However, for the remainder of the paper, we will let $\util(\cdot)$ be defined as in Equation~\eqref{eq:policy-utility} and simply refer to it as ``the utility" unless otherwise specified, since this is one of the most common objectives in causal inference and its choice will help illustrate the key ideas in our paper.

\hfill

\noindent \textbf{Notation and conventions.} Let $[n] := \set{1,...,n}$ and $[n]_0 := \set{0,...,n}$. Let $\Omega$ denote the sample space of the base probability measure. All random variables (including the data $(X_i, Y_i)_{i=1}^n$) and $\sigma$-algebras are defined with respect to $\Omega$. We will abbreviate ``almost surely" as ``a.s." For random variables $A$ and $B$, we say that $A \leq B$ almost surely conditional on an event $E$ if $\Prb\paren{\set{A \leq B} \cap E} = \Prb(E)$, and similarly with ``$\leq$" signs replaced with ``$=$" signs. We let $\Phi(\cdot)$ denote the standard normal CDF. We will write $(x)_+ = \max\set{0, x}$, sometimes use the wedge notation for minimum, i.e. $a \wedge b = \min\set{a, b}$, and will let the minimum of an empty set be $\infty$. We assume that all functions are measurable throughout. We will sometimes use language such as ``choose a function $\score(\cdot)$ based on $\F$" where $\F$ is a $\sigma$-algebra. Formally, what we mean by this is that $\score(\cdot)$ is an $\F$-measurable random function.

\section{Methodology}
\label{section:methodology}

\subsection{Overview of our approach}
\label{section:approach-overview}

The challenge of controlled subgroup selection is that the data needs to be used to both (i) identify the candidate subgroup and (ii) perform statistical inference on that subgroup. Data splitting accomplishes this under extremely weak assumptions, and we improve upon data splitting by revealing information in an incremental fashion, allowing us to leverage more information before definitively selecting any particular subgroup. Algorithm~\ref{alg:chiseling-testing} summarizes the full approach. Below, we also give a high-level overview of our main ideas.

\hfill

\noindent \textbf{High-level overview.} Chiseling proceeds by initializing the region under consideration as the full population $\reg = \mathcal{X}$ and interweaving two basic operations: (1) a step that shrinks $\reg$, and (2) a step that chooses a fraction of $\alpha$ to ``spend" on testing $H_0: \mean(\reg) \leq \cutoff$, stopping and reporting $\reg$ if a rejection is made.

The key idea behind (1) is that as long as the shrinkage directions are chosen based on data outside of $\reg$, the subset of data that lies in $\reg$ is \emph{untarnished} (defined in Section~\ref{section:thought-experiment}) in the sense that it is an i.i.d. sample of units from $\reg$. This makes it simple to test properties of $\reg$ even though $\reg$ is a complex, data-dependent object. The key idea behind (2) is that it is possible to characterize the dependence between the tests that we design for $H_0: \mean(\reg) \leq \cutoff$ as $\reg$ evolves throughout the procedure. This lets us non-conservatively test multiple regions until a rejection is made via \emph{conditionally valid tests} (introduced in Section~\ref{section:general-testing-framework}).

A feature of our approach is its immense flexibility: Type I error is controlled regardless of how the analyst shrinks the region (any machine learning method can be used) or allocates $\alpha$, as long as decisions are made only based on information allowed by the protocol. Throughout, we will suggest guidelines to navigate this flexibility, including ways to construct interpretable regions (Section~\ref{section:interpretable-regions}).

\begin{algorithm}[t]
    \caption{Chiseling}
    \label{alg:chiseling-testing}
    \hspace*{\algorithmicindent} \textbf{Input:} \text{dataset $\D = \paren{X_i, Y_i}_{i=1}^n$, cutoff $\cutoff$, and nominal Type I error level $\alpha$}
    \begin{algorithmic}[1]
    \State Initialize region $\reg_0 \gets \mathcal{X}$ and revealed information $\F_0 \gets \set{\emptyset, \Omega}$
    \For{$t = 0,...,\tmax$}
        \State Choose level $\allocalpha_t \in [0,1]$ based on $\F_t$, ensuring valid $\alpha$-budget \Comment{Definition~\ref{def:alpha-budget}}
        \State Perform conditionally valid level-$\allocalpha_t$ test $\gentest_t$ of $\genhyp_t : \mean(\reg_t) \leq \cutoff$ \Comment{Definition~\ref{def:cond-valid-test-seq}}
        \If{$\gentest_t$ rejects} \label{line:chisel-testing-if}
            \State \textbf{Return:} $\reg_t$ \label{line:chisel-testing-mid-return}
        \EndIf \label{line:chisel-testing-endif}
        \State Obtain next region $\reg_{t+1}$ by shrinking $\reg_t$ based on $\F_t$ \Comment{Algorithm~\ref{alg:chiseling}}
        \State See revealed data points: $\F_{t+1} \gets \sigma(\F_t, (X_i, Y_i)_{i : X_i \in \reg_t \setminus \reg_{t+1}})$
    \EndFor
    \State \textbf{If Line~\ref{line:chisel-testing-mid-return} never executes, return:} $\emptyset$ \label{line:chisel-testing-return}
    \end{algorithmic}
\end{algorithm}

\subsection{Untarnished subgroups}
\label{section:thought-experiment}

When a subgroup $\reg$ is fixed \textit{a priori}, inferences about $\reg$ are straightforward because $(X_i, Y_i)_{i : X_i \in \reg}$ behaves like an i.i.d. sample from the distribution of $(X, Y) \mid X \in \reg$. Thus, any standard inferential procedure can be applied to the subgroup sample $(X_i, Y_i)_{i : X_i \in \reg}$ to yield inferences about $\reg$. For instance, the sample mean of $(Y_i)_{i : X_i \in \reg}$ is unbiased for $\Ec{Y}{X \in \reg}$, and the usual confidence intervals derived from the central limit theorem (CLT) are asymptotically valid. Our primary insight is that the key property that the distribution of points inside $\reg$ is not ``distorted" can also hold for a \emph{data-dependent region}, conditional on having selected the region, if we construct it carefully. We begin with a useful definition.

\begin{definition}[Untarnished subgroup]
\label{def:untarnished-subgroup}
Let $\reg \subseteq \mathcal{X}$ be a (possibly data-dependent) subgroup, and let $\F$ be a $\sigma$-algebra. We say that $\reg$ is \emph{untarnished} with respect to $\F$ if $\reg$ and $n(\reg) := \abs{\set{i : X_i \in \reg}}$ are $\F$-measurable, and the distribution of the subsample $(X_i, Y_i)_{i : X_i \in \reg}$ conditional on $\F$ is the same as that of an independent sample of $n(\reg)$ data points drawn i.i.d. from the distribution of $(X, Y) \mid X \in \reg$.
\end{definition}

\begin{figure}[t]
  \centering
  \includegraphics[width=\linewidth]{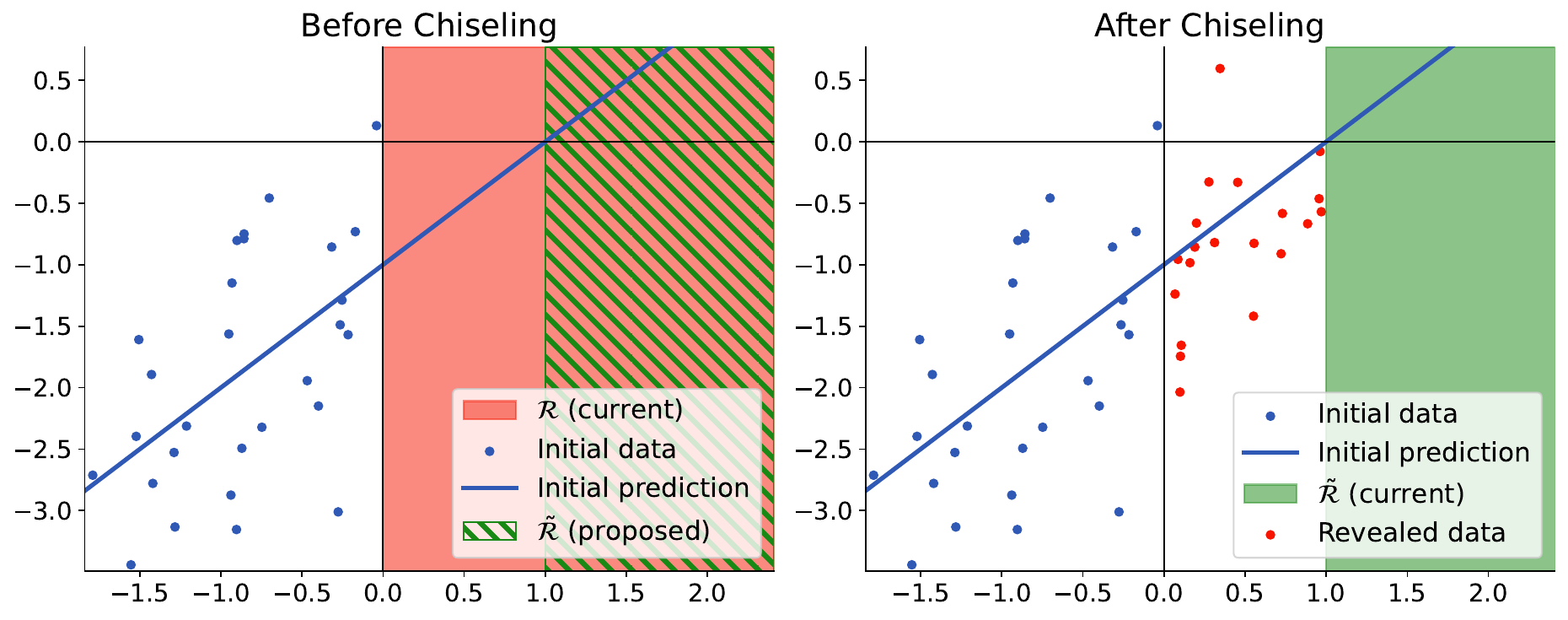}
  \caption{A thought experiment using a toy example: the blue line represents the least squares fit through the initially revealed blue points. After shrinking the red region $\reg = (0,\infty)$ to the green region $\tilde{\reg} = (1,\infty)$, the red points are revealed.}
  \label{fig:thought-experiment}
\end{figure}

Formalizing what we just discussed, an \textit{a priori} fixed $\reg$ is untarnished with respect to $\F := \sigma(n(\reg))$. Regions selected by data splitting can also be thought of as untarnished (see Section~\ref{section:simple-demo}). To see how else untarnishedness arises, consider the following thought experiment. Let $\reg$ be fixed \textit{a priori}. Since the data points outside of $\reg$ are conditionally independent of the data points inside of it, $\reg$ is untarnished with respect to $\F := \sigma((X_i, Y_i)_{i : X_i \not\in \reg})$. Thus, even upon scrutinizing $(X_i, Y_i)_{i : X_i \not\in \reg}$, an analyst may still treat $(X_i, Y_i)_{i : X_i \in \reg}$ like an i.i.d. sample from $(X, Y) \mid X \in \reg$. For instance, in Figure~\ref{fig:thought-experiment}, the blue points constitute our state of knowledge after revealing the points outside the non-data-dependent region $\reg = (0, \infty) \subseteq \mathbb{R}$. Studying them, we may be dissatisfied with $\reg$. After all, if the optimal subgroup $\set{x \in \mathbb{R} : \E[Y \mid X = x] > 0}$ is our target, a simple linear regression (the solid blue line) suggests that $\tilde{\reg} := (1, \infty)$ is better than $\reg$. Our key insight is this: we may switch the focus of our inference to $\tilde{\reg}$, and $\tilde{\reg}$ remains untarnished even though it was determined in a data-dependent way, because it was only determined using points outside of $\reg$. Thus, the further subsetted sample $(X_i, Y_i)_{i : X_i \in \tilde{\reg}}$ may be used as if it were an i.i.d. sample from $\tilde{\reg}$.

In this example, untarnishedness does \emph{not} hinge on the correctness of the linear extrapolation. But we will be rewarded with a better region and generally more powerful inferences if our modeling assumptions are accurate. Furthermore, $\tilde{\reg}$ is untarnished with respect to $\tilde{\F} := \sigma(\F, (X_i, Y_i)_{i : X_i \in \reg \setminus \tilde{\reg}})$ because the newly excluded points (the red points) are conditionally independent of the points inside of $\tilde{\reg}$. Thus, if we remain dissatisfied after inspecting $(X_i, Y_i)_{i : X_i \not\in \tilde{\reg}}$, for instance finding that our new least squares fit has shifted substantially, we may repeat the above process to further shrink $\tilde{\reg}$. But since we cannot reverse this process, it seems prudent to shrink as little as possible at each step in order to avoid overshrinking. Indeed, our construction in the next subsection shrinks infinitesimally until new data is revealed.

We call our approach \emph{chiseling}, in analogy with how a sculptor reduces their starting material to produce a final artwork, refining their artistic choices as they interact with the medium, and moreover, \emph{committing} to any reductions they make, since chiseled material cannot be restored to the main body. Later, we will explore why the chiseling framework is beneficial, including a simple numerical demonstration in Section~\ref{section:simple-demo}. But first, the next subsection describes in more detail how chiseling is operationalized and some of its properties.

\subsection{Chiseling subroutine: region shrinking}
\label{section:chiseling-properties}

The level sets of functions allow us to conveniently parameterize region shrinking. Let $\reg$ be the current region and untarnished with respect to $\F$. For any function $\score : \mathcal{X} \to \mathbb{R}$, let $\tilde{\reg}(z) := \reg \cap \set{x \in \mathcal{X} : \score(x) > z}$. It helps to imagine $\tilde{\reg}(z)$ shrinking from $\reg$ to $\emptyset$ as $z$ ``sweeps" continuously from $-\infty$ to $\infty$. Where should $z$ stop? There is no loss of generality in letting $z$ sweep no further than $\min_{i : X_i \in \reg} \score(X_i)$, the level required to reveal at least one new data point, since to shrink more we may simply repeat this procedure. But also, we may wish to cap $z$ at some value $\threshlim$ even if no data points fall in $\reg \setminus \tilde{\reg}(c)$ (e.g. to ensure we only reveal points with negative predicted treatment effects; see Section~\ref{section:simple-demo}). Thus, we define the next region as
\begin{equation}
\label{eq:region-sweep}
\begin{aligned}
\tilde{\reg} := \reg \cap \set*{x \in \mathcal{X} : \score(x) > \threshlim \wedge \paren*{\min_{i : X_i \in \reg} \score(X_i)}}.
\end{aligned}
\end{equation}
We allow $\score(\cdot)$ and $\threshlim$ to be chosen in an arbitrary fashion by the analyst as long as they are chosen based on $\F$ (i.e. they are $\F$-measurable). We also allow the choice $\threshlim = \infty$, which removes the role of $\threshlim$ from the above construction. Essentially, this shrinks $\reg$ along the contours defined by $\score(\cdot)$ until either new data is revealed or $\threshlim$ is reached. For maximizing utility in causal inference, a natural idea is to let $\score(\cdot)$ be an estimate of the CATE using the information in $\F$, as then units with the lowest predicted CATEs are revealed first, in principle leaving more of the optimal region intact. We will soon see that this choice is empirically effective (Section~\ref{section:simple-demo}), though further constraints may be imposed if the analyst desires more interpretable regions (Section~\ref{section:interpretable-regions}).

\begin{algorithm}[ht]
    \caption{Chiseling subroutine (region shrinking)}
    \label{alg:chiseling}
    \hspace*{\algorithmicindent} \textbf{Input:} \text{dataset $\D = \paren{X_i, Y_i}_{i=1}^n$, current region $\reg$, current information $\F$}
    \begin{algorithmic}[1]
    \State Analyst chooses $\score : \mathcal{X} \to \mathbb{R}$ and $\threshlim \in \mathbb{R} \cup \set{\infty}$ based on $\F$
    \State \textbf{Return:} $\tilde{\reg} := \reg \cap \set{x \in \mathcal{X} : \score(x) > \threshlim \wedge \paren{\min_{i : X_i \in \reg} \score(X_i)}}$
    \end{algorithmic}
\end{algorithm}

Algorithm~\ref{alg:chiseling} summarizes the construction we have just described, which is a subroutine of Algorithm~\ref{alg:chiseling-testing}. We call $\score(\cdot)$ the \emph{scoring function} and $\threshlim$ the \emph{cap}, and we say that Algorithm~\ref{alg:chiseling} \emph{chisels} $\reg$ to produce a new region $\tilde{\reg} \subseteq \reg$.\footnote{While we call our entire shrinking and testing procedure (Algorithm~\ref{alg:chiseling-testing}) chiseling, we will also refer to the act of shrinking a region (via Algorithm~\ref{alg:chiseling}) as chiseling.} The following theorem formalizes the property described in the thought experiment visualized by Figure~\ref{fig:thought-experiment}.

\begin{theorem}[Chiseling preserves untarnishedness]
\label{theorem:untarnished-chiseling}
Suppose that $\reg$ is untarnished with respect to $\F$ (Definition~\ref{def:untarnished-subgroup}). Let $\tilde{\reg}$ be the output of Algorithm~\ref{alg:chiseling} applied to $(\D, \reg, \F)$, and define $\tilde{\F} := \sigma(\F, (X_i, Y_i)_{i : X_i \in \reg \setminus \tilde{\reg}})$. Then $\tilde{\reg}$ is untarnished with respect to $\tilde{\F}$.
\end{theorem}

A proof is given in Appendix~\ref{appendix:proofs-interactive-selection}. We now describe how to chain Algorithm~\ref{alg:chiseling} to produce a nested sequence of regions $(\reg_t)_{t=0}^{\infty}$ while accumulating information the analyst can use for selection. Section~\ref{section:general-testing-framework} will describe how to select one of these regions while controlling Type I error. We assume without loss of generality that $\reg_0 = \mathcal{X}$, the entire covariate space, and let $\F_0 := \set{\emptyset, \Omega}$, the trivial $\sigma$-algebra. Thus, $\reg_1 \subseteq \reg_0$ must be obtained by chiseling $\reg_0$ using only prior information (see Section~\ref{section:init-chiseling} for a discussion of this point). Then recursively define for $t = 1,2,...$,
\begin{equation}
\begin{aligned}
\label{eq:formalized-chiseling}
\reg_t &:= \text{ the output of Algorithm~\ref{alg:chiseling} applied to $(\D, \reg_{t-1}, \F_{t-1})$},\\
\F_t &:= \sigma(\F_{t-1}, (X_i, Y_i)_{i : X_i \in \reg_{t-1} \setminus \reg_t}).
\end{aligned}
\end{equation}
Observing that $\reg_0$ is trivially untarnished with respect to $\F_0$ since it is not data-dependent, the first part of the following corollary follows directly from recursively applying Theorem~\ref{theorem:untarnished-chiseling}. The second part is proved in Appendix~\ref{appendix:proofs-distribution-subsamples}.
\begin{corollary}[Sequentially chiseled regions]
\label{corollary:distribution-subsamples}
For any fixed $t$, $\reg_t$ is untarnished with respect to $\F_t$. If $\genstop$ is an a.s. finite stopping time with respect to $(\F_t)_{t=0}^{\infty}$, then $\reg_{\genstop}$ is untarnished with respect to $\F_{\genstop}$, the corresponding stopped $\sigma$-algebra.
\end{corollary}

That is, the subsample $(X_i, Y_i)_{i : X_i \in \reg_t}$ behaves like an i.i.d. sample of size $n(\reg_t)$ from $(X, Y) \mid X \in \reg_t$, conditionally on $\F_t$, even if $t$ is a stopping time $\genstop$. While Corollary~\ref{corollary:distribution-subsamples} applies to an infinite sequence of regions, practically it is only necessary to apply chiseling a finite number of times to produce a finite sequence of regions. Thus, unless stated otherwise we will fix an $\tmax$ and restrict our attention to the first $\tmax + 1$ steps and the sequence $(\reg_t, \F_t)_{t=0}^{\tmax}$.

\begin{remark}[Auxiliary randomness]
Corollary~\ref{corollary:distribution-subsamples} still holds if the analyst leverages auxiliary randomness (randomness that is independent of the data) when making decisions such as selecting $\score(\cdot)$ and $\threshlim$. For instance, the analyst may leverage machine learning methods with inherent randomness such as random forests or stochastic gradient descent. In this case, we simply condition on the auxiliary randomness and omit it from our discussion.
\end{remark}

In summary, as long as we only shrink the current region, we may alter the direction of this shrinkage at any time, in any way, and particularly in light of new information conferred by data points that are excluded from the region. Corollary~\ref{corollary:distribution-subsamples} ensures that $\reg_t$ is untarnished with respect to $\F_t$, and thus the samples within it are suitable for use in standard inferential procedures. For any stopping time $\nu$, Corollary~\ref{corollary:distribution-subsamples} paves the way for inferences on $\reg_{\genstop}$. In fact, something much more general is possible: in Section~\ref{section:general-testing-framework}, we will show that it is possible to tightly ``split" one's $\alpha$ error budget over many regions in the sequence $(\reg_t)_{t=0}^{\tmax}$, so that instead of testing $H_0: \mean(\reg_{\genstop}) \leq 0$, regions may be tested successively until a rejection is made. Still, even the simplest approach of using samples in $(X_i, Y_i)_{i : X_i \in \reg_{\genstop}}$ for inference on $\reg_{\genstop}$ yields immediate improvements over data splitting, as we highlight in the next subsection.

\subsection{Chiseling is more efficient than data splitting}
\label{section:simple-demo}

To understand why chiseling improves upon data splitting, we first describe data splitting in a way that reveals a direct comparison. Let $p \in (0,1)$ and let $\mathcal{M}$ denote a random subset of $[n]$ of size $\ceil{(1 - p)n}$ so that there are $\floor{pn}$ data points in $\D_{\mathrm{train}} := (X_i, Y_i)_{i \not\in \maskind}$. A standard use of data splitting would build an estimate $\hat{\mean}(\cdot)$ of $\mean(\cdot)$ via machine learning on $\D_{\mathrm{train}}$, define $\reg_{\mathrm{ds}} := \set{x \in \mathcal{X} : \hat{\mean}(x) > \cutoff}$, and test $H_0 : \mean(\reg_{\mathrm{ds}}) \leq \cutoff$ by running a one-sided $t$-test using the independent test subsample $(X_i, Y_i)_{i \in \maskind : X_i \in \reg_{\mathrm{ds}}}$. But another way to view data splitting is by unraveling the construction of $\reg_{\mathrm{ds}}$ into a shrinking sequence of upper level sets of $\hat{\mean}(\cdot)$ that terminates at $\reg_{\mathrm{ds}}$. In particular, if we consider the process $\mathcal{X} = \reg_0 \supseteq \reg_1 \supseteq ... \supseteq \reg_{\mathrm{ds}}$ where
\begin{equation*}
\begin{aligned}
\reg_t = \set*{x \in \mathcal{X} : \hat{\mean}(x) > \cutoff \wedge \paren*{\min_{i \in \maskind : X_i \in \reg_{t-1}} \hat{\mean}(X_i)}}
\end{aligned}
\end{equation*}
then we see that $\reg_{\mathrm{ds}}$ is simply a \emph{special case} of chiseling where in Algorithm~\ref{alg:chiseling} we use $\score(\cdot) = \hat{\mean}(\cdot)$---fit only using $\D_{\mathrm{train}}$---and $\threshlim = \cutoff$ at every shrinking step, and where $(X_i, Y_i)_{i \in \maskind}$ is treated as the dataset. Formally, chiseling can also be seen to generalize the full data splitting procedure (including fitting $\hat{\mean}(\cdot)$ to $\D_{\mathrm{train}}$) via an argument that adds an independent, auxiliary dimension to the covariates and chisels in that dimension (see Appendix~\ref{appendix:chiseling-generalizes-ds}).

From here, it is easy to see how data splitting can be improved. Corollary~\ref{corollary:distribution-subsamples} tells us that at the $t$th stage, the data that is excluded from $\reg_t$ can be used without invalidating the $t$-test that is run on the final region.\footnote{Later, once we have defined conditionally valid tests, one will see that running a $t$-test on the final region is a particular instantiation of Algorithm~\ref{alg:chiseling-testing}.} Thus, chiseling can follow the exact same steps as above with just one change: at each stage, refit $\hat{\mean}(\cdot)$ to \emph{all} the data that is available, which includes $\D_{\mathrm{train}}$ and the points revealed thus far by shrinking. This implies that, given the same $\D_{\mathrm{train}}$, chiseling will tend to find better regions than data splitting since it iteratively improves its estimate of $\mean(\cdot)$ using more data. As a consequence, chiseling can find better regions with smaller split proportions $p$, leaving larger sample sizes available for testing.

\begin{figure}[t]
  \centering
  \includegraphics[width=\linewidth]{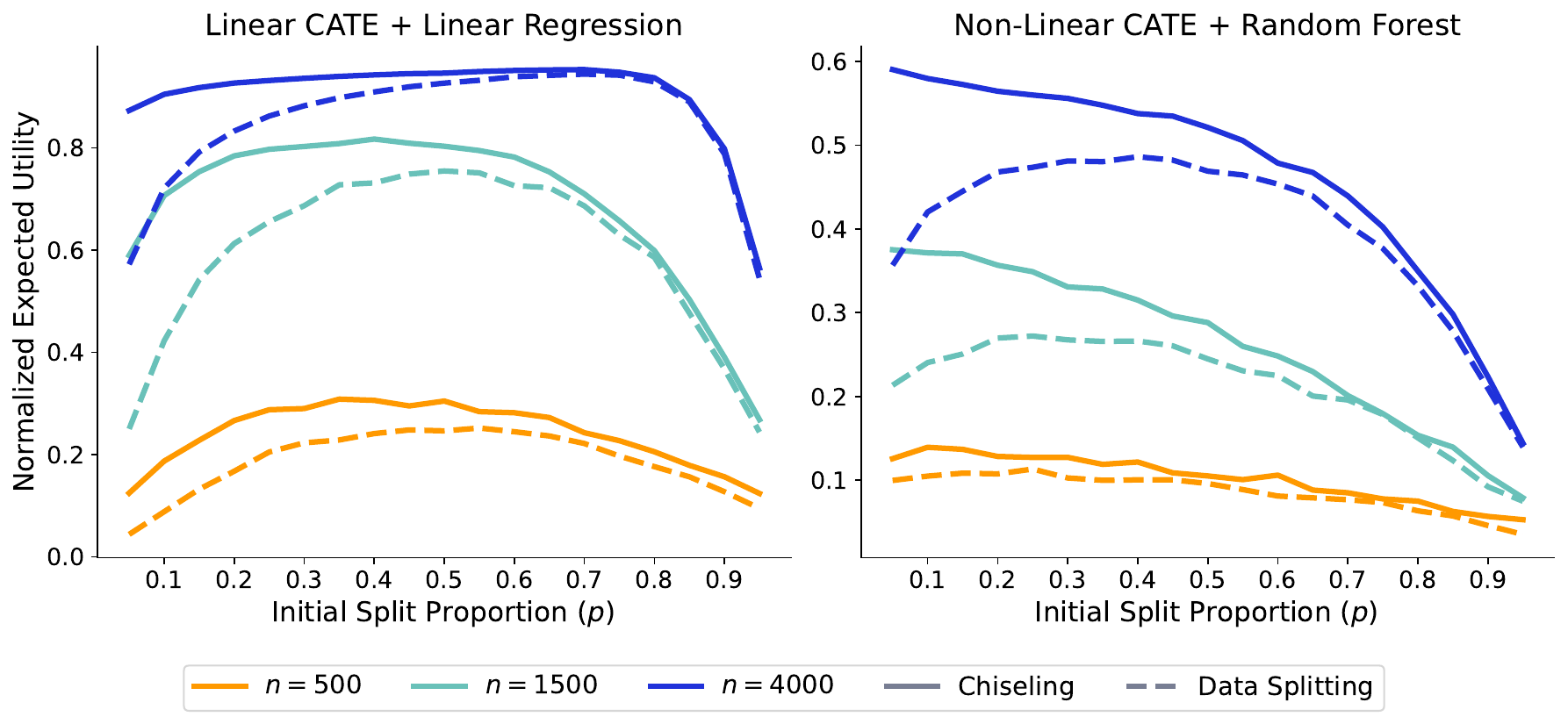}
  \caption{Normalized expected utility of region reported by chiseling versus data splitting, controlling for the same starting information (via initial split proportion) and machine learning method in two simulated RCTs with three different sample sizes $n$.
  }
  \label{fig:simple-demo-utility}
\end{figure}

Figure~\ref{fig:simple-demo-utility} compares the expected utility of this simple variant of chiseling (see Appendix~\ref{appendix:chiseling-to-boundary} for a precise description) to data splitting across a range of initial split proportions $p$ in two simulated RCTs: one where the CATE follows a linear model and the other where the CATE is nonlinear (see Appendix~\ref{appendix:simulation-details} for details). As expected, for every $n$ and $p$, chiseling dominates data splitting, and the difference across split proportions $p$ can be quite substantial, e.g., for $n=1500$ in the second panel, the maximum of chiseling's expected utility curve is 38\% higher than that of the data splitting curve. Furthermore, we see that there are typically many choices for $p$ where, for most or all $p' \in (0,1)$, chiseling with a $p$ initial split outperforms data splitting with a $p'$ initial split.

While leveraging an initial split allows us to directly compare to data splitting, chiseling can also be applied without an initial split if there is useful prior information---we discuss this possibility in Section~\ref{section:init-chiseling}. But first, we situate chiseling in an even more general sequential testing framework that gives the analyst more flexibility in how regions are tested.

\subsection{General framework: sequential testing of chiseled regions}
\label{section:general-testing-framework}

As there is uncertainty in our estimate of the optimal region, we may prefer not to overcommit to testing just a single region, but rather spread power over multiple candidate regions. Though multiple testing can be addressed via the Bonferroni correction, this will be conservative because the data points in $\reg_t$ greatly overlap with those in $\reg_{t+1}$, creating strong positive dependence between successive tests.\footnote{Nonetheless, the Bonferroni correction constitutes a valid and general approach, and has some other interesting uses that we discuss in Appendix~\ref{appendix:point-est-cis}.} Instead, we introduce a framework based on constructing a sequence of hypothesis tests $\phi_t$ which are \emph{conditionally valid} for null hypotheses $\genhyp_t$ at level $\alpha_t$. In this paper, $\genhyp_t$ will be the hypothesis that $\reg_t$, the $t$th region produced by chiseling, satisfies $\mean(\reg_t) \le \cutoff$.

We first present a very general version of this idea. As we will allow the error allocated to each adaptively determined null hypothesis to be chosen in an online manner, we need the following definition.

\begin{definition}[$\alpha$-budget]
\label{def:alpha-budget}
For a fixed $\alpha$, we call $(\allocalpha_t)_{t=0}^{m}$ a \emph{valid $\alpha$-budget} with respect to a filtration $(\G_t)_{t=0}^{m}$ if $\allocalpha_t$ is $\G_t$-measurable for all $t=0,...,m$ and $1 - \prod_{t=0}^{m} (1 - \allocalpha_t) \leq \alpha$ almost surely. If additionally, $1 - \prod_{t=0}^{m} (1 - \allocalpha_t) = \alpha$ almost surely, then we call it an \emph{exact $\alpha$-budget}.
\end{definition}

When choosing $\allocalpha_t$ in an online manner, Definition~\ref{def:alpha-budget} can always be satisfied by choosing $\allocalpha_t \leq 1 - (1 - \alpha) \prod_{s=0}^{t-1} (1 - \allocalpha_s)^{-1}$ based on $\G_t$. 

Next, let us formalize a null hypothesis $\genhyp$ as both a proposition about the data distribution and a random variable that evaluates to $0$ if that proposition is true (we are under the null) and $1$ otherwise (we are under the alternative). For instance, when $\reg$ is random the hypothesis $\genhyp : \mean(\reg) \leq \cutoff$ corresponds to the random variable $\genhyp = \indic\set{\mean(\reg) > \cutoff}$. We formalize a test $\gentest$ for $\genhyp$ as a binary random variable that evaluates to $1$ if $\genhyp$ is rejected; thus, a false rejection occurs if and only if $\genhyp = 0$ and $\gentest = 1$.

\begin{definition}[Conditionally valid testing sequence]
\label{def:cond-valid-test-seq}
$(\genhyp_t, \allocalpha_t, \gentest_t)_{t=0}^{m}$ is a \emph{conditionally valid testing sequence} with respect to a filtration $(\G_t)_{t=0}^{m}$ if for all $t=0,...,m$ we have that $\genhyp_t$ and $\allocalpha_t \in [0,1]$ are $\G_t$-measurable and the following holds:
\begin{equation}\label{eq:cond-valid-test-seq}
\Prbc{\gentest_t = 1}{\sigma(\G_t, (\gentest_s)_{s < t})} \leq \allocalpha_t \text{ a.s. on the event } \set{\genhyp_t = 0 \text{ and } \max_{s < t} \gentest_{s} = 0}.
\end{equation}
Additionally, we call $(\genhyp_t, \allocalpha_t, \gentest_t)_{t=0}^{m}$ a \emph{conditionally exact testing sequence} with respect to $(\G_t)_{t=0}^{m}$ if Equation~\eqref{eq:cond-valid-test-seq} holds with equality for all $t=0,...,m$.
\end{definition}

In words, Definition~\ref{def:cond-valid-test-seq} asks that the random variable $\gentest_t$ specifies a valid level-$\allocalpha_t$ test of the null hypothesis $\genhyp_t$ \emph{conditional} on all the information in $\G_t$ as well as the fact that no previous test rejected. A simple lemma shows that conditionally valid testing sequences control the overall Type I error rate. Though the result is intuitive, its proof contains subtleties and is presented in Appendix~\ref{appendix:proofs-interactive-testing}.

\begin{lemma}[Sequential error control]
\label{lemma:sequential-error-control}
Let $(\genhyp_t, \allocalpha_t, \gentest_t)_{t=0}^{m}$ be a conditionally valid testing sequence and $(\allocalpha_t)_{t=0}^{m}$ be a valid $\alpha$-budget with respect to a filtration $(\G_t)_{t=0}^{m}$. Define the first rejected index $\tsel := \min\set{t : \gentest_t = 1}$ with $\tsel = \infty$ if the set is empty, and let $\genhyp_{\infty} = 1$ by convention. Then rejecting $\genhyp_{\tsel}$ controls the Type I error rate:
\begin{equation}
\label{eq:sequential-error-control}
\begin{aligned}
\Prb\paren{\genhyp_{\tsel} = 0} \leq \alpha.
\end{aligned}
\end{equation}
Furthermore, if $(\genhyp_t, \allocalpha_t, \gentest_t)_{t=0}^{m}$ is conditionally exact, $(\allocalpha_t)_{t=0}^{m}$ is an exact $\alpha$-budget, and $\genhyp_t = 0$ almost surely for all $t=0,...,m$ (i.e. we are under the global null), then Equation~\eqref{eq:sequential-error-control} holds with equality.
\end{lemma}

The above lemma ensures that $\genhyp_{\tsel}$, the rejected hypothesis, is a true null with probability no greater than $\alpha$. We now connect this result to our subgroup selection problem. By convention, let $\reg_{\infty} = \emptyset$ (corresponding to no rejection). Now letting $\genhyp_t := \indic\set{\mean(\reg_t) > \cutoff}$, Lemma~\ref{lemma:sequential-error-control} implies that an appropriately designed filtration $(\G_t)_{t=0}^{\tmax}$, sequence of levels $(\allocalpha_t)_{t=0}^{\tmax}$, and collection of tests $\paren{\gentest_t}_{t=0}^{\tmax}$ yields the desired error control,
\begin{equation*}
\Prb(\mean(\reg_{\tsel}) \leq \cutoff) \leq \alpha,
\end{equation*}
where we recall that $\tsel := \min\set{t : \gentest_t = 1}$. That is, by rejecting the first (largest) region in our sequence of chiseled regions using conditionally valid tests, we satisfy Type I error control.\footnote{In Appendix~\ref{appendix:multiple-testing}, we describe a variant of chiseling that rejects multiple regions while controlling the family-wise error rate.}

At this point we have specified Algorithm~\ref{alg:chiseling-testing} up to the design of the conditionally valid tests $(\gentest_t)_{t=0}^{\tmax}$, which we will do in the following subsection, and strategies for shrinking the regions and designing $\alpha$-budgets, which we discuss in Section~\ref{section:practical-considerations}.

\subsection{Design of the tests \texorpdfstring{$\gentest_t$ and main validity results}{}}
\label{section:general-test-design}

Constructing conditionally valid tests may seem quite challenging. For our problem, however, conditional tests can be readily constructed by leveraging the untarnishedness of chiseled regions. Since by Corollary~\ref{corollary:distribution-subsamples} the sample mean of points in a chiseled region $\reg_t$ would be unbiased for $\mean(\reg_t)$ given $\F_t$ if we did not condition on the previous tests $(\gentest_s)_{s < t}$, arguably the most natural test simply thresholds the sample mean of points in the region, i.e.
\begin{equation*}
\begin{aligned}
\gentest_t := \indic\set{\meanest_t > \critval_t}
\end{aligned}
\end{equation*}
where\footnote{If ever $\meanest_t$ or $\critval_t$ is undefined, we may either let $\gentest_t = 0$, which is trivially conditionally valid, or let $\gentest_t \sim \text{Bern}(\allocalpha_t)$ independently, which is trivially conditionally exact.}
\begin{equation*}
\begin{aligned}
\meanest_t := \frac{1}{\un_t} \sum_{i : X_i \in \reg_t} Y_i \quad \text{ and } \quad \un_t := \abs{\set{i : X_i \in \reg_t}}.
\end{aligned}
\end{equation*}
The challenge is to choose $\critval_t$ so that the tests are part of a conditionally valid testing sequence, with the main difficulty arising from the fact that conditioning on $\set{\max_{s < t} \gentest_s = 0}$ creates bias. Fortunately, it is tractable to characterize this bias. It turns out that $\max_{s < t} \gentest_s = 0$ if and only if $\meanest_t \leq \trunc_t$ where
\begin{equation}
\label{eq:truncation-definition}
\begin{aligned}
\trunc_t := \min_{s < t} \set*{ \frac{1}{\un_t} \paren*{ \un_s \critval_s - \sum_{i : X_i \in \reg_s \setminus \reg_t} Y_i } }.
\end{aligned}
\end{equation}
See Appendix~\ref{appendix:conditioning-truncation-equivalence} for a detailed derivation. 

First, when $Y$ is binary, we can derive a sequence of exact tests that demonstrates our main idea (we return to the case of general $Y$ in a moment).\footnote{For causal inference with binary outcomes, Equation~\eqref{eq:ipw-transform} does not produce binary $Y$. One may either use the asymptotic variant described later, or use a finite-sample valid extension to binary potential outcomes described in Appendix~\ref{appendix:binary-potential-outcomes}.} We make use of the fact that prior to truncation, $\un_t \meanest_t \sim \text{Binomial}(\un_t, \mean(\reg_t))$, which is stochastically dominated by $\text{Binomial}(\un_t, \cutoff)$ under the null. This leads to the following definition and theorem (see Appendix~\ref{appendix:critval-motivation} for a more detailed motivation).

\begin{definition}[Binary $Y$ critical values]
\label{def:binary-critval}
For $t = 0,...,\tmax$, let
\begin{equation*}
\begin{aligned}
\critval_t := \frac{\qtbinom\paren{1 - \allocalpha_t; \un_t, \cutoff, \un_t \trunc_t}}{\un_t}
\end{aligned}
\end{equation*}
where $\trunc_t$ is defined by Equation~\eqref{eq:truncation-definition}, and $\qtbinom\paren{q; \tilde{n}, \tilde{\mu}, \tilde{\trunc}}$ denotes the $q$th quantile of a $\text{Binomial}(\tilde{n}, \tilde{\mu})$ truncated to be no greater than $\tilde{\trunc}$ (see Appendix~\ref{appendix:generating-qtbinom} for a precise definition).
\end{definition}

\begin{theorem}[Validity and tightness for binary $Y$]
\label{theorem:binary-test-validity}
Assume $Y$ is binary and let $\reg_{\tsel}$ be the region produced by Algorithm~\ref{alg:chiseling-testing} using tests of the form $\gentest_t := \indic\set{\meanest_t > \critval_t}$ where $\critval_t$ is defined in Definition~\ref{def:binary-critval}. Then
\begin{equation*}
\begin{aligned}
\Prb\paren*{ \mean(\reg_{\tsel}) \leq \cutoff } \leq \alpha.
\end{aligned}
\end{equation*}
If in addition $\mean(X) = \cutoff$ a.s. (i.e., we are at the boundary of the global null) and $1 - \prod_{t=0}^{\tmax} (1 - \allocalpha_t) = \alpha$ almost surely, then the Type I error guarantee is tight in the sense that
\begin{equation*}
\begin{aligned}
\Prb\paren*{ \mean(\reg_{\tsel}) \leq \cutoff } = \alpha.
\end{aligned}
\end{equation*}
\end{theorem}

A proof is given in Appendix~\ref{appendix:proofs-binary-validity}. In it, we use the important fact that stochastic domination for binomials is preserved under right truncations (Lemma~\ref{lemma:truncated-binomial-stochastic-dominance}), which is not true for general distributions. Of particular note is the fact that Theorem~\ref{theorem:binary-test-validity} places no restrictions or assumptions on the choice of $\score(\cdot)$, $\threshlim$, and $\allocalpha_t$ beyond what is already stated in Algorithms~\ref{alg:chiseling-testing} and~\ref{alg:chiseling}, providing the user enormous latitude in how they make such choices in order to maximize the utility of the returned subgroup $\reg_{\tsel}$ without concern of violating Type I error. Furthermore, the only distributional assumptions we have made are that the dataset $(X_i, Y_i)_{i=1}^n$ consists of i.i.d. data points and that $Y$ is binary, yet the validity guarantee is non-asymptotic and tight. It is worth noting that here $\critval_t$, which is based on the null distribution truncated at $\trunc_t$, is deterministically no larger than the $1 - \allocalpha_t$ quantile of $\text{Binomial}(\un_t, \cutoff)$, and potentially much smaller if $\trunc_t$ is small. The latter is what the critical value would be if we based it on the \emph{untruncated} null distribution (the null distribution conditional only on $\F_t$). Thus, in our case conditioning makes it easier for each test to reject compared to not conditioning.

When the support of $Y$ is unrestricted, we rely on the fact that when $\un_t$ is large, $\meanest_t$ is approximately Gaussian with approximately known variance. We also suppose without loss of generality that $\cutoff = 0$ in the remainder of this subsection, since for any other $\cutoff$ we may simply subtract it from $Y$ to obtain an equivalent problem where $\cutoff = 0$. The following critical values are motivated by a Gaussian approximation.

\begin{definition}[General $Y$ critical values]
\label{def:general-critval}
In increasing order of $t$ for $t = 0,...,\tmax$, let
\begin{equation*}
\begin{aligned}
\critval_t := \paren*{\frac{\Phi^{-1}((1 - \allocalpha_t) \cdot \Phi(\sqrt{\un_t} \cdot \hsigsq_t^{-1/2} \trunc_t))}{\sqrt{\un_t} \cdot \hsigsq_t^{-1/2}}}_+ \quad \text{ where } \quad \hsigsq_t := \frac{1}{\un_t} \sum_{i : X_i \in \reg_t} (Y_i - \meanest_t)^2
\end{aligned}
\end{equation*}
and $\trunc_t$ is defined by Equation~\eqref{eq:truncation-definition}. The numerator of $\critval_t$ is the $(1 - \allocalpha_t)$th quantile of a standard normal truncated to be no greater than $\sqrt{\un_t} \cdot \hsigsq_t^{-1/2} \trunc_t$.
\end{definition}

To precisely state our asymptotic validity guarantee for Algorithm~\ref{alg:chiseling-testing} with general outcomes, we must define our asymptotics. Let $(X_i, Y_i)_{i=1}^{\infty}$ be an infinite stream of i.i.d. samples from the distribution of $(X, Y)$, and let $\reg_{\tsel}^{(n)}$ be the output of Algorithm~\ref{alg:chiseling-testing} obtained by treating the first $n$ data points as the entire dataset $\D$. Here we also allow $\tmax$, the maximum number of chiseling steps, to vary with $n$. Our asymptotics will be taken as $n \to \infty$. As with all CLT-based inference, we need some very basic moment assumptions about the data.
\begin{assumption}[Moment conditions]
\label{assumption:dgp-regularity}
$\E[|Y|^4]<\infty$ and $\Var(Y \mid X) > 0$ a.s.
\end{assumption}

Because our tests are based on normal approximations, we need some algorithmic restrictions in order to keep the total approximation error under control. For instance, we need to ensure that the minimum sample size associated to any test with nonzero $\allocalpha_t$ is growing. We also need to ensure that any nonzero $\allocalpha_t$ is not too small, since otherwise the inversion from $\allocalpha_t$ to quantile is too unstable, and we do rely on the accuracy of the quantiles themselves since they figure into the truncation levels.

\begin{constraint}[Chiseling constraints]
\label{constraint:chisel-cons}
Fix $\minprop, \alphamin \in (0, 1)$. Algorithm~\ref{alg:chiseling-testing} obeys the following constraints for all $t$:
\begin{enumerate}
\item \emph{Minimum sample size:} If $\allocalpha_t > 0$, then $\un_t / n \geq \minprop$.
\item \emph{Tail error restriction:} If $\allocalpha_t > 0$, then $\allocalpha_t \geq \alphamin$.
\end{enumerate}
\end{constraint}
Note that the tail error restriction in Constraint~\ref{constraint:chisel-cons} implies that there is a $\maxreg$ such that the number of tested regions is bounded, i.e. $\abs{\set{t : \allocalpha_t > 0}} \leq \maxreg$. We emphasize that Constraint~\ref{constraint:chisel-cons} is not an assumption but a constraint on how Algorithm~\ref{alg:chiseling-testing} is instantiated. It is entirely within the analyst's control to satisfy it irrespective of any unknowns about the problem. We suggest practical ways to fulfill Constraint~\ref{constraint:chisel-cons} and demonstrate that these choices achieve excellent empirical performance in Section~\ref{section:simulations}. We can now state our validity guarantee for Algorithm~\ref{alg:chiseling-testing} for general outcomes.

\begin{theorem}[Validity for general $Y$]
\label{theorem:general-test-validity}
Suppose $(X, Y)$ satisfies Assumption~\ref{assumption:dgp-regularity} and that Algorithm~\ref{alg:chiseling-testing} satisfies Constraint~\ref{constraint:chisel-cons}. When the sample size is $n$, let $\reg_{\tsel}^{(n)}$ be the region produced by Algorithm~\ref{alg:chiseling-testing} using tests of the form $\gentest_t := \indic\set{\meanest_t > \critval_t}$ where $\critval_t$ is defined in Definition~\ref{def:general-critval}. Then
\begin{equation*}
\begin{aligned}
\limsup_{n \to \infty} \Prb\paren*{ \mean(\reg_{\tsel}^{(n)}) \leq 0 } \leq \alpha.
\end{aligned}
\end{equation*}
\end{theorem}

A proof is given in Appendix~\ref{appendix:proofs-general-validity}. We note that Theorem~\ref{theorem:general-test-validity} does not come with an exactness guarantee. This is a byproduct of the clipping at $0$ in Definition~\ref{def:general-critval}, which is necessary due to an asymptotic subtlety requiring us to disallow $\critval_t$ from diverging to $-\infty$ (see Appendix~\ref{appendix:clipping-asymptotics} for an extended discussion). Nevertheless, we find chiseling to be essentially exact in our simulations (Section~\ref{section:simulations-type1-error}).

Although we state Theorem~\ref{theorem:general-test-validity} with a pointwise (i.e. fixed $(X, Y)$ distribution) guarantee for simplicity of exposition, in Appendix~\ref{appendix:proofs-general-validity} we prove validity in a more general triangular array setting that subsumes the pointwise case, which allows the distribution of $(X, Y)$ to also change with $n$. In particular, we emphasize that the asymptotic validity of Algorithm~\ref{alg:chiseling-testing} does not depend on the dimension of $X$ at all (it could even be infinite), nor on any structural assumption about $Y \mid X$ or the consistency of any estimator of $\Ec{Y}{X}$, nor on how the analyst shrinks the regions or allocates the $\allocalpha_t$ as long as they satisfy the mild Constraint~\ref{constraint:chisel-cons}. In fact, we can construct and work with a generalized procedure that relaxes the constraint that the minimum $\un_t$ is proportional to $n$, allowing it to grow to $\infty$ at any rate and thus permitting the discovery of regions of arbitrarily small probability mass; this general result's statement is quite a bit more technical, so we defer it to Appendix~\ref{appendix:proofs-general-validity}.

For causal problems, the AIPW estimator is generally preferred to the IPW estimator because it is more statistically efficient. Even though the AIPW transformation does not preserve the i.i.d. structure of the resulting sample since it typically utilizes data-dependent nuisance functions, we nonetheless prove that the AIPW estimator can be validly leveraged to perform chiseling in RCTs. As the setup is more verbose, we defer a detailed discussion and proofs to Appendix~\ref{appendix:proofs-aipw}.

\section{Practical considerations}
\label{section:practical-considerations}

In Section~\ref{section:simple-demo}, we discussed the choice of $\score(\cdot)$ and $\cutoff$. We now address some additional practical considerations: the initialization of chiseling, how to set the levels $(\allocalpha_t)_{t=0}^{\tmax}$, subgroup selection when non-responders are not harmed, computational efficiency, and the (optional) construction of interpretable regions. We refrain from making any universal prescriptions since chiseling is a flexible framework that an analyst can leverage in creative ways to validly incorporate domain knowledge. However, we hope that our discussion can serve as a helpful guide to practitioners in a breadth of scenarios.

\subsection{Initializing chiseling}
\label{section:init-chiseling}

When domain knowledge or structural assumptions are completely absent, chiseling may be initialized by splitting off a proportion of the sample (as in Section~\ref{section:simple-demo}). Thus, our simulations in Section~\ref{section:simulations} will leverage such initial data splits to kickstart chiseling. In practice, however, a researcher with sufficient understanding of their problem can initialize more powerfully and non-randomly. For instance, known directions of effect modification, mechanistic models, and predictive models trained on observational data or data from earlier clinical trial phases may inform the choice of initial shrinking direction. As new data is revealed by shrinking, the analyst is free to incorporate this information into $\score(\cdot)$.

When an initial split---let us say of proportion $p$---is necessary, how large should one take it to be? For the same reason that it is not easy to answer this for data splitting, it is not easy to answer this for chiseling. However, we make two observations. First, chiseling is effective even when $p$ is small because chiseling will continue to learn on data beyond the initial split. Second, any method that relies on an initial split proportion $p$ can be run at multiple $p_1,...,p_k$ with Type I error rates $\alpha^{(1)} + ... + \alpha^{(k)} = \alpha$, then combined using the Bonferroni correction. In our case, if $p_1 \leq ... \leq p_k$ and $\reg^{(1)},...,\reg^{(k)}$ are the regions reported by the $k$ different analyses, it is natural to report $\reg^{(j^*)}$ where $j^* := \max\set{j : \reg^{(j)} \neq \emptyset}$, since among nonempty regions $\reg^{(j^*)}$ is learned using the most initial training data. We call this \emph{Bonferroni aggregation} and we use it to aggregate results across different $p$ for chiseling, data splitting, and other competing methods in Section~\ref{section:simulations} (we find $k=3$, split proportions $\set{0.2,0.5,0.8}$, and equal $\alpha$ splitting to be an effective default choice). Surprisingly, this strategy can sometimes outperform any single choice of $p$ run at the full level $\alpha$, which we discuss in more detail in Appendix~\ref{appendix:multi-chiseling}.

\subsection{Setting the levels \texorpdfstring{$\allocalpha_t$}{}}
\label{section:setting-alpha}

Suppose we run Algorithm~\ref{alg:chiseling-testing} by letting at step $t$ the scoring function $\score(\cdot)$ be a machine learning estimate of the conditional mean function $\mean(x) = \E[Y \mid X = x]$ based on $\F_t$ and the cap $\threshlim$ be set equal to $\cutoff$. Suppose that after $\genstop$ shrinking steps the cap prevents chiseling from shrinking the current region any further. It makes sense to set $\allocalpha_t = 0$ for all $t < \genstop$ since at those stages we still estimate that $\reg_t$ contains points with negative conditional mean. In Section~\ref{section:simple-demo}, our simulations tested $H_0 : \mean(\reg_{\genstop}) \leq \cutoff$ at the full level $\alpha$, i.e. $\allocalpha_{\genstop} = \alpha$. But this may not be optimal, for example, because of estimation uncertainty. A natural way to accommodate multiple comparisons is to allocate $\alpha$ ``equally" between stage $\nu$ and some future stages.

Consider the following strategy. Initialize $\alpha_{\mathrm{spent}} = 0$ and choose some minimum sample size $n_{\min} > 0$ and initial split proportion $p \in (0,1)$. If it is necessary to enforce Constraint~\ref{constraint:chisel-cons}, take $\alphamin > 0$, but let $\alphamin = 0$ otherwise. Initialize chiseling with a $p$ fraction of the full sample as in Section~\ref{section:simple-demo} and for $t < \nu$ let $\allocalpha_t = 0$, while at each stage $t \geq \nu$, (1) set $\alpha_{\mathrm{budget}} \gets \paren*{\frac{\un_{\genstop} - \un_t}{\un_{\genstop} - n_{\min}}} \times \alpha$ if $\un_t \geq n_{\min}$ and $\alpha$ otherwise, (2) calculate $\allocalpha_t' \gets 1 - \frac{1 - \alpha_{\mathrm{budget}}}{1 - \alpha_{\mathrm{spent}}}$ and test at level $\allocalpha_t \gets \allocalpha_t' \cdot \indic\set{\allocalpha_t' \geq \alphamin}$, then (3) update $\alpha_{\mathrm{spent}} \gets 1 - (1 - \alpha_{\mathrm{spent}})(1 - \allocalpha_t)$. Essentially, this sets the remaining $\alpha$ proportional to the remaining sample size. During these stages, continue refitting the conditional mean $\hat{\mean}(\cdot)$ and shrinking the region using $\score(\cdot) = \hat{\mean}(\cdot)$ and $\threshlim = \infty$ so that new data is always revealed. Sometimes, it may be reasonable to spend a bit of the error budget on testing $\reg_0 = \mathcal{X}$, i.e. let $\allocalpha_0 > 0$ (see Section~\ref{section:practical-no-harm}). In this case, initialize $\alpha_{\mathrm{spent}} \gets \allocalpha_0$ and calculate $\alpha_{\mathrm{budget}} \gets \allocalpha_0 + \paren*{\frac{\un_{\genstop} - \un_t}{\un_{\genstop} - n_{\min}}} \times (\alpha - \allocalpha_0)$ for $t \geq \genstop$, which generalizes the above.

This strategy is intentionally simple. While more sophisticated $\alpha$ allocation strategies can lead to more powerful procedures, we find throughout Section~\ref{section:simulations} that even this simple approach dramatically outperforms data splitting, and moreover improves over the variant of chiseling in Section~\ref{section:simple-demo} that spends all of $\alpha$ at step $\genstop$. We leave an exploration of more sophisticated strategies to future work.

\subsection{Subgroup selection when non-responders are not harmed}
\label{section:practical-no-harm}

In some applications including certain clinical RCTs, it may be implausible for the CATE to ever be negative, e.g., since such trials are only approved if there is a strong prior belief that the treatment will not bring harm to any individuals. Though subgroup selection can still be used to discover units with an effect much larger than zero (i.e. by taking $\cutoff > 0$), it can bring value even when $\cutoff = 0$ and $\mean(x)\ge 0$ for all $x\in\mathcal{X}$, so that $\mathcal{X}$ is an optimal subgroup. This is because some units may have small or zero treatment effects and thus excluding them from the subgroup can boost the power of a test of significance (and hence the expected utility of the reported region). We will see a numerical demonstration of this in Section~\ref{section:simulation-nonneg-rct}.

When $\mathcal{X}$ is believed to be optimal, however, it is also natural to compare chiseling to the procedure that tests $H_0: \mean(\mathcal{X}) \leq 0$ using a one-sided $t$-test, reporting $\mathcal{X}$ if a rejection is made and $\emptyset$ otherwise; we will refer to this procedure as the `global $t$-test' for short. The global t-test always tests an optimal region, and when treatment effects are positive and homogeneous, it is essentially an optimal procedure. Luckily, we can instantiate chiseling in such a way that even when effects are homogeneous, chiseling is not much worse than the global $t$-test, but when a sufficient fraction of units have zero or nearly zero treatment effects (but are still all nonnegative, so $\mathcal{X}$ remains optimal), chiseling outperforms the global $t$-test. This is possible by tuning $\allocalpha_0$; setting $\allocalpha_0 = 0$ puts no stock in testing $\mathcal{X}$, while setting $\allocalpha_0 = \alpha$ reduces to just testing $\mathcal{X}$ and is essentially identical to the global $t$-test procedure.\footnote{At $t=0$, the asymptotic test we construct in Section~\ref{section:general-test-design} reduces to running a one-sided $Z$-test on $\mathcal{X}$, which is asymptotically equivalent to the one-sided $t$-test.} Anything in-between trades off between these extremes, and in Appendix~\ref{appendix:inferiority-t-test} we show that one can bound the worst case loss in expected utility of chiseling when compared to the $t$-test as a function of $\allocalpha_0$. For instance, if $\mean(\cdot)$ is nonnegative everywhere and $\allocalpha_0 = \alpha / 2$, then the normalized expected utility of chiseling cannot be less than that of the $t$-test by more than $12.5$ percentage points (asymptotically), regardless of how chiseling is otherwise instantiated, but can potentially be much greater (Section~\ref{section:simulation-nonneg-rct}).

\subsection{Computational efficiency}

While updating $\hat{\mean}(\cdot)$, the current estimate of $\mean(\cdot)$, each time a single new data point is revealed is ideal from the standpoint of statistical efficiency, it may not be practical if the machine learning method is computationally expensive and/or the dataset is large. Luckily, some machine learning methods can be efficiently updated upon observing a single new data point, e.g. rank-one updating ridge regression or online stochastic gradient descent, but since chiseling is interactive it is easy to enforce computational constraints as needed. For instance by only updating $\hat{\mean}(\cdot)$ every $B > 1$ calls to Algorithm~\ref{alg:chiseling}, or ``warm starting" fitting procedures by initializing parameters at previous estimates.

\subsection{Constructing interpretable regions}
\label{section:interpretable-regions}

In practice, investigators may prefer subgroups that are easy to interpret so that it is easier to assess the plausibility of the discovered subgroups, in addition to being easier to communicate to stakeholders. There are many different criteria for what makes a subgroup interpretable, and each different criteria will necessitate a different way of using chiseling to produce a satisfactorily interpretable region. In this subsection, we consider one common standard for interpretability: regions that are axis-aligned hyperrectangles. That is, we desire regions of the form $\reg = (a_1, b_1) \times ... \times (a_d, b_d) \subseteq \mathbb{R}^d$, so that checking whether $x \in \reg$ reduces to checking whether each of $d$ covariates individually lies in an interval. If desired, a notion of sparsity may be incorporated by encouraging $(a_j, b_j) = (-\infty, \infty)$ for most $j$.

Let $x^{(j)}$ denote the $j$th element of a $x \in \mathbb{R}^d$. We have presented chiseling as shrinking along the upper level sets of a function $\score(\cdot)$, and in order to ensure the upper level sets of $\score(\cdot)$ are hyperrectangular, it suffices to consider the form $\score(x) = \min_{j=1,...,d} \set{f_j(x^{(j)})}$, where each $f_j : \mathbb{R} \to \mathbb{R} \cup \set{\infty}$ is any function whose upper level sets are intervals (we call such a function ``unimodal"). This is because for any $z \in \mathbb{R}$,
\begin{equation*}
\begin{aligned}
\score(x) > z \iff \forall j \,\, f_j(x^{(j)}) > z \iff \forall j \,\, x^{(j)} \in \set{x' \in \mathbb{R} : f_j(x') > z},
\end{aligned}
\end{equation*}
thus implying that $\set{x \in \mathbb{R}^d : \score(x) > z}$ is the product of intervals. Setting $f_j(x^{(j)}) = \infty$ for most $j$ facilitates sparsity. In fact, any $\score(\cdot)$ with hyperrectangular upper level sets has this form (Appendix~\ref{appendix:hyperrectangular-shrinking}), so this is a fully general formulation. The intersection of hyperrectangles remains a hyperrectangle, so using scoring functions of the above form in each shrinking step will result in a sequence of chiseled regions that are hyperrectangles. The endpoints of any region reported by chiseling are straightforward to calculate by simply keeping track of the $2d$ coordinate endpoints of each hyperrectangle at each stage of Algorithm~\ref{alg:chiseling-testing} and intersecting them.

The individual functions $f_j(\cdot)$ need to be learned from the data. A simple idea, showcased in Section~\ref{section:real-data}, is to let $f_j(\cdot)$ be an estimate of $\E[Y \mid X^{(j)}]$ using all of the data available thus far that also enforces the structural requirement of unimodality. Examples of unimodal regressions include isotonic regression and constrained quadratic regression. A further challenge is that in this case $\score(\cdot)$ is no longer an estimate of $\hat{\mean}(\cdot)$, and hence there is no special reason to begin testing at the point where $\score(\cdot)$ crosses $0$. Instead, a naive workaround (which in Section~\ref{section:real-data} we find to be effective) is to run the strategy outlined in Section~\ref{section:setting-alpha} but with $\genstop = 0$ so that $\alpha$ is allocated ``equally" over the entire sequence of chiseled regions.

\section{Numerical studies}
\label{section:simulations}

In this section, we verify using numerical experiments that chiseling possesses valid Type I error control, which is exact or nearly exact at the boundary of the global null, and that regions reported by chiseling have higher expected utility than those reported by methods with comparable guarantees. In particular, we consider the following methods:
\begin{itemize}
    \item \textbf{$p$-chiseling}. The version of chiseling described in Section~\ref{section:setting-alpha} with initial split proportion $p$.
    \item \textbf{$p$-data splitting.} Data splitting with initial split proportion $p$ as described in Section~\ref{section:simple-demo}.
    \item \textbf{$p$-simultaneous data splitting.} An approach that leverages data splitting but then produces simultaneously valid tests for a nested collection of regions. This is representative of the methods of \textcite{Bonetti2004, Song2004, li2023statisticalperformanceguaranteesubgroup}, which are all based on a common principle. In particular, this method uses a training set of proportion $p$ to define $10$ nested regions where the largest is the estimated optimal region, e.g., the one that $p$-data splitting would test. It reports the largest rejected region. See Appendix~\ref{appendix:simultaneous-data-splitting} for details.
\end{itemize}
Because the parameter $p$ may be difficult to tune, where appropriate we will (1) show results as a function of $p$, and (2) show results for the Bonferroni aggregation (described in Section~\ref{section:init-chiseling}) across $p \in \set{0.2, 0.5, 0.8}$, letting the target level be $\alpha / 3$ in each of the three instances.

We will also consider the \textbf{global $t$-test} procedure from Section~\ref{section:practical-no-harm}, an \textbf{oracle data splitting} approach which has access to twice the sample size and runs $0.5$-data splitting, and an \textbf{oracle simultaneous data splitting} approach which has access to twice the sample size and runs $0.5$-simultaneous data splitting. Effectively, the oracle methods tell us what utility to expect if we were able to use the entire sample both for subgroup selection and for testing in a valid way, and thus provide approximate upper bounds on the achievable utility.

Throughout this section, we only consider the version of chiseling described in Section~\ref{section:setting-alpha}, which requires the multiple testing framework we have developed, as opposed to the version described in Section~\ref{section:simple-demo}, which tests only a single region. While we consistently find that the latter approach dominates data splitting for any given $p$ (Figure~\ref{fig:simple-demo-utility}), we showcase the former because we find that it is uniformly more powerful than the latter (see Appendix~\ref{appendix:simulation-multiple-vs-single}).

\hfill

\noindent \textbf{High-level details.} Throughout, $n$ will denote the sample size and $p$ is synonymous with the initial split proportion. We let $\alpha = 0.05$, $\alphamin = 1 - (1 - \alpha)^{1/40}$, and $n_{\min} = 30$. All expected utilities are normalized by the utility of the optimal region. While maximizing expected utility is our primary goal, we report power curves for every setting in Appendix~\ref{appendix:simulation-power}. Simulations are repeated at least 2,500 times, and all standard errors are less than $1\%$. All additional details can be found in Appendix~\ref{appendix:simulation-details}.

\subsection{Type I error and exactness}
\label{section:simulations-type1-error}

Our simulations confirm that, at the boundary of the global null, the Type I error of chiseling is essentially exact across a diverse array of data generating processes and choices of machine learning algorithm within chiseling. We confirm the quality of the asymptotic approximation when the IPW transformation is used to construct the outcomes $Y$, as well as for when a misspecified AIPW transformation is used. To save space, we relegate figures and additional details to Appendix~\ref{appendix:empirical-type1-error-global-null}, but we broadly report that the quality of our asymptotic approximation is high and that the Type I error is essentially exact even when the minimum sample size corresponding to any test is as small as $30$. Next, we confirm that the Type I error of chiseling is controlled when we are not under the global null. In each of the simulation settings in the subsequent sections (none of which are under the global null), we also calculate the Type I error $\Prb(\mean(\reg_{\tsel}) \leq \cutoff)$ and find that the maximum Type I error across any setting is no greater than $1\%$.

\subsection{Binary regression}
\label{section:simulation-binary-regression}

\begin{figure}[ht]
  \centering
  \includegraphics[width=\linewidth]{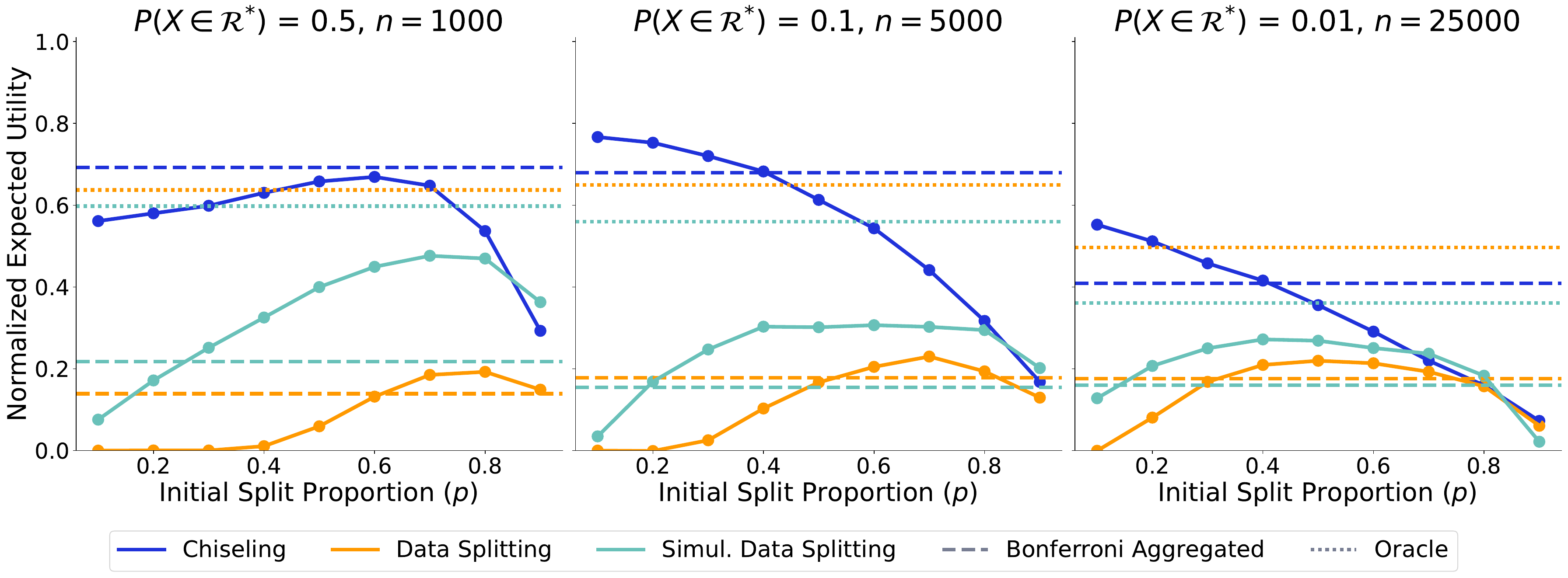}
  \caption{Normalized expected utility in a simulated binary regression problem as the probability mass of the optimal subgroup varies from $50\%$ to $1\%$. Solid horizontal lines correspond to using Bonferroni to aggregate across initial split proportions $\in \set{0.2,0.5,0.8}$ for the method with the corresponding color. The machine learning method used is $\ell^2$-penalized logistic regression.}
  \label{fig:binary-regression-utility}
\end{figure}

We consider a binary regression problem where $\cutoff = 0.9$, $d = 100$, $Y$ follows a logistic regression model in $X$, and the covariates are mildly correlated. The machine learning algorithm employed by every method is $l^2$-penalized logistic regression with scikit-learn's default regularization settings, and we vary the probability mass of the optimal subgroup ($\Prb(X \in \reg^*)$ where $\reg^*$ is the optimal subgroup) from $50\%$ to $1\%$ and tune $n$ so that the expected utility is nontrivial. Our findings are displayed in Figure~\ref{fig:binary-regression-utility}. Because in this case $\Prb(Y = 1) < 0.9$ in every setting, we do not report results for the global $t$-test (its expected utility is negative). For the data splitting methods, an exact binomial test is used in place of a $t$-test. We find that the expected utility of chiseling dominates that of the other methods across essentially all initial split proportions and optimal subgroup probability masses, and that the Bonferroni aggregated variant of chiseling dominates the Bonferroni aggregated variants of data splitting and simultaneous data splitting. Comparing the best performance of each non-oracle method across initial split proportions, we see that the peak expected utility for chiseling exceeds that of the next-best method by $40\text{--}150\%$. In some situations, chiseling even outperforms the oracles which utilize twice the sample size, which is remarkable but by no means impossible since chiseling leverages a multiple testing framework that is different than that used by the oracles.

\subsection{Heterogeneous randomized control trial}
\label{section:simulation-heterogeneous-rct}

\begin{figure}[ht]
  \centering
  \includegraphics[width=\linewidth]{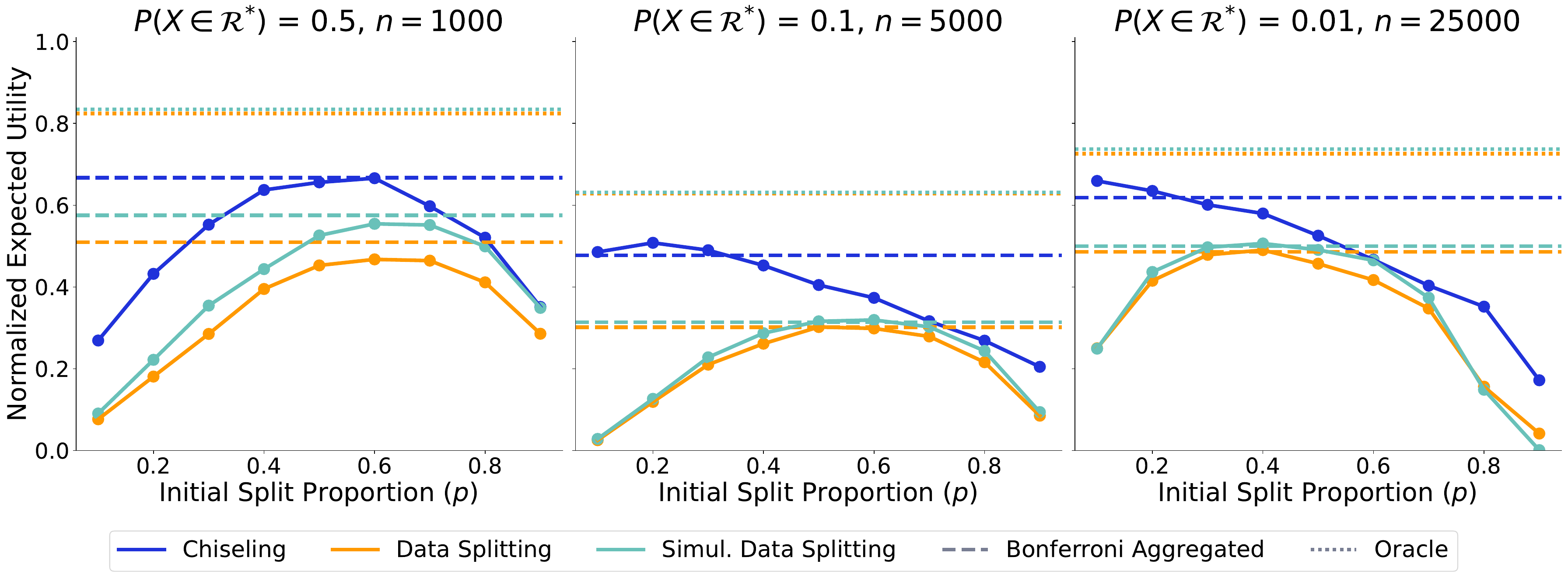}
  \caption{Normalized expected utility in a simulated heterogeneous RCT as the probability mass of the optimal subgroup varies from $50\%$ to $1\%$. Solid horizontal lines correspond to using Bonferroni to aggregate across initial split proportions $\in \set{0.2,0.5,0.8}$ for the method with the corresponding color. The machine learning method used is cross-validated lasso regression.}
  \label{fig:heterogeneous-linear-rct-utility}
\end{figure}

We consider a simulated RCT where $\cutoff = 0$, $d = 100$, the CATE is linear in $X$, and the covariates are mildly correlated. The treatments are assigned using a fair coin flip, and we apply the IPW transformation to convert each instance to a regression problem. The machine learning algorithm we use here is the lasso where the penalty is selected via cross-validation. Like in the previous subsection, we vary the probability mass of the optimal subgroup from $50\%$ to $1\%$, and we do not report results for the global $t$-test because the full population $\mathcal{X}$ is under the null. We find that the expected utility of chiseling dominates that of the other methods across all initial split proportions and optimal subgroup probability masses, and that the Bonferroni aggregated variant of chiseling dominates the Bonferroni aggregated variants of data splitting and simultaneous data splitting (Figure~\ref{fig:heterogeneous-linear-rct-utility}). Comparing the best performance of each method across initial split proportions, we see that the peak expected utility for chiseling exceeds that of the next-best method by $20\text{--}59\%$.

\subsection{RCT with misspecified machine learning model}
\label{section:simulation-rct-misspecified}

We empirically investigate how misspecification of the machine learning model affects the relative performances of the subgroup selection procedures. Recall that the methods we consider are valid despite the misspecification. We let the CATE be non-linear in $X$ but use linear regression with a ridge penalty selected using cross-validation as the base machine learning method. Across a variety of optimal subgroup probability masses ranging from $75\%$ to $25\%$, we find that the expected utility of chiseling dominates that of other methods, including when results are aggregated using Bonferroni. As the expected utility curves do not look substantially qualitatively different from those seen thus far, we relegate the plots to Appendix~\ref{appendix:simulation-additional-results} (Figures~\ref{fig:kang-schafer-utility} and~\ref{fig:kang-schafer-power}). In these simulations, the peak expected utility for chiseling exceeds that of the next-best method by $13\text{--}40\%$.

\subsection{RCT with no or small negative treatment effects}
\label{section:simulation-nonneg-rct}

We consider a simulated clinical RCT where the treatment does not bring substantial harm to any individuals. Moreover, $\cutoff = 0$, so that we wish to include units with any positive CATE while being agnostic about units with zero CATE.

\begin{figure}[ht]
  \centering
  \includegraphics[width=\linewidth]{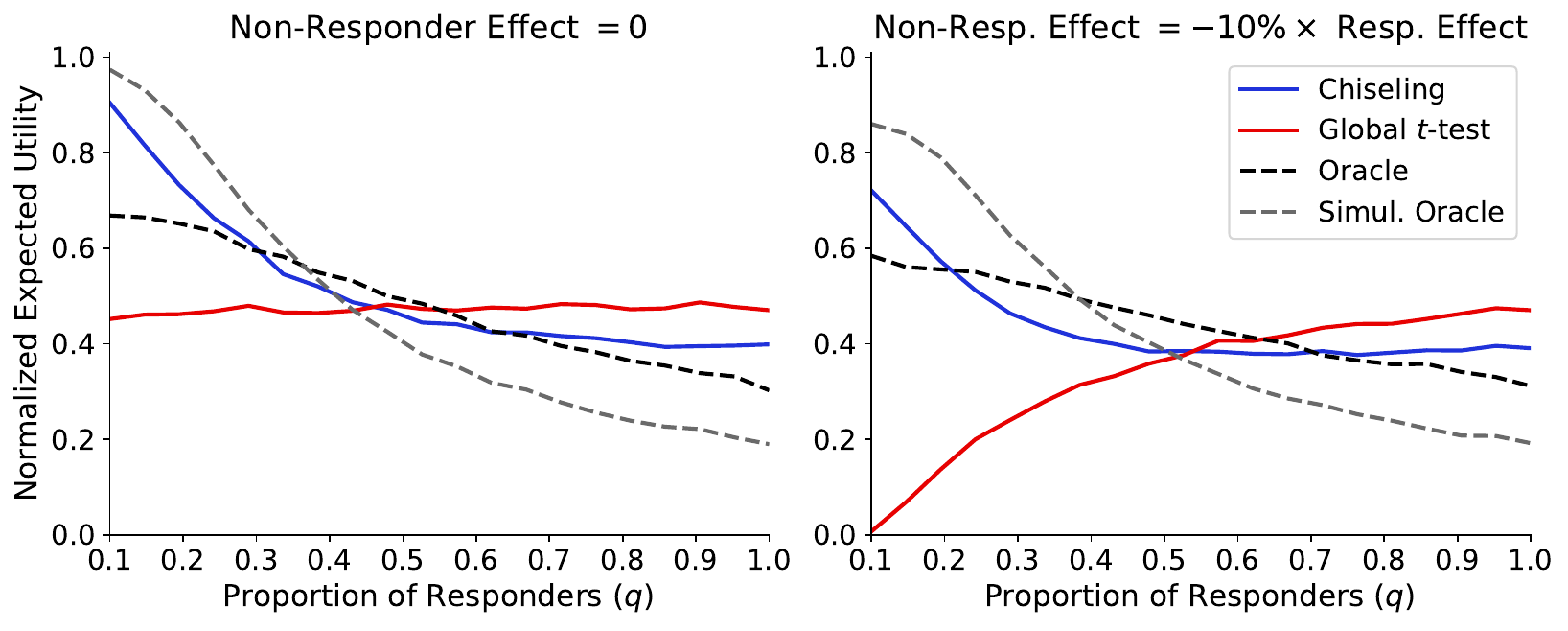}
  \caption{Normalized expected utility in a simulated RCT where the treatment effect for responders is $\tau$, and where non-responders either have $0$ treatment effect (left panel) or treatment effect equal to $-0.1\tau$ (right panel). Sample size is $n = 1000$ and $\tau = 0.1 / q$ where $q$ is the proportion of responders. The machine learning method used is linear regression. We initialize chiseling with $p = 0.1$ and $\alpha_0 = \alpha / 2$.}
  \label{fig:linear-nonneg-rct-utility}
\end{figure}

First, suppose that the CATE only ever takes on values in $\set{0, \tau}$ for some constant $\tau$ so that $\mathcal{X}$ is an optimal subgroup. We choose the CATE to be linear in $X \in \mathbb{R}^5$ and restrict the support of $X$ so that the CATE lies in $\{0,\tau\}$ (see Appendix~\ref{appendix:simulation-details} for details; we choose this stylized setting to make the distribution of the CATEs maximally interpretable). Chiseling uses linear regression. We vary $q$, the proportion of units with CATE equal to $\tau$. As per Section~\ref{section:practical-no-harm}, we let $\allocalpha_0 = \alpha / 2$ and $p = 0.1$ so that asymptotically the normalized expected utility of chiseling can not be more than $12.5$ percentage points smaller than that of the global $t$-test for any $q$ or $\tau$. The left panel of Figure~\ref{fig:linear-nonneg-rct-utility} empirically confirms this at the extreme where $q = 1$. But when the proportion of responders is $q = 0.1$, chiseling doubles the expected utility of the global $t$-test. Also, the point at which it becomes more beneficial to use chiseling is roughly the same as the point at which the oracle methods outperform the global $t$-test, suggesting that chiseling only struggles when subgroup selection is intrinsically difficult.

The second panel in Figure~\ref{fig:linear-nonneg-rct-utility} shows the same as the above, but where the CATE takes on values in $\set{-0.1\tau, \tau}$; that is, the treatment is $10$ times more effective for responders than it is harmful for non-responders. It paints the same picture but naturally yields greater potential gains over the global $t$-test. For instance, when $q = 10\%$, chiseling achieves a normalized expected utility of over $70\%$ while the global $t$-test has zero expected utility (since $\mean(\mathcal{X})=0$).

\section{Real data application}
\label{section:real-data}

\begin{figure}[ht]
  \centering
  \includegraphics[width=\linewidth]{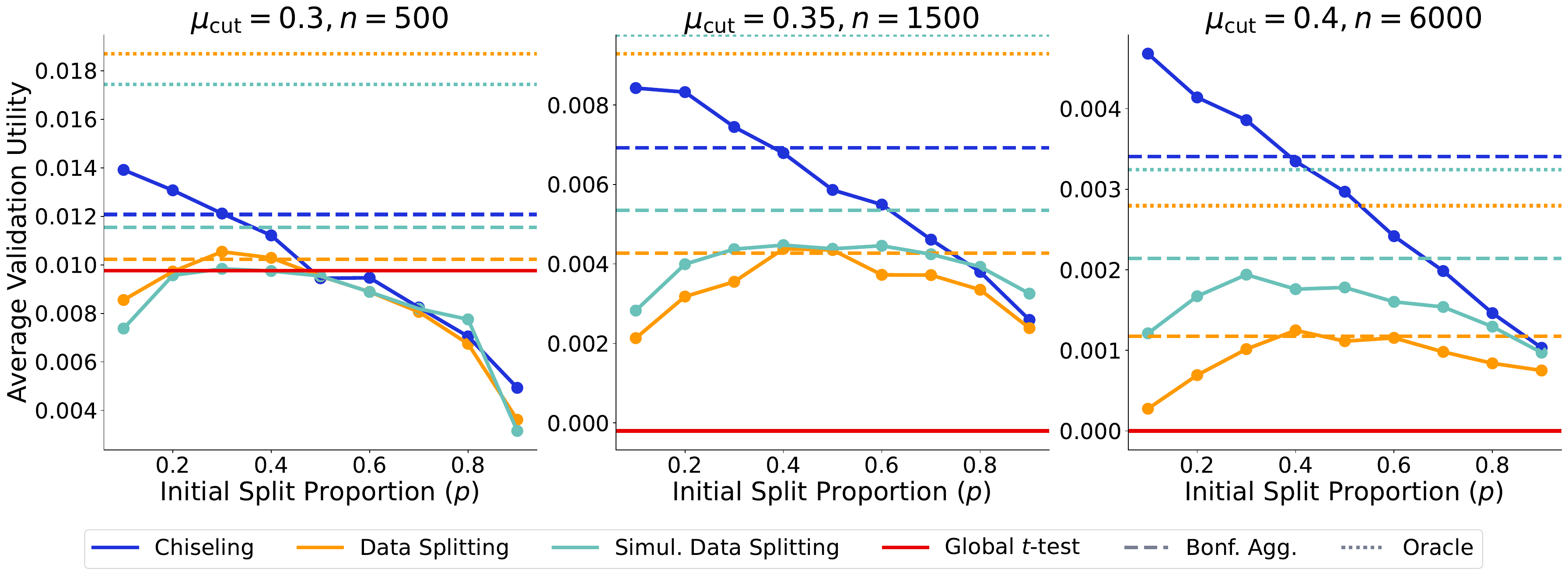}
  \caption{Average validation utilities on the GSS data across different testing cutoffs $\cutoff$.}
  \label{fig:bart-utility}
\end{figure}

In the mid-1980s, the General Social Survey (GSS) incorporated an experiment into its survey of American adults in order to assess whether the wording of a question could influence expressed support for government spending. Survey takers were randomized into either receiving a question which described the spending program as ``welfare" or one which described the spending program as ``assistance to the poor." Many analyses have replicated the famous $\approx 0.35$ increase in probability of support for the spending program when the wording is changed from ``welfare" to ``assistance to the poor." The experiment was reanalyzed in \textcite{Green2012} using Bayesian Additive Regression Trees in order to understand the heterogeneity in the effect across different demographic characteristics and survey responses.

Imitating the analysis of \textcite{Green2012}, we take 36,501 survey data points from the GSS and apply minimal preprocessing (impute non-responses, drop missing data, and one-hot encode non-ordinal categorical variables). This produces $d = 152$ features. We then test the ability of methods to detect subgroups with effects greater than $\cutoff \in \set{0.3, 0.35, 0.4}$. In order to assess the average utility of the reported subgroups, we repeatedly split the dataset into a training set which is used to run the subgroup selection methods and a validation set which is used to estimate the utilities of the reported subgroups. Note that for each method, both selection and testing are performed strictly on the training set. We appropriately set the size of the training set, which we refer to as $n$, for each setting so that the expected utility is nontrivial. We apply a simple AIPW transformation with only an intercept and no covariates, and we use the T-learner \parencite{Knzel2019} with scikit-learn's default implementation of the random forest classifier as the base learner.

We see in Figure~\ref{fig:bart-utility} that the average utility of chiseling dominates that of the other methods in all settings (results for power and expected probability mass may be found in Appendix~\ref{appendix:simulation-additional-results}). Comparing the best performance of each method across initial split proportions, we see that the peak expected utility for chiseling exceeds that of the next-best method by $32\text{--}141\%$. When $\cutoff = 0.4$, chiseling outperforms the oracle methods, though we caution that the repeated train/validation splits are not independent and hence we cannot produce reliable standard errors for these plots when $n$ is a sizable fraction of 36,501.

Next, we run chiseling on the full 36,501 data points with $\cutoff = 0.45$. To construct an interpretable subgroup, we use the approach described in Section~\ref{section:interpretable-regions} with coordinate-wise isotonic regression (direction adaptively chosen using Spearman's rank correlation) to discover hyperrectangular regions. We initialize chiseling with 1,000 randomly revealed data points. Chiseling discovers a subgroup that includes roughly $4\%$ of the sample. In this subgroup, the empirical difference between treatment and control means is $0.50$. We examine in Appendix~\ref{appendix:simulation-additional-results} (Table~\ref{table:bart-feature-ranks}) the ten most important features as measured by exclusivity of that feature's range in the discovered subgroup, and find strong agreement with previous analyses such as \textcite{Green2012}. Further investigation suggests these ten features are the dominant drivers of the effect size in the discovered subgroup; see Figure~\ref{fig:bart-interpret} for details.

\section{Discussion}
\label{section:discussion}

We presented a novel method for subgroup selection that maintains the leanness of data splitting's assumptions but is able to utilize the given sample much more efficiently. It consists of two primary methodological insights: that a specific way of shrinking a subgroup leaves it untarnished, and that we can exactly (or essentially exactly) sequentially test multiple regions while shrinking. In its most basic form, chiseling information-theoretically dominates data splitting for subgroup selection, and we showed that these gains are realized empirically via natural instantiations of chiseling.

The chiseling framework is extremely flexible, and while this paper provided some guidance and default choices for its use, we ultimately believe its flexibility is its key strength, allowing users to incorporate any and all qualitative and quantitative knowledge however they want without having to worry about validity. We list here a few directions for future research on leveraging and further expanding this flexibility.

\begin{itemize}
    \item \emph{Can we better shrink the regions?} We recommended shrinking regions along an estimate $\hat{\mean}(\cdot)$ of $\mean(\cdot)$, but more sophisticated ideas could be superior. For instance, \textcite{spiess2023findingsubgroupssignificanttreatment} suggest that power can be gained by considering how the variance of the outcome varies across the covariate space, favoring regions with moderate effects and small variances to those with large effects and outsized variances.
    \item \emph{Can we better allocate $\alpha$?} We suggested a simple strategy for allocating $\alpha$, but it may not be the most powerful choice. Relatedly, how should one set $\alpha$ when $\hat{\mean}(\cdot)$ is grossly misspecified? While $\hat{\mean}(\cdot)$ may still be effective for shrinking, the value of $\hat{\mean}(\cdot)$ cannot be trusted to indicate the optimal region threshold. Ideas from the calibration literature could prove useful.
    \item \emph{Can we better initialize chiseling?} Can the initial split proportion $p$ be chosen adaptively, or even done away with entirely? If prior information exists, how can it be effectively incorporated into the initialization? For an idea along these lines, see Appendix~\ref{appendix:two-team-cross-screening} for an initialization inspired by the ``two team cross-screening" approach of \textcite{roy2025explorationconfirmationreplicationobservational}.
    \item \emph{How should chiseling be used to maximize other utility functions?} The right way to use chiseling will differ based on the utility one wishes to maximize, and in particular may be different if one wishes to maximize power or expected probability mass.
    \item \emph{Can we give inferences for other estimands?} The structure of the naive mean estimator yields tractable conditioning events that we take advantage of in our sequential testing framework. We also showed that one can utilize sophisticated transformations such as the AIPW transformation with known propensities, but additional work is required to extend our approach to the observational setting. It would be interesting to see if estimands that are not based on an average of transformed outcomes (e.g. conditional medians or M-estimands) can be addressed in our framework.
    \item \emph{Can we broaden the inferential targets?} Chiseling, as we have developed in this paper, is a method for rejecting one region at a fixed cutoff, but there may be situations where an analyst wishes to reject multiple regions, or to pair the selected subgroup with a point estimate or tight confidence interval for its effect. Though we discuss some multiple testing extensions in Appendix~\ref{appendix:multiple-testing} and methods for point estimates and simultaneous confidence intervals in Appendix~\ref{appendix:point-est-cis}, there is more work to be done.
    \item \emph{Can we expand the action space of chiseling?} Currently, there are two basic actions one can take at any point during chiseling: reveal a random point from the current region (see Appendix~\ref{appendix:chiseling-generalizes-ds}) or shrink the region. What other actions can be added to this set? For instance, are there techniques that would allow one to restore erroneously excised regions of the covariate space? Instead of revealing a point uniformly at random, can we reveal it according to a weight function? And can we reveal points based not only on $X$ but also on $Y$ while still permitting valid inference?
\end{itemize}

\section*{Acknowledgments}

The authors would like to thank Alan Chung, Kosuke Imai, Zeyang Jia, Michael Lingzhi Li, Yash Nair, and Kostas Sechidis for helpful discussions regarding this work. NC and LJ were partially supported by DMS-2045981. NC was partially supported by a Graduate Research Fellowship from the National Science Foundation. AS was partially supported by the Two Sigma Graduate Fellowship Fund, the Citadel GQS PhD Fellowship, and a Graduate Research Fellowship from the National Science Foundation.

\FloatBarrier

\printbibliography

\FloatBarrier
\newpage

\begin{appendices}

\startcontents[appendix]
\section*{Appendix Contents}
\printcontents[appendix]{}{1}[2]{}

\setcounter{figure}{0}
\renewcommand{\thefigure}{\thesection\arabic{figure}}

\section{Extended literature review}
\label{appendix:extended-lit-review}

\subsection{Subgroup learning}

There is a vast literature on works that seek to learn subgroups from the data, but which do not by themselves propose ways of evaluating that subgroup without data splitting, or otherwise only provide weak guarantees on the learned subgroup. This includes the literature on policy learning, which is distinguished enough that we discuss it separately in the next subsection. While these works do not focus on the inferential criteria that we consider here, essentially any one of them has potentially important insight to offer about one of the key tuning parameters in chiseling: how to shrink the region.

In this space, tree-based methods have been enormously popular since they naturally form partitions of the covariate space into relatively interpretable subgroups \parencite{Su2009, Lipkovich2011, Lipkovich2014,Athey2016,ladhania2020learningtestingsubgroupsheterogeneous, Zhang2021, Yang2021, Cai2022, Zhang2023, huang2025distillingheterogeneoustreatmenteffects}. The aspects of subgroup learning covered in these works include: splitting criteria, for instance, distinguishing the goals of minimizing the generalization mean-squared error (MSE) and the expected empirical MSE of ``honest" estimates calculated on a held out test set; resampling based heuristics for reducing overfitting; variable screening steps; the use of more efficient (doubly robust) estimators of the treatment effect to inform splits; and the incorporation of constraints on subgroup structure. Non tree-based procedures include ``bump hunting" \parencite{FRIEDMAN1999, Kehl2006}, covariate coarsening via optimal transport \parencite{zhang2024coarsepersonalization}, and mixture modeling \parencite{Ding2016}. Furthermore, some works such as \textcite{Foster2011, Bertsimas2019, Zhao2023} validate their approaches using informal principles and simulations but are not accompanied by rigorous inferential guarantees.

Lastly, we highlight two works where a subgroup is designed with the specific goal of optimizing the power to replicate a significant subgroup finding on an external dataset. \textcite{Talisa2021} evaluate a subgroup selection procedure favorably if the selected subgroup yields a significant finding on a holdout set. \textcite{spiess2023findingsubgroupssignificanttreatment} initiate a formal study of this problem, and make the interesting observation that the best powered subgroup may not necessarily be the one that maximizes the policy value. This is because certain regions of the covariate space with strong average effects may be offset by high treatment effect variance, which can deteriorate the power of a significance test. Under a local asymptotic scaling, they write down a formal objective that they develop methods to optimize for. It is also noteworthy that they consider, but do not investigate, a possible extension of their method that, rather than optimizing power, optimizes an expected utility very similar to what we have considered in this work.

\subsection{Policy learning}

The policy learning literature is broadly distinguished by a focus on regret minimization in a decision-theoretic framework. A now-classical goal in policy learning is to minimize
\begin{equation*}
\begin{aligned}
\mathrm{regret} := \mathcal{V}(\pi^*) - \E[\mathcal{V}(\hat{\pi}_n)],
\end{aligned}
\end{equation*}
where $\mathcal{V}(\pi) := \E[Y(\pi(X))]$, $\pi^*$ is a maximizer of $\mathcal{V}(\pi)$ over a class of policies $\Pi$, and $\hat{\pi}_n$ is the (random, data-dependent) policy that is obtained from a random sample of $n$ data points. \textcite{Manski2004} calls $\hat{\pi}_n$ a \textit{statistical treatment rule} (STR). Generally, an STR is judged favorably relative to a class of distributions on $(Y(1), Y(0), W, X)$ if it is \textit{admissible} and \textit{minimax} with regards to regret minimization. In the literature, particular focus has been devoted to establishing the minimaxity of various STRs. \textcite{Manski2004} established regret bounds over a restricted set $\Pi$, which were subsequently refined and generalized in a number of works \parencite{Hirano2009, Stoye2009, Kitagawa2018}. Often, regret bounds are provided in terms of complexity measures of the policy class $\Pi$ such as the Vapnik--Chervonenkis dimension. An important insight that arises in this literature is that for regret minimization under misspecified parametric models, one should base decisions on direct maximization of the empirical welfare rather than using the plug-in policy that thresholds an estimate of $\E[Y(1) - Y(0) \mid X]$, since the latter may not converge to the welfare maximizing policy due to the misspecification. This is an aspect that we have not explored in our work, but which may offer insights about how to best leverage chiseling.

Additional work in this vein includes the following. \textcite{Qian2011} use a very large basis expansion of the covariate space to linearly model the conditional mean function $\E[Y \mid X, T]$. \textcite{Zhao2012} convert the policy maximization problem into a weighted classification problem where each unit is weighted according to its outcome. \textcite{Imai2013} leverages the support vector machine to produce a sparse, parsimonious model for treatment effect heterogeneity across treatments and subgroups. \textcite{VanderWeele2019} describe the population optimal treatment allocation rules under various constraints such as a constraint on the treated proportion or cost/side effect constraints. \textcite{Luedtke2020} show that the regret of empirical risk minimization can decay at rates faster than $n^{-1/2}$ for fixed data generating distributions. \textcite{Athey2021} leverage ideas from semiparametric efficiency theory, using estimates of doubly robust ``scores" as pseudo-outcomes and plugging these into an empirical risk minimization framework. We also refer the reader to the ``Related Work" section of \textcite{Athey2021} for a survey of developments along the theme of regret minimization.

Note that none of the above discuss inference. Indeed, the ethos of a decision-theoretic solution is generally to obviate a direct need for inference by focusing on the expected gain or cost of each action. There are, however, some exceptions. For instance, some papers do consider inference for the value of the optimal policy or the policy at hand \parencite{Zhang2012Robust, Luedtke2016, ponomarev2024lowerconfidencebandoptimal}, usually either with the requirement that the policy is learned efficiently or via data splitting. Works such as \textcite{chernozhukov2025policylearningconfidence, andrews2025certifieddecisions} are still situated in a decision-theoretic framework but discuss the role of uncertainty in decision making: the former absorbs the uncertainty quantification into the objective by utilizing losses that reflect risk aversion, while the latter characterizes the role of confidence sets in decision making.

\subsection{Sequential, online, and adaptive hypothesis testing}

Sequential hypothesis testing was introduced in \textcite{Wald1945} to address the challenge of assessing significance in an experiment that continues making observations until a significant conclusion is reached. For testing a simple null against a simple alternative, \textcite{Wald1945} introduced the sequential probability ratio test, which is optimal in the sense of minimizing the expected number of observations needed to yield a significant conclusion. In clinical trials, Wald's work inspired new protocols for repeatedly assessing evidence for treatment efficacy throughout an experiment while controlling Type I error \parencite{Armitage1969, Haybittle1971, POCOCK1977, OBrien1979, Lan1983, Demets1994}. At a high level, these methods work by comparing an accumulating test statistic to different rejection boundaries at different time points. In particular, \textcite{Lan1983} set testing boundaries according to an object they call the $\alpha$-spending function, which dictates how much of the error budget should have been utilized by each time point. We refer the reader to the introduction of \textcite{Bartroff2013} for a history of the field.

Generally, sequential hypothesis testing considers the calibration of accumulating evidence for a single, fixed hypothesis. In online hypothesis testing, the goal is to make a determination about whether to accept or reject different hypotheses based on evidence that arrives over time. The goal might be to control a notion of error such as the family-wise error or the false discovery rate (FDR). This paradigm dates as far back as \textcite{Foster2008}, who developed a method for controlling a quantity known as the \textit{marginal false discovery rate} uniformly over all finite stopping times. Subsequently, more powerful methods were developed, and the framework was generalized to accommodate different error rates such as the classical false discovery rate and the FWER \parencite{GSell2015, javanmard2015onlinecontrolfalsediscovery, Tian2021}. We refer the reader to \textcite{Robertson2023} for a survey of recent developments.

The label \textit{adaptive} is somewhat loosely used to describe any method that selects certain hyperparameters of a procedure---especially those that are traditionally fixed \textit{a priori}---in a data-dependent way. For instance, some FDR-controlling methods boost power by estimating the true proportion of nulls from the $p$-values themselves \parencite{Benjamini2000, Storey2002}. An \textit{interactive} testing procedure is a special kind of adaptive procedure, but where the adaptivity is driven by a human analyst who interacts with the data---often with arbitrary flexibility---as opposed to a pre-specified adaptive strategy. We have already surveyed interactive hypothesis testing in Section~\ref{section:related-work} and described its relationship to our work. We re-emphasize that while our work is similar in spirit to existing interactive approaches, it addresses a different problem which requires substantially different technical tools to solve. For instance, the proofs of our main ideas do not rely on properties of martingales, while---with the exception of \textcite{Duan2020FWER}---all of the interactive methods we are aware of rely on some special property of martingales such as the optional stopping theorem or Ville's inequality. \textcite{Duan2020FWER} prove validity of their FWER-controlling procedure by directly characterizing the distribution of false discoveries when the null $p$-values are independent.

Lastly, we consider some works that are conceptually related but which do not fit neatly into the above categories. \textcite{Lai2014} consider the goal of rejecting the null hypothesis of no treatment effect for the full population if possible, and otherwise selecting a subgroup (from a fixed, \textit{a priori} collection of subgroups) with the largest generalized likelihood ratio test statistic and testing the hypothesis of no treatment effect for that subgroup, conditionally on the selection. \textcite{wager2024sequentialvalidationtreatmentheterogeneity} sequentially split the data in order to sequentially construct and combine independent tests for treatment effect heterogeneity, with each subsequent test leveraging more data to learn the test statistic but using an independent fold to test. \textcite{Fraser1951} construct prediction sets of a desired coverage level using a procedure that iteratively and interactively cuts out regions of the sample space in a fashion that is reminiscent of chiseling. Indeed, a few of our intermediate lemmas recapitulate results in this work, though our inferential goal is quite different. The focus of \textcite{Fraser1951} is on characterizing the probability mass of the resulting region while ours is on leveraging the distribution of the samples within it.

\subsection{Parametric and Bayesian approaches}

\textcite{Kovalchik2013} leverage a parametric ``proportional interactions" model which assumes a structural relationship between the treatment response surface and the control response surface. \textcite{Wan2024} construct confidence level sets for linear and generalized linear models by leveraging the known distribution of $\hat{\beta}$, the estimated regression coefficients. Though not precisely a parametric approach, \textcite{Ma2017} suppose that there are $K$ subgroups that partition the data, and that units in the same subgroup have the same subject-specific intercept in a linear model with unobserved latent heterogeneity. They describe an optimization-based approach for recovering the subgroup structure and give conditions on the data-generating process (including $K$ and the gap in subgroup effect between different partitions) under which the latent heterogeneity can be consistently recovered. Naturally, the validity of these approaches hinges on the validity of the model specification.

Bayesian approaches obviate selective inference concerns since the inferences are given conditionally on the data, on average over the prior. As with all Bayesian inference, validity depends upon having a well-specified prior and correctly interpreting the inference as an update to the prior in light of the data. A Bayesian model that has seen popular use in the assessment of treatment effect heterogeneity is the Bayesian additive regression tree, or BART, which also has a number of extensions specifically to the field of causal inference (\cite{Chipman2010, Hill2011, Hahn2020}). BART is a fully Bayesian method that yields a posterior over the space of regression surfaces $\E[Y \mid X]$. Methods that consider Bayesian approaches more specifically for subgroup analysis include \textcite{Berger2014, Schnell2016, Dixon1991}. These methods either tend to be parametric, require the covariate factors to be dichotomous, and/or do not model interactions (or model a few, pre-specified interactions). An exception is \textcite{oganisian2020bayesiannonparametriccosteffectivenessanalyses}, which leverages an Enriched Dirichlet Process prior to model the space of ``cost-effectiveness profiles." This nonparametric prior naturally gives rise to clusters, which they interpret as subgroups. Some works investigate the intersection of Bayesian modeling and decision-theoretic aspects (\cite{Schnell2017}). Finally, some works incorporate Bayesian ideas for testing and model selection, but ultimately evaluate their methodology according to frequentist criteria such as Type I error control (\cite{Sivaganesan2010}). These methods typically do not yield rigorous frequentist inferential guarantees.

\subsection{Conformal inference}

Conformal inference \parencite{vovk2005} provides distribution-free inferential guarantees by leveraging the exchangeability of the training and test samples, and providing guarantees that are unconditional over \textit{both} training and test samples. The literature on conformal inference is now vast. We highlight three specific works that are most related to ours.

In causal inference, \textcite{Lei2021} construct conformalized prediction intervals for the individual treatment effect,
\begin{equation*}
\begin{aligned}
\Prb(Y_{n+1}(1) - Y_{n+1}(0) \in \hat{C}(X_{n+1})) \geq 1 - \alpha
\end{aligned}
\end{equation*}
where $\hat{C}$ is a function of samples $\set{Y_i, T_i, X_i}_{i=1}^n$ drawn from a randomized experiment, and $(Y_{n+1}(1), Y_{n+1}(0), X_{n+1})$ is an additional tuple drawn from the same distribution that the experiment is sampled from. We reiterate that the probability is taken over both $\set{Y_i, T_i, X_i}_{i=1}^n$ and $(Y_{n+1}(1), Y_{n+1}(0), X_{n+1})$. We also observe the inherent hardness of this problem, since the individual treatment effect is unidentifiable without assumptions on the joint distribution of the potential outcomes. \textcite{Medarametla2021} provide conformalized inferences for the conditional median $\mathrm{Median}(Y \mid X = X_{n+1})$. This marks an interesting departure from previous works in conformal inference since the target of inference is a parameter of the distribution (which is never directly observed in the training sample) rather than the outcome itself. \textcite{Jin2023} specifically consider the task of selecting a proportion of units that satisfy a false discovery rate guarantee using ideas from conformal inference.

The guarantee that conformal inference provides is different from what we consider in this work, and in general they are difficult to compare. For one, a valid prediction interval does not directly imply anything about the expected welfare gain if one were to treat according to that prediction interval (for instance, treating all units whose prediction intervals exceed $0$). Moreover, the coverage guarantees in conformal inference are unconditional over the training set, meaning that the realized population miscoverage rate can be far from the unconditional miscoverage rate (which is the nominal level by construction) unless certain algorithmic stability assumptions are satisfied \parencite{liang2024algorithmicstabilityimpliestrainingconditional}.

\FloatBarrier
\section{Simulation details and additional results}
\label{appendix:simulations}

\subsection{Type I error and exactness}
\label{appendix:empirical-type1-error-global-null}

The implementation of chiseling we consider here is the one discussed in Section~\ref{section:setting-alpha}. Let $f(x) = \arctan((X_1 + ... + X_5) / \sqrt{5})$ and in the case of RCTs always consider i.i.d. treatment indicators $W \sim \text{Bern}(0.5)$. We consider the following base simulation settings.
\begin{enumerate}
    \item \textbf{Simulation Setting 1.} $X \in \mathbb{R}^{50}$ is a mean-centered correlated normal with covariance $\Sigma_{ij} = 0.2^{\abs{i - j}}$. The machine learning algorithm used is ridge regression with leave-one-out cross-validation to select the penalty. We let $\epsilon \sim \text{Expo}(1) - 1$.
    \item \textbf{Simulation Setting 2.} $X \in \mathbb{R}^{100}$ consists of independent Rademacher entries. The machine learning algorithm used is a multi-layer perceptron (neural network). We let $\epsilon \sim t_5$, the $t$-distribution with $5$ degrees of freedom.
    \item \textbf{Simulation Setting 3.} $X \in \mathbb{R}^{150}$ consists of independent $\text{Expo}(1) - 1$ entries. The machine learning algorithm used is random forests. We let $\epsilon \mid X \sim \mathcal{N}(0, 1 + f(X)^2)$ (i.e. the errors are heteroscedastic).
\end{enumerate}
See Appendix~\ref{appendix:ml-methods} for more details about the machine learning algorithms. Within each base setting, we consider three sub-settings. First, we consider the \textbf{binary} case where $Y \sim \text{Bern}(0.5)$ independently of $X$ (and $\cutoff = 0.5$). Next, we let $\cutoff = 0$ and $Y'(1) = Y'(0) = f(X) + \epsilon$. We will consider both the \textbf{IPW transformation} and the \textbf{AIPW transformation}. The AIPW transformation is fit using $5$-fold cross-fitting and linear regression on the first five coordinates, which is in particular misspecified.

\begin{figure}[ht]
  \centering
  \includegraphics[width=\linewidth]{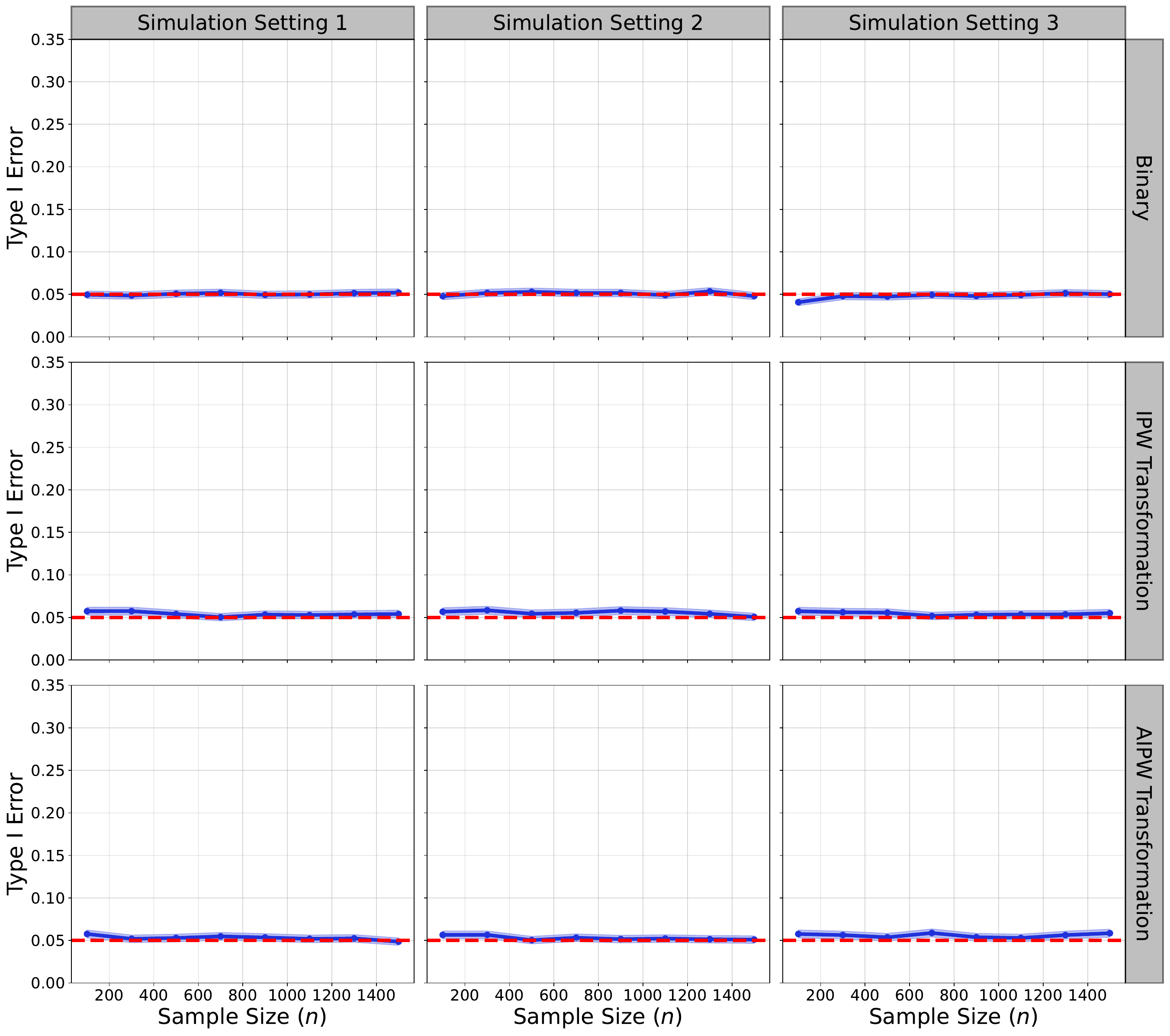}
  \caption{Type I error of chiseling under the global null across different data generating processes and machine learning methods (target level is $\alpha = 0.05$, given by the dashed red line). Error band shown reflects $\pm$ twice the standard error. A minimum sample size of $n_{\min} = 30$, not scaling with $n$, is enforced in order to showcase the robustness of chiseling to small $n_{\min}$.}
  \label{fig:null-dgps-type1-error}
\end{figure}

In the first row of Figure~\ref{fig:null-dgps-type1-error}, we empirically confirm the theoretical exactness of chiseling when $Y$ is binary (using the exact test for binary outcomes designed in Section~\ref{section:general-test-design}). Next, we run the asymptotic variant of chiseling while only enforcing a minimum sample size of $n_{\min} = 30$ in order to assess the robustness of our asymptotic approximation. We find that even with such a small minimum sample size, our asymptotic approximation is of high quality and the Type I error is essentially exact at level $\alpha$ (second and third rows of Figure~\ref{fig:null-dgps-type1-error}). Note, importantly, that the asymptotic regime is determined not by the sample size $n$ but by the minimum sample size $n_{\min}$. Because in these experiments we are not scaling $n_{\min}$ with $n$, we are not approaching an asymptotic regime as $n$ grows. Nonetheless, the Type I error is robust, and the asymptotics can only improve as $n_{\min}$ is increased (for instance, requiring it to be $5\%$ of $n$).

\FloatBarrier

\subsection{Simulation details}
\label{appendix:simulation-details}

\subsubsection{\texorpdfstring{$p$}{p}-simultaneous data splitting}
\label{appendix:simultaneous-data-splitting}

The dataset is split into a training set of size $\floor{pn}$, with the rest set aside for testing. Using the training dataset, a machine learning model is used to estimate $\hat{\mean}(\cdot)$ and $\reg_1 := \set{x : \hat{\mean}(x) > \cutoff}$. We then define $\reg_2,...,\reg_{10}$ by thresholding $\hat{\mean}(\cdot)$ at different cutoffs, where the cutoffs are chosen so that $\reg_{10}$ is the smallest region with at least $n_{\min}$ samples and the number of samples in $\reg_j \setminus \reg_{j-1}$ for $j=2,...,10$ is as even as possible. Though we define these regions using the test data, we will treat them as fixed. This approximation is reasonable since they are only defined using $X$.

When the outcomes are continuous, we let $\hat{\theta}_j$ denote estimates of $\theta_j := \mean(\reg_j)$ obtained by taking the simple mean of points in the region. We then calculate the standard error $\hat{\sigma}_j$ for each $\hat{\theta}_j$ by bootstrapping the test sample. We form the statistics
\begin{equation*}
\begin{aligned}
T := \max_j \set*{ \frac{\hat{\theta}_j - \theta_j}{\hat{\sigma}_j} }
\end{aligned}
\end{equation*}
and in particular $B$ bootstrapped versions of the above $T^{(1)},...,T^{(B)}$. The lower simultaneous confidence bounds are then given by $\hat{\theta}_j - \hat{\sigma}_j q$ for $j=1,...,10$ where $q$ is the $1 - \alpha$ quantile of $\set{T^{(1)},...,T^{(B)}}$. We report $\reg_j$ for the smallest $j$ such that $\hat{\theta}_j - \hat{\sigma}_j q > \cutoff$. The validity of this construction follows from standard theory regarding the validity of the bootstrap for $Z$-estimators (e.g. Chapter 10 of \textcite{Kosorok2008}) once we recognize that $(\hat{\theta}_1,...,\hat{\theta}_{10})$ is the solution to a set of $Z$-estimating equations.

When the outcomes are binary, we are not aware of any obvious method that provides simultaneous testing for the means of nested regions that is valid in finite samples. We find that the asymptotic approximation can fail in small samples, so instead we consider the following approach. We form the natural statistic
\begin{equation*}
\begin{aligned}
S_j := \sqrt{n(\reg_j)} \cdot \frac{\hat{\theta}_j - \cutoff}{\sqrt{\hat{\theta}_j (1 - \hat{\theta}_j)}}
\end{aligned}
\end{equation*}
where $n(\reg_j)$ is the number of test samples in $\reg_j$. If $\hat{\theta}_j = 1$ we let $S_j = \infty$, and if the right-hand side is otherwise undefined we let it equal $0$. This is the $Z$-statistic for testing $H_j : \theta_j \leq \cutoff$ where we have imputed some edge cases. We will reject $H_j$ for all $S_j > c$. How should we set $c$? We condition on $X$ (which in particular fixes the regions) and desire
\begin{equation*}
\begin{aligned}
\Prb_{\mathcal{P}}(\exists j : S_j > c \text{ and } \mathcal{P} \in H_j \mid X) \leq \alpha
\end{aligned}
\end{equation*}
for all distributions $\mathcal{P}$. Let $c(\mathcal{P})$ be the smallest $c$ satisfying the above for distribution $\mathcal{P}$. Ideally, we would take $c := \sup_{\mathcal{P}} c(\mathcal{P})$, which would ensure the finite-sample validity of our test. As an \emph{anti-conservative} approximation, we let $\tilde{c} := c(\mathcal{P}_0)$ where $\mathcal{P}_0$ is the distribution where $Y_1,...,Y_n$ are i.i.d. $\text{Bern}(\cutoff)$ and independent of $X$. Clearly, $\tilde{c} \leq c$, and so this approximation rejects strictly more often than the finite-sample valid test. Furthermore, we can easily calculate $\tilde{c}$ by simulations. This lets us \emph{upper bound} the expected utility and power of the finite-sample variant of $p$-simultaneous data splitting, and this is what we show throughout our simulations. Though it is an upper bound, we actually expect it to be quite close to the power of the finite-sample valid variant, as $\mathcal{P}_0$ is specifically chosen with the heuristic that it is at least close to maximizing $c(\mathcal{P})$ (it is at the boundary of the global null). To the extent that it fails to maximize $c(\mathcal{P})$, it creates an inflation in expected utility and power. Lastly, we note that in this variant, it can be quite important to randomize the quantile since the test statistics are highly discrete. Thus, we in fact let $\tilde{c}$ be the randomized quantile such that $\Prb_{\mathcal{P}_0}(\exists j : S_j > \tilde{c} \mid X) = \alpha$ exactly.

\subsubsection{Machine learning methods}
\label{appendix:ml-methods}

We use off-the-shelf implementations from scikit-learn of $\ell^2$-penalized logistic regression, lasso with $5$-fold cross-validation, ridge regression with leave-one-out cross-validation, multi-layer perception regression (i.e. neural network), and random forest regression and classification. The only modifications we make are that for the linear model-based methods we standardize $X$ (and $Y$, where appropriate) prior to fitting, and that for ridge regression we increase the grid of penalty parameters to $20$ evenly spaced values between $10^{-5}$ and $10^5$ in $\log$-space. Unless stated otherwise, when the IPW transformation is used to convert a causal problem into a regression problem, the resulting pseudo-outcomes are directly regressed onto the covariates. For computational reasons, when chiseling we generally do not refit the machine learning model each time a new data point is revealed, but only in batches of $1\%$, $5\%$ or $10\%$ depending on the computational cost of the simulation setting.

\subsubsection{Data generating processes}

Throughout, when the simulation is an RCT we let $Y'(1) = \mean(X) + Y'(0)$ where we will specify $\mean(\cdot)$ and $Y'(0)$. The treatment indicator will always be i.i.d. $W \sim \text{Bern}(0.5)$, and we always work with the regression formulation with IPW outcomes, i.e. $Y = 2 W Y'(1) - 2 (1 - W) Y'(0)$. Machine learning methods are directly regressed on the IPW outcomes.

\hfill

\noindent \textbf{DGPs for Section~\ref{section:simple-demo}.} For the linear case, we let $X \sim \mathcal{N}(0, I_d)$, $\mean(X) = X^T \beta$, and $Y(0) \sim \mathcal{N}(0,1)$. We let $d = 20$ and $\beta \propto (1,...,1)$ with $\norm{\beta} = 0.5$. For the non-linear case, we let $X$ and $\mean(X)$ follow the data generating process of \textcite{Kang2007}, to which we defer for additional details. The noise term in \textcite{Kang2007} is $\mathcal{N}(0,1)$, but we inflate the noise so that $Y(0) \sim \mathcal{N}(0, 80^2)$. Furthermore, we set the intercept so that $\E[Y(1)] = 0$. Note that the \textcite{Kang2007} example also has non-constant propensities, which we ignore since we are imitating an RCT.

\hfill

\noindent \textbf{DGP for Section~\ref{section:simulation-binary-regression}.} We let $X \sim \mathcal{N}(0, \Sigma)$ where $\Sigma \in \mathbb{R}^{d \times d}$ and $\Sigma_{ij} = 0.2^{\abs{i - j}}$. We let $\Prb(Y = 1 \mid X) = \mean(X) = \text{logit}(\tau + X^T \beta)$. We let $d = 100$, $\beta \propto (1,1,1,1,1,0,...,0)$. Letting $\norm{\beta} = \theta$, we consider $(\theta, \tau) \in \set{(2,2.2), (1.3, 0.243), (1.5, -1.936)}$ so that the probability masses of the optimal regions are $0.5$, $0.1$, and $0.01$ respectively.

\hfill

\noindent \textbf{DGP for Section~\ref{section:simulation-heterogeneous-rct}.} We let $X \sim \mathcal{N}(0, \Sigma)$ where $\Sigma \in \mathbb{R}^{d \times d}$ and $\Sigma_{ij} = 0.2^{\abs{i - j}}$. We let $\mean(X) = \tau + X^T \beta$ and $Y(0) \sim \text{Expo}(1) - 1$. We let $d = 100$, $\beta \propto (1,1,1,1,1,0,...,0)$. Letting $\norm{\beta} = \theta$, we consider $(\theta, \tau) \in \set{(0.45,0), (0.4, -0.601), (0.8, -2.201)}$ so that the probability masses of the optimal regions are $0.5$, $0.1$, and $0.01$ respectively.

\hfill

\noindent \textbf{DGP for Section~\ref{section:simulation-rct-misspecified}.} We let $\cutoff = 0$ and again consider the DGP from \textcite{Kang2007}. We now let $Y(0) \sim \mathcal{N}(0, 125^2)$ and vary $\tau \in \set{24.45, 0, -24.35}$ so that the probability masses of the optimal subgroups are $0.75, 0.5$, and $0.25$ respectively. Here, the CATE is nonlinear in $X$ but we use a linear model to learn the subgroups.

\hfill

\noindent \textbf{DGP for Section~\ref{section:simulation-nonneg-rct}.} First we draw $Z \sim \text{Bern}(q)$. Then we draw $L \in \mathbb{R}^5$ from the distribution of $A \mid A_1 + ... + A_5 = 0$ where $A \sim \mathcal{N}(0, I_5)$. Then we define $X_j = \tau Z + \tau_0 (1 - Z) + L_j$ for $j=1,...,5$ and let $\mean(X) = (X_1 + ... + X_5) / 5 = \tau Z + \tau_0 (1 - Z)$. In the first simulation setting, we let $\tau_0 = 0$. In the second, we let $\tau_0 = -0.1 \tau$. We let $\tau = 0.1 / q$. The sample size is $n = 1000$.

\subsection{Additional results}
\label{appendix:simulation-additional-results}

\FloatBarrier

\begin{figure}[ht]
  \centering
  \includegraphics[width=\linewidth]{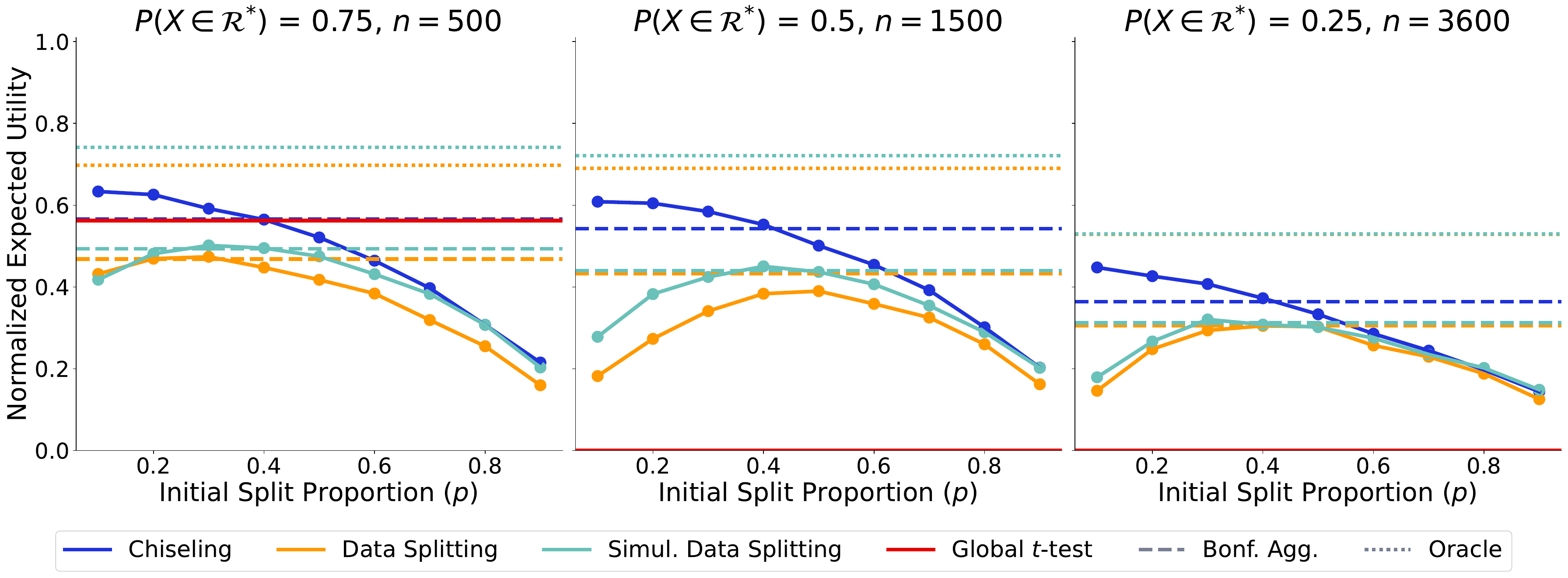}
  \caption{Results for Section~\ref{section:simulation-rct-misspecified}. Normalized expected utility in a simulated RCT where the CATE is non-linear. Solid horizontal lines correspond to using Bonferroni to aggregate across initial split proportions $\in \set{0.2, 0.5, 0.8}$ for the method with the corresponding color. The machine learning method used is ridge regression with leave-one-out cross-validation, which is misspecified.}
  \label{fig:kang-schafer-utility}
\end{figure}

\begin{figure}[ht]
  \centering
  \includegraphics[width=\linewidth]{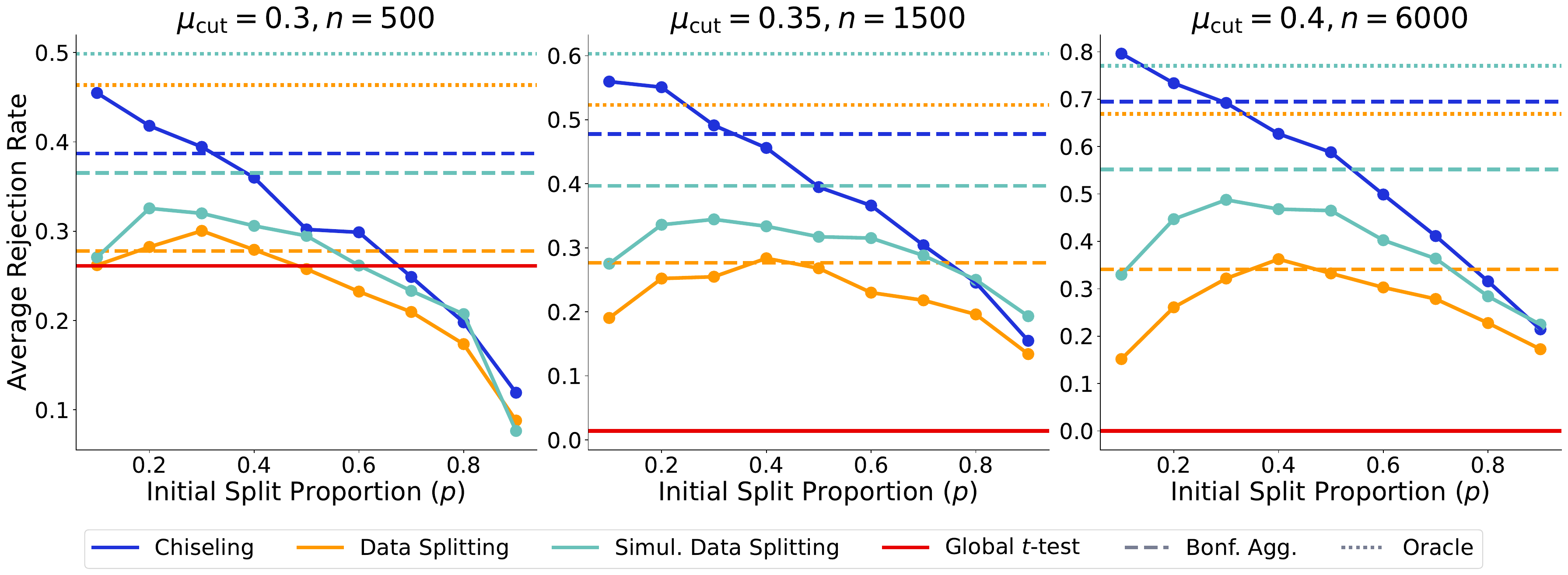}
  \caption{Results for Section~\ref{section:real-data}. The analogue of Figure~\ref{fig:bart-utility} but for the average rejection rate across repeated train/validation splits.}
  \label{fig:bart-power}
\end{figure}

\begin{figure}[ht]
  \centering
  \includegraphics[width=\linewidth]{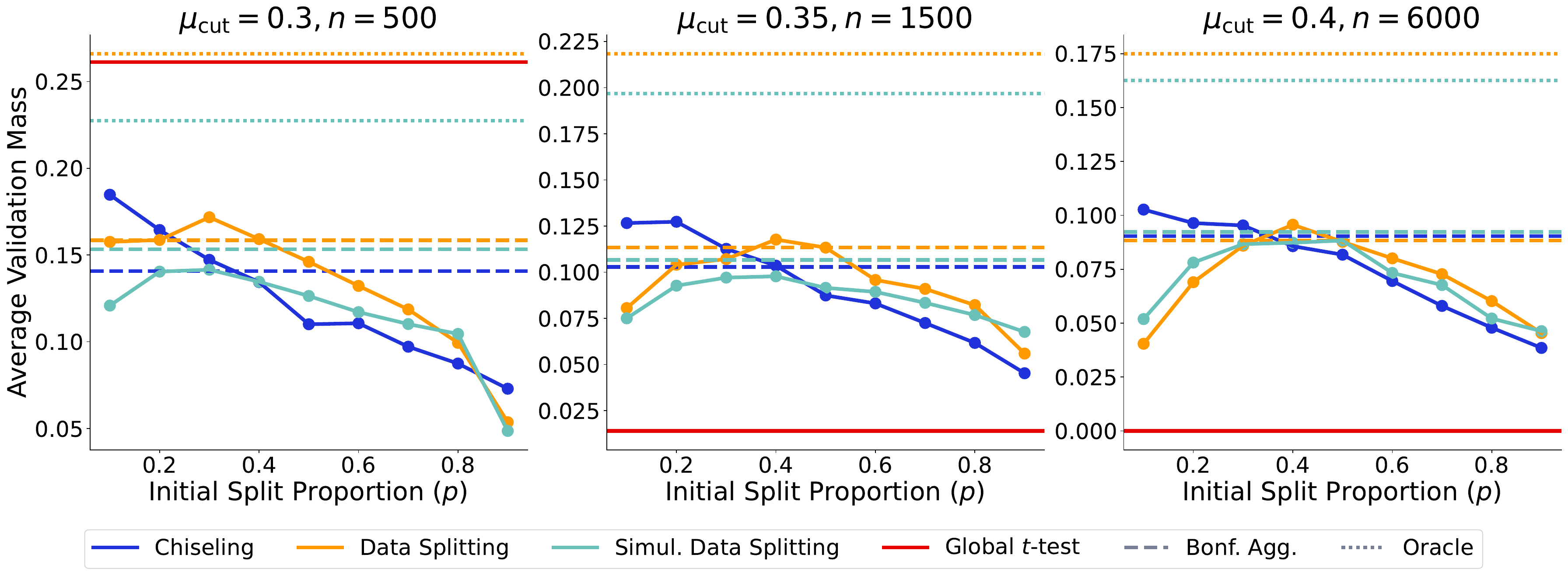}
  \caption{Results for Section~\ref{section:real-data}. The analogue of Figure~\ref{fig:bart-utility} but for the average probability mass of the rejected region (estimated using the validation set) across repeated train/validation splits.}
  \label{fig:bart-size}
\end{figure}

\begin{table}[htbp]
\centering
\caption[GSS feature ranks]{\textbf{Top 10 features determining chiseled subgroup membership in GSS dataset.}
``Rate" refers to the proportion of units who would be included in the subgroup based on that feature alone, and features are ranked by rate. Some features are redundant because they represent different codings of the same survey item.}
\begin{tabular}{@{}r l p{3.4cm} p{6.2cm} r@{}}
\toprule
\textbf{Rank} & \textbf{Feature ID} & \textbf{Feature Description} & \textbf{Inclusion Criteria} & \textbf{Rate} \\
\midrule
1 & partyid & Political party affiliation & Neither, no response, independent, close to Republican, not very strong Republican, strong Republican, or other party & 53.2\% \\ \cmidrule{1-5}
2 & year & GSS year for respondent & 1998 or earlier & 53.5\% \\ \cmidrule{1-5}
3 & polviews & Political ideology & Moderate, slightly conservative, conservative, or extremely conservative & 68.3\% \\ \cmidrule{1-5}
4 & income & Total family income & $\geq\!\$15{,}000$ & 68.9\% \\ \cmidrule{1-5}
5 & income\_num & Total family income & $\geq\!\$15{,}000$ & 68.9\% \\ \cmidrule{1-5}
6 & racdif3 & Differences due to education & Did not answer ``yes" to ``On the average African-Americans have worse jobs, income, and housing than white people. Do you think these differences are because most African-Americans don't have the chance for education that it takes to rise out of poverty?" & 72.6\% \\ \cmidrule{1-5}
7 & marital & Marital status & Married, widowed, or divorced & 73.7\% \\ \cmidrule{1-5}
8  & polviews\_num & Political ideology & Moderate, slightly conservative, conservative, or extremely conservative & 76.1\% \\ \cmidrule{1-5}
9  & racdif1 & Differences due to discrimination & Did not answer ``yes" to ``On the average African-Americans have worse jobs, income, and housing than white people. Do you think these differences are mainly due to discrimination?" & 78.1\% \\ \cmidrule{1-5}
10 & race & Race of respondent & White & 79.1\% \\
\bottomrule
\end{tabular}
\label{table:bart-feature-ranks}
\end{table}

\begin{figure}[ht]
  \centering
  \includegraphics[width=\linewidth]{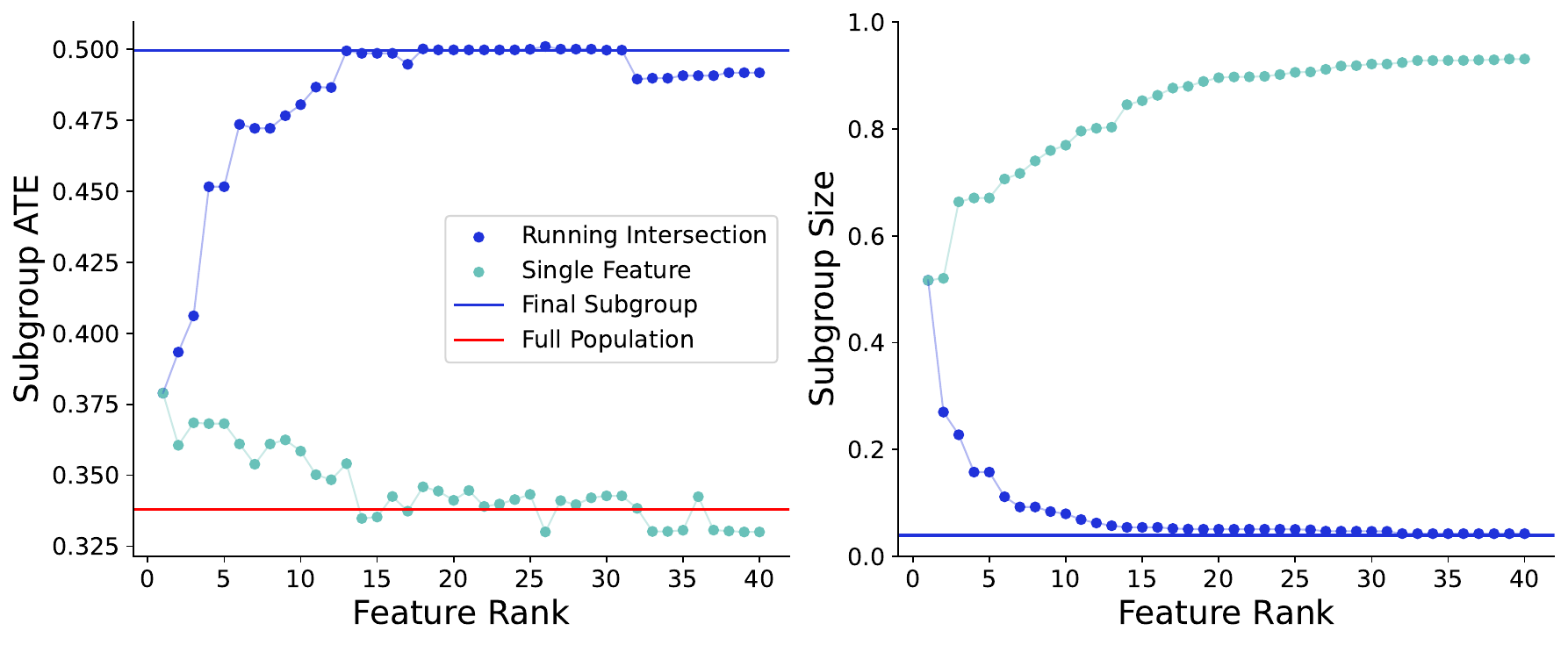}
  \caption{The rejected region has the form $[l_1, u_1] \times ... \times [l_d, u_d]$, and feature $j$'s importance is given by the fraction of samples falling in $[l_j, u_j]$. \textbf{Left panel:} the estimated subgroup ATE within subgroups defined by single feature inclusion criteria, as well as the estimated subgroup ATE within the subgroup defined by the running intersection of inclusion criteria (both in decreasing order of feature importance) for the first $40$ features. Intersecting all $d = 152$ inclusion criteria yields the subgroup discovered by chiseling. We also show the full population ATE for comparison. \textbf{Right panel:} the same as the left panel but where subgroup ATE is replaced with subgroup probability mass (i.e. the proportion of units belonging to the subgroup).}
  \label{fig:bart-interpret}
\end{figure}

\FloatBarrier

\subsection{Multiple testing chiseling versus single test chiseling}
\label{appendix:simulation-multiple-vs-single}

\FloatBarrier

\begin{figure}[ht]
  \centering
  \includegraphics[width=\linewidth]{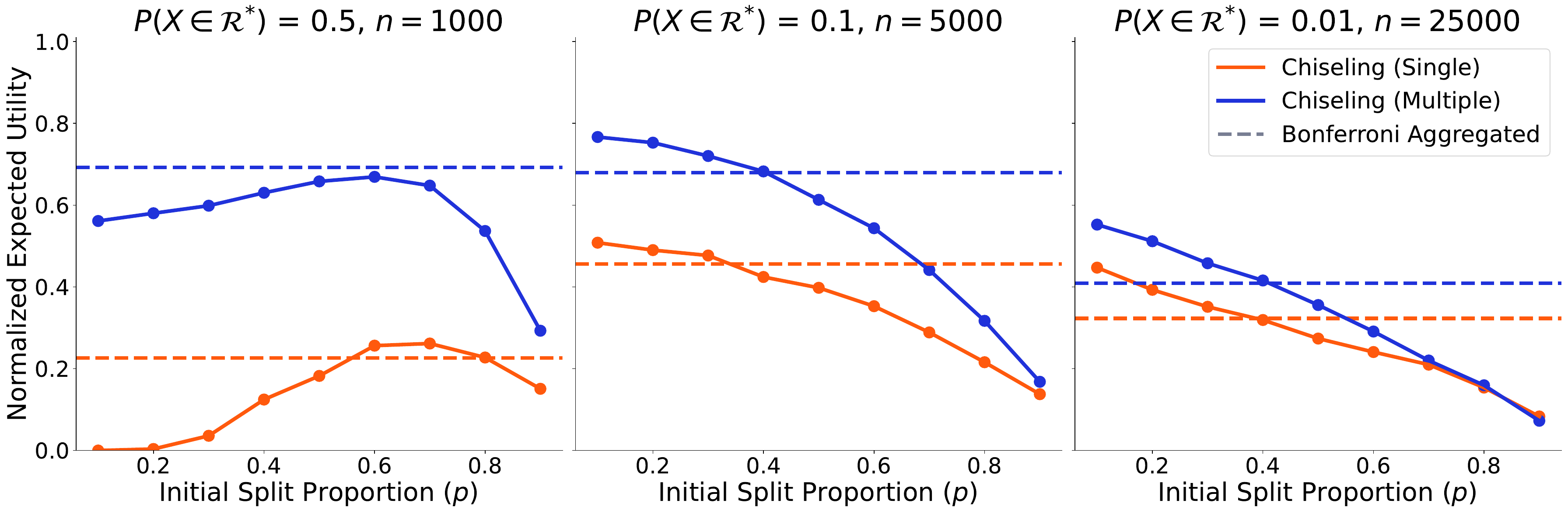}
  \caption{A comparison of chiseling with multiple testing (Section~\ref{section:setting-alpha}) versus chiseling with a single test (Section~\ref{section:simple-demo}) in the simulation setting of Section~\ref{section:simulation-binary-regression}.}
\end{figure}

\begin{figure}[ht]
  \centering
  \includegraphics[width=\linewidth]{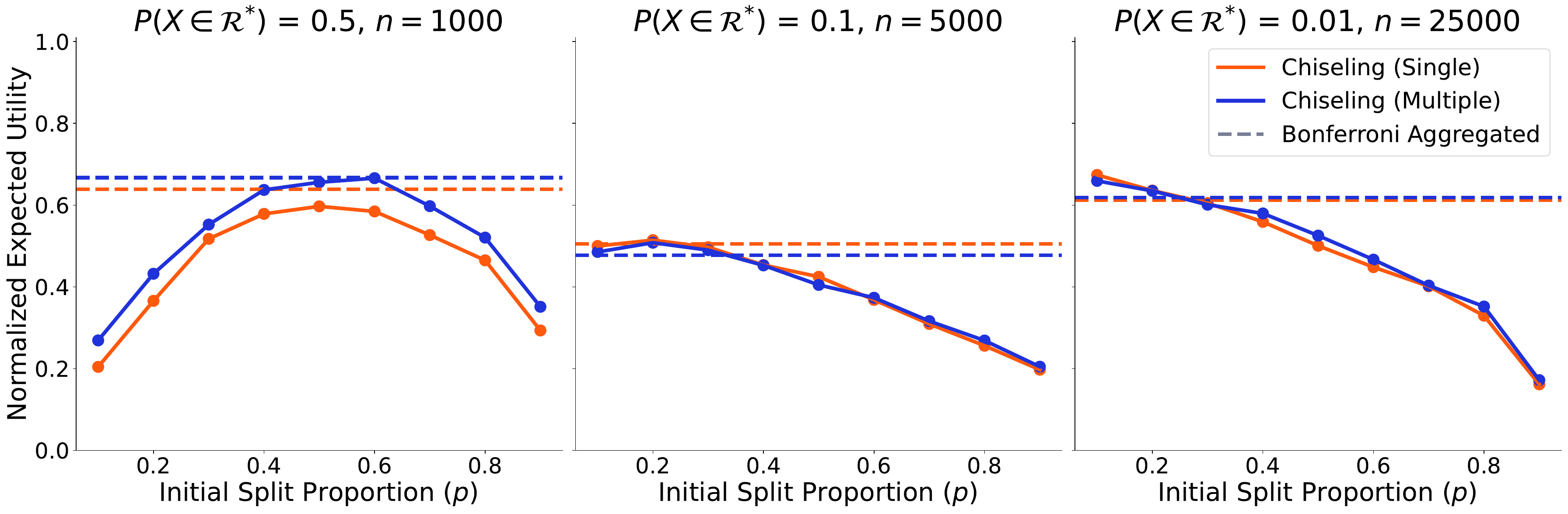}
  \caption{A comparison of chiseling with multiple testing (Section~\ref{section:setting-alpha}) versus chiseling with a single test (Section~\ref{section:simple-demo}) in the simulation setting of Section~\ref{section:simulation-heterogeneous-rct}.}
\end{figure}

\begin{figure}[ht]
  \centering
  \includegraphics[width=\linewidth]{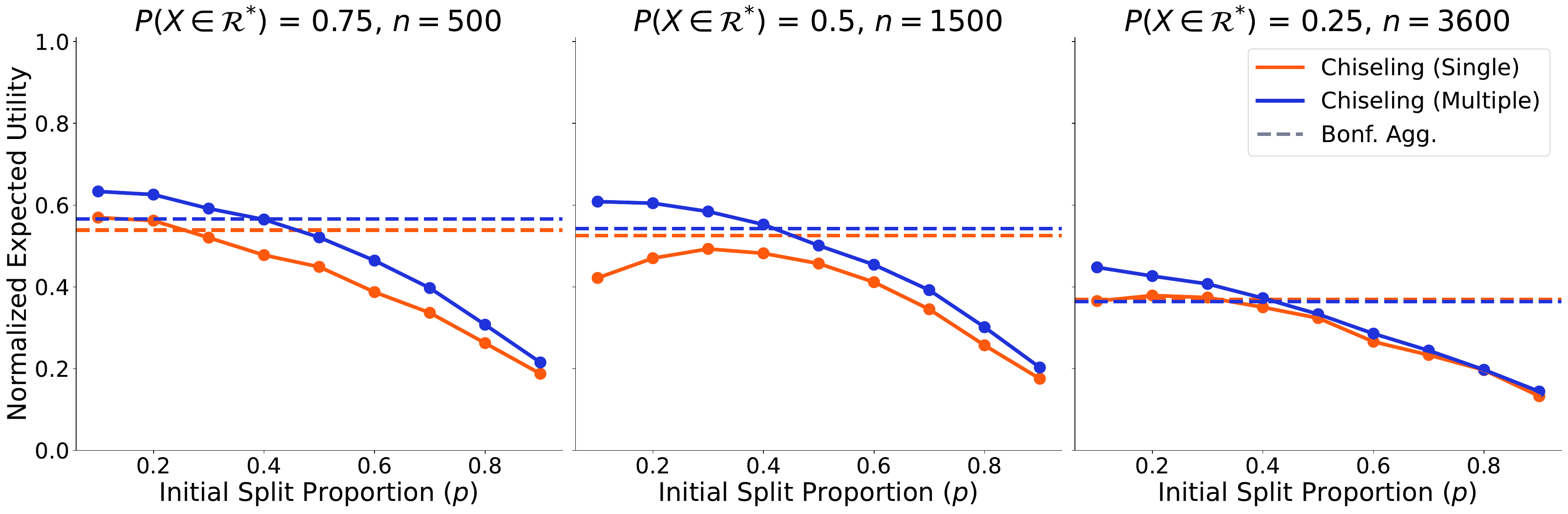}
  \caption{A comparison of chiseling with multiple testing (Section~\ref{section:setting-alpha}) versus chiseling with a single test (Section~\ref{section:simple-demo}) in the simulation setting of Section~\ref{section:simulation-rct-misspecified}.}
\end{figure}

\FloatBarrier

\subsection{Power}
\label{appendix:simulation-power}

\FloatBarrier

\begin{figure}[ht]
  \centering
  \includegraphics[width=\linewidth]{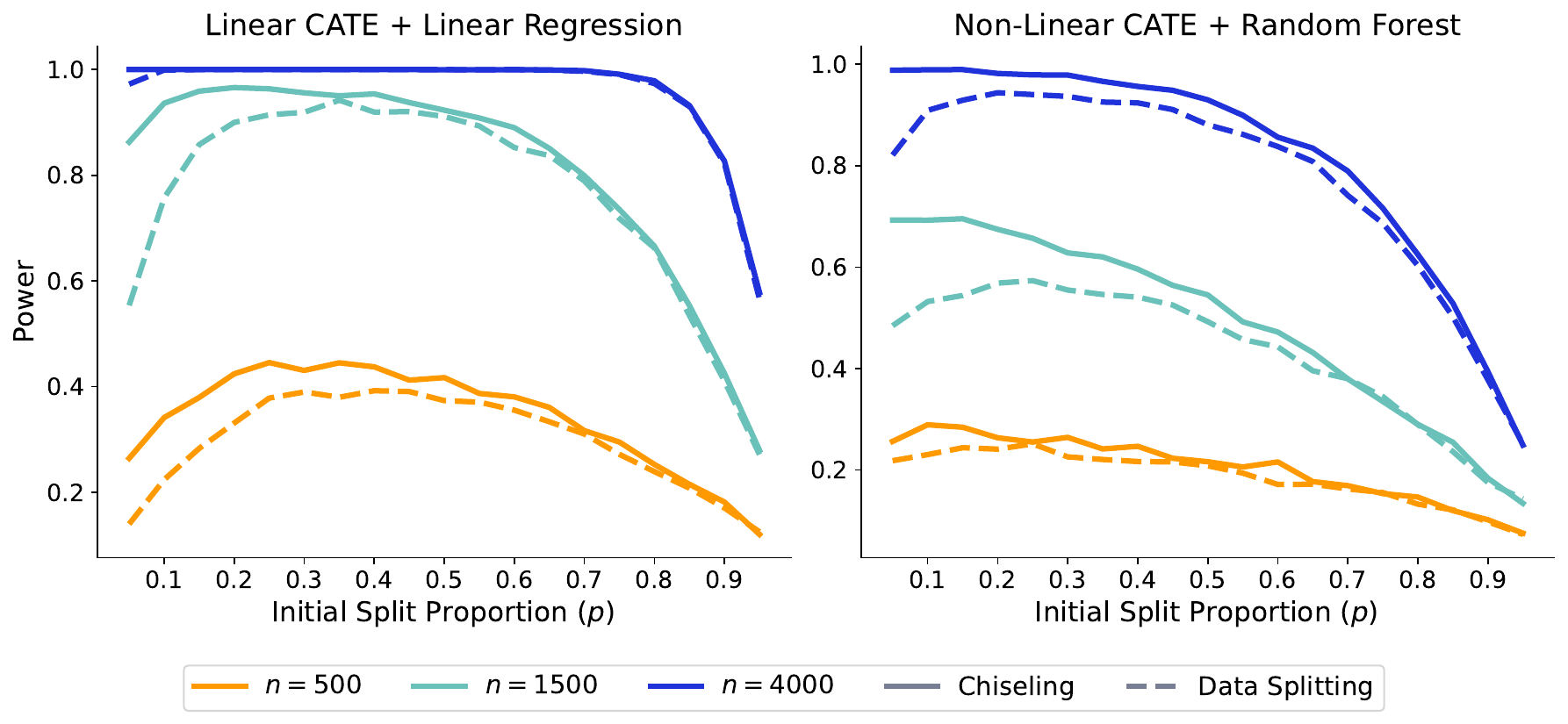}
  \caption{Results for Section~\ref{section:simple-demo}. The analogue of Figure~\ref{fig:simple-demo-utility} for power.}
  \label{fig:simple-demo-power}
\end{figure}

\begin{figure}[ht]
  \centering
  \includegraphics[width=\linewidth]{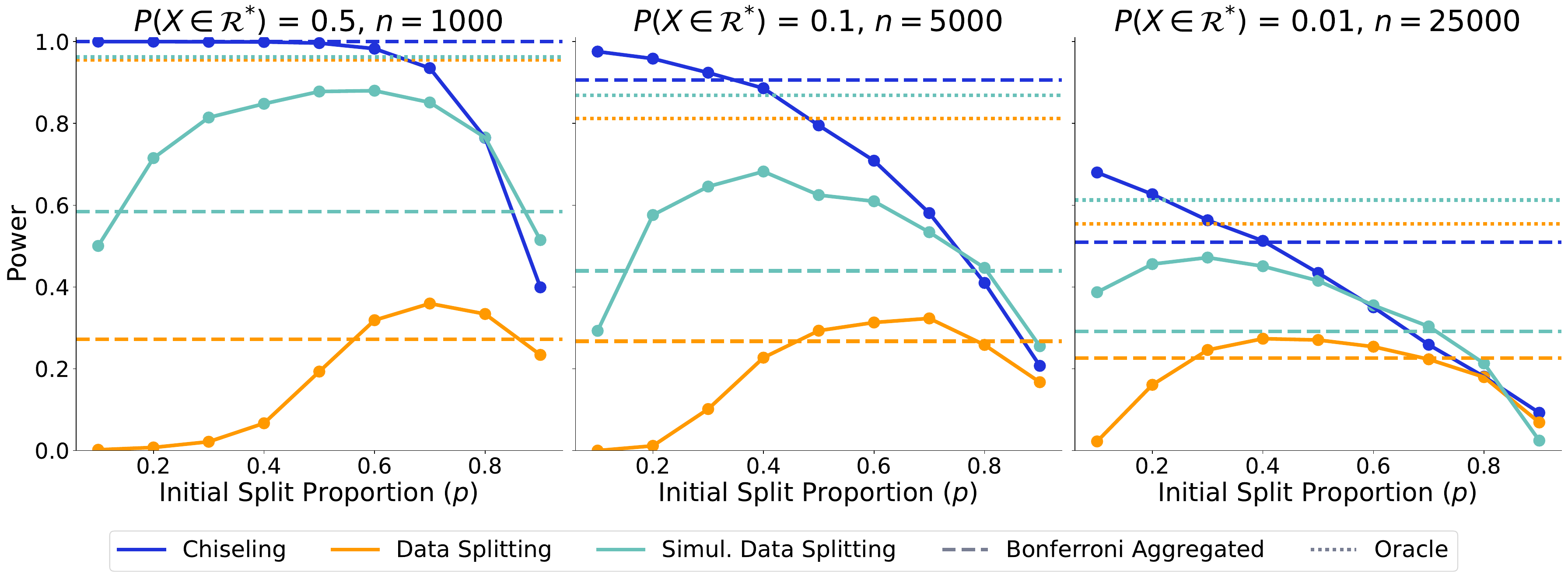}
  \caption{Results for Section~\ref{section:simulation-binary-regression}. The analogue of Figure~\ref{fig:binary-regression-utility} for power.}
  \label{fig:binary-regression-power}
\end{figure}

\begin{figure}[ht]
  \centering
  \includegraphics[width=\linewidth]{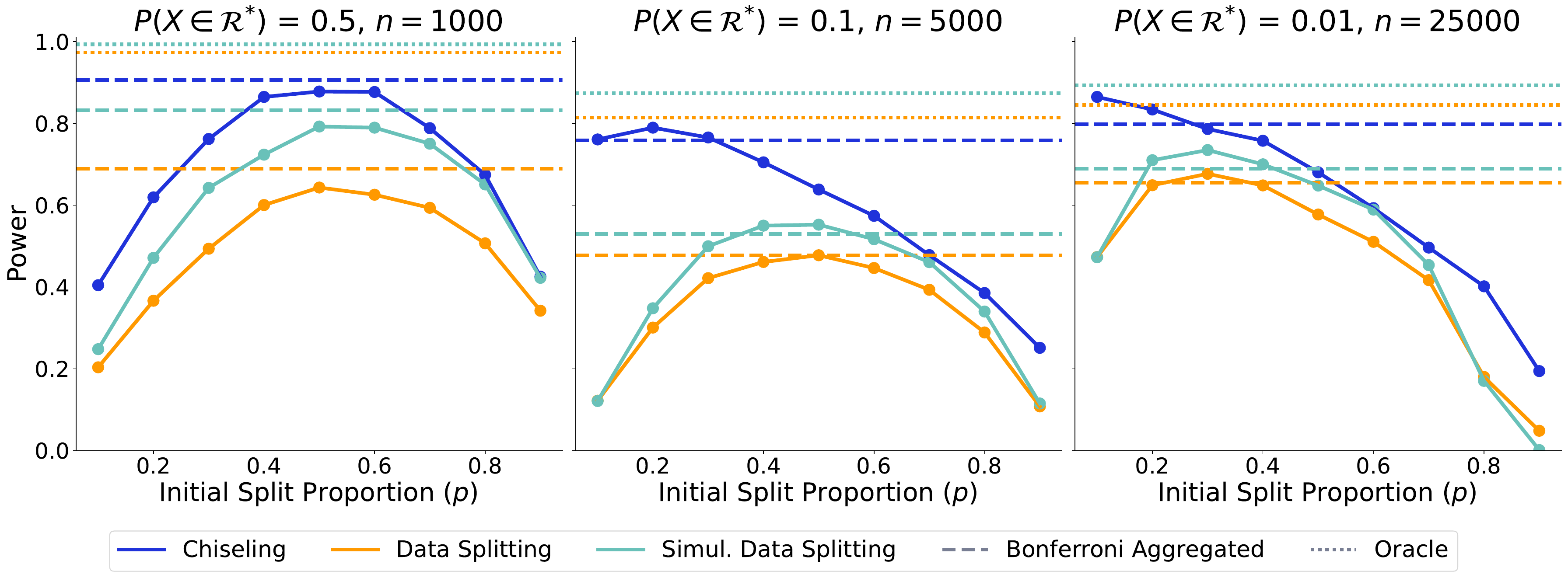}
  \caption{Results for Section~\ref{section:simulation-heterogeneous-rct}. The analogue of Figure~\ref{fig:heterogeneous-linear-rct-utility} for power.}
  \label{fig:heterogeneous-linear-rct-power}
\end{figure}

\begin{figure}[ht]
  \centering
  \includegraphics[width=\linewidth]{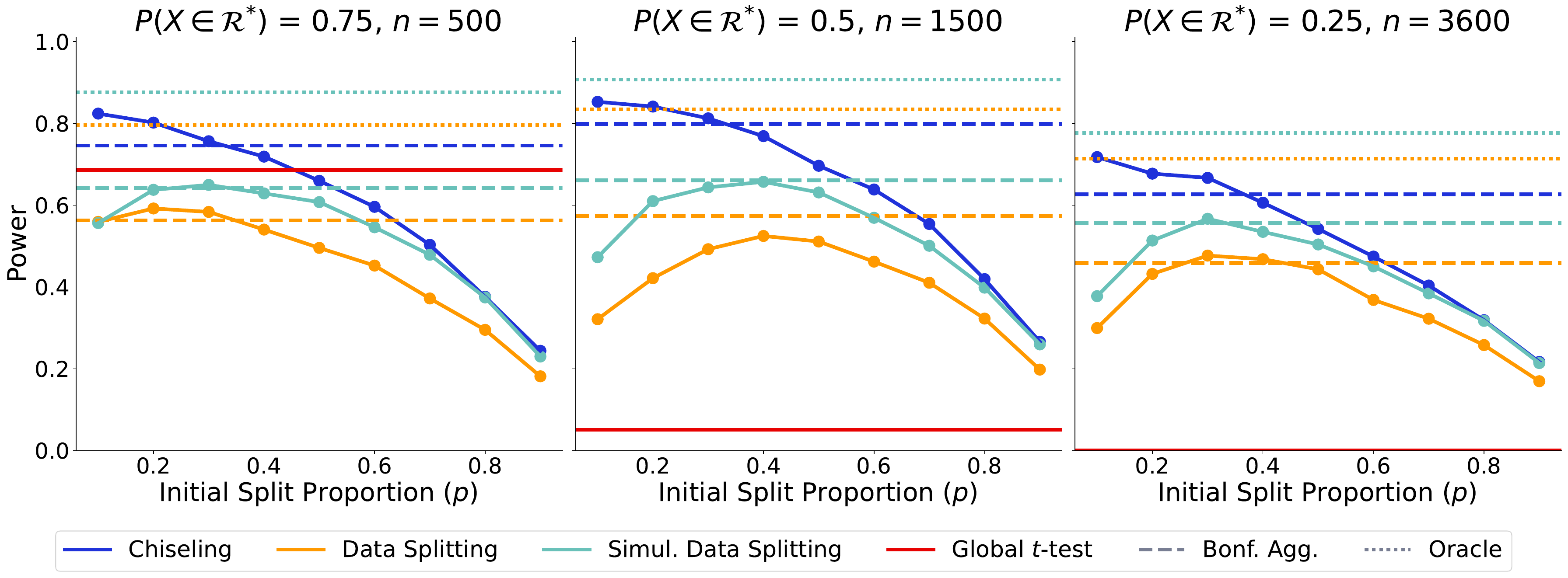}
  \caption{Results for Section~\ref{section:simulation-rct-misspecified}. The analogue of Figure~\ref{fig:kang-schafer-utility} for power.}
  \label{fig:kang-schafer-power}
\end{figure}

\begin{figure}[ht]
  \centering
  \includegraphics[width=\linewidth]{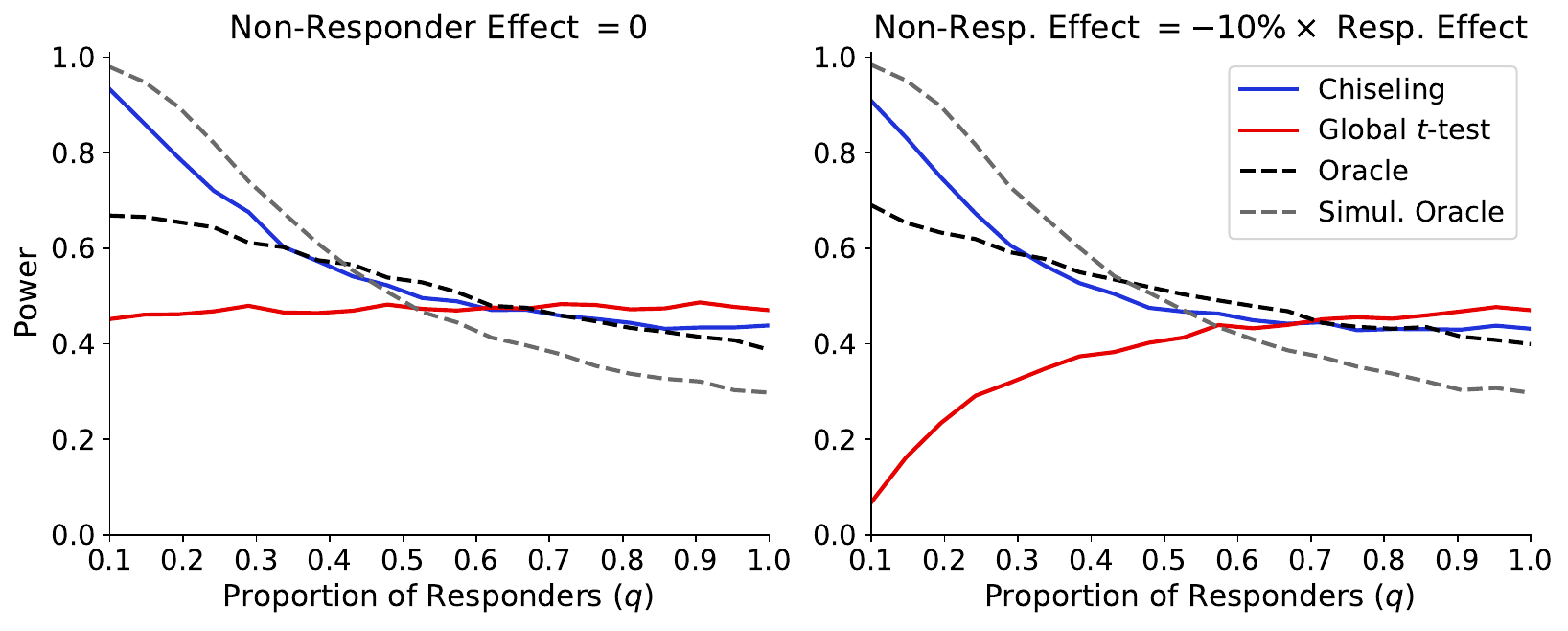}
  \caption{Results for Section~\ref{section:simulation-nonneg-rct}. The analogue of Figure~\ref{fig:linear-nonneg-rct-utility} for power.}
  \label{fig:linear-nonneg-rct-power}
\end{figure}

\FloatBarrier

\FloatBarrier

\section{Extended discussion}

\subsection{Relationship between utility and policy value}
\label{appendix:policy-max-equivalence}

In the policy learning literature, it is common to let $\cutoff = 0$ (to protect against discovering non-efficacious treatment policies) and to desire policies with high \emph{policy value}. This is equivalent to Equation~\eqref{eq:policy-utility} since
\begin{equation*}
\begin{aligned}
\E[Y\indic\set{X \in \reg}] &= \E[(Y'(1) - Y'(0)) \indic\set{X \in \reg}]\\
&= \E[Y'(1) \indic\set{X \in \reg} + Y'(0) \indic\set{X \not\in \reg}] - \E[Y'(0)]\\
&= \E[Y'(\pi(X))] - \E[Y'(0)]
\end{aligned}
\end{equation*}
where $\pi(X) = \indic\set{X \in \reg}$ is the policy that treats everyone in $\reg$. Thus, Equation~\eqref{eq:policy-utility} is equivalent to the policy value $\E[Y'(\pi(X))]$ up to the constant factor $\E[Y'(0)]$. Thus, one maximizer of the policy value is the policy that treats units whose CATEs are positive, i.e., the units in $\reg^*$.

\subsection{Querying regions}

For $t \geq 1$, let $\score_t(\cdot)$ and $\threshlim_t$ denote what is chosen by the $t$th application of Algorithm~\ref{alg:chiseling} used to generate $(\reg_t)_{t=0}^{\infty}$. Let $\pthresh_t := \threshlim_t \wedge \min_{i : X_i \in \reg_{t-1}} \score_t(X_i)$. For any $x \in \mathcal{X}$, one may check whether $x \in \reg_t$ by simply checking whether $\score_s(x) > \pthresh_s$ for all $0 < s \leq t$. If the scoring functions $\score_t(\cdot)$ are computationally feasible to evaluate, then it is feasible to check whether any $x$ belongs to $\reg_t$ or not. In the context of causal inference, this makes the treatment rule ``treat a person with covariates $x$ if $x \in \reg_t$" one that is computationally feasible to use in practice.

\subsection{Total order}
\label{appendix:total-order}

In Algorithm~\ref{alg:chiseling}, we have chosen to parameterize the shrinking of regions by the upper level sets of a real-valued function as this is a simple, intuitive, and useful choice. However, it is not the most general parameterization. In general, $\score(\cdot)$ may take values in any set that is given a total order---a binary relation that is reflexive, transitive, antisymmetric, and connected. For example, the analyst may choose $\score : \mathcal{X} \to \mathbb{R}^2$ and give $\mathbb{R}^2$ the lexicographic order. Then in Algorithm~\ref{alg:chiseling} we may replace minima with lexicographic minima and comparisons such as $>$ with lexicographic comparisons; Theorem~\ref{theorem:untarnished-chiseling} and Corollary~\ref{corollary:distribution-subsamples} will still hold.

\subsection{Chiseling generalizes data splitting}
\label{appendix:chiseling-generalizes-ds}

Suppose without loss of generality that the covariates have the form $(X, U) \in \mathcal{X} \times [0,1]$ where $U$ is an independent standard uniform random variable (appending an extra feature if necessary). Then for any region of the form $\reg = \reg' \times I$ for $\reg' \subseteq \mathcal{X}$ and $I \subseteq [0,1]$, we have $\Ec{Y}{(X, U) \in \reg} = \Ec{Y}{X \in \reg'}$. Thus, guarantees on the former translate to guarantees on the latter. Now as long as either
\begin{equation*}
\begin{aligned}
\score(x, u) = f'(x) \text{ for some } f' \quad \text{ or } \quad \score(x, u) = u
\end{aligned}
\end{equation*}
in each application of Algorithm~\ref{alg:chiseling}, we will have that all $\reg_t$ are of the form $\reg_t' \times (u,1]$ for some $\reg_t' \subseteq \mathcal{X}$. The former choice is used to shrink the region within the covariate space of interest $\mathcal{X}$, while the latter is used to reveal a point uniformly at random from the current region. Splitting off $k$ random data points is thus equivalent to chiseling with $\score(x, u) = u$ successively $k$ times. Moreover, this choice may be used to reveal points randomly from the current region at \emph{any} point during the procedure.

We note that here since $\reg_t'$ are the regions of interest, though the ambient space is $\mathcal{X} \times [0,1]$ we should redefine utility as $\util(\reg') = \E[(Y - \cutoff)\indic\set{X \in \reg'}]$ for $\reg' \subseteq \mathcal{X}$ (i.e. we do not marginalize over $U$).

\subsection{Tiebreaking}
\label{appendix:tiebreaking}

If multiple data points in the region have the same $\score(\cdot)$ value, then they will all be revealed by Algorithm~\ref{alg:chiseling}. If the analyst wishes for each application of Algorithm~\ref{alg:chiseling} to reveal no more than one data point, then this can be accomplished by randomly breaking ties in $\score(\cdot)$. In fact, this can be formalized by adding an additional auxiliary dimension of independent standard uniforms (as in Appendix~\ref{appendix:chiseling-generalizes-ds}) and letting $\tilde{\score}(x, u) = (\score(x), u)$, now using lexicographic comparisons to order points (Appendix~\ref{appendix:total-order}). Then this will reveal points in order of $\score(x)$ but break ties using $u$. However, it results in a region that depends in a more complex way on the auxiliary random feature, and hence in the context of causal inference corresponds to a treatment rule that stochastically treats a fraction of units at the boundary of the region. Whether this is useful in practice or not is case-dependent. However, we will use this tiebreaking construction in the proof of Lemma~\ref{lemma:dist-smallest-mass}.

\subsection{Chiseling to the boundary}
\label{appendix:chiseling-to-boundary}

Algorithm~\ref{alg:chisel-to-boundary} is the concrete algorithm that produces the regions used to compare chiseling to data splitting in Section~\ref{section:simple-demo}. It also corresponds to the first step of the strategy described in Section~\ref{section:setting-alpha}. Note that Line~\ref{line:chisel-to-boundary-stopping-time} corresponds to a stopping time, so that if we let $\genstop$ count the number of times Algorithm~\ref{alg:chiseling} is called, then Corollary~\ref{corollary:distribution-subsamples} indeed applies to the output $\reg_{\genstop}$.

\begin{algorithm}[h]
    \caption{Chiseling to the boundary using machine learning}
    \label{alg:chisel-to-boundary}
    \hspace*{\algorithmicindent} \textbf{Input:} dataset $\D = (X_i, Y_i)_{i=1}^n$, region $\reg$, split proportion $p \in (0,1)$, ML method
    \begin{algorithmic}[1]
    \State $\regend \gets \reg, \maskindend \gets$ random subset of $[n]$ of size $\ceil{(1 - p)n}$
    \While{$\abs{\maskindend} > 0$}
        \State Refit $\hat{\mean}(\cdot)$ to $(X_i, Y_i)_{i \not\in \maskindend}$ using ML method to predict $\mean(x) = \Ec{Y}{X = x}$ \label{line:refit-cond-mean}
        \State Obtain $\regmid \subseteq \regend$ by chiseling using Algorithm~\ref{alg:chiseling} with $\score(\cdot) = \hat{\mean}(\cdot)$ and $\threshlim = \cutoff$ \label{line:chisel-boundary-implicit-call}
        \State \textbf{if} $\abs{\set{i \in \maskindend : X_i \in \regend \setminus \regmid}} = 0$ \textbf{then} quit loop after \textbf{update} \label{line:chisel-to-boundary-stopping-time}
        \State \textbf{update} $(\regend, \maskindend) \gets (\regmid, \maskindend \setminus \set{i \in \maskindend : X_i \in \regend \setminus \regmid})$
    \EndWhile
    \State \textbf{Return:} $\regend$, $\maskindend$
    \end{algorithmic}
\end{algorithm}

\subsection{Derivation of conditioning--truncation equivalence}
\label{appendix:conditioning-truncation-equivalence}

Recall the notation from Section~\ref{section:general-test-design}. Then for $s < t$,
\begin{equation*}
\begin{aligned}
\un_s \meanest_s - \un_t \meanest_t = \sum_{i : X_i \in \reg_s} Y_i - \sum_{i : X_i \in \reg_t} Y_i = \sum_{i : X_i \in \reg_s \setminus \reg_t} Y_i
\end{aligned}
\end{equation*}
which, by rearranging, implies that
\begin{equation*}
\begin{aligned}
\meanest_t = \frac{1}{\un_t} \paren*{n_s \meanest_s - \sum_{i : X_i \in \reg_s \setminus \reg_t} Y_i}.
\end{aligned}
\end{equation*}
Then,
\begin{equation*}
\begin{aligned}
\meanest_s \leq \critval_s \iff \meanest_t \leq \frac{1}{\un_t} \paren*{n_s \critval_s - \sum_{i : X_i \in \reg_s \setminus \reg_t} Y_i}.
\end{aligned}
\end{equation*}
Therefore, $\max_{s < t} \gentest_s = 0$ if and only if
\begin{equation*}
\begin{aligned}
\meanest_s \leq \critval_s \text{ for all $s < t$} \iff \meanest_t \leq \min_{s < t} \set*{ \frac{1}{\un_t} \paren*{n_s \critval_s - \sum_{i : X_i \in \reg_s \setminus \reg_t} Y_i}} \iff \meanest_t \leq \trunc_t.
\end{aligned}
\end{equation*}

\subsection{Truncated binomial quantiles}
\label{appendix:generating-qtbinom}

Recall in Definition~\ref{def:binary-critval} that we defined $\qtbinom\paren{q; \tilde{n}, \tilde{\mu}, \tilde{\trunc}}$ to be the $q$th quantile of $\text{Binom}(\tilde{n}, \tilde{\mu})$ truncated to be no greater than $\tilde{\trunc}$. However, due to the discreteness of the binomial distribution, an exact $q$th quantile may not exist. In this case, $\qtbinom\paren{q; \tilde{n}, \tilde{\mu}, \tilde{\trunc}}$ returns an exogenously randomized quantile. Concretely, let $Z \sim \text{Binom}(\tilde{n}, \tilde{\mu})$. Then let
\begin{equation*}
\begin{aligned}
z^*_{\text{upper}} &:= \inf_{z \in \mathbb{N}} \set*{ \Prb(Z \leq z \mid Z \leq \tilde{\trunc}) > q }, \\
z^*_{\text{lower}} &:= \sup_{z \in \mathbb{N}} \set*{ \Prb(Z \leq z \mid Z \leq \tilde{\trunc}) \leq q } .
\end{aligned}
\end{equation*}
Then define
\begin{equation*}
\qtbinom\paren{q; \tilde{n}, \tilde{\mu}, \tilde{\trunc}} :=
\begin{cases}
z^*_{\text{upper}} & \text{ with probability }\, \frac{q - \Prb(Z \leq z^*_{\text{lower}} \mid Z \leq \tilde{\trunc})}{\Prb(Z \leq z^*_{\text{upper}} \mid Z \leq \tilde{\trunc}) - \Prb(Z \leq z^*_{\text{lower}} \mid Z \leq \tilde{\trunc})},\\
z^*_{\text{lower}} & \text{ otherwise.}
\end{cases}
\end{equation*}
In other words, $\qtbinom\paren{q; \tilde{n}, \tilde{\mu}, \tilde{\trunc}}$ is a random variable that we always consider to be completely independent of all other random variables. Under this definition,
\begin{equation*}
\begin{aligned}
\Prb\paren*{ Z \leq \qtbinom\paren{q; \tilde{n}, \tilde{\mu}, \tilde{\trunc}} \mid Z \leq \tilde{\trunc} } = q.
\end{aligned}
\end{equation*}
In Algorithm~\ref{alg:chiseling-testing}, we imagine that each time this function is queried, it produces a random output, but that once the output is observed, its value is fixed. Though verbose, all of these quantities can be readily and efficiently calculated given access to the binomial CDF. Lastly, note that if $\Prb(Z \leq z^*_{\text{lower}} \mid Z \leq \tilde{\trunc}) = q$, then $\qtbinom\paren{q; \tilde{n}, \tilde{\mu}, \tilde{\trunc}} = z^*_{\text{lower}}$ with probability $1$. So deterministic versions of our methods can be implemented by choosing $q$ so that this equality can be achieved. In practice, if $\allocalpha_t$ is the desired error level, we recommend rounding $\allocalpha_t$ to the nearest value such that a deterministic quantile will be reported.

\subsection{Derivation of critical values}
\label{appendix:critval-motivation}

Let us consider the case of binary $Y$. As per Lemma~\ref{lemma:sequential-error-control} and Appendix~\ref{appendix:conditioning-truncation-equivalence}, we need to construct $\critval_t$ so that
\begin{equation*}
\begin{aligned}
\Prb(\meanest_t > \critval_t \mid \F_t, \set{\mean(\reg_t) \leq 0, \meanest_t \leq \trunc_t}) \leq \allocalpha_t \quad \text{ a.s.}
\end{aligned}
\end{equation*}
Conditional on $\F_t$ only, the binary nature of $Y$ means that $\un_t \meanest_t \sim \text{Binomial}(\un_t, \mean(\reg_t))$. When $\reg_t$ is null (i.e. $\mean(\reg_t) \leq \cutoff$), then $\text{Binomial}(\un_t, \mean(\reg_t))$ is stochastically dominated by $\text{Binomial}(\un_t, \cutoff)$. But we must additionally condition on $\meanest_t \leq \trunc_t$, which is equivalent to $\un_t \meanest_t \leq \un_t \trunc_t$. In general stochastic domination is not necessarily preserved under right truncations, but in the case of binomials it is preserved (Lemma~\ref{lemma:truncated-binomial-stochastic-dominance}). This motivates letting $\un_t \critval_t$ be the $1 - \allocalpha_t$ quantile of an appropriately truncated binomial distribution, which yields the critical value in Definition~\ref{def:binary-critval}.

Note that since $\sum_{i : X_i \in \reg_s \setminus \reg_t} Y_i$ and $\un_s$ are $\F_t$-measurable for all $s < t$, then an inductive argument implies that $\trunc_t$ is $\sigma(\F_t, (U_s)_{s < t})$-measurable for all $t$, where $U_t$ is some auxiliary random variable, independent of everything, used to generate the $t$th randomized quantiles outputted by $\qtbinom\paren{q; \tilde{n}, \tilde{\mu}, \tilde{\trunc}}$ which are used to define $\critval_t$ (see Appendix~\ref{appendix:generating-qtbinom} for more details on the randomized quantiles). This measurability implies that the truncation level $\trunc_t$ can be calculated without peeking inside $\reg_t$. This important property allows us to characterize the distribution of the sample means under the null without requiring any information from inside the region, thus preventing any further distortion of the distribution of the masked sample.

The story for general $Y$ is similar, except that $\trunc_t$ is no longer $\F_t$-measurable due to the fact that $\hsigsq_t$ depends on data from inside the region. If we replaced $\hsigsq_t$ with $\sigma^2(\reg_t)$ where $ \sigma^2(\reg) := \Var(Y \mid X \in \reg)$, then $\trunc_t$ would in fact be $\F_t$-measurable, and this makes it easier to characterize the null distribution of each test statistic conditional on having not previously rejected. In Appendix~\ref{appendix:proofs-general-validity}, we prove validity for an oracle version of our procedure which uses $\sigma^2(\reg_t)$ in place of $\hsigsq_t$, and then prove that these two procedures produce the same output with probability converging to $1$ asymptotically.

\subsection{Clipped critical value and non-exactness}
\label{appendix:clipping-asymptotics}

As mentioned in Section~\ref{section:general-test-design}, in Definition~\ref{def:general-critval} we clip $\critval_t$ to be no smaller than $0$. This is to address an asymptotic subtlety: even if $\sup_{z}\abs{\Prb(Z_n \leq z) - \Phi(z)} \to 0$, we may not have $\sup_{z \leq u_n} \abs{\Prb(Z_n \leq z) / \Prb(Z \leq u_n) - \Phi(z) / \Phi(u_n)} \to 0$ if $u_n$ diverges to $-\infty$ too fast. Here, $u_n$ is just a sequence of real numbers, but the story would be the same if it were a sequence of random variables. The point is that even if one random variable is well-approximated by another, that approximation may fail when conditioning on extremely rare events. It is rare though not impossible to observe an extremely negative value for $\trunc_t$, which may cause problems for the Gaussian approximation deep in the left tail. Thus, we address this by essentially limiting full use of the truncation information, e.g. clipping the critical value $\critval_t$ at $0$. It is possible that a more careful analysis could relax this clipping.

Empirically, however, we do not observe significant loss in exactness from this choice (Section~\ref{section:simulations-type1-error}). We should not expect to, either, as at the boundary of the global null we should observe $\critval_t = 0$ only very rarely. This is by design, since the point of the correction is to address the failure of the Gaussian approximation under rare events. Another way to see this is by noticing, from Equation~\eqref{eq:truncation-definition}, that $\trunc_t$ is highly negative only when $(Y_i)_{i \not\in \reg_t}$ contains substantially many positive values. Under the global null that $\E[Y \mid X] = 0$ almost surely, it would be rare for chiseling to consecutively reveal a substantial quantity of positive $Y$ as chiseling can only reveal points based on $X$ which is not predictive of the mean of $Y$.

Lastly, we note that while technically the asymptotic subtlety could be addressed by clipping $\critval_t$ at any value, the clipping of $\critval_t$ at zero could be considered a methodological advantage, as it prevents us from rejecting $H_0 : \mean(\reg_t) \leq 0$ when $\meanest_t$ is negative, which would result in the unusual (though not necessarily invalid from a frequentist standpoint) conclusion that the population region mean is significantly positive while the empirical region mean is negative. Again, however, this would only happen extremely rarely even in the absence of any clipping.

\subsection{Two team cross-screening}
\label{appendix:two-team-cross-screening}

\textcite{roy2025explorationconfirmationreplicationobservational} propose a procedure called ``two team cross-screening," which is like data splitting except that instead of splitting the dataset randomly, it is split based on some meaningful covariate. For instance, $\D_1$ may consist of all units below the age of $45$ while $\D_2$ may consist of all units at or above the age of $45$. $\D_1$ is used to plan an analysis for $\D_2$, and vice versa; if a hypothesis is selected in both planning phases and rejected in both datasets, then the finding is deemed highly credible.

This idea can be used to initialize chiseling without auxiliary randomness. We first observe that splitting on a meaningful covariate partitions $\mathcal{X}$ into disjoint sets $\mathcal{X}_1$ and $\mathcal{X}_2$, where the data in $\D_1$ is drawn from the former and the data in $\D_2$ is drawn from the latter. Next, we note that if $\reg_1 \subseteq \mathcal{X}_1$ and $\reg_2 \subseteq \mathcal{X}_2$, then $\mean(\reg_1 \cup \reg_2)$ is a convex combination of $\mean(\reg_1)$ and $\mean(\reg_2)$, and hence
\begin{equation*}
\begin{aligned}
\mean(\reg_1) > 0 \text{ and } \mean(\reg_2) > 0 \implies \mean(\reg_1 \cup \reg_2) > 0.
\end{aligned}
\end{equation*}
This holds even when $\reg_1$ or $\reg_2$ are equal to $\emptyset$, where we recall that $\mean(\emptyset) > 0$ is a true proposition. Thus, an idea is to use $\D_2$ to initialize chiseling to be run on $\D_1$ with level $\alpha / 2$, producing $\reg_1 \subseteq \mathcal{X}_1$, and vice versa to produce $\reg_2 \subseteq \mathcal{X}_2$. Then
\begin{equation*}
\begin{aligned}
\Prb(\mean(\reg_1) \leq 0 \text{ or } \mean(\reg_2) \leq 0) \leq \Prb(\mean(\reg_1) \leq 0) + \Prb(\mean(\reg_2) \leq 0) \leq \alpha / 2 + \alpha / 2 = \alpha.
\end{aligned}
\end{equation*}
So reporting the union of the regions $\reg_1 \cup \reg_2 \subseteq \mathcal{X}$ controls Type I error.

This works best when the covariate being split on does not have an interacting effect with the other covariates, so that learned relationships in $\D_1$ can be largely transported to $\D_2$ and vice versa. Even when non-interaction is satisfied, however, this initialization can be less powerful than randomly initializing a single instance of chiseling due to the reduction of the overall significance levels for each instance of chiseling to $\alpha / 2$, as well as the reduction in sample size for each instance.

\subsection{Combining chiseling across different initial split proportions}
\label{appendix:multi-chiseling}

\FloatBarrier

\begin{figure}[ht]
  \centering
  \includegraphics[width=\linewidth]{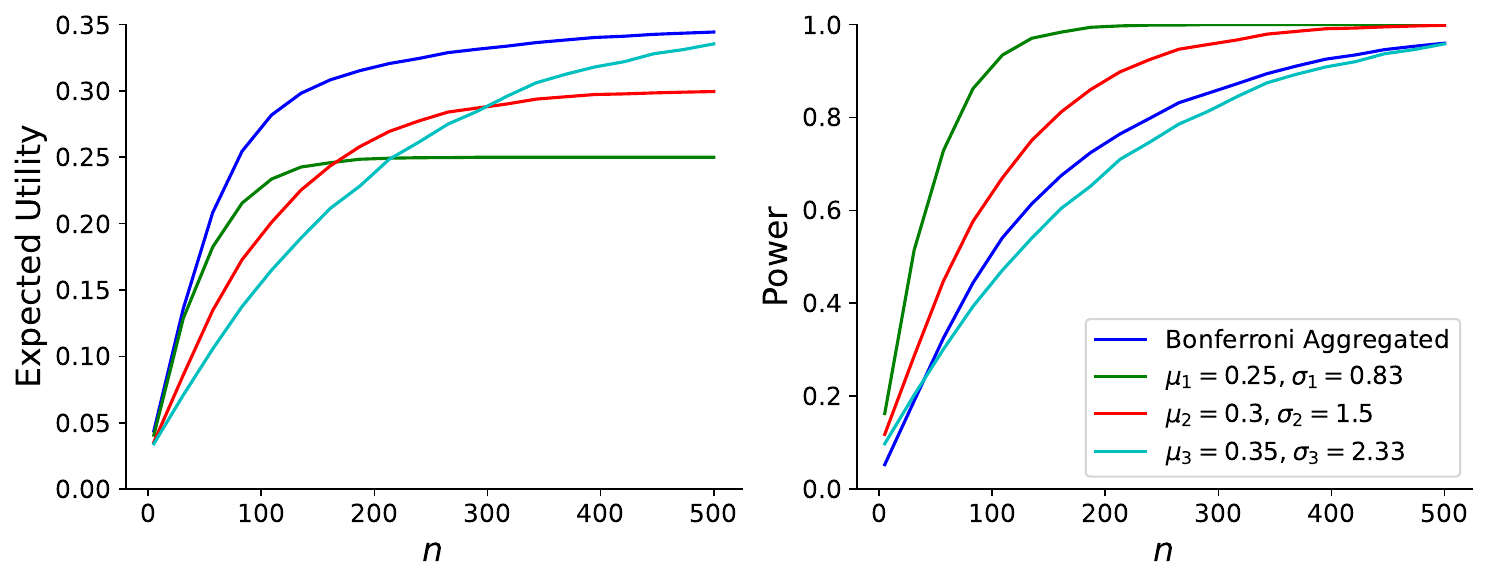}
  \caption{Stylized empirical demonstration of the effectiveness of Bonferroni aggregation.}
  \label{fig:bonferroni-aggregation-demo}
\end{figure}

In Section~\ref{section:init-chiseling}, we described Bonferroni aggregation, which combines the results of chiseling (or any method that relies on an initial split proportion $p$) across $p_1 \leq ... \leq p_k$ by simply running the analyses at adjusted levels $\alpha^{(1)} + ... + \alpha^{(k)} = \alpha$. First, we make a remark regarding implementation: a valid but less effective way to realize Bonferroni combination is as follows. Consider, for instance, data splitting. Randomly shuffle and order the data points, and use the first $p_1$ proportion to train and the last $1 - p_1$ proportion to test, then use the first $p_2$ proportion to train and the last $1 - p_2$ proportion to test, and so on, finally aggregating the results of these $k$ procedures via Bonferroni combination. This is clearly valid, but it is ineffective because there is high positive correlation between the different analyses, which makes the Bonferroni correction conservative. However, if we reshuffle the data before \emph{each} analysis so that the $p_j \cdot n$ training points for the $j$th analysis are not simply a subset of the $p_{j + 1} \cdot n$ training points for the $(j + 1)$st analysis, this makes the analyses more independent and alleviates some of Bonferroni's conservativeness. This is what we do in our simulations and we recommend it in practice.

Next, we discuss why Bonferroni aggregation is effective. Consider data splitting for simplicity. Typically, the utility of the tested subgroup will be higher for $p_2$-data splitting than for $p_1$-data splitting if $p_1 < p_2$. However, the power may sometimes (but not always) be lower for $p_2$ because the testing sample size is smaller, which brings down the expected utility. Suppose we have $p_1 < p_2 < p_3$. We can stylize the behavior of data splitting as follows: let $\mu_1 < \mu_2 < \mu_3$ and $\mu_1 / \sigma_1 > \mu_2 / \sigma_2 > \mu_3 / \sigma_3$. The $\mu_j$ represent the mean of the tested subgroups discovered by $p_j$-data splitting, which are increasing in $j$ but such that the signal size $\mu_j / \sigma_j$ is decreasing in $j$. We let $Y_{ij} \sim \mathcal{N}(\mu_j, \sigma_j^2)$ independently for $i=1,...,n$ and $j=1,2,3$. The individual analyses test $H_j : \mu_j \leq 0$ at level $\alpha$ using the $z$-test and receive utility $\mu_j$ if a rejection is made and $0$ otherwise. The Bonferroni aggregated analysis tests $H_1,H_2,H_3$ simultaneously at levels $\alpha / 3$, then receives utility $\mu_j$ for the largest $j$ such that $H_j$ rejected and $0$ if no rejection is made.

We can understand the behavior in a limiting setting where $\alpha \to 0$ and the signal sizes are scaled so that the testing problem is non-trivial. Let $R_j(\alpha)$ denote the indicator of the event that we can reject $H_j$ at level $\alpha$ and let $q_j(\alpha) = \E[R_j(\alpha)]$. As $\alpha \to 0$, the worst-case power loss between running the $z$-test at level $\alpha$ and level $\alpha / 3$ converges to $0$ (see Appendix~\ref{appendix:inferiority-t-test} for a related discussion). This means that for $\alpha$ in a sufficiently small neighborhood of $0$, we have that the expected utility of Bonferroni aggregation is
\begin{equation*}
\begin{aligned}
\E[\max\set{\mu_1 R_1(\alpha / 3), \mu_2 R_2(\alpha / 3), \mu_3 R_3(\alpha / 3)}] \geq \E[\mu_j R_j(\alpha / 3)] = \mu_j q_j(\alpha / 3) \approx \mu_j q_j(\alpha) 
\end{aligned}
\end{equation*}
for all $j=1,2,3$. Thus, the expected utility of Bonferroni aggregation is no worse than that of any individual test when $\alpha$ is sufficiently small. In fact, it can be much larger, since the mean of the maximum is generally larger than the maximum of the individual means. We verify in a toy example that even when $\alpha = 0.05$ the difference can be quite stark (Figure~\ref{fig:bonferroni-aggregation-demo}).

In practice, the improvements we observe are not as dramatic because the tests are not independent but positively correlated. However, we do find Bonferroni aggregation to be surprisingly effective and we believe this stylized example hints at why.

\FloatBarrier

\subsection{Weak non-inferiority of chiseling to the global \texorpdfstring{$t$}{t}-test}
\label{appendix:inferiority-t-test}

\FloatBarrier

\begin{figure}[ht]
  \centering
  \includegraphics[width=\linewidth]{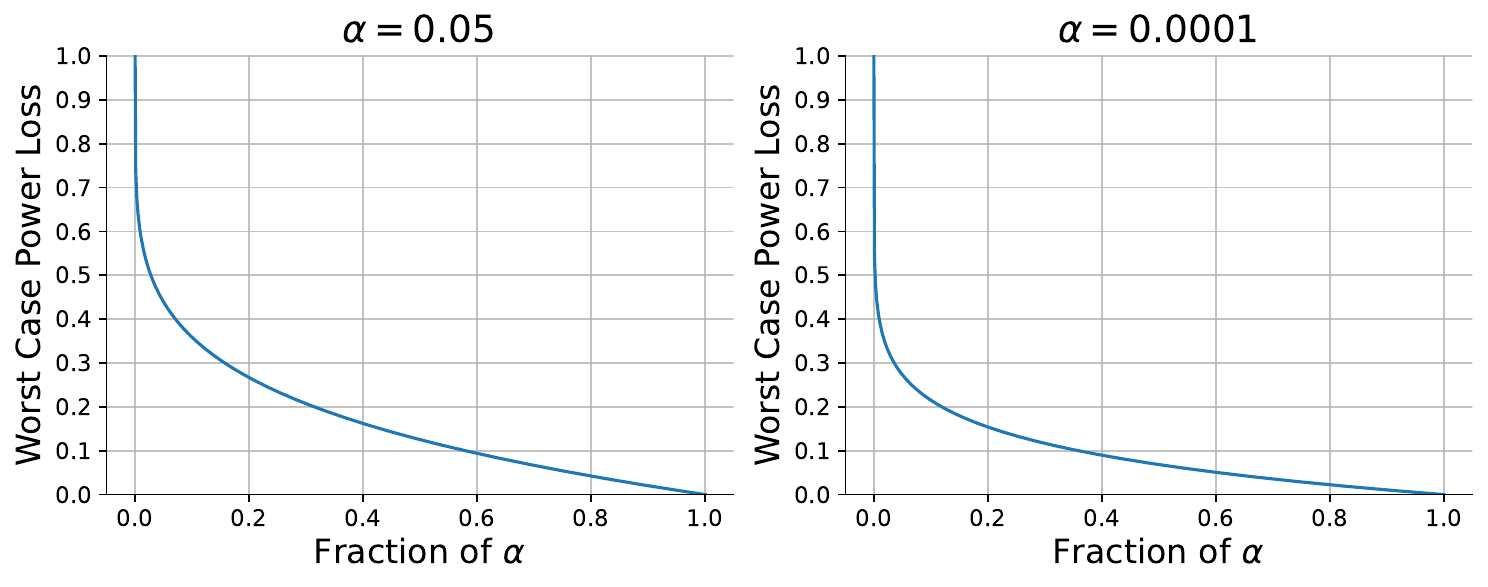}
  \caption{Asymptotic worst case power loss of running a $t$-test with a fraction of $\alpha$ compared to running the $t$-test at level $\alpha$.}
  \label{fig:t-test-non-inferiority}
\end{figure}

Suppose without loss of generality that $\Var(Y) = 1$ and $\cutoff = 0$. Under local asymptotic scaling where $\mean(\mathcal{X}) = \tau / \sqrt{n}$, the asymptotic power of the $t$-test for testing $H_0: \mean(\mathcal{X}) \leq 0$ at level $\alpha$ is given by $\Phi(\tau - z_{1 - \alpha})$ where $z_{1 - \alpha}$ is the $1 - \alpha$ quantile of a standard normal (see, for instance, \textcite{spiess2023findingsubgroupssignificanttreatment}). Then if chiseling sets $\allocalpha_0 = c \alpha$ for $c \in [0,1]$ the asymptotic difference in power between chiseling and the $t$-test can be no greater than
\begin{equation*}
\begin{aligned}
L(c; \alpha) := \sup_{\tau \in \mathbb{R}} \Phi(\tau - z_{1 - \alpha}) - \Phi(\tau - z_{1 - c \alpha}).
\end{aligned}
\end{equation*}
For instance, $L(0.5; 0.05) \approx 0.125$, while $L(0.5; 0.0001) \approx 0.068$. $L(c; \alpha)$ is decreasing in $c$ and increasing in $\alpha$ (see the end of this subsection). This is the worst case loss in power, attained when the effect is completely homogeneous, but of course chiseling can achieve much greater power if the remaining $(1 - c)\alpha$ of the error can be effectively utilized to learn a subgroup, as we have seen in Section~\ref{section:simulation-nonneg-rct}. The analyst may resort to curves such as those in Figure~\ref{fig:t-test-non-inferiority} in order to select an acceptable level of non-inferiority relative to the $t$-test, trading this off with the strength of their prior confidence in being able to detect subgroups.

When $\E[Y \mid X] \geq 0$ almost surely, then this bound on the power difference immediately translates to a bound on the difference in normalized expected utility. Let $A$ be the event that chiseling rejects $\reg_0 = \mathcal{X}$ and let $B$ be the event that the global $t$-test rejects. Then
\begin{equation*}
\begin{aligned}
\E[\util(\reg_{\tsel})] = \E[\util(\reg_{\tsel}) \mid A] \Prb(A) + \E[\util(\reg_{\tsel}) \mid A^{\comp}] \Prb(A^{\comp}) \geq \util(\mathcal{X}) \Prb(A) = \util(\reg^*) \Prb(A)
\end{aligned}
\end{equation*}
where $\reg^*$ is an optimal utility region (because in this case $\mathcal{X}$ is an optimal utility region). The inequality follows from the fact that $\util(\reg_{\tsel}) \geq 0$ almost surely since $\E[Y \mid X] \geq 0$. Then as the expected utility of the $t$-test is $\util(\reg^*) \Prb(B)$,
\begin{equation*}
\begin{aligned}
\frac{\util(\reg^*) \Prb(B) - \E[\util(\reg_{\tsel})]}{\util(\reg^*)} \leq \frac{\util(\reg^*) \Prb(B) - \util(\reg^*)\Prb(A)}{\util(\reg^*)} = \Prb(B) - \Prb(A) \leq L(c; \alpha).
\end{aligned}
\end{equation*}

\begin{proof}[Proof that $L(c; \alpha)$ is increasing in $c$ and decreasing in $\alpha$]
Fix $c$ and $\alpha$. It suffices to consider $c \in (0,1)$. Let $\phi$ be the density of a standard normal. Setting the derivative of $\Phi(\tau - z_{1 - \alpha}) - \Phi(\tau - z_{1 - c\alpha})$ with respect to $\tau$ equal to $0$, we get
\begin{equation*}
\begin{aligned}
\phi(\tau - z_{1 - \alpha}) - \phi(\tau - z_{1 - c\alpha}) = 0.
\end{aligned}
\end{equation*}
Since $\phi$ is symmetric and strictly decreasing away from $0$, and since $z_{1 - \alpha} \neq z_{1 - c\alpha}$, this holds if and only if $\tau - z_{1 - \alpha} = -(\tau - z_{1 - c\alpha})$, i.e. $\tau = \frac{z_{1 - \alpha} + z_{1 - c \alpha}}{2}$. So we have
\begin{equation*}
\begin{aligned}
L(c; \alpha) = \Phi\paren*{\frac{z_{1 - c\alpha} - z_{1 - \alpha}}{2}} - \Phi\paren*{\frac{- z_{1 - c\alpha} + z_{1 - \alpha}}{2}} = 2\Phi\paren*{\frac{z_{1 - c\alpha} - z_{1 - \alpha}}{2}} - 1.
\end{aligned}
\end{equation*}
Since $\Phi$ is increasing, it suffices to show that $z_{1 - c\alpha} - z_{1 - \alpha}$ is decreasing in $c$ and increasing in $\alpha$. The former is easy to see, so we focus on the latter. By the inverse function rule, the derivative of this quantity with respect to $\alpha$ is given by
\begin{equation*}
\begin{aligned}
\frac{-c}{\phi(z_{1 - c\alpha})} + \frac{1}{\phi(z_{1 - \alpha})} = \frac{1}{\alpha} \paren*{ \frac{1 - \Phi(z_{1 - \alpha})}{\phi(z_{1 - \alpha})} - \frac{1 - \Phi(z_{1 - c\alpha})}{\phi(z_{1 - c\alpha})}}.
\end{aligned}
\end{equation*}
Since $z_{1 - c\alpha} > z_{1  - \alpha}$, it suffices to show that $\frac{1 - \Phi(x)}{\phi(x)}$ is decreasing in $x$. But this is simply the Mills ratio, and it is a well-known fact that this quantity is strictly decreasing on $\mathbb{R}$ (see, for instance, \textcite{Baricz2008}).
\end{proof}

\FloatBarrier

\subsection{Shrinking along hyperrectangular contours}
\label{appendix:hyperrectangular-shrinking}

Let $\score : \mathbb{R}^d \to \mathbb{R}$ be any function. Define $A_z := \set{x \in \mathbb{R}^d : \score(x) > z}$. We show that $A_z$ is hyperrectangular for all $z$ if and only if $\score(x) = \min_{j=1,...,d} \set{f_j(x^{(j)})}$ for some univariate functions $f_j$ whose upper level sets are intervals. The reverse direction was shown in Section~\ref{section:interpretable-regions}, so we focus on the forward direction. By assumption, $A_z = \prod_{j=1}^d I_j(z)$ where $I_j(z)$ are intervals such that $I_j(z) \supseteq I_j(z')$ for all $z \leq z'$. Define $f_j(x^{(j)}) = \sup\set{z \in \mathbb{R} : x^{(j)} \in I_j(z)}$. Then
\begin{equation*}
x^{(j)} \in \set{x' : f_j(x') > z} \iff f_j(x^{(j)}) > z  \iff \exists z' > z \text{ s.t. } x^{(j)} \in I_j(z') \implies x^{(j)} \in I_j(z).
\end{equation*}
To reverse the last implication, suppose without loss of generality that $j = 1$. Note that $x^{(1)} \in I_1(z)$ implies that there exists $u \in \mathbb{R}^{d-1}$ such that $\score(x^{(1)}, u) > z$, and thus there exists $z'$ such that $\score(x^{(1)}, u) > z' > z$. But this implies that $x^{(1)} \in I_1(z')$, so the last implication may be reversed, and we have that the upper level sets of $f_j(\cdot)$ are intervals. Then finally note that
\begin{equation*}
\begin{aligned}
\score(x) > z \iff \forall j \,\, x^{(j)} \in I_j(z) \iff \forall j \,\, f_j(x^{(j)}) > z \iff \min_{j=1,..,d} f_j(x^{(j)}) > z.
\end{aligned}
\end{equation*}
The upper level sets of $\score(\cdot)$ and $\min_{j=1,..,d} f_j(x^{(j)})$ are identical for all $z$ and hence the two functions are equal. This last fact can be quickly seen by contradiction: for functions $f$ and $g$, suppose the upper level sets of $f$ and $g$ agree but $f(x') < g(x')$ for some $x'$. Then there is a $z$ such that $f(x') < z < g(x')$, but then $x' \not\in \set{x : f(x) > z}$ while $x' \in \set{x : g(x) > z}$, a contradiction.

\subsection{Rejecting multiple regions}
\label{appendix:multiple-testing}

We show that the chiseling procedure we have developed can be used to reject multiple regions while controlling the family-wise error rate (FWER) as long as it is used in an appropriate way. The construction is essentially an application of the closed testing principle. There are, however, some subtleties which arise from the adaptive nature of our problem.

Imagine the following procedure: the analyst proceeds through the chiseling algorithm as usual, producing a sequence of regions $\reg_0,...,\reg_{\tmax}$. Let us suppose that the analyst does not run any tests (and hence does not terminate until after the last region $\reg_{\tmax}$ has been defined). Instead, at stage $t$, the analyst picks a value $v_t(S) \in [0,1]$ for every $S \subseteq [t]_0$ such that $t \in S$, and sets $v_t(S) = 0$ for every other $S \subseteq [t]_0$, We require that the choices satisfy $\prod_{s \leq t} 1 - v_s(S \cap [s]_0) \geq 1 - \alpha$ for all such $S$. We call $v_t(S)$ the \emph{hypothetical $\alpha$ budgets}. After defining all the hypothetical $\alpha$ budgets, we can go back and do the following: for each stage $t$, let $\phi_t = 1$ if, for every $S \subseteq [t]_0$ such that $t \in S$, chiseling run with $\allocalpha_s = v_s(S \cap [s]_0)$ for $s \leq t$ rejects \emph{some} region $\reg_s$ with $s \leq t$. Then
\begin{equation*}
\begin{aligned}
\Prb(\exists t \text{ s.t. } \phi_t = 1 \text{ and } \mean(\reg_t) \leq \cutoff) \leq \alpha
\end{aligned}
\end{equation*}
when the tests are finite-sample valid, and
\begin{equation*}
\begin{aligned}
\limsup_{n \to \infty} \Prb(\exists t \text{ s.t. } \phi_t = 1 \text{ and } \mean(\reg_t) \leq \cutoff) \leq \alpha
\end{aligned}
\end{equation*}
when the tests are asymptotically valid. We prove this below and then sketch out a simple and illuminating example. In general this procedure can be computationally intensive for the same reasons that closed testing is computationally intensive (i.e. needing to test every intersection null), but computational shortcuts are possible if structure is imposed on $v_t(S)$.

\begin{proof}
The analyst defines, in the course of this modified variant of chiseling, a set of hypothetical $\alpha$ budgets. We will consider an oracle that runs the usual variant of chiseling by picking out and abiding by a single hypothetical $\alpha$ budget corresponding to testing all the true nulls. The oracle knows $\mathcal{N}_t := \set{s \leq t : \mean(\reg_s) \leq \cutoff} \subseteq [t]_0$ at stage $t$ since $\reg_s$ is $\F_t$-measurable for $s \leq t$. We let the oracle set $\allocalpha_t = v_t(\mathcal{N}_t)$. By construction,
\begin{equation*}
\begin{aligned}
\prod_{t=0}^{\tmax} 1 - \allocalpha_t = \prod_{t=0}^{\tmax} 1 - v_t(\mathcal{N}_t) = \prod_{t=0}^{\tmax} 1 - v_t(\mathcal{N}_{\tmax} \cap [t]_0) \geq 1 - \alpha
\end{aligned}
\end{equation*}
so this is a valid $\alpha$-budget. Now we define the oracle multiple rejection set $\mathcal{V}_{\mathrm{oracle}} := \mathcal{N}_m^{\comp}$ if the oracle instance of chiseling does not reject (i.e. reports the empty region), and $\mathcal{V}_{\mathrm{oracle}} := [\tmax]_0$ otherwise. In words, the oracle rejects all the false nulls (and makes no errors) if its instance of chiseling accepts, and otherwise rejects every hypothesis. Hence, the oracle FWER is bounded by the probability that the oracle instance of chiseling rejects. Since the oracle is only testing null regions and chiseling is valid, the probability of this event is no greater than $\alpha$, and thus the oracle FWER is bounded by $\alpha$.

Now we argue that $\mathcal{V} := \set{t : \phi_t = 1} \subseteq \mathcal{V}_{\mathrm{oracle}}$ deterministically. Suppose $t \in \mathcal{V}$. If $\mean(\reg_t) > \cutoff$, then $t \in \mathcal{V}_{\mathrm{oracle}}$ since $\mathcal{V}_{\mathrm{oracle}}$ always contains all the false nulls. If $\mean(\reg_t) \leq \cutoff$, then the oracle instance of chiseling must have rejected at some stage $s \leq t$. This is because $t \in \mathcal{N}_t$ and the oracle $\alpha$-budget up to stage $t$ corresponds to $(v_0(\mathcal{N}_0),...,v_t(\mathcal{N}_t)) = (v_0(\mathcal{N}_t \cap [0]_0),...,v_t(\mathcal{N}_t \cap [t]_0))$, but every hypothetical instance of chiseling that uses $\alpha$-budget $(v_0(S \cap [0]_0),...,v_t(S \cap [t]_0))$ rejects for any $S \subseteq [t]_0$ as long as $t \in S$; this is by definition of $t$ belonging to $\mathcal{V}$. Then $\mathcal{V}_{\mathrm{oracle}} = [\tmax]_0$ and trivially includes $t$. Thus,
\begin{equation*}
\begin{aligned}
\Prb(\exists t \in \mathcal{V} \text{ s.t. } \mean(\reg_t) \leq \cutoff) \leq \Prb(\exists t \in \mathcal{V}_{\mathrm{oracle}} \text{ s.t. } \mean(\reg_t) \leq \cutoff) \leq \alpha
\end{aligned}
\end{equation*}
when the tests are finite-sample valid. The first inequality is always true regardless of the validity of the chiseling. Thus, when the tests are asymptotic we can simply put a $\limsup$ in front of the probabilities.
\end{proof}

\noindent \textbf{Example: rejecting two regions.} Suppose we wish to test $\reg_0 = \mathcal{X}$ and one additional chiseled region. At a high level, the closed testing construction allows us to do following: we pick some $\allocalpha_0$, and then use chiseling to shrink and stop at one region $\reg_{\genstop}$ (e.g. Algorithm~\ref{alg:chisel-to-boundary}). If we do not reject $H_0 : \mean(\reg_0) \leq \cutoff$ at level $\allocalpha_0$, we apply the sequential testing framework developed in this paper and test $H_{\genstop} : \mean(\reg_{\genstop}) \leq \cutoff$ at level $\allocalpha_{\genstop} = 1 - (1 - \alpha) / (1 - \allocalpha_0)$ conditionally on having accepted the previous test, rejecting if possible. If we do reject $H_0$, we report that, and then we are entitled to test $H_{\genstop}$ \emph{unconditionally} at level $\alpha$ and also report that if it rejects. The intuition is that when we do reject $H_0$, there are only two scenarios: $H_0$ is null, in which case we have already committed a Type I error and additional rejections are ``free," or (2) $H_0$ is not null, in which case there was no need to test it and we could have, in the first place, saved all of our $\alpha$ for testing $H_{\genstop}$. Note that here we have implicitly set, for every $t \not\in \set{0, \genstop}$, $v_t(S) = 0$ for every $S \subseteq [t]_0$, which renders the test feasible to execute.

\subsection{Point estimates and confidence intervals}
\label{appendix:point-est-cis}

Beyond testing, a standard analysis would provide point estimates and confidence intervals.

\hfill

\noindent \textbf{Point estimates.} $\meanest_t = \frac{1}{\un_t} \sum_{i : X_i \in \reg_t} Y_i$ is the natural estimate for $\mean(\reg_t)$. In fact, Corollary~\ref{corollary:distribution-subsamples} guarantees that $\meanest_t$ is unbiased for $\mean(\reg_t)$ conditionally on $\F_t$ as long as there is at least one point in $\reg_t$. If we take the convention that an undefined value---call it $\nan$---is equal to any number, i.e. $\nan - x = 0$ for all $x$, then by applying the tower rule to each coordinate we have
\begin{equation*}
\begin{aligned}
\E[(\meanest_0,...,\meanest_{\tmax}) - (\mean(\reg_0),...,\mean(\reg_{\tmax}))] = (0,...,0).
\end{aligned}
\end{equation*}
Though the convention that $\nan - x = 0$ for all $x$ appears odd, it is not artificial. It simply reflects the idea that if we have no estimate to report, then we are neither overestimating nor underestimating the truth. It is thus appropriate to call $\nan$ a trivially unbiased estimate. We note, of course, that the above unbiasedness does not hold \emph{conditionally} on having made a rejection, for instance using Algorithm~\ref{alg:chiseling-testing}. But neither is any ordinarily unbiased point estimate $\hat{\theta}$ for $\theta$ conditional on having either accepted or rejected $H_0: \theta \leq 0$, say, using a $t$-test. Thus, it may not be unreasonable to interpret $\meanest_t$ similarly to how one would interpret an unbiased estimate in the context of a larger analysis that includes hypothesis tests (i.e. as unbiased prior to testing).

In fact, the natural unbiased estimates have some reasonable consistency properties. For example, suppose that $p \in (0,1)$, $Y$ is bounded, and $\tmax = o(e^{pn})$.\footnote{$\tmax = o(e^{pn})$ is easily satisfied; note that if every $k$ applications of chiseling reveals at least $1$ data point, then we may take $\tmax = kn$ as all subsequent regions will be $\emptyset$.} Then
\begin{equation*}
\begin{aligned}
\lim_{n \to \infty} \Prb\paren*{\sup_{t : \un_t \geq pn} \abs{\meanest_t - \mean(\reg_t)} > \epsilon } = 0 \quad \text{ for all } \epsilon > 0.
\end{aligned}
\end{equation*}
The result follows from applying Hoeffding's inequality conditionally. For some constant $C$, the following holds for all $t$:
\begin{equation*}
\begin{aligned}
\Prbc{ \abs{\meanest_t - \mean(\reg_t)} > \epsilon  }{ \F_t } \cdot \indic\set{ \un_t \geq pn } \leq C e^{-\un_t} \cdot \indic\set{ \un_t \geq pn } \leq C e^{-pn} \quad \text{a.s.}
\end{aligned}
\end{equation*}
Then
\begin{equation*}
\begin{aligned}
\Prb\paren*{\sup_{t : \un_t \geq pn} \abs{\meanest_t - \mean(\reg_t)} > \epsilon } \leq \sum_{t=0}^{\tmax} \Prb\paren*{ \abs{\meanest_t - \mean(\reg_t)} > \epsilon \text{ and } \un_t \geq pn } \leq C (\tmax + 1) e^{-pn} \to 0
\end{aligned}
\end{equation*}
by observing that the summands in the second expression are simply the expectations of the left-hand side of the previous display. It is probably possible to weaken the condition that $Y$ is bounded, but this requires more technical effort.

\hfill

\begin{algorithm}[ht]
    \caption{Generic multiple testing via chiseling}
    \label{alg:chisel-bonferroni}
    \hspace*{\algorithmicindent} \textbf{Input:} \text{dataset $\D = \paren{X_i, Y_i}_{i=1}^n$ and nominal family-wise error rate $\alpha$}
    \begin{algorithmic}[1]
    \State Initialize region $\reg_0 \gets \mathcal{X}$ and revealed information $\F_0 \gets \set{\emptyset, \Omega}$
    \For{$t = 0,...,\tmax$}
        \State Choose a hypothesis $\genhyp_t$ and $\allocalpha_t \leq \alpha - \sum_{s=0}^{t-1} \allocalpha_s$ based on $\F_t$
        \State Perform a valid level $\allocalpha_t$ test $\gentest_t$ of $\genhyp_t$ conditional on $\F_t$ \label{line:bonf-valid-test}
        \State Obtain $\reg_{t+1} \subseteq \reg_t$ by chiseling $\reg_t$ based on $\F_t$ using Algorithm~\ref{alg:chiseling}
        \State See revealed data points: $\F_{t+1} \gets \sigma(\F_t, (X_i, Y_i)_{i : X_i \in \reg_t \setminus \reg_{t+1}})$
    \EndFor
    \State \textbf{Return:} hypotheses $(\genhyp_t)_{t=0}^{\tmax}$ and rejections $(\gentest_t)_{t=0}^{\tmax}$
    \end{algorithmic}
\end{algorithm}

\noindent \textbf{Confidence intervals.} First, Algorithm~\ref{alg:chisel-bonferroni} describes a generic multiple testing wrapper for chiseling based on the Bonferroni correction. Note that in the final line we return the hypotheses $(\genhyp_t)_{t=0}^{\tmax}$ but here we should think of the hypotheses as being the adaptively determined null propositions about the data that are being tested by the analyst, not the actual values of $(\genhyp_t)_{t=0}^{\tmax}$ which are of course unobservable. The family-wise error control of Algorithm~\ref{alg:chisel-bonferroni} follows directly from the union bound:
\begin{equation*}
\begin{aligned}
\Prb\paren{\exists t \text{ s.t. } \genhyp_t = 0 \text{ and } \gentest_t = 1} &\leq \sum_{t=0}^{\tmax} \Prb\paren{\genhyp_t = 0, \gentest_t = 1} = \sum_{t=0}^{\tmax} \E\bkt{\Prb\paren{\genhyp_t = 0, \gentest_t = 1 \mid \F_t}}\\
&\leq \sum_{t=0}^{\tmax} \E\bkt{\allocalpha_t} = \E\bkt*{\sum_{t=0}^{\tmax} \allocalpha_t} \leq \E[\alpha] = \alpha.
\end{aligned}
\end{equation*}

We clarify Line~\ref{line:bonf-valid-test} of Algorithm~\ref{alg:chisel-bonferroni}. This is different from Definition~\ref{def:cond-valid-test-seq} which requires validity conditional on the outcomes of the previous tests $(\gentest_s)_{s < t}$. Here, we only require validity conditional on $\F_t$. Generally, we can achieve this with little fuss using Corollary~\ref{corollary:distribution-subsamples}. For instance, imagine if $\genhyp_t = \indic\set{\mathrm{median}(\reg_t) \neq 0}$ (i.e. $\genhyp_t : \mathrm{median}(\reg_t) = 0$) where $\mathrm{median}(\reg)$ is the median of $Y \mid X \in \reg$. Then for every $t$ we may simply let $\gentest_t$ be the sign test applied to the samples $(Y_i)_{i : X_i \in \reg_t}$ with level $\allocalpha_t$.

Now we note that we may replace $\genhyp_t$ with an estimand $\theta_t$, $\gentest_t$ with a $1 - \allocalpha_t$ confidence interval $\mathcal{C}_t$ for $\theta_t$ that is valid conditional on $\F_t$, and the return statement with the collection of selected estimands and confidence intervals. Then the same union bound argument implies
\begin{equation*}
\begin{aligned}
\Prb(\exists t \text{ s.t. }\theta_t \not\in \mathcal{C}_t) \leq \alpha.
\end{aligned}
\end{equation*}
In other words, we have simultaneous coverage. To give a concrete example: we may let $\theta_t := \mathrm{median}(\reg_t)$ and $\mathcal{C}_t$ be the $1 - \allocalpha_t$ confidence interval for the median obtained by inverting the sign test using the samples $(Y_i)_{i : X_i \in \reg_t}$.

Note, of course, that this may be much less powerful than what we describe in the main text because we are not accounting for the dependence among tests. However, it is more general, precisely because it is valid under the worst-case dependence structure. What is lost in power is gained here in flexibility.

Lastly, if we replace Line~\ref{line:bonf-valid-test} with ``construct a valid level-$\alpha$ confidence interval $\mathcal{C}_t$ for $\theta_t$ conditional on $\F_t$," then we control the expected miscoverage rate:
\begin{equation*}
\begin{aligned}
\E\bkt*{ \frac{1}{\tmax + 1} \sum_{t=0}^{\tmax} \indic\set{\theta_t \not\in \mathcal{C}_t} } \leq \alpha.
\end{aligned}
\end{equation*}
This is the analogue of providing pointwise coverage. For instance, if $\theta_t = \mean(\reg_t)$ and we set aside asymptotics for the moment, then $(\mathcal{C}_0,...,\mathcal{C}_{\tmax})$ is precisely like a pointwise confidence band for the process traced out by $(\mean(\reg_0),...,\mean(\reg_{\tmax}))$. While pointwise coverage is a weak statistical guarantee, it can be useful in exploratory contexts.

\subsection{Finite-sample approach for binary potential outcomes}
\label{appendix:binary-potential-outcomes}

Suppose that our outcomes $Y$ are binary, that we additionally measure a binary covariate $W$, and that we are interested in finding a region $\reg$ such that there is a positive contrast, i.e. that $\E[Y \mid W = 1, X \in \reg] > \E[Y \mid W = 0, X \in \reg]$. If $W$ is a vector of randomized treatment assignments and $Y$ is the outcome of the randomized experiment, then this is equivalent to a region where the average treatment effect is positive, i.e. this would reduce to $\E[Y(1) \mid X \in \reg] > \E[Y(0) \mid X \in \reg]$. Alternatively, since the outcomes are binary and the subgroup means are rates, it may be natural to desire regions such that the subgroup rate ratio is at least some value, i.e.
\begin{equation*}
\begin{aligned}
\frac{\E[Y \mid W = 1, X \in \reg]}{\E[Y \mid W = 0, X \in \reg]} > \text{ some fixed threshold}.
\end{aligned}
\end{equation*}
We will see that the ideas presented here can be readily extended to testing properties of arbitrary functions of $\E[Y \mid W = 1, X \in \reg]$ and $\E[Y \mid W = 0, X \in \reg]$, but for simplicity we will focus on the case of positive contrasts.

Let us use the notation $\bin_j(\reg) = \E[Y \mid W = j, X \in \reg]$ for $j = 0, 1$. Our situation would be very similar to the binary regression setting of Section~\ref{section:general-test-design} if not for the appearance of nuisance parameters. Note that even in the non-adaptive setting where $\reg$ is fixed a priori, there are nuisances in testing $H_0: \bin_1(\reg) \leq \bin_0(\reg)$ as we care about the single unknown contrast but not the two unknown means. For a general test statistic $\ts$, it becomes necessary to search over the $1 - \alpha$ quantiles of $\ts$ under all nulls that are consistent with $H_0$ in order to properly calibrate the null distribution. This entails a two-dimensional grid search. The goal of this section is to illustrate that this grid search is readily incorporated into our adaptive testing framework, where now the algorithm must perform a grid search at each stage where the analyst wishes to perform a test.

Define $\un_{t,1} = \sum_{i=1}^n W_i \indic\set{X_i \in \reg}$ and $\un_{t,0} = \sum_{i=1}^n (1 - W_i) \indic\set{X_i \in \reg}$, and suppose that the analyst is shown $\un_{0,1}$ and $\un_{0,0}$ at the outset, i.e. we condition on the sample sizes of the treated and control groups, and let $\F_0 = \sigma(\un_{0,1}, \un_{0,0})$. Then $\un_{t,1}$ and $\un_{t,0}$ are $\F_t$-measurable. Let us suppose that at stage $t$ the analyst selects, using the information in $\F_t$, a test statistic $\ts_t$, which by abuse of notation we will suppose is both a function from $[\un_{t,1}]_0 \times [\un_{t,0}]_0$ to $\mathbb{R}$, and also the value it attains when applied to the following statistics of the masked data,
\begin{equation}
\label{eq:binary-contrast-test-stat}
\begin{aligned}
\ts_t := \ts_t\paren*{ \sum_{i=1}^n W_i Y_i \indic\set{X_i \in \reg_t}, \sum_{i=1}^n (1 - W_i)Y_i \indic\set{X_i \in \reg_t} }.
\end{aligned}
\end{equation}
Note that the actual values that $\ts_t$ outputs do not matter so much as the ranking it induces on the elements of $[\un_{t,1}]_0 \times [\un_{t,0}]_0$. Ultimately, rejecting for large values of $\ts_t$ will carve out a subset of $[\un_{t,1}]_0 \times [\un_{t,0}]_0$ corresponding to rejections, and \textit{not} rejecting will tell us that the above statistics of the masked data are \textit{not} in this subset. We can show that this information propagates cleanly from stage to stage. At stage $t$, we recursively define the \textit{allowed pairs} $\apairs_t$ and critical values $\critval_t$ at stage $t$ via the following procedure. Begin with $\apairs_0 := [\un_{0,1}]_0 \times [\un_{0,0}]_0$, and recursively define the following for $t = 0,...,\tmax$:
\begin{equation}
\label{eq:binary-contrast-crit-val}
\begin{aligned}
\critval_t &:= \sup_{(p_1, p_0) \in [0,1]^2 : p_1 \leq p_0} \resbinquant(1 - \allocalpha_t; (\un_{t,1}, \un_{t,0}), (p_1, p_0), \apairs_t)(\ts_t),\\
\interR_{t,1} &:= \sum_{i=1}^n W_i Y_i \indic\set{X_i \in \reg_t \setminus \reg_{t+1}}, \\
\interR_{t,0} &:= \sum_{i=1}^n (1 - W_i) Y_i \indic\set{X_i \in \reg_t \setminus \reg_{t+1}}, \\
\apairs_{t+1} &:= \set*{ \paren*{j - \interR_{t,1}, k - \interR_{t,0}} : (j, k) \in \apairs_t \text{ and } \ts_t(j, k) \leq \critval_t } \cap \paren*{[\un_{t+1,1}]_0 \times [\un_{t+1,0}]_0}\\
\end{aligned}
\end{equation}
where, for a function $f : [n_1]_0 \times [n_0]_0 \to \mathbb{R}$, we define the \textit{restricted binomial statistic quantile} $\resbinquant(\quant; (n_1, n_0), (p_1, p_0), \apairs)(f)$ as
\begin{equation*}
\begin{aligned}
\resbinquant(\quant; (n_1, n_0), (p_1, p_0), \apairs)(f) := \inf\set*{ z \in \mathbb{R} : \Prb(f(B_1, B_2) \leq z \mid (B_1, B_2) \in \apairs) \geq \quant },
\end{aligned}
\end{equation*}
where $B_1$ and $B_2$ are drawn independently from $\text{Binom}(n_1, p_1)$ and $\text{Binom}(n_0, p_0)$, respectively. It is possible to allow randomized versions of $\resbinquant$ as in Appendix~\ref{appendix:generating-qtbinom}, but we will not discuss this.

The complexity of evaluating $\resbinquant(\quant; (n_1, n_0), (p_1, p_0), \apairs)(f)$ is essentially $O(n_1 \cdot n_0)$, assuming $f$ takes constant time to evaluate. Though the additional search over $(p_1, p_0)$ makes evaluating $\critval_t$ more expensive, it is still within the capabilities of modern personal computers, especially when the sample sizes are modest. For $n_1 = n_0 = 100$, it takes a matter of seconds to calculate $\resbinquant(\quant; (n_1, n_0), (p_1, p_0), \apairs)(f)$ over 20,000 pairs of $(p_1, p_0) \in [0,1]^2$ on a personal computer. There may be further computational improvements. For instance, it may be possible that for certain choices of test statistics, the solution to the optimization is always attained on the diagonal $p_1 = p_0$.

Though these definitions may appear convoluted, the motivation for these constructions is the same as for our other tests. As we show at the end of this section,
\begin{equation}
\label{eq:binary-po-restriction-equivalence}
\begin{aligned}
\ts_s \leq \critval_s \text{ for } s=1,...,t-1 \iff \paren*{ \sum_{i=1}^n W_i Y_i \indic\set{X_i \in \reg_t}, \sum_{i=1}^n (1 - W_i) Y_i \indic\set{X_i \in \reg_t} } \in \apairs_t.
\end{aligned}
\end{equation}
Conditionally on $\F_t$, the pair on the right-hand side is distributed as $\text{Binom}(\un_{t,1}, p_1)$ in the first element and as $\text{Binom}(\un_{t,0}, p_0)$ in the second element, for some $p_1 \leq p_0$ under the null. Knowing that we have not yet rejected simply reduces the support to $\apairs_t$, and this motivates the definition of $\resbinquant$. The proof of the following theorem follows that of Appendix~\ref{appendix:proofs-binary-validity} exactly except we replace truncation by the restriction of pairs given by Equation~\eqref{eq:binary-po-restriction-equivalence}. As the proof is otherwise redundant, we omit it.

\begin{theorem}[Validity of test for binary contrasts]
\label{theorem:binary-contrast-validity}
Under the setting described in this section, let $\reg_{\tsel}$ be the region produced by Algorithm~\ref{alg:chiseling-testing} using any test statistic of the form described by Equation~\eqref{eq:binary-contrast-test-stat}, and using Equation~\eqref{eq:binary-contrast-crit-val} to compute the critical values. Then
\begin{equation*}
\begin{aligned}
\Prb\paren*{ \bin_1(\reg_{\tsel}) \leq \bin_0(\reg_{\tsel}) } \leq \alpha.
\end{aligned}
\end{equation*}
\end{theorem}

To implement this test, it is natural to let the test statistic $\ts_t$ be the standardized difference in means, which can be calculated from $\sum_{i=1}^n W_i Y_i \indic\set{X_i \in \reg_t}$, $\sum_{i=1}^n (1 - W_i)Y_i \indic\set{X_i \in \reg_t}$ and the information in $\F_t$, though other statistics may also be reasonable. We leave an empirical exploration to future work.

\begin{remark}[Targeted unmasking]
Since the relative sample sizes $\un_{t,1}$ and $\un_{t,0}$ are ancillary to the parameters of interest in this setting, whenever the analyst wishes to reveal a point at random, she may opt to \textit{selectively reveal} whether this point should be from the treated group or the control group (as opposed to revealing a point from either group at random).
\end{remark}

\begin{proof}[Proof of Equation~\eqref{eq:binary-po-restriction-equivalence}]
Define
\begin{equation*}
\begin{aligned}
\binsuft_t &:= \sum_{i=1}^n W_i Y_i \indic\set{X_i \in \reg_t} \quad \text{ and } \quad \binsufc_t &:= \sum_{i=1}^n (1 - W_i) Y_i \indic\set{X_i \in \reg_t}.
\end{aligned}
\end{equation*}
We show the equivalence in Equation~\eqref{eq:binary-po-restriction-equivalence} by induction. The case $t = 0$ is vacuously true. Suppose it is true for $t - 1$. Then the inductive hypothesis yields
\begin{equation*}
\begin{aligned}
\ts_s \leq \critval_s \text{ for } s=1,...,t-2 \iff (\binsuft_{t-1}, \binsufc_{t-1}) \in \apairs_{t-1}
\end{aligned}
\end{equation*}
which implies
\begin{equation*}
\begin{aligned}
\ts_s \leq \critval_s \text{ for } s=1,...,t-1 \iff (\binsuft_{t-1}, \binsufc_{t-1}) \in \apairs_{t-1} \text{ and } \ts_{t-1} \leq \critval_{t-1}.
\end{aligned}
\end{equation*}
Thus, it suffices to show that the right-hand side of the above is equivalent to $(\binsuft_{t}, \binsufc_{t}) \in \apairs_{t}$, which is the right-hand side of Equation~\eqref{eq:binary-po-restriction-equivalence}.

Recall that $\ts_{t-1} = \ts_{t-1}(\binsuft_{t-1}, \binsufc_{t-1})$, that $\binsuft_{t-1} - \interR_{t-1,1} = \binsuft_{t}$, and that $\binsufc_{t-1} - \interR_{t-1,0} = \binsufc_{t}$. Hence, the right-hand side of the above implies that $(\binsuft_{t}, \binsufc_{t}) = (j - \interR_{t-1,1}, k - \interR_{t-1,0})$ for some $(j, k) \in \apairs_{t-1}$ that also satisfies $\ts_{t-1}(j, k) \leq \critval_{t-1}$. Then since we additionally have $(\binsuft_{t}, \binsufc_{t}) \in \paren*{[\un_{t,1}]_0 \times [\un_{t,0}]_0}$ by construction, this yields
\begin{equation*}
\begin{aligned}
(\binsuft_{t-1}, \binsufc_{t-1}) \in \apairs_{t-1} \text{ and } \ts_{t-1} \leq \critval_{t-1} \implies (\binsuft_{t}, \binsufc_{t}) \in \apairs_{t}
\end{aligned}
\end{equation*}
by the definition of $\apairs_{t}$. To show the reverse implication, note that $(\binsuft_{t}, \binsufc_{t}) \in \apairs_{t}$ implies that $(\binsuft_{t}, \binsufc_{t}) = (j - \interR_{t-1,1}, k - \interR_{t-1,0})$ for some $(j, k) \in \apairs_{t-1}$ that also satisfies $\ts_{t-1}(j, k) \leq \critval_{t-1}$. But this uniquely identifies $j = \binsuft_{t-1}$ and $k = \binsufc_{t-1}$, so we get the reverse implication. This completes the proof by induction.
\end{proof}

\section{Proofs of main results}

First, we describe some additional notation, conventions, and basic facts that will be used throughout the proofs. For a sequence of random variables $X_n$, write $X_n \stackrel{p}{\to} \infty$ to mean that $\Prb(X_n > B) \to 1$ for every $B \in \mathbb{R}$. It will sometimes be useful for us to condition on both a $\sigma$-algebra and an event. We formalize what this means. Let $X$ be a random variable with finite first moment, $A$ an event, and $\mathcal{F}$ a $\sigma$-algebra. We use the notation
\begin{equation*}
\begin{aligned}
\E[X \mid \mathcal{F}, A] = \frac{\Ec{X \indic\set{A}}{\mathcal{F}}}{\Ec{\indic\set{A}}{\mathcal{F}}}
\end{aligned}
\end{equation*}
where we leave $0/0$ undefined. We note that
$\E[X \mid \mathcal{F}, A]$ is $\F$-measurable, and that
\begin{equation*}
\begin{aligned}
\E[X \mid \mathcal{F}, A] = \E[X \mid \mathcal{F}, \indic\set{A}] \quad \text{ a.s. conditional on } A.
\end{aligned}
\end{equation*}
For completeness, the above fact is proved as Lemma~\ref{lemma:measure-theory-conditional-prob} in Appendix~\ref{appendix:technical-proofs}. In particular, we note that $\E[X \mid \mathcal{F}, A]$ is well-defined and finite almost surely conditional on $A$. In our proofs, we will only ever consider the values of such quantities conditional on $A$, and hence our particular convention for $0/0$ will not figure into our arguments.

Let $Z$ be a discrete random variable. In general, $\E[X \mid Z]$ and $\E[X \mid h(Z)]$ are not equal unless $h$ is invertible. However, they are equal for almost all $\omega$ belonging to $\set{\omega \in \Omega : h^{-1}(h(Z(\omega))) \text{ is a singleton}}$. That is, whenever $h(Z)$ takes on a value from which $Z$ can be perfectly deduced, the two are interchangeable. For completeness, this is proved as Lemma~\ref{lemma:equivalence-local-invertible} in Appendix~\ref{appendix:technical-proofs}.

\subsection{Properties of chiseling}

\subsubsection{Preliminary results}

Though the following result is somewhat obvious, we offer a proof in Appendix~\ref{appendix:proof-distribution-min-cond} for completeness. It simply states that conditional on the clipped minimum of an i.i.d. real-valued sample, the points that do not attain the clipped minimum are i.i.d. and have distributions truncated to be above the minimum.

\begin{lemma}[Distribution conditional on minimum]
\label{lemma:distribution-min-cond}
Let $S_1,...,S_m$ be i.i.d. real-valued random variables. Let $\threshlim \in \mathbb{R} \cup \set{\infty}$ be fixed and define $\pthresh := \threshlim \wedge \min_{i=1,...,m} S_i$ and $\I := \set{i : S_i > \pthresh}$. Then the distribution of $(S_i)_{i \in \I}$ conditional on $\pthresh$ and $\I$ is the same as $\abs{\I}$ i.i.d. draws from the distribution of $S \mid S > \pthresh$.
\end{lemma}

Now we isolate the essential idea of the proof of Theorem~\ref{theorem:untarnished-chiseling} as a standalone lemma.

\begin{lemma}[General thresholding]
\label{lemma:general-thresholding}
Let $(Z_1,...,Z_m)$ be i.i.d. samples from any distribution taking values in $\mathcal{Z}$ and let $Z$ be an additional independent copy from the same distribution. Let $\threshlim \in \mathbb{R} \cup \set{\infty}$ and $g : \mathcal{Z} \to \mathbb{R}$ be fixed. Define random variables $\pthresh := \threshlim \wedge \paren{\min_{i=1,...,m} g(Z_i)}$ and $m' := \abs{\set{i : g(Z_i) > \pthresh}}$. Also define $\mathbf{Z}_{\mathrm{out}} := (Z_i)_{i : g(Z_i) \leq \pthresh}$ and $\mathbf{Z}_{\mathrm{in}} := (Z_i)_{i : g(Z_i) > \pthresh}$. Then $\pthresh$ and $m'$ are $\sigma(\mathbf{Z}_\mathrm{out})$-measurable, and the distribution of $\mathbf{Z}_{\mathrm{in}}$ given $\mathbf{Z}_\mathrm{out}$ is the same as $m'$ i.i.d. draws from the distribution of $Z \mid g(Z) > \pthresh$.
\end{lemma}

\begin{proof}
By construction, $\pthresh$ is equal to $g$ applied to any element of $\mathbf{Z}_{\mathrm{out}}$ if it is non-empty, and equal to $\threshlim$ otherwise. Also, $m'$ is equal to $m$ minus the number of elements in $\mathbf{Z}_{\mathrm{out}}$. Hence, $\pthresh$ and $m'$ are $\sigma(\mathbf{Z}_{\mathrm{out}})$-measurable. Define the random set $\I := \set{i : g(Z_i) > \pthresh}$. Note that, by a symmetry argument, conditional on $\mathbf{Z}_{\mathrm{out}}$, $\I$ is uniformly distributed over subsets of $\set{1,...,m}$ of size $m'$ independently of $\mathbf{Z}_{\mathrm{in}}$. Thus $\mathbf{Z}_{\mathrm{in}}$ is independent of $\I$ given $\mathbf{Z}_{\mathrm{out}}$.

Since $\pthresh$ and $\I$ are functions of $(Z_1,...,Z_m)$, they have a distribution implied by the distribution of $(Z_1,...,Z_m)$. We describe a way of sampling $(Z_1,...,Z_m)$ that reveals the underlying conditional independence.
\begin{enumerate}
    \item Jointly sample the pair $(\pthresh, \I)$ from its marginal distribution.
    \item For $i \not\in \I$, set $G_i = \pthresh$. Then independently sample $Z_i$ from the distribution of $Z \mid g(Z) = G_i$
    \item For $i \in \I$, independently sample $G_i$ from the distribution of $g(Z) \mid g(Z) > \pthresh$. Then independently sample $Z_i$ from the distribution of $Z \mid g(Z) = G_i$.
\end{enumerate}
In a moment, we will show that this produces $(Z_1,...,Z_m)$ from the correct distribution. But first we observe that from this the lemma follows. This is because $\mathbf{Z}_{\mathrm{out}} = (Z_i)_{i \not\in \I}$ is realized in Step~2, while $\mathbf{Z}_{\mathrm{in}}$ is precisely $(Z_i)_{i \in \I}$, whose conditional distribution given $(\mathbf{Z}_{\mathrm{out}}, \pthresh, \I)$ is specified in Step~3. There are $m'$ elements in $(Z_i)_{i \in \I}$, and each is independently obtained by first sampling $G$ from $g(Z) \mid g(Z) > \pthresh$, and then sampling $Z$ from $Z \mid g(Z) = G$. But this is the same as sampling $Z$ from $Z \mid g(Z) > \pthresh$. This gives us the desired form for the distribution of $\mathbf{Z}_{\mathrm{in}}$ given $(\mathbf{Z}_{\mathrm{out}}, \pthresh, \I)$, but earlier we showed that $\sigma(\mathbf{Z}_{\mathrm{out}}, \pthresh, \I) = \sigma(\mathbf{Z}_{\mathrm{out}}, \I)$ and that $\mathbf{Z}_{\mathrm{in}}$ is independent of $\I$ given $\mathbf{Z}_{\mathrm{out}}$. So this gives the same characterization for the distribution of $\mathbf{Z}_{\mathrm{in}}$ given $\mathbf{Z}_{\mathrm{out}}$, and it has the desired form.

Now we show that $(Z_1,...,Z_m)$ sampled in this way produces samples from the correct distribution. Since each $Z_i$ is independently sampled from the distribution of $Z \mid g(Z) = G_i$, it suffices to show that $(G_1,...,G_m)$ is an i.i.d. sample from the distribution of $(g(Z_1),...,g(Z_m))$. To see this, note that under the correct distribution, (i) $g(Z_i) = \pthresh$ for $i \not\in \I$ surely, and (ii) conditional on $(\pthresh, \I)$, the correct distribution of the subvector $(g(Z_i))_{i \in \I}$ is that it is an i.i.d. draw of $\abs{\I}$ points from the distribution of $g(Z) \mid g(Z) > \pthresh$ by Lemma~\ref{lemma:distribution-min-cond} (letting $S_i := g(Z_i)$). Then noting that in our sampler $(\pthresh, \I)$ has the correct distribution by assumption, and that both of the aforementioned properties (i) and (ii) are respected by the way our sampler constructs $(G_1,...,G_m)$ conditionally on $(\pthresh, \I)$, we conclude that $(G_1,...,G_m)$ indeed follows the distribution of $(g(Z_1),...,g(Z_m))$.
\end{proof}

Now we may establish Theorem~\ref{theorem:untarnished-chiseling}. In essence, it is simply a re-expression of Lemma~\ref{lemma:general-thresholding}, but applied conditionally on $\F$.

\subsubsection{Proof of Theorem~\ref{theorem:untarnished-chiseling}}
\label{appendix:proofs-interactive-selection}

\begin{proof}
By assumption, $(X_i, Y_i)_{i : X_i \in \reg}$ is an i.i.d. sample from the distribution of $(X, Y) \mid X \in \reg$ conditionally on $\F$. Let $\threshlim$ and $\score$ be as in Algorithm~\ref{alg:chiseling} and let $g(x, y) = \score(x)$. Note that $\threshlim$ and $\score$ are fixed conditionally on $\F$. Now, we may apply Lemma~\ref{lemma:general-thresholding} to $(X_i, Y_i)_{i : X_i \in \reg}$ conditionally on $\F$. First note that by definition,
\begin{equation*}
\begin{aligned}
\set{i : X_i \in \reg \text{ and } g(X_i, Y_i) \leq \pthresh} &= \set{i : X_i \in \reg\setminus\tilde{\reg}}\\
\set{i : X_i \in \reg \text{ and } g(X_i, Y_i) > \pthresh} &= \set{i : X_i \in \tilde{\reg}}
\end{aligned}
\end{equation*}
where we have used the fact that $g(X_i, Y_i) = \score(X_i)$. Then in Lemma~\ref{lemma:general-thresholding}, $\mathbf{Z}_{\mathrm{out}}$ corresponds to $(X_i, Y_i)_{i : X_i \in \reg\setminus\tilde{\reg}}$, while $\mathbf{Z}_{\mathrm{in}}$ corresponds to $(X_i, Y_i)_{i : X_i \in \tilde{\reg}}$. Lemma~\ref{lemma:general-thresholding}, applied conditionally on $\F$, thus allows us to conclude the following:
\begin{enumerate}
    \item $\pthresh = \threshlim \wedge \min_{i : X_i \in \reg} \score(X_i)$ is $\sigma((X_i, Y_i)_{i : X_i \in \reg\setminus\tilde{\reg}})$-measurable conditionally on $\F$, and is thus $\tilde{\F}$-measurable. Since $\tilde{\reg}$ is a function of $\reg$, $\score$, $\threshlim$, and $\pthresh$, all of which are $\tilde{\F}$-measurable, $\tilde{\reg}$ is also $\tilde{\F}$-measurable.
    \item Similarly, $n(\tilde{\reg})$ corresponds to $m'$ in the statement of Lemma~\ref{lemma:general-thresholding}, and is thus $\tilde{\F}$-measurable.
    \item The distribution of $(X_i, Y_i)_{i : X_i \in \tilde{\reg}}$ conditional on $\tilde{\F}$ is $n(\tilde{\reg})$ i.i.d. draws from the distribution of $(X, Y) \mid \set{X \in \reg \text{ and } \score(X) > \pthresh}$. But $X \in \reg$ and $\score(X) > \pthresh$ if and only if $X \in \tilde{\reg}$, so this is equivalent to the distribution of $(X, Y) \mid X \in \tilde{\reg}$, as desired.
\end{enumerate}
\end{proof}

\subsubsection{Proof of Corollary~\ref{corollary:distribution-subsamples}}
\label{appendix:proofs-distribution-subsamples}

\begin{proof}
The first part of the following corollary follows directly from recursively applying Theorem~\ref{theorem:untarnished-chiseling}. For the second part, write $Z := (X_i, Y_i)_{i : X_i \in \reg_{\genstop}}$, and $Z_t := (X_i, Y_i)_{i : X_i \in \reg_t}$. Let $h : \mathbb{R}^n \to \mathbb{R}$ be bounded and measurable. We will allow $h(\cdot)$ to be applied to a vector $z$ of length less than $n$ by simply padding $z$ with an appropriate number of zeros. Then
\begin{equation*}
\begin{aligned}
\E[h(Z) \mid \F_{\genstop}] &= \sum_{t=0}^{\infty} \E[h(Z) \mid \F_t] \indic\set{\genstop = t}\\
&= \sum_{t=0}^{\infty} \E[h(Z) \indic\set{\genstop = t} \mid \F_t]\\
&= \sum_{t=0}^{\infty} \E[h(Z_t) \indic\set{\genstop = t} \mid \F_t]\\
&= \sum_{t=0}^{\infty} \E[h(Z_t) \mid \F_t] \indic\set{\genstop = t}.
\end{aligned}
\end{equation*}
The first equality follows from Lemma~\ref{lemma:stopped-sigma-algebra}, and the remainder follows from noting that $\indic\set{\genstop = t}$ is $\F_t$-measurable (as it is a stopping time) and that $h(Z) \indic\set{\genstop = t} = h(Z_t) \indic\set{\genstop = t}$. Thus, for almost all $\omega \in \Omega$,
\begin{equation*}
\begin{aligned}
\E[h(Z) \mid \F_{\genstop}](\omega) = \sum_{t=0}^{\infty} \E[h(Z_t) \mid \F_t](\omega) \indic\set{\genstop(\omega) = t} = \E[h(Z_{\genstop(\omega)}) \mid \F_{\genstop(\omega)}](\omega).
\end{aligned}
\end{equation*}
The second equality follows from noting that the sum is nonzero only when $t = \genstop(\omega)$. Since conditional on $\F_{\genstop(\omega)}$, $Z_{\genstop(\omega)}$ is almost surely of length $m(\omega) := \set{\abs{i : X_i(\omega) \in \reg_{\genstop(\omega)}}}$, for any fixed $\omega$ it suffices to consider functions $h : \mathbb{R}^{m(\omega)} \to \mathbb{R}$, where now we drop the zero padding convention. In particular, the above holds for $h(z) := \indic\set{z \in B}$ for measurable $B \subseteq \mathbb{R}^{m(\omega)}$. But this shows that the distribution of $Z$ given $\F_{\genstop}$ at $\omega$ is the same as the distribution of $Z_t$ given $\F_t$ at $\omega$ when we let $t = \genstop(\omega)$. The latter we have already characterized as having the distribution of $\abs{\set{i : X_i(\omega) \in \reg_t(\omega)}}$ i.i.d. draws from $(X, Y) \mid X \in \reg_t(\omega)$. Simply plugging in $t = \genstop(\omega)$ yields the conclusion.
\end{proof}

\subsection{Error control for interactive testing}
\label{appendix:proofs-interactive-testing}

We prove a generalization of Lemma~\ref{lemma:sequential-error-control} which allows for slight discrepancies between the achieved level and the target levels of the conditional tests. This will be useful when we turn to asymptotic arguments later, and Lemma~\ref{lemma:sequential-error-control} follows as a special case.

\begin{lemma}[Approximate sequential error control]
\label{lemma:approx-sequential-error-control}
Let $(\genhyp_t, \tilde{\allocalpha}_t, \gentest_t)_{t=0}^m$ be a conditionally valid testing sequence with respect to a filtration $(\G_t)_{t=0}^m$, and suppose $\tilde{\allocalpha}_t = \allocalpha_t + \errordelta_t$ where $\allocalpha_t \in [0,1]$ and $\errordelta_t$ are also $\G_t$-measurable. We will think of $\allocalpha_t$ as the target levels and $\tilde{\allocalpha}_t$ as the achieved levels. Define the first rejected index $\tsel := \min\set{t : \gentest_t = 1}$ with $\tsel = \infty$ if the set is empty, and let $\genhyp_{\infty} = 1$ by convention. If the target levels satisfy $1 - \prod_{t=0}^m (1 - \allocalpha_t) \leq \alpha$ almost surely, then rejecting $\genhyp_{\tsel}$ approximately controls the Type I error rate:
\begin{equation}
\label{eq:approx-sequential-error-control}
\begin{aligned}
\Prb\paren{\genhyp_{\tsel} = 0} \leq \alpha + \E[\erroreps_0]
\end{aligned}
\end{equation}
where $\erroreps_0$ is a linear combination of $(\errordelta_t)_{t=0}^m$ with random coefficients bounded between $0$ and $1$. Furthermore, if $(\genhyp_t, \tilde{\allocalpha}_t, \gentest_t)_{t=0}^m$ is conditionally exact, $1 - \prod_{t=0}^m (1 - \allocalpha_t) = \alpha$, and $\genhyp_t = 1$ almost surely for all $t=0,...,m$ (i.e. we are under the global null), then Equation~\eqref{eq:approx-sequential-error-control} holds with equality.
\end{lemma}

Note that Lemma~\ref{lemma:sequential-error-control} is an immediate corollary of Lemma~\ref{lemma:approx-sequential-error-control} by letting $\errordelta_t = 0$ for all $t$, whence $\E[\erroreps_0] = 0$. Also note the following asymptotic corollary which places explicit growth conditions on the approximation errors $(\errordelta_t)_{t=0}^m$. To state the asymptotics, we need to define a sequence of conditionally valid testing sequences indexed by $n$. That is, let $\paren{\genhyp_t^{(n)}, \tilde{\allocalpha}_t^{(n)}, \gentest_t^{(n)}}_{t=0}^{m(n)}$ be a conditionally valid testing sequence with respect to $(\G_t^{(n)})_{t=0}^{m(n)}$ for each $n \geq 1$, and note that the total number of tests $m(n)$ can vary with $n$. We will also write $\tilde{\allocalpha}_t^{(n)} = \allocalpha_t^{(n)} + \errordelta_t^{(n)}$. Let $\tsel^{(n)}$ be the first rejected index for the $n$th conditionally valid testing sequence, and by slight abuse of notation write $\genhyp^{(n)}_{\tsel} = \genhyp_{\tsel^{(n)}}^{(n)}$.

\begin{corollary}[Asymptotic sequential error control]
\label{corollary:asymptotic-sequential-error-control}
Suppose that $\paren{\genhyp_t^{(n)}, \tilde{\allocalpha}_t^{(n)}, \gentest_t^{(n)}}_{t=0}^{m(n)}$, $(\G_t^{(n)})_{t=0}^{m(n)}$, and $(\allocalpha_t^{(n)}, \errordelta_t^{(n)})_{t=0}^{m(n)}$ satisfy the conditions of Lemma~\ref{lemma:approx-sequential-error-control} for each $n$. Furthermore, suppose that $\abs{\errordelta_0^{(n)}} + ... + \abs{\errordelta_{m(n)}^{(n)}} \stackrel{L^1}{\to} 0$ as $n \to \infty$. Then
\begin{equation*}
\begin{aligned}
\limsup_{n \to \infty} \Prb(\genhyp_{\tsel}^{(n)} = 0) \leq \alpha.
\end{aligned}
\end{equation*}
In the case where we satisfy the exactness conditions in Lemma~\ref{lemma:approx-sequential-error-control}, we have
\begin{equation*}
\begin{aligned}
\lim_{n \to \infty} \Prb(\genhyp_{\tsel}^{(n)} = 0) = \alpha.
\end{aligned}
\end{equation*}
\end{corollary}

Corollary~\ref{corollary:asymptotic-sequential-error-control} follows directly from Lemma~\ref{lemma:approx-sequential-error-control} by noting that since $\erroreps_0$ is a linear combination of $(\errordelta_t)_{t=0}^m$ with random coefficients bounded between $0$ and $1$, then (introducing an $n$ superscript on $\erroreps_0$) we have the bound
\begin{equation*}
\begin{aligned}
\abs{\E\bkt{\erroreps_0^{(n)}}} \leq \E\bkt{\abs{\erroreps_0^{(n)}}} \leq \E\bkt{\abs{\errordelta_0^{(n)}} + ... + \abs{\errordelta_{m(n)}^{(n)}}}.
\end{aligned}
\end{equation*}
Both the limit and the limit superior of the right-hand side go to $0$ as $n \to \infty$ by assumption, and the result follows.

Finally, we prove the centerpiece of this section.

\begin{proof}[Proof of Lemma~\ref{lemma:approx-sequential-error-control}]
First note that we may suppose without loss of generality that $1 - \prod_{t=0}^m (1 - \allocalpha_t) = \alpha$. If not, we can simply define an additional hypothesis $\genhyp_{m+1} = 0$, test $\gentest_{m+1} = 0$, and target level $\allocalpha_{m+1} = 1 - (1 - \alpha) \prod_{t=0}^m (1 - \allocalpha_t)^{-1}$ with no approximation error, i.e. $\errordelta_{m+1} = 0$ and $\tilde{\allocalpha}_{m+1} = \allocalpha_{m+1}$. Then since this additional test accepts almost surely, including it will not change the value of $\Prb(\genhyp_{\tsel} = 0)$. Thus, we may apply the proof below to $(\genhyp_t, \tilde{\allocalpha}_t, \gentest_t)_{t=0}^{m+1}$, which is still a conditionally valid testing sequence, and is conditionally exact if $(\genhyp_t, \tilde{\allocalpha}_t, \gentest_t)_{t=0}^m$ was conditionally exact (because then $\allocalpha_{m + 1} = 0$ almost surely).

Define $\accind_t := 1 - \gentest_t (1 - \indic\set{\tsel < t})$ and $\safeind_t := 1 - (1 - \genhyp_t) (1 - \accind_t)$. $\accind_t$ is $1$ if the $t$th hypothesis is not reported ($A$ for ``accept") either because $\gentest_t$ did not reject or because there was an earlier rejection, and $\safeind_t$ represents no mistake at stage $t$ ($S$ for ``safe"), which happens if either we don't report the $t$th hypothesis as a rejection or if the $t$th hypothesis is not null (i.e. $\genhyp_t = 1$, we are under the alternative). Also define $\GAug_t := \sigma(\G_t, (\gentest_s)_{s < t})$. Note that since $\tsel < t$ if and only if $\max_{s < t} \gentest_s = 1$, then $\accind_t, \safeind_t$ are $\GAug_{t+1}$-measurable.

First we will show that $\E[\safeind_t \mid \GAug_t] \geq 1 - \allocalpha_t - \errordelta_t$ almost surely for all $t$. We can see this by cases.
\begin{enumerate}
\item \textbf{Case 1:} Condition on $\set{\accind_0 = ... = \accind_{t-1} = 0} \cap \set{\genhyp_t = 0} \in \GAug_t$, i.e. that there has not been a rejection yet and the null hypothesis is true. Conditional on this, $\safeind_t = \accind_t = 1 - \gentest_t$ by construction, and hence $\E[\safeind_t \mid \GAug_t] \geq 1 - \tilde{\allocalpha}_t = 1 - \allocalpha_t - \errordelta_t$ by the definition of conditionally valid testing sequences (note that here we are conditioning on both $\genhyp_t = 0$ and $\max_{s < t} \gentest_s = 0$ which is implied by $\accind_0 = ... = \accind_{t-1} = 0$).
\item \textbf{Case 2:} Condition on $\set{\accind_s = 1 \text{ for some } s < t} \cup \set{\genhyp_t = 1} \in \GAug_t$, i.e. that there has been a rejection or the null hypothesis is false. Conditional on this, we either have that $\tsel < t$ and hence $\accind_t = 1$, or $\genhyp_t = 1$. Either way, we see that $\safeind_t = 1$ almost surely, and in particular $\E[\safeind_t \mid \GAug_t] = 1 \geq 1 - \tilde{\allocalpha}_t = 1 - \allocalpha_t - \errordelta_t$ since $\tilde{\allocalpha}_t \in [0,1]$.
\end{enumerate}
Since these two cases are exhaustive, we conclude that $\E[\safeind_t \mid \GAug_t] \geq 1 - \allocalpha_t - \errordelta_t$ almost surely for all $t$.

Now, define $\erroreps_m := \errordelta_m$ and recursively define for $t = m - 1,...,0$,
\begin{equation*}
\begin{aligned}
\erroreps_t := \errordelta_t (1 - \alpha) \prod_{s=0}^t (1 - \allocalpha_s)^{-1} + \safeind_t \erroreps_{t+1}.
\end{aligned}
\end{equation*}
Note that $\erroreps_t$ is a linear combination of $\errordelta_t,...,\errordelta_m$ with random coefficients bounded between $0$ and $1$. Also define $\erroreps_{m+1} = 0$ as a convention. We will show that
\begin{equation}
\begin{aligned}
\label{eq:approx-abstract-error-control-inductive-relation}
\E\bkt*{\safeind_0...\safeind_t \paren*{(1 - \alpha)\prod_{s=0}^t (1 - \allocalpha_s)^{-1} - \erroreps_{t+1}}} \geq \E\bkt*{\safeind_0...\safeind_{t-1} \paren*{(1 - \alpha) \prod_{s=0}^{t-1} (1 - \allocalpha_s)^{-1} - \erroreps_t}}
\end{aligned}
\end{equation}
for all $t = 0,...,m$, where we use the convention that empty products evaluate to $1$. To see this, note that
\begin{equation}
\label{eq:approx-abstract-error-control-main-comp}
\begin{aligned}
& \E\bkt*{\safeind_0...\safeind_t \paren*{(1 - \alpha) \prod_{s=0}^t (1 - \allocalpha_s)^{-1} - \erroreps_{t+1}}}\\
= & \E\bkt*{\safeind_0...\safeind_{t-1} \Ec*{\paren*{\safeind_t (1 - \alpha) \prod_{s=0}^t (1 - \allocalpha_s)^{-1} - \safeind_t \erroreps_{t+1}}}{\GAug_t} }\\
= & \E\bkt*{\safeind_0...\safeind_{t-1} \paren*{ \Ec*{\safeind_t}{\GAug_t} (1 - \alpha) \prod_{s=0}^t (1 - \allocalpha_s)^{-1} - \Ec*{\safeind_t \erroreps_{t+1}}{\GAug_t} } }\\
\geq & \E\bkt*{\safeind_0...\safeind_{t-1} \paren*{ (1 - \allocalpha_t - \errordelta_t) (1 - \alpha) \prod_{s=0}^t (1 - \allocalpha_s)^{-1} - \Ec*{\safeind_t \erroreps_{t+1}}{\GAug_t} } }\\
= & \E\bkt*{\safeind_0...\safeind_{t-1} \Ec*{(1 - \allocalpha_t - \errordelta_t) (1 - \alpha) \prod_{s=0}^t (1 - \allocalpha_s)^{-1} - \safeind_t \erroreps_{t+1}}{\GAug_t} }\\
= & \E\bkt*{\safeind_0...\safeind_{t-1} \Ec*{(1 - \alpha) \prod_{s=0}^{t-1} (1 - \allocalpha_s)^{-1} - \errordelta_t (1 - \alpha) \prod_{s=0}^t (1 - \allocalpha_s)^{-1} - \safeind_t \erroreps_{t+1}}{\GAug_t} }\\
= & \E\bkt*{\safeind_0...\safeind_{t-1} \Ec*{(1 - \alpha) \prod_{s=0}^{t-1} (1 - \allocalpha_s)^{-1} - \erroreps_t}{\GAug_t} }\\
= & \E\bkt*{\safeind_0...\safeind_{t-1} \paren*{(1 - \alpha) \prod_{s=0}^{t-1} (1 - \allocalpha_s)^{-1} - \erroreps_t} }.
\end{aligned}
\end{equation}
To go to the second line of the above, we use the fact that $\safeind_0,...,\safeind_{t-1}$ are $\GAug_t$-measurable and hence can be moved out of the inner conditional expectation. The third line follows from linearity of expectation and the fact that $\allocalpha_s$ is $\GAug_t$-measurable for $s \leq t$. The fourth line follows from $\E[\safeind_t \mid \GAug_t] \geq 1 - \allocalpha_t - \errordelta_t$ as we showed above. The fifth line follows from moving the central term back inside of the conditional expectation, which we can do since it consists of $\GAug_t$-measurable quantities. The sixth line follows from algebraic manipulations and the seventh line follows from the definition of $\erroreps_t$. The last line follows from moving the $\safeind_0...\safeind_{t-1}$ back inside of the conditional expectation and applying the law of total expectation. This establishes Equation~\eqref{eq:approx-abstract-error-control-inductive-relation}.

Finally, note that the right-hand side of Equation~\eqref{eq:approx-abstract-error-control-inductive-relation} is simply the left-hand side but with $t - 1$ substituted for $t$. Thus, by forming the chain of inequalities implied by Equation~\eqref{eq:approx-abstract-error-control-inductive-relation} beginning with $t = m$ and ending with $t = 0$, we obtain
\begin{equation*}
\begin{aligned}
\E[\safeind_0...\safeind_m] \geq \E[(1 - \alpha) - \erroreps_0] = 1 - \alpha - \E[\erroreps_0].
\end{aligned}
\end{equation*}
Since $\Prb(\genhyp_{\tsel} = 0) = 1 - \E[\safeind_0...\safeind_m]$, this proves the first part of the lemma.

Now we turn our attention to the second part of the lemma. Recall that we have already assumed without loss of generality that $1 - \prod_{t=0}^m (1 - \allocalpha_t) = \alpha$. We additionally assume that the testing sequence is conditionally exact and $\genhyp_t = 0$ a.s. for $t = 0,...,m$. We argue that under these additional assumptions, the inequality in Equation~\eqref{eq:approx-abstract-error-control-inductive-relation} can be turned into an equality. In particular, we argue that the sole inequality in Equation~\eqref{eq:approx-abstract-error-control-main-comp} can be turned into an equality. The reason is that now $\safeind_0...\safeind_{t-1} \E[\safeind_t \mid \GAug_t] = \safeind_0...\safeind_{t-1} (1 - \allocalpha_t - \errordelta_t)$ almost surely. We again proceed by cases. Conditional on $\accind_s = 0$ for some $s < t$, both sides of the expression are $0$ since $\genhyp_s = 0$ for all $s$ and thus $\safeind_s = 0$ for some $s < t$. Conditional on the complement, $\safeind_0...\safeind_{t-1} = 1$ and the expression reduces to $\E[\safeind_t \mid \GAug_t] = 1 - \allocalpha_t - \errordelta_t$. But this holds by the definition of conditional exactness. Using $\safeind_0...\safeind_{t-1}\E[\safeind_t \mid \GAug_t] = \safeind_0...\safeind_{t-1}(1 - \allocalpha_t - \errordelta_t)$ lets us manipulate the third line into the fourth line of Equation~\eqref{eq:approx-abstract-error-control-main-comp} with equality. Hence, Equation~\eqref{eq:approx-abstract-error-control-inductive-relation} holds with equality, and the remainder of the proof is the same with ``$\geq$" signs replaced with ``$=$" signs.
\end{proof}

\subsection{Validity results for binary outcomes}

\subsubsection{Proof of Theorem~\ref{theorem:binary-test-validity}}
\label{appendix:proofs-binary-validity}

\begin{proof}
We will make use of Lemma~\ref{lemma:sequential-error-control}. Let $(\reg_t, \allocalpha_t, \F_t)_{t=0}^{\tmax}$ be defined as in Algorithm~\ref{alg:chiseling-testing}. Let $\genhyp_t := \indic\set{\mean(\reg_t) > \cutoff}$, and let $\gentest_t$ and $\critval_t$ be defined as in the theorem statement. Let $\G_t := \sigma(\F_t, (\qaux_s)_{s < t})$ where $\qaux_t$ is the auxiliary random variable (for instance, a standard uniform independent of everything else) used to sample from $\qtbinom\paren{1 - \allocalpha_t; \un_t, \cutoff, \un_t \trunc_t}$ in the definition of $\critval_t$; in particular, $\critval_t$ is $\sigma(\F_t, (\qaux_s)_{s \leq t})$-measurable and $\trunc_t$ is $\G_t$-measurable. Then by Lemma~\ref{lemma:sequential-error-control} it suffices to show that $(\genhyp_t, \allocalpha_t, \gentest_t)_{t=0}^{\tmax}$ is conditionally valid, and conditionally exact under the additional assumptions (see Definition~\ref{def:cond-valid-test-seq}).

Let $t$ be fixed. We need to show that
\begin{equation*}
\begin{aligned}
\Prb(\meanest_t \leq \critval_t \mid \sigma(\G_t, (\gentest_s)_{s < t})) \geq 1 - \allocalpha_t \quad \text{ conditional on } \quad \set{\genhyp_t = 0 \text{ and } \max_{s < t} \gentest_s = 0}.
\end{aligned}
\end{equation*}
Implicitly condition on $\set{\genhyp_t = 0 \text{ and } \max_{s < t} \gentest_s = 0}$ throughout. We rewrite the left-hand side,
\begin{equation*}
\begin{aligned}
\Prb(\meanest_t \leq \critval_t \mid \G_t, (\gentest_s)_{s < t}) &= \Prb(\meanest_t \leq \critval_t \mid \G_t, \max_{s < t} \gentest_s)\\
&= \Prb(\meanest_t \leq \critval_t \mid \G_t, \max_{s < t} \gentest_s = 0)\\
&= \frac{\Prb(\meanest_t \leq \critval_t, \max_{s < t} \gentest_s = 0 \mid \G_t)}{\Prb(\max_{s < t} \gentest_s = 0 \mid \G_t)}\\
&= \frac{\Prb(\meanest_t \leq \critval_t, \meanest_t \leq \trunc_t \mid \G_t)}{\Prb(\meanest_t \leq \trunc_t \mid \G_t)}\\
&= \frac{\Prb(\meanest_t \leq \critval_t \mid \F_t, (\qaux_s)_{s < t})}{\Prb(\meanest_t \leq \trunc_t \mid \F_t, (\qaux_s)_{s < t})}.
\end{aligned}
\end{equation*}
The first equality follows from the fact that $\max_{s < t} \gentest_s = 0 \iff \gentest_1 = ... = \gentest_{t-1} = 0$, so we may apply Lemma~\ref{lemma:equivalence-local-invertible}. The second equality follows from Lemma~\ref{lemma:measure-theory-conditional-prob}, and the third equality is by definition. The fourth equality follows from the fact that $\max_{s < t} \gentest_s = 0$ if and only if $\meanest_t \leq \trunc_t$ (recalling the calculations in Appendix~\ref{appendix:conditioning-truncation-equivalence}). The last equality follows from expanding $\G_t$ and the fact that $\critval_t \leq \trunc_t$ almost surely by definition. Now consider a fixed $\omega \in \set{\genhyp_t = 0 \text{ and } \max_{s < t} \gentest_s = 0}$. Then
\begin{equation}
\label{eq:binary-test-validity-numerator}
\begin{aligned}
\Prb(\meanest_t \leq \critval_t \mid \F_t, (\qaux_s)_{s < t})(\omega) = \Prb(\un_t \meanest_t \leq \un_t \critval_t \mid \F_t, (\qaux_s)_{s < t})(\omega) = \Prb(Z \leq C')
\end{aligned}
\end{equation}
where $Z \sim \text{Binom}(\un_t(\omega), \mean(\reg_t(\omega)))$ and $C'$ is the independently randomized output of $\qtbinom(1 - \allocalpha_t(\omega); \un_t(\omega), \mean(\reg_t(\omega)), \un_t(\omega) \trunc_t(\omega))$. This follows from noticing a few things. First,  $\allocalpha_t, \un_t, \reg_t$, and $\trunc_t$ are $\sigma(\F_t, (\qaux_s)_{s < t})$-measurable and hence we can refer to their fixed values conditionally on $\sigma(\F_t, (\qaux_s)_{s < t})$ at $\omega$ unambiguously; this is what the notation $\allocalpha_t(\omega)$ means, and so on. Next, $\un_t \meanest_t$ is the sum of the $Y$ values of points that fall in $\reg_t(\omega)$, and Corollary~\ref{corollary:distribution-subsamples} tells us that conditionally on $\sigma(\F_t, (\qaux_s)_{s < t})$, each of these points is i.i.d. Bernoulli with rate $\mean(\reg_t(\omega))$ (note that the $\qaux_s$ are independent of everything and thus do not affect the application of Corollary~\ref{corollary:distribution-subsamples}). Thus, $\un_t \meanest_t$ is distributed as $\text{Binom}(\un_t(\omega), \mean(\reg_t(\omega)))$ conditionally on $\sigma(\F_t, (\qaux_s)_{s < t})$ at $\omega$. Next, $\un_t \critval_t$ is by definition the independently randomized output of $\qtbinom(1 - \allocalpha_t(\omega); \un_t(\omega), \mean(\reg_t(\omega)), \un_t(\omega) \trunc_t(\omega))$  given $\sigma(\F_t, (\qaux_s)_{s < t})$ at $\omega$, and in particular it is conditionally independent of $\un_t \meanest_t$. With these observations, we see that the right-hand side of Equation~\eqref{eq:binary-test-validity-numerator} is merely a re-expression of the middle term.

Following the same logic, we obtain that
\begin{equation*}
\begin{aligned}
\Prb(\meanest_t \leq \trunc_t \mid \F_t, (\qaux_s)_{s < t})(\omega) = \Prb(Z \leq M')
\end{aligned}
\end{equation*}
where $M' = \trunc_t(\omega)$. Putting this all together, we obtain that
\begin{equation*}
\begin{aligned}
\Prb(\meanest_t \leq \critval_t \mid \G_t)(\omega) = \frac{\Prb(Z \leq C')}{\Prb(Z \leq M')} = \Prb(Z \leq C' \mid Z \leq M')
\end{aligned}
\end{equation*}
for almost all $\set{\genhyp_t = 0 \text{ and } \max_{s < t} \gentest_s = 0}$, where we have again used the fact that $C' \leq M'$ a.s. Recalling that $\genhyp_t(\omega) = 0 \iff \mean(\reg_t(\omega)) \leq \cutoff$, then by stochastic domination for truncated binomials (Lemma~\ref{lemma:truncated-binomial-stochastic-dominance}) we have that the above continues simplifying as
\begin{equation}
\label{eq:binary-test-validity-domination}
\begin{aligned}
\dots = \Prb(Z \leq C' \mid Z \leq M') \geq \Prb(Z' \leq C' \mid Z' \leq M')
\end{aligned}
\end{equation}
where $Z' \sim \text{Binom}(\un_t(\omega), \cutoff)$. But $\Prb(Z' \leq C' \mid Z' \leq M') = 1 - \allocalpha_t(\omega)$ by definition, and thus we conclude that $\Prb(\meanest_t \leq \critval_t \mid \G_t) \geq 1 - \allocalpha_t$ a.s. conditionally on $\set{\genhyp_t = 0 \text{ and } \max_{s < t} \gentest_s = 0}$ as desired. This shows conditional validity, and as mentioned an application of the first part of Lemma~\ref{lemma:sequential-error-control} completes the first part of the theorem.

To show conditional exactness under the additional assumptions, we note that $\mean(X) = \cutoff$ a.s. implies that $\mean(\reg_t(\omega)) = \cutoff$ almost surely. Thus, the inequality in Equation~\eqref{eq:binary-test-validity-domination} can be replaced with an equality, allowing us to conclude that $\Prb(\meanest_t \leq \critval_t \mid \G_t) = 1 - \allocalpha_t$ a.s. conditional on $\set{\genhyp_t = 0 \text{ and } \max_{s < t} \gentest_s = 0}$. Then we have conditional exactness, and an application of the second part of Lemma~\ref{lemma:sequential-error-control} establishes the second part of the theorem.
\end{proof}

\subsection{Validity results for general outcomes}
\label{appendix:proofs-general-validity}

In Section~\ref{section:general-test-design}, we stated a simpler version of our asymptotic results for expositional clarity. Here, we will prove a more general set of claims. We briefly motivate the generalization. First of all, we relax the asymptotic conditions that are stated in the main text. Moreover, this relaxation allows us to state uniform guarantees (or equivalently, guarantees over a triangular array where the base distribution may change from row to row). Finally, we relax the requirement that the minimum sample size allocated to any non-trivial test must be a constant proportion $\minprop$ of $n$. The reason this last relaxation is appealing is that the original condition essentially precludes the discovery of a region $\reg$ whose probability mass $\Vol(\reg) := \Prb(X \in \reg)$ is much less than $\minprop$, since $\Vol(\reg_t) \approx \un_t / n$. This is generally not of major concern in key application areas such as RCTs, since the study populations are typically designed with a strong prior belief of efficacy for most if not all units in the population. Moreover, RCT sample sizes are typically modest so that extremely small subgroups will typically lack sufficient sample sizes to justify approximate inference via CLT. Nonetheless, there may be situations where one hopes to identify an extremely small subgroup from a very large dataset, perhaps arising in observational causal inference or regression settings.

We generalize both the construction of the tests and the asymptotic conditions; these are given in Appendices~\ref{appendix:formal-oracle-empirical} and~\ref{appendix:formalizing-asymptotic-conditions} respectively. Our generalized claims are stated at the end of the latter, and we devote Appendices~\ref{appendix:implied-properties}--\ref{appendix:proof-empirical-validity} to proving them. We end with a section describing primitive conditions---including those given in Section~\ref{section:general-test-design}---which imply our generalized conditions (Appendix~\ref{appendix:primitives}), and a somewhat stylized discussion of why we expect our generalization to be powerful in practice (Appendix~\ref{appendix:spacing-tests}).

Finally, we introduce some notation that we will use throughout this section. Recall that for a fixed $\reg \subseteq \mathcal{X}$ we have defined $\Vol(\reg) = \Prb(X \in \reg)$. For any fixed $\reg \subseteq \mathcal{X}$ and positive integer $k$, recall that $\mean(\reg) = \E[Y \mid X \in \reg]$ and define
\begin{equation}
\label{eq:region-moment-functions}
\begin{aligned}
\Vol(\reg) &:= \Prb(X \in \reg),\\
\Mom{k}(\reg) &:= \E[\abs{Y}^k \mid X \in \reg],\\
\CMom{k}(\reg) &:= \E[\abs{Y - \mean(\reg)}^k \mid X \in \reg],\\
\CStd{k}(\reg) &:= \sqrt{\Var(Y^k \mid X \in \reg)}.
\end{aligned}
\end{equation}
We will also write $\sigma(\reg) = \CStd{1}(\reg)$ for convenience.

\subsubsection{Formalizing the oracle and empirical tests}
\label{appendix:formal-oracle-empirical}

The proof strategy we will take is to describe an \emph{oracle} test which standardizes the critical values using the true variance of the mean estimate. Because the true variance of $\meanest_t$ is a property of the region $\reg_t$, we have that this true variance is $\F_t$-measurable and thus easier to analyze. We then show that the test which standardizes via the empirical variance produces the same output as this oracle test with probability tending to $1$ asymptotically. We devote this section to setting up the proof framework.

Let $(\reg_t, \allocalpha_t, \F_t)_{t=0}^{\tmax}$ be as defined in Algorithm~\ref{alg:chiseling-testing}. For a fixed region $\reg \subseteq \mathcal{X}$ define $\sigma^2(\reg) := \Var(Y \mid X \in \reg)$, and define
\begin{equation*}
\begin{aligned}
\sigsq_t := \sigma^2(\reg_t).
\end{aligned}
\end{equation*}
We will now use the variables $\trunc_t$ and $\critval_t$ slightly differently from how they are presented in Definition~\ref{def:general-critval}, but which are more convenient to work with. First, define the following \emph{oracle intermediate statistics} for $s \leq t$:
\begin{equation}
\label{eq:oracle-intermediate-statistics}
\begin{aligned}
\interR_{s,t} &:= \frac{1}{\sqrt{\sigsq_s} \cdot \sqrt{\un_s}} \sum_{i : X_i \in \reg_s \setminus \reg_t} Y_i \quad \text{ and } \quad \interv_{s,t} &:= \sqrt{\frac{\sigsq_t}{\sigsq_s}} \sqrt{\frac{\un_t}{\un_s}}.
\end{aligned}
\end{equation}
Define the \textit{empirical intermediate statistics} $\hat{\interR}_{s,t}$ and $\hat{\interv}_{s,t}$ identically to the above, but with $\sigsq_s$ replaced with $\hsigsq_s$ and $\sigsq_t$ replaced with $\hsigsq_t$. These quantities track various statistics that will be used to relate the test statistic at stage $s$ to the test statistic at stage $t$. We use the notation $\qtnorm(q; \trunc) := \Phi^{-1}(q \cdot \Phi(\trunc))$ to denote the $q$th quantile of $\mathcal{N}(0,1)$ truncated to be no greater than $\trunc$. We will also define $\Phi(x; \trunc) = \Phi(x) / \Phi(\trunc)$ for $x \leq \trunc$ and $\Phi(x; \trunc) = 1$ otherwise. This is the CDF of a $\mathcal{N}(0,1)$ truncated to be no greater than $\trunc$, so that $\Phi(\Phi^{-1}(q; \trunc); \trunc) = q$. We define two versions of the test statistic and critical values: an \textit{oracle} version which has access to the random variables $\sigsq_t$ and an \textit{empirical} version that uses $\hsigsq_t$ in place of $\sigsq_t$ wherever the latter appears. We also describe these tests in greater generality than what is described in the main text and will elaborate on the connection momentarily. Throughout, let $\minprop \in (0,1)$ be fixed.

\begin{definition}[Oracle normal test statistic and critical values]
\label{def:oracle-normal-test-stat-crit-val}
Define $\ts_t := \sqrt{\un_t} \cdot \sigsq_t^{-1/2} \meanest_t$. Also, recursively define
\begin{equation*}
\begin{aligned}
\trunc_t &:= \min_{s < t : \un_t / \un_s \geq \minprop} \set*{ \frac{\critval_s - \interR_{s,t}}{\interv_{s,t}} } \quad \text{ and } \quad \critval_t &:= \max\set{ 0, \Phi^{-1}(1 - \allocalpha_t; \trunc_t) }.
\end{aligned}
\end{equation*}
where the minimum is $\infty$ if the set is empty (i.e. when $t = 0$). The oracle tests are defined as $\gentest_t := \indic\set{\ts_t > \critval_t}$.
\end{definition}

\begin{definition}{(Empirical normal test statistics and critical values)}
\label{def:empirical-normal-test-stat-crit-val}
Define $\hat{\ts}_t := \sqrt{\un_t} \cdot \hsigsq_t^{-1/2} \meanest_t$. Also, recursively define
\begin{equation*}
\begin{aligned}
\hat{\trunc}_t &:= \min_{s < t : \un_t / \un_s \geq \minprop} \set*{ \frac{\hat{\critval}_s - \hat{\interR}_{s,t}}{\hat{\interv}_{s,t}} } \quad \text{ and } \quad \hat{\critval}_t &:= \max\set{ 0, \Phi^{-1}(1 - \allocalpha_t; \hat{\trunc}_t) }
\end{aligned}
\end{equation*}
where the minimum is $\infty$ if the set is empty (i.e. when $t = 0$). The empirical tests are defined as $\hat{\gentest}_t := \indic\set{\ts_t > \hat{\critval}_t}$.
\end{definition}

Note that by construction, $\Phi(\critval_t; \trunc_t) \geq 1 - \allocalpha_t$, and the same is true when $(\critval_t, \trunc_t)$ is replaced with $(\hat{\critval}_t, \hat{\trunc}_t)$. Also, $\sigsq_t, \interR_{s,t}, \interv_{s,t}, \trunc_t$, and $\critval_t$ are $\F_t$-measurable for all $s \leq t$ (though the same is not true of the hatted quantities). This important property will allow us to more easily establish the Type I error control for the oracle test.

Now we connect this to what is described in the main text. Suppose Constraint~\ref{constraint:chisel-cons} were enforced with the same $\minprop$ as defined in this section. Then under Constraint~\ref{constraint:chisel-cons}, it is the case that if $\allocalpha_t > 0$, then $\un_t / \un_s \geq \minprop$ for all $s < t$. In this case, the behavior of Algorithm~\ref{alg:chiseling-testing} is not affected by dropping the condition $\un_t / \un_s \geq \minprop$ from the definition of $\trunc_t$ in Definition~\ref{def:oracle-normal-test-stat-crit-val} and $\hat{\trunc}_t$ in Definition~\ref{def:empirical-normal-test-stat-crit-val}. From here, the rationale for these definitions is the same as in Appendix~\ref{appendix:conditioning-truncation-equivalence}. Since
\begin{equation*}
\begin{aligned}
\ts_t = \frac{\ts_s - \interR_{s,t}}{\interv_{s,t}}
\end{aligned}
\end{equation*}
for $s \leq t$, then
\begin{equation}
\begin{aligned}
\label{eq:oracle-truncation-equivalence}
\ts_s \leq \critval_s \text{ for all } s < t &\iff \ts_t = \frac{\ts_s - \interR_{s,t}}{\interv_{s,t}} \leq \frac{\critval_s - \interR_{s,t}}{\interv_{s,t}} \text{ for all } s < t\\
&\iff \ts_t \leq \min_{s < t} \set*{\frac{\critval_s - \interR_{s,t}}{\interv_{s,t}}} = \trunc_t.
\end{aligned}
\end{equation}
The same is true when $(\critval_t, \trunc_t)$ is replaced with $(\hat{\critval}_t, \hat{\trunc}_t)$. In fact, a bit of straightforward algebraic manipulation shows that the empirical test we have described here is equivalent to the test we described in the main body as long as Constraint~\ref{constraint:chisel-cons} is enforced, in the sense that $\gentest_t$ (based on Definition~\ref{def:general-critval} and used in Theorem~\ref{theorem:general-test-validity}) and $\hat{\gentest}_t$ (from Definition~\ref{def:empirical-normal-test-stat-crit-val}) are equal almost surely. Thus, the validity guarantees we will give for these tests subsume the validity guarantees stated in the main text.

When Constraint~\ref{constraint:chisel-cons} is not enforced, the rationale for these definitions is similar. As a matter of fact, Equation~\eqref{eq:oracle-truncation-equivalence} still holds up to the last line, but where now instead of having the final quantity equal to $\trunc_t$ we have
\begin{equation*}
\begin{aligned}
\min_{s < t} \set*{\frac{\critval_s - \interR_{s,t}}{\interv_{s,t}}} \leq \min_{s < t : \un_t / \un_s \geq \minprop} \set*{\frac{\critval_s - \interR_{s,t}}{\interv_{s,t}}} = \trunc_t.
\end{aligned}
\end{equation*}
That is, we truncate more conservatively, which results in a more conservative critical value $\critval_t$ and, consequently, a more conservative test, but which does not affect validity. This generalization was designed to overcome a technical difficulty in our proofs; namely, that it is somewhat difficult to characterize the effect of truncation for tests that utilize vastly different sample sizes.\footnote{In fact, this is only a difficulty for the empirical tests, which are what we ultimately care about. It is interesting to note that we can prove validity for a version of the oracle test that uses all the truncation information, suggesting that the primary difficulty is in controlling the error of the variance estimation.} Thus, these truncation levels are designed to utilize only truncation information from tests whose sample sizes are of the same order as the current sample size. This will allow us to weaken the requirement in Constraint~\ref{constraint:chisel-cons} that the minimum sample size is a constant proportion of $n$, allowing chiseling to validly discover regions of arbitrarily small probability mass. In Appendix~\ref{appendix:spacing-tests}, we give some heuristics to explain why we do not expect to lose substantial power from discarding distal truncation information. Lastly, we note that in some cases this more general construction does not discard distal truncation information at all---for instance, when Constraint~\ref{constraint:chisel-cons} is satisfied as we have seen, or when we enforce a similar constraint that the non-trivial tests (i.e. where $\allocalpha_t > 0$) are not spaced too far apart.

\begin{remark}[Undefined edge cases]
As in the main text, if ever a quantity is undefined, any test that depends upon it will be assumed to \emph{accept}. Under assumptions that we will soon state, for $t$ where $\allocalpha_t > 0$ this edge case will happen with probability going to $0$, and hence for simplicity of exposition we will simply assume that all of the above quantities are well-defined almost surely. This will already be the case if the distribution of $Y$ has no discrete atoms and $n$ is sufficiently large, and it is straightforward (though burdensome) to amend our proofs to handle undefined quantities. By, in essence, ``intersecting" every event that appears in any probability statement with the event
\begin{equation*}
E := \set{ \textup{all quantities are well-defined at stages } t \textup{ such that } \allocalpha_t > 0 },
\end{equation*}
all of our arguments may be carried out identically, and the same results follow upon noticing that $\Prb(E) \to 1$ asymptotically.
\end{remark}

\subsubsection{Formalizing the asymptotic conditions}
\label{appendix:formalizing-asymptotic-conditions}

We will prove a more general version of Theorem~\ref{theorem:general-test-validity} under more general conditions. We first state a high-level set of conditions that will suffice to imply validity. These conditions are somewhat abstract, so in Appendix~\ref{appendix:primitives} we will describe primitive conditions (including the primitives stated in Section~\ref{section:general-test-design}) that suffice to imply these high-level conditions.

Consider a triangular array of data points $(X_{i,n}, Y_{i,n})_{i=1}^n$ for $n=1,2,...$ where the $n$th row consists of $n$ i.i.d. data points drawn from the distribution of $(X^{(n)}, Y^{(n)})$. We allow the distribution of the pair of random variables to vary with $n$. Though $Y^{(n)}$ will always be real-valued, we suppose $X^{(n)} \in \mathcal{X}^{(n)}$ where $\mathcal{X}^{(n)}$ may vary with $n$. We imagine that chiseling is applied separately to each row of the triangular array. As the number of steps $\tmax$ may vary with $n$, we write $\tmax(n)$. For each $n$, this produces chiseled regions $(\reg_t^{(n)})_{t=0}^{\tmax(n)}$, levels $(\allocalpha_t^{(n)})_{t=0}^{\tmax(n)}$, and a selected region which we denote $\reg_{\tsel}^{(n)}$. Subsequently, wherever appropriate, quantities will be understood to depend on $n$ unless stated otherwise (for instance, the critical values $\critval_t$ now also depend on $n$ and should be implicitly understood as $\critval_t^{(n)}$). For readability, we will often suppress dependence on $n$ in the proofs where it is not confusing, only reintroducing the dependence when formality is required (such as in the statements of theorems). In particular, we will not write probability and expectation operators ($\Prb$ and $\E$) with explicit dependence on $n$; this also applies to the functions $\Vol(\cdot), \Mom{k}(\cdot), \CMom{k}(\cdot)$, and $\CStd{k}(\cdot)$. The following assumption, constraint, and condition completes the description of our most general asymptotic setup.\footnote{A word on nomenclature: we have designated the first of these to be an assumption as what it imposes is essentially unverifiable. The second is a constraint since the analyst can satisfy it by choice. The last requires that our sample sizes are sufficiently large; as this is not exactly unverifiable, nor exactly satisfiable by choice, we call it a condition.}

\begin{assumption}[General moment conditions]
\label{assumption:gen-dgp-regularity}
Define
\begin{equation*}
\begin{aligned}
\LFourMom^{(n)} &:= \sup_{t \in [\tmax(n)]_0 : \allocalpha_t^{(n)} > 0} \Mom{4}(\reg_t^{(n)}),\\
\LInvVar^{(n)} &:= \sup_{t \in [\tmax(n)]_0 : \allocalpha_t^{(n)} > 0} \sigma^{-2}(\reg_t^{(n)}).
\end{aligned}
\end{equation*}
Then $\LFourMom^{(n)}$ and $\LInvVar^{(n)}$ are bounded in probability.
\end{assumption}

\begin{constraint}[Tail error restriction]
\label{constraint:tail-error-restriction}
Fix $\alphamin \in (0, 1)$. Algorithm~\ref{alg:chiseling-testing} satisfies the following for all $n$ almost surely:
\begin{equation*}
\begin{aligned}
\text{If } \allocalpha_t^{(n)} > 0, \text{ then } \allocalpha_t^{(n)} \geq \alphamin.
\end{aligned}
\end{equation*}
\end{constraint}

\begin{condition}[Diverging sample size]
\label{condition:diverging-sample-size}
Define
\begin{equation*}
\begin{aligned}
\nmin^{(n)} &:= \min\set{ \un_t^{(n)} \textup{ for } t=0,...,m(n) : \allocalpha_t^{(n)} > 0}.
\end{aligned}
\end{equation*}
Then $\nmin^{(n)} \stackrel{p}{\to} \infty$.
\end{condition}

Recall that Constraint~\ref{constraint:tail-error-restriction} implies that there exists a constant $\maxreg$ such that
\begin{equation}
\label{eq:maxreg-implication}
\begin{aligned}
\abs{\set{t = 0,...,m(n) : \allocalpha_t^{(n)} > 0}} \leq \maxreg \text{ a.s. for all } n.
\end{aligned}
\end{equation}
For instance, it suffices to let $\maxreg$ be such that $1 - (1 - \alphamin)^{\maxreg} \geq \alpha$. For convenience, we will simply treat the redundant Equation~\eqref{eq:maxreg-implication} as a part of Constraint~\ref{constraint:tail-error-restriction} throughout the appendix, and we will refer to it as the ``bounded number of tested regions" constraint. In fact, Constraint~\ref{constraint:tail-error-restriction} is not exactly required to prove validity of the oracle test (Proposition~\ref{prop:oracle-subgroup-mean-validity} below); rather, proofs for the oracle test only rely on Equation~\eqref{eq:maxreg-implication}. However, Constraint~\ref{constraint:tail-error-restriction} is required for subsequent results concerning the empirical tests.

We are finally prepared to state the results we prove in this section.

\begin{prop}[Error control for oracle test]
\label{prop:oracle-subgroup-mean-validity}
When the sample size is $n$, let $\reg_{\tsel}^{(n)}$ be the region produced by Algorithm~\ref{alg:chiseling-testing} using tests of the form $\gentest_t^{(n)} := \indic\set{\ts_t^{(n)} > \critval_t^{(n)}}$ as specified by Definition~\ref{def:oracle-normal-test-stat-crit-val}. Then under Assumption~\ref{assumption:gen-dgp-regularity}, Constraint~\ref{constraint:tail-error-restriction}, and Condition~\ref{condition:diverging-sample-size},
\begin{equation*}
\begin{aligned}
\limsup_{n \to \infty} \Prb\paren*{ \mean(\reg_{\tsel}^{(n)}) \leq 0 } \leq \alpha.
\end{aligned}
\end{equation*}
\end{prop}

A proof is given in Appendix~\ref{appendix:proof-oracle-validity}. Of course, the oracle test cannot be realized in practice since it depends on the unknown variances $\sigsq_t$. The next lemma connects the oracle test to the empirical test.

\begin{lemma}[Oracle/empirical algorithm convergence.]
\label{lemma:oracle-empirical-convergence}
Let $\reg_{\tsel_{\mathrm{oracle}}}^{(n)}$ be the region outputted by Algorithm~\ref{alg:chiseling-testing} using the tests defined in Definition~\ref{def:oracle-normal-test-stat-crit-val}. Let $\reg_{\tsel_{\mathrm{empirical}}}^{(n)}$ be the region outputted by Algorithm~\ref{alg:chiseling-testing} using the tests defined in Definition~\ref{def:empirical-normal-test-stat-crit-val} applied to the same dataset as the former. Note that in both instances we assume that we select the same $\score(\cdot)$, $\threshlim$, and $\allocalpha_t$ in Algorithms~\ref{alg:chiseling-testing} and~\ref{alg:chiseling}. Then under Assumption~\ref{assumption:gen-dgp-regularity}, Constraint~\ref{constraint:tail-error-restriction}, and Condition~\ref{condition:diverging-sample-size},
\begin{equation*}
\begin{aligned}
\lim_{n \to \infty} \Prb\paren*{ \reg_{\tsel_{\mathrm{oracle}}}^{(n)} = \reg_{\tsel_{\mathrm{empirical}}}^{(n)} } = 1.
\end{aligned}
\end{equation*}
\end{lemma}

A proof is given in Appendix~\ref{appendix:proof-oracle-empirical-conv}. Hence, the empirical test inherits error control from the oracle test.

\begin{theorem}[Error control for empirical test]
\label{theorem:empirical-subgroup-mean-validity}
When the sample size is $n$, let $\reg_{\tsel}^{(n)}$ be the region produced by Algorithm~\ref{alg:chiseling-testing} using tests of the form $\hat{\gentest}_t^{(n)} := \indic\set{\ts_t^{(n)} > \hat{\critval}_t^{(n)}}$ as defined in Definition~\ref{def:empirical-normal-test-stat-crit-val}. Then under Assumption~\ref{assumption:gen-dgp-regularity}, Constraint~\ref{constraint:tail-error-restriction}, and Condition~\ref{condition:diverging-sample-size},
\begin{equation*}
\begin{aligned}
\limsup_{n \to \infty} \Prb\paren*{ \mean(\reg_{\tsel}^{(n)}) \leq 0 } \leq \alpha.
\end{aligned}
\end{equation*}
\end{theorem}

A proof is given in Appendix~\ref{appendix:proof-empirical-validity}. Finally, we reiterate how this connects to the test and guarantee described in Section~\ref{section:general-test-design}. As noted in Appendix~\ref{appendix:formal-oracle-empirical}, the test described in Definition~\ref{def:empirical-normal-test-stat-crit-val} almost surely reports the same rejections as the test described in Definition~\ref{def:general-critval} as long as Constraint~\ref{constraint:chisel-cons} is satisfied. Moreover, we will show in Appendix~\ref{appendix:primitives} that Assumption~\ref{assumption:dgp-regularity} and Constraint~\ref{constraint:chisel-cons} imply Assumption~\ref{assumption:gen-dgp-regularity}, Constraint~\ref{constraint:chisel-cons}, and Condition~\ref{condition:diverging-sample-size}. Hence, the guarantee of Theorem~\ref{theorem:empirical-subgroup-mean-validity} is more general than that of Theorem~\ref{theorem:general-test-validity}.

In the next section, we establish some intermediate properties that will be useful for proving the above results.

\subsubsection{Implications of Assumption~\ref{assumption:gen-dgp-regularity}, Constraint~\ref{constraint:tail-error-restriction}, and Condition~\ref{condition:diverging-sample-size}}
\label{appendix:implied-properties}

We state three properties that are implied by Assumption~\ref{assumption:gen-dgp-regularity}, Constraint~\ref{constraint:tail-error-restriction}, and Condition~\ref{condition:diverging-sample-size}, which will be helpful in deriving the asymptotic validity of the oracle and empirical tests for subgroup means. Note that Properties~\ref{property:var-ratio-bound}--\ref{property:consist-var} will not be used until Appendix~\ref{appendix:supporting-proofs}. In the definitions below, we take the convention that the supremum over an empty set is $0$.

\begin{property}[Uniformly consistent approximations]
\label{property:consist-approx}
Define
\begin{equation*}
\begin{aligned}
\approxerr^{(n)} := \sup_{t \in [\tmax(n)]_0 : \allocalpha_t^{(n)} > 0} \sup_{x \in \mathbb{R}} \abs*{ \Prb\paren*{\frac{\sqrt{\un_t^{(n)}}}{\sqrt{\sigsq_t^{(n)}}} \paren*{\meanest_t^{(n)} - \mean(\reg_t^{(n)})} \leq x \bigcond \F_t^{(n)}} - \Phi(x) }.
\end{aligned}
\end{equation*}
Then $\approxerr^{(n)} \stackrel{p}{\to} 0$.
\end{property}

\begin{property}[Variance ratio bound]
\label{property:var-ratio-bound}
Define
\begin{equation*}
\begin{aligned}
\vratbound^{(n)} := \sup_{s, t \in [\tmax(n)]_0 : \allocalpha_s^{(n)}, \allocalpha_t^{(n)} > 0} \frac{\un_s^{(n)} \sigsq_s^{(n)}}{\un_t^{(n)} \sigsq_t^{(n)}} \cdot \indic\set{\minprop \leq \un_s / \un_t \leq \minprop^{-1}}
\end{aligned}
\end{equation*}
Then $\vratbound^{(n)}$ is bounded in probability.
\end{property}

\begin{property}[Uniformly consistent variance estimates]
\label{property:consist-var}
Define
\begin{equation*}
\begin{aligned}
\vestbound^{(n)} := \sup_{t \in [\tmax(n)]_0 : \allocalpha_t^{(n)} > 0} \abs*{ \frac{\hsigsq_t^{(n)}}{\sigsq_t^{(n)}} - 1 }.
\end{aligned}
\end{equation*}
Then $\vestbound^{(n)} \stackrel{p}{\to} 0$.
\end{property}

Now we state the lemma.

\begin{lemma}[Implied convergence properties]
\label{lemma:implied-convergence-properties}
Assumption~\ref{assumption:gen-dgp-regularity}, Constraint~\ref{constraint:tail-error-restriction}, and Condition~\ref{condition:diverging-sample-size} imply Properties~\ref{property:consist-approx},~\ref{property:var-ratio-bound}, and~\ref{property:consist-var}.
\end{lemma}

A proof is given in Appendix~\ref{proof:lemma-implied-convergence-properties}. With these, we may establish some results.

\subsubsection{Proof of Proposition~\ref{prop:oracle-subgroup-mean-validity}}
\label{appendix:proof-oracle-validity}

\begin{proof}
The strategy we take will be nearly identical to the proof of Theorem~\ref{theorem:binary-test-validity}, except that we must keep track of an additional, vanishing error term. As a reminder, the conditions of the proposition imply, via Lemma~\ref{lemma:implied-convergence-properties}, that we have Property~\ref{property:consist-approx}. It suffices to show that that we satisfy the conditions of the asymptotic abstract error control result (Corollary~\ref{corollary:asymptotic-sequential-error-control}).

Let $(\reg_t, \allocalpha_t, \F_t)_{t=0}^{\tmax}$ be defined as in Algorithm~\ref{alg:chiseling-testing} and note that we are suppressing dependence on $n$. Let $\genhyp_t := \indic\set{\mean(\reg_t) > 0}$, and let $\gentest_t$ be defined as in the proposition statement. To apply Corollary~\ref{corollary:asymptotic-sequential-error-control}, we will let $\G_t := \F_t$. Define the following random quantities:
\begin{equation*}
\begin{aligned}
\tilde{\trunc}_t &:= \min_{s < t} \set*{ \frac{\critval_s - \interR_{s,t}}{\interv_{s,t}} }, & \text{(true trunc. level)}\\
W_t &:= \frac{\sqrt{\un_t}}{\sqrt{\sigsq_t}} \mean(\reg_t), & \text{(rescaled mean)}\\
Z_t &:= \ts_t - W_t = \frac{\sqrt{\un_t}}{\sqrt{\sigsq_t}}\paren{\meanest_t - \mean(\reg_t)}, & \text{($Z$-stat.)}\\
F_t(x; y) &:= \Prb\paren*{ Z_t \leq x \bigcond \F_t, Z_t \leq y}, & \text{(trunc. dist. of $Z$-stat.)}\\
\errordelta_t &:= \paren*{\sup_{x \geq 0, y \in \mathbb{R}} \abs*{ F_t(x; y) - \Phi(x; y) }} \cdot \indic\set{\allocalpha_t > 0}, & \text{(normal approx. error)}\\
\truncapproxerr &:= \sup_{t \in [\tmax]_0 : \allocalpha_t > 0} \sup_{x \geq 0, y \in \mathbb{R}} \abs*{ F_t(x; y) - \Phi(x; y) }. & \text{(max approx. error)}
\end{aligned}
\end{equation*}
First, we show that
\begin{equation*}
\begin{aligned}
\Prb(\ts_t \leq \critval_t \mid \sigma(\F_t, (\gentest_s)_{s < t}) ) \geq 1 - \allocalpha_t - \errordelta_t \quad \text{ conditional on } \quad \set{\genhyp_t = 0 \text{ and } \max_{s < t} \gentest_s = 0}.
\end{aligned}
\end{equation*}
We rewrite the left-hand side. Implicitly condition on $\set{\genhyp_t = 0 \text{ and } \max_{s < t} \gentest_s = 0}$ throughout. Making use of the equivalence described in Equation~\eqref{eq:oracle-truncation-equivalence}, the first few steps of the expansion are identical to the analogous expansion in the proof of Theorem~\ref{theorem:binary-test-validity}, so we can skip ahead to deduce the first equality below, and then continue expanding:
\begin{equation*}
\begin{aligned}
\Prb(\ts_t \leq \critval_t \mid \sigma(\F_t, (\gentest_s)_{s < t}) ) &= \Prb(\ts_t \leq \critval_t \mid \F_t, \ts_t \leq \tilde{\trunc}_t)\\
&= \Prb(\ts_t - W_t \leq \critval_t - W_t \mid \F_t, \ts_t - W_t \leq \tilde{\trunc}_t - W_t)\\
&= \Prb(Z_t \leq \critval_t - W_t \mid \F_t, Z_t \leq \tilde{\trunc}_t - W_t).
\end{aligned}
\end{equation*}
Next, note that since $\critval_t, \tilde{\trunc}_t,$ and $W_t$ are $\F_t$-measurable, we may make the formal substitution
\begin{equation*}
\begin{aligned}
F_t(\critval_t - W_t; \tilde{\trunc}_t - W_t) = \Prb(Z_t \leq \critval_t - W_t \mid \F_t, Z_t \leq \tilde{\trunc}_t - W_t),
\end{aligned}
\end{equation*}
and hence
\begin{equation*}
\begin{aligned}
& \abs{\Prb(Z_t \leq \critval_t - W_t \mid \F_t, Z_t \leq \tilde{\trunc}_t - W_t) - \Phi(\critval_t - W_t; \tilde{\trunc}_t - W_t)}\\
= & \abs{F_t(\critval_t - W_t; \tilde{\trunc}_t - W_t) - \Phi(\critval_t - W_t; \tilde{\trunc}_t - W_t)}\\
\leq & \sup_{x \geq 0, y \in \mathbb{R}} \abs{F_t(x; y) - \Phi(x; y)} \cdot \indic\set{\allocalpha_t > 0}\\
= & \errordelta_t.
\end{aligned}
\end{equation*}
To reach the third line of the above, we use the fact that $\critval_t \geq 0$ almost surely and that $W_t \leq 0$, so that $\critval_t - W_t \geq 0$ almost surely (conditional on $\genhyp_t = 0$). We also use the fact that when $\allocalpha_t = 0$ then $\critval_t \geq \trunc_t \geq \tilde{\trunc}_t$ almost surely and hence the difference in the second line is $0$. Putting the above together,
\begin{equation*}
\begin{aligned}
& \Prb(\ts_t \leq \critval_t \mid \sigma(\F_t, (\gentest_s)_{s < t}) )\\
= & \Phi(\critval_t - W_t; \tilde{\trunc}_t - W_t)\\
& + \Prb(Z_t \leq \critval_t - W_t \mid \F_t, Z_t \leq \tilde{\trunc}_t - W_t) - \Phi(\critval_t - W_t; \tilde{\trunc}_t - W_t)\\
\geq & \Phi(\critval_t - W_t; \tilde{\trunc}_t - W_t)\\
& - \abs{\Prb(Z_t \leq \critval_t - W_t \mid \F_t, Z_t \leq \tilde{\trunc}_t - W_t) - \Phi(\critval_t - W_t; \tilde{\trunc}_t - W_t)}\\
\geq &  \Phi(\critval_t - W_t; \tilde{\trunc}_t - W_t) - \delta_t\\
\geq & \Phi(\critval_t ; \tilde{\trunc}_t) - \delta_t\\
\geq & \Phi(\critval_t ; \trunc_t) - \delta_t\\
\geq & 1 - \allocalpha_t - \delta_t
\end{aligned}
\end{equation*}
where to reach the third to last line we recall that in this case $W_t \leq 0$ almost surely, so we may apply a stochastic dominance lemma for truncated normals (Lemma~\ref{lemma:truncated-normal-stochastic-dominance}). The second to last line follows from the fact that $\Phi(x; y)$ is decreasing in its second argument and $\tilde{\trunc}_t \leq \trunc_t$. The last line follows since $\Phi(\critval_t; \trunc_t) \geq 1 - \allocalpha_t$ by construction.

To apply Corollary~\ref{corollary:asymptotic-sequential-error-control}, it remains to be shown that the sum of the errors vanishes in $L^1$. Note that $\abs{\errordelta_t} \leq \truncapproxerr$ almost surely for all $t$, and so
\begin{equation*}
\begin{aligned}
\sum_{t=0}^{\tmax} \abs{\errordelta_t} &= \sum_{t=0}^{\tmax} \abs{\errordelta_t} \cdot \indic\set{\allocalpha_t > 0}\\
&\leq \sum_{t=0}^{\tmax} \truncapproxerr \cdot \indic\set{\allocalpha_t > 0}\\
&= \truncapproxerr \cdot \sum_{t=0}^{\tmax} \indic\set{\allocalpha_t > 0}\\
&\leq \truncapproxerr \cdot \maxreg\\
&\stackrel{p}{\to} 0.
\end{aligned}
\end{equation*}
In the above, we applied Constraint~\ref{constraint:chisel-cons} to reach the second to last line. Furthermore, Property~\ref{property:consist-approx} implies that $\truncapproxerr$ converges to $0$ in probability (Lemma~\ref{lemma:trunc-cdf-conv}), which allows us to conclude the last line. Since $\truncapproxerr \leq 1$ almost surely, by dominated convergence we can conclude that $\sum_{t=0}^{\tmax} \abs{\errordelta_t} \stackrel{L^1}{\to} 0$. Thus, we may apply Corollary~\ref{corollary:asymptotic-sequential-error-control}, and the proposition follows.
\end{proof}

\subsubsection{Proof of Lemma~\ref{lemma:oracle-empirical-convergence}}
\label{appendix:proof-oracle-empirical-conv}

Before we proceed to the proof of Lemma~\ref{lemma:oracle-empirical-convergence}, we need to show that certain quantities converge in probability. Define the following quantities that we will use throughout:
\begin{equation}
\label{eq:orac-emp-shifted-quant}
\begin{aligned}
d_t &:= \critval_{t} - \sqrt{\frac{\un_{t}}{\sigsq_{t}}} \cdot \mean(\reg_{t}), & \text{(critical value minus normalized mean)}\\
\hat{d}_t &:= \sqrt{\frac{\hsigsq_{t}}{\sigsq_{t}}} \cdot \hat{\critval}_{t} - \sqrt{\frac{\un_{t}}{\sigsq_{t}}} \cdot \mean(\reg_{t}), & \text{(empirical variant of $d_t$)}\\
Z_t &:= \ts_{t} - \sqrt{\frac{\un_{t}}{\sigsq_{t}}} \cdot \mean(\reg_{t}) & \text{($Z$-stat.)}
\end{aligned}
\end{equation}
and note that equivalently,
\begin{equation*}
\begin{aligned}
Z_t = \frac{1}{\sqrt{\sigsq_{t}} \cdot \sqrt{\un_{t}}} \sum_{i=1}^n (Y_i - \mean(\reg_{t})) \cdot \indic\set{X_i \in \reg_{t}}.
\end{aligned}
\end{equation*}

\begin{lemma}[Uniform convergence of shifted critical values]
\label{lemma:uniform-conv-shift-crit-val}
Under Assumption~\ref{assumption:gen-dgp-regularity}, Constraint~\ref{constraint:tail-error-restriction}, and Condition~\ref{condition:diverging-sample-size}, we have
\begin{equation*}
\begin{aligned}
\sup_{t \in [\tmax(n)]_0 : \allocalpha_t^{(n)} > 0} \abs{d_t^{(n)} - \hat{d}_t^{(n)}} \stackrel{p}{\to} 0
\end{aligned}
\end{equation*}
where here we take the convention that the supremum over an empty set is $0$.
\end{lemma}

A proof is given in Appendix~\ref{proof:lemma-uniform-conv-shift-crit-val}. We now prove the lemma.

\begin{proof}[Proof of Lemma~\ref{lemma:oracle-empirical-convergence}]
Since the sequence of regions produced by both the oracle and empirical algorithm are assumed to be the same (as they make the same decisions regarding $\score(\cdot)$ and $\threshlim$ in Algorithm~\ref{alg:chiseling}), it suffices to show that their first rejections coincide with probability going to $1$. In particular, it suffices to show that the probability that $\gentest_t \neq \hat{\gentest}_t$ for some $t \in [\tmax]_0$ vanishes.

Note that
\begin{equation*}
\begin{aligned}
& \gentest_t \neq \hat{\gentest}_t \text{ for some } t \in [\tmax]_0 \\
\implies & \indic\set{\ts_{t} \leq \critval_{t}} \neq \indic\set{\hat{\ts}_{t} \leq \hat{\critval}_{t}} \text{ and } \allocalpha_t > 0 \text{ for some } t \in [\tmax]_0.
\end{aligned}
\end{equation*}
Next, note that since $\sqrt{\frac{\hsigsq_t}{\sigsq_t}} \cdot \hat{\ts}_t = \ts_t$ for all $t$, we have
\begin{equation*}
\begin{aligned}
\indic\set{\ts_{t} \leq \critval_{t}} \neq \indic\set{\hat{\ts}_{t} \leq \hat{\critval}_{t}} &\iff \min\set*{\critval_{t}, \sqrt{\frac{\hsigsq_{t}}{\sigsq_{t}}} \cdot \hat{\critval}_{t}} < \ts_{t} \leq \max\set*{\critval_{t}, \sqrt{\frac{\hsigsq_{t}}{\sigsq_{t}}} \cdot \hat{\critval}_{t}}\\
&\iff \min\set{d_t, \hat{d}_t} < Z_t \leq \max\set{d_t, \hat{d}_t}
\end{aligned}
\end{equation*}
where we recall the definition of $d_t$, $\hat{d}_t$, and $Z_t$ from Equation~\eqref{eq:orac-emp-shifted-quant}. For a fixed $\delta > 0$ that we will specify later, let $E := \set{\sup_{t \in [\tmax]_0 : \allocalpha_t > 0} \abs{d_t - \hat{d}_t} < \delta}$. Then for each $t \in [\tmax]_0$,
\begin{equation*}
\begin{aligned}
\Prb(\min\set{d_t, \hat{d}_t} < Z_t \leq \max\set{d_t, \hat{d}_t}, \allocalpha_t > 0, E) &\leq \Prb(d_t - \delta < Z_t \leq d_t + \delta, \allocalpha_t > 0).
\end{aligned}
\end{equation*}
Now, conditional on $\set{\allocalpha_t > 0}$ and recalling that $\allocalpha_t$ is $\F_t$-measurable,
\begin{equation*}
\begin{aligned}
& \Prb(d_t - \delta < Z_t \leq d_t + \delta \mid \F_t)\\
= & \Prb(Z_t \leq d_t + \delta \mid \F_t) - \Prb(Z_t \leq d_t - \delta \mid \F_t)\\
= & \Prb(Z_t \leq d_t + \delta \mid \F_t) - \Phi(d_t + \delta) - \Prb(Z_t \leq d_t - \delta \mid \F_t) + \Phi(d_t - \delta)\\
& + \Phi(d_t + \delta) - \Phi(d_t - \delta)\\
\leq & \abs*{\Prb(Z_t \leq d_t + \delta \mid \F_t) - \Phi(d_t + \delta)} + \abs*{\Prb(Z_t \leq d_t - \delta \mid \F_t) - \Phi(d_t - \delta)}\\
& + \abs*{\Phi(d_t + \delta) - \Phi(d_t - \delta)}\\
\leq & \sup_{x \in \mathbb{R}} \abs*{\Prb(Z_t \leq x \mid \F_t) - \Phi(x)} + \sup_{x \in \mathbb{R}} \abs*{\Prb(Z_t \leq x \mid \F_t) - \Phi(x)}\\
& + \abs*{\Phi(d_t + \delta) - \Phi(d_t - \delta)}\\
\leq & 2\approxerr + 2\delta.
\end{aligned}
\end{equation*}
To reach the fifth line of the above, we use the fact that $d_t$ is $\F_t$-measurable and is hence a constant conditional on $\F_t$. To reach the last line, we recall the definition on $\approxerr$ from Property~\ref{property:consist-approx}, recalling that it is implied by Lemma~\ref{lemma:implied-convergence-properties}. We also use the fact that $\Phi$ is $1$-Lipschitz to bound $\abs*{\Phi(d_t + \delta) - \Phi(d_t - \delta)} \leq 2\delta$. Thus for all $t \in [\tmax]_0$,
\begin{equation}
\label{eq:orac-emp-conv-case-center}
\begin{aligned}
& \Prb(\min\set{d_t, \hat{d}_t} < Z_t \leq \max\set{d_t, \hat{d}_t}, \allocalpha_t > 0, E)\\
\leq & \E[\Prb(d_t - \delta < Z_t \leq d_t + \delta, \allocalpha_t > 0 \mid \F_t)]\\
= & \E[\Prb(d_t - \delta < Z_t \leq d_t + \delta \mid \F_t) \indic\set{ \allocalpha_t > 0}]\\
\leq & \E[(2\approxerr + 2\delta) \cdot \indic\set{\allocalpha_t > 0}].
\end{aligned}
\end{equation}
Now let $\epsilon > 0$. Choose $\delta = \epsilon / (8 \maxreg)$. Since $\approxerr \stackrel{p}{\to} 0$ by Property~\ref{property:consist-approx} and $\approxerr \leq 1$ almost surely, then dominated convergence implies that $\lim_{n \to \infty} \E[\approxerr] = 0$. In particular, we may choose $N_1$ such that $\E[\approxerr] < \epsilon / (8 \maxreg)$ for all $n > N_1$. By Lemma~\ref{lemma:uniform-conv-shift-crit-val}, we can choose $N_2$ such that $\Prb(E) > 1 - \epsilon / 2$ for all $n > N_2$. Then for all $n > \max\set{N_1, N_2}$,
\begin{equation*}
\begin{aligned}
& \Prb(\gentest_t \neq \hat{\gentest}_t \text{ for some } t \in [\tmax]_0)\\
\leq & \Prb(\min\set{d_t, \hat{d}_t} < Z_t \leq \max\set{d_t, \hat{d}_t} \text{ and } \allocalpha_t > 0 \text{ for some } t \in [\tmax]_0)\\
\leq & \Prb(E^{\comp}) + \Prb(\min\set{d_t, \hat{d}_t} < Z_t \leq \max\set{d_t, \hat{d}_t} \text{ and } \allocalpha_t > 0 \text{ for some } t \in [\tmax]_0, E)\\
\leq &  \epsilon / 2 + \sum_{t=0}^{\tmax} \Prb(\min\set{d_t, \hat{d}_t} < Z_t \leq \max\set{d_t, \hat{d}_t} \text{ and } \allocalpha_t > 0, E)\\
\leq & \epsilon / 2 + \sum_{t=0}^{\tmax} \E\bkt{(2 \approxerr + 2 \delta) \cdot \indic\set{\allocalpha_t > 0}}\\
= & \epsilon / 2 + \E\bkt*{(2 \approxerr + 2 \delta) \cdot \sum_{t=0}^{\tmax} \indic\set{\allocalpha_t > 0}}\\
\leq & \epsilon / 2 + 2\maxreg \E[\approxerr] + 2 \maxreg \delta\\
\leq & \epsilon / 2 + \epsilon / 4 + \epsilon / 4\\
= & \epsilon.
\end{aligned}
\end{equation*}
Note that we have applied Constraint~\ref{constraint:tail-error-restriction} to conclude that $\sum_{t=0}^{\tmax} \indic\set{\allocalpha_t > 0} \leq \maxreg$ almost surely. As discussed before, this establishes the result.
\end{proof}

\subsubsection{Proof of Theorem~\ref{theorem:empirical-subgroup-mean-validity}}
\label{appendix:proof-empirical-validity}

\begin{proof}
Let $\tsel_{\mathrm{oracle}}$ and $\tsel_{\mathrm{empirical}}$ be defined as in Lemma~\ref{lemma:oracle-empirical-convergence}. Then
\begin{equation*}
\begin{aligned}
& \limsup_{n \to \infty} \Prb\paren{\mean(\reg_{\tsel_{\mathrm{empirical}}}) \leq 0}\\
\leq & \limsup_{n \to \infty} \Prb\paren{\mean(\reg_{\tsel_{\mathrm{empirical}}}) \leq 0, \reg_{\tsel_{\mathrm{oracle}}} = \reg_{\tsel_{\mathrm{empirical}}}} + \Prb\paren*{\reg_{\tsel_{\mathrm{oracle}}} \neq \reg_{\tsel_{\mathrm{empirical}}}}\\
= & \limsup_{n \to \infty} \Prb\paren{\mean(\reg_{\tsel_{\mathrm{empirical}}}) \leq 0, \reg_{\tsel_{\mathrm{oracle}}} = \reg_{\tsel_{\mathrm{empirical}}}}\\
= & \limsup_{n \to \infty} \Prb\paren{\mean(\reg_{\tsel_{\mathrm{oracle}}}) \leq 0, \reg_{\tsel_{\mathrm{oracle}}} = \reg_{\tsel_{\mathrm{empirical}}}}\\
\leq & \limsup_{n \to \infty} \Prb\paren{\mean(\reg_{\tsel_{\mathrm{oracle}}}) \leq 0}\\
\leq & \alpha
\end{aligned}
\end{equation*}
where the third line follows from Lemma~\ref{lemma:oracle-empirical-convergence} and the last line follows from Proposition~\ref{prop:oracle-subgroup-mean-validity}.
\end{proof}

\subsubsection{Sufficient primitive conditions}
\label{appendix:primitives}

We describe primitive sufficient conditions where our theory applies.

\hfill

\noindent \textbf{Pointwise validity.} First, we consider the setting described in Section~\ref{section:general-test-design}. The pointwise asymptotics described in that section can be embedded into our triangular array setting by letting each row be drawn i.i.d. from the same distribution. Also, Constraint~\ref{constraint:chisel-cons} directly implies Constraint~\ref{constraint:tail-error-restriction} and Condition~\ref{condition:diverging-sample-size}. It remains to show that Assumption~\ref{assumption:dgp-regularity} and Constraint~\ref{constraint:chisel-cons} imply Assumption~\ref{assumption:gen-dgp-regularity}. For the first part,
\begin{equation*}
\begin{aligned}
\sup_{t : \allocalpha_t > 0} \Mom{4}(\reg_t) \leq \sup_{t : \allocalpha_t > 0} \frac{\E[\abs{Y}^4]}{\Vol(\reg_t)} \leq \E[\abs{Y}^4] \cdot \sup_{t : \un_t / n \geq \minprop} \Vol(\reg_t)^{-1}.
\end{aligned}
\end{equation*}
The expectation of the supremum on the right-hand side is bounded by a universal constant that only depends on $\minprop$ (Corollary~\ref{corollary:l1-bound-inverse-volume}). Therefore, introducing dependence on $n$, the right-hand side is bounded in probability by Markov's inequality. As for the inverse variance part of Assumption~\ref{assumption:gen-dgp-regularity}, this follows directly from Lemma~\ref{lemma:bound-inverse-variance}.

\hfill

\noindent \textbf{Uniform validity.} Suppose that we still enforce Constraint~\ref{constraint:chisel-cons} but now we allow the distributions to vary with $n$ in the triangular array. If we suppose that there exists constants $L, B > 0$ such that
\begin{equation*}
\begin{aligned}
\E[\abs{Y^{(n)}}^4] \leq L \quad \text{ and } \quad \Var(Y^{(n)} \mid X^{(n)}) \geq B \text{ a.s.}
\end{aligned}
\end{equation*}
for all $n$, then the fourth moment part of Assumption~\ref{assumption:gen-dgp-regularity} follows from an identical calculation as in the pointwise case, while the inverse variance part follows immediately from the fact that all region inverse variances are bounded almost surely.

\hfill

\noindent \textbf{Relaxing $\minprop$ constraint.} Lastly, suppose that we enforce Constraint~\ref{constraint:tail-error-restriction} and Condition~\ref{condition:diverging-sample-size} directly. This only requires that the minimum sample size diverges to $\infty$ at any rate. Then a variety of weak distributional assumptions suffice to imply Assumption~\ref{assumption:gen-dgp-regularity}. For instance, simply requiring that for some $M, B > 0$,
\begin{equation*}
\begin{aligned}
\abs{Y^{(n)}} \leq M \quad \text{ and } \quad \Var(Y^{(n)} \mid X^{(n)}) \geq B \text{ a.s.}
\end{aligned}
\end{equation*}
for all $n$ suffices. A weaker and more sophisticated sufficient condition is that for some $M, L, B > 0$,
\begin{equation*}
\begin{aligned}
\abs{\E[Y^{(n)} \mid X^{(n)}]} &\leq M,\\
\E[(Y^{(n)} - \E[Y^{(n)} \mid X^{(n)}])^4 \mid X^{(n)}] &\leq L,\\
\Var(Y^{(n)} \mid X^{(n)}) &\geq B
\end{aligned}
\end{equation*}
almost surely for all $n$. Note that we have relaxed $\minprop$ in the sense of getting rid of Constraint~\ref{constraint:chisel-cons}, but $\minprop$ still appears in the definitions of the tests, e.g. Definition~\ref{def:empirical-normal-test-stat-crit-val}.

\subsubsection{Discussion of distal truncation information}
\label{appendix:spacing-tests}

Our generalized test ignores distal truncation information (distal in the sense that once the sample size of an earlier test becomes incomparable to the current sample size, that truncation information is ignored). This was designed in order to overcome some technical difficulties in our proofs; however, we expect this choice to have a limited effect in practice. To get a heuristic sense for the features driving the asymptotic behavior, consider the following calculations. Let $(X, Y)$ be a fixed distribution and let $\reg^{(1)}, \reg^{(2)},... \subseteq \mathcal{X}$ be a fixed sequence of regions. Let $\rho_n := \Prb(X \in \reg^{(n)})$. We suppose that $\rho_n \to 0$ while $n \rho_n \to \infty$. Define $k_n := \sum_{i=1}^n \indic\set{X_i \in \reg^{(n)}}$ and note that $k_n \stackrel{p}{\to} \infty$ since
\begin{equation*}
\begin{aligned}
\E[k_n] = n \rho_n, \quad \Var(k_n) = n \rho_n (1 - \rho_n) \leq n
\end{aligned}
\end{equation*}
so that for all $\delta > 0$ Chebyshev's inequality yields
\begin{equation*}
\begin{aligned}
\Prb(k_n \leq n \rho_n -  \delta \sqrt{n}) \leq \frac{1}{\delta^2} \quad \text{ for all } n.
\end{aligned}
\end{equation*}
Then we simply note that $n\rho_n - \delta \sqrt{n} \to \infty$ for any $\delta > 0$. Also, $k_n / n$ concentrates around $\rho_n$, and since the latter is going to $0$ we have $k_n / n \stackrel{p}{\to} 0$.

Now define $\theta_0 := \mean(\mathcal{X})$, $\sigma^2_0 := \Var(Y)$, $\theta_n := \mean(\reg^{(n)})$, and $\sigma^2_n := \Var(Y \mid X \in \reg^{(n)})$. Then define
\begin{equation*}
\begin{aligned}
Z_n := \frac{1}{\sigma_0 \sqrt{n}} \sum_{i=1}^n (Y_i - \theta_0) \quad \text{ and } \quad W_n := \frac{1}{\sigma_n \sqrt{k_n}} \sum_{i : X_i \in \reg^{(n)}} (Y_i - \theta_n).
\end{aligned}
\end{equation*}
For simplicity suppose that $Y$ is bounded and $\Var(Y \mid X) \geq b$ a.s. for some $b > 0$; then both of the above converge to standard normal distributions. More crucially, we can show that $Z_n$ and $W_n$ are asymptotically independent, which we do at the end of this section. This result suggests why we do not expect to lose much from ignoring very distal truncation information in Definitions~\ref{def:oracle-normal-test-stat-crit-val} and~\ref{def:empirical-normal-test-stat-crit-val}. Tests that utilize comparable sample sizes pass along their truncation information, but for tests with incomparable sample sizes the truncation information is asymptotically irrelevant. Intuitively, this is because $W_n$ contributes a vanishingly small fluctuation to $Z_n$. To say something slightly more concrete: Suppose that $\theta_0 = 0$ and $\cutoff = 0$, and that we run the oracle version of Algorithm~\ref{alg:chiseling-testing} to test $H_0: \theta_0 \leq 0$, then immediately chisel to $\reg^{(n)}$ and test $\theta_n$. If we do not reject at the first stage, then we pass along the information that $Z_n \leq \critval_0$ to the test of $H_0: \theta_n \leq 0$, which is based on the distribution of $W_n$. But conditioning on $Z_n \leq \critval_0$ asymptotically does not affect the distribution of $W_n$, and thus in principle we do not sacrifice much by ignoring the truncation information, as we do in Definitions~\ref{def:oracle-normal-test-stat-crit-val} and~\ref{def:empirical-normal-test-stat-crit-val} once $k_n$ becomes sufficiently small relative to $n$.

\begin{proof}[Proof that $Z_n$ and $W_n$ are asymptotically independent]
Without loss of generality we may assume $\theta_0 = 0$ by subtracting off $\theta_0$ from $Y$. Then by straightforward algebra we may write
\begin{equation*}
\begin{aligned}
Z_n = \underbrace{\frac{\sigma_n \sqrt{k_n}}{\sigma_0 \sqrt{n}} \paren*{W_n + \frac{\sqrt{k_n}}{\sigma_n} \theta_n}}_{\delta_n} + \frac{\bar{\sigma}_n \sqrt{n - k_n}}{\sigma_0 \sqrt{n}} \times \underbrace{\frac{1}{\bar{\sigma}_n \sqrt{n - k_n}} \sum_{i : X_i \not\in \reg^{(n)}} (Y_i - \bar{\theta}_n)}_{Z_n^*} + \underbrace{\frac{n - k_n}{\sigma_0 \sqrt{n}} \bar{\theta}_n}_{\epsilon_n}
\end{aligned}
\end{equation*}
where $\bar{\theta}_n := \E[Y \mid X \not\in \reg^{(n)}]$ and $\bar{\sigma}_n^2 := \Var(Y \mid X \not\in \reg^{(n)})$. Conditionally on $W_n$ and $k_n$, we have that $\delta_n$ and $\epsilon_n$ are constants while $Z_n^*$ is approximately standard normal. In particular, the Berry-Esseen theorem gives
\begin{equation*}
\begin{aligned}
\Prb(Z_n \leq z \mid W_n, k_n) = \Phi\paren*{ (z - \delta_n - \epsilon_n) \cdot \frac{\sigma_0 \sqrt{n}}{\bar{\sigma}_n \sqrt{n - k_n}} } + \beta_n
\end{aligned}
\end{equation*}
where $\beta_n$ is the approximation error, a random variable such that $\abs{\beta_n} \leq K / \sqrt{n - k_n}$ almost surely for some universal constant $K$ (since $Y$ is bounded with lower bounded conditional variance). Then $\beta_n \to 0$ in probability. Furthermore, $\bar{\sigma}_n \to \sigma_0$ and $\sqrt{n / (n - k_n)} \stackrel{p}{\to} 1$. It remains to show that $\delta_n + \epsilon_n \stackrel{p}{\to} 0$; once we've shown this, it follows by Slutsky's theorem and the continuous mapping theorem that
\begin{equation*}
\begin{aligned}
\Prb(Z_n \leq z \mid W_n, k_n) \stackrel{p}{\to} \Phi(z)
\end{aligned}
\end{equation*}
and hence $Z_n$ is asymptotically independent of $W_n$ and $k_n$.

To see that $\delta_n + \epsilon_n \stackrel{p}{\to} 0$, we expand
\begin{equation*}
\begin{aligned}
\delta_n + \epsilon_n &= \frac{\sigma_n \sqrt{k_n}}{\sigma_0 \sqrt{n}} W_n + \frac{k_n}{\sigma_0 \sqrt{n}} \theta_n + \frac{n - k_n}{\sigma_0 \sqrt{n}} \bar{\theta}_n\\
&= \frac{\sigma_n \sqrt{k_n}}{\sigma_0 \sqrt{n}} W_n + \frac{1}{\sigma_0} \cdot \sqrt{n} \cdot \frac{k_n}{n} \theta_n + \frac{1}{\sigma_0} \cdot \sqrt{n} \cdot \frac{n - k_n}{n} \bar{\theta}_n.
\end{aligned}
\end{equation*}
The first term is easy to deal with since $W_n$ is of constant order, $\sigma_n / \sigma_0$ is bounded, and $\sqrt{k_n / n} \stackrel{p}{\to} 0$. To deal with the last two terms, we first note that
\begin{equation*}
\begin{aligned}
\rho_n \theta_n + (1 - \rho_n) \bar{\theta}_n &= \Prb(X \in \reg^{(n)}) \E[Y \mid X \in \reg^{(n)}] + \Prb(X \not\in \reg^{(n)}) \E[Y \mid X \not\in \reg^{(n)}]\\
&= \E[Y]\\
&= 0.
\end{aligned}
\end{equation*}
Then
\begin{equation*}
\begin{aligned}
\frac{1}{\sigma_0} \cdot \sqrt{n} \cdot \frac{k_n}{n} \theta_n + \frac{1}{\sigma_0} \cdot \sqrt{n} \cdot \frac{n - k_n}{n} \bar{\theta}_n = \frac{\theta_n}{\sigma_0} \cdot \underbrace{\sqrt{n} \paren*{\frac{k_n}{n} - \rho_n}}_{A_1} + \frac{\bar{\theta}_n}{\sigma_0} \cdot \underbrace{\sqrt{n} \paren*{\frac{n - k_n}{n} - (1 - \rho_n)}}_{A_2}.
\end{aligned}
\end{equation*}
Now $\theta_n / \sigma_0$ and $\bar{\theta}_n / \sigma_0$ are bounded, so it suffices to show that $A_1$ and $A_2$ converge to $0$ in probability. Note that $A_1 = -A_2$, that $\E[A_1] = 0$, and that
\begin{equation*}
\begin{aligned}
\Var(A_1) = n \cdot \frac{\rho_n (1 - \rho_n)}{n} = \rho_n (1 - \rho_n) \to 0.
\end{aligned}
\end{equation*}
Thus, $A_1$ concentrates around its mean and hence both $A_1$ and $A_2$ converge to $0$ in probability. Combining this with the above yields $\delta_n + \epsilon_n \stackrel{p}{\to} 0$ as desired.
\end{proof}

\subsection{Validity results for AIPW}
\label{appendix:proofs-aipw}

The IPW estimator is less efficient than the AIPW estimator, which reduces the variance of the ATE estimate by eliminating some nuisance variation. Moreover, in an observational study, it is often necessary to estimate the propensity scores in order to conduct inference on the ATE. Here, we describe how one can make use of the augmented IPW (AIPW) estimator in our framework. We describe the general construction and rigorously prove validity for the setting where the propensities are known (i.e. AIPW estimation in RCTs). We anticipate that the proof for the general case when propensities are also estimated from the data is similar, though we leave it for future work.

\subsubsection{Setup}
\label{appendix:aipw-setup}

We use slightly different notation from the main text. Let $(Y'_i(1), Y'_i(0))$ denote potential outcomes for unit $i$ and suppose we observe $(X_i, W_i, Y'_i)_{i=1}^n$ where $Y'_i := W_i Y'_i(1) + (1 - W_i) Y'_i(0)$ is the observed outcome, $W_i$ is a binary exposure indicator, and $X_i$ is a vector of pre-exposure covariates belonging to $\mathcal{X}$. If $g_1(\cdot)$ and $g_0(\cdot)$ are any functions and $e(x) := \Prb(W = 1 \mid X = x)$, then define the transformed outcomes
\begin{equation*}
\begin{aligned}
Y_{i,1} &:= g_1(X_i) + \frac{W_i(Y'_i - g_1(X_i))}{e(X_i)},\\
Y_{i,0} &:= g_0(X_i) - \frac{(1 - W_i)(Y'_i - g_0(X_i))}{1 - e(X_i)},
\end{aligned}
\end{equation*}
and $Y_i := Y_{i,1} - Y_{i,0}$. Then
\begin{equation*}
\begin{aligned}
\mean(\reg) := \E[Y'(1) - Y'(0) \mid X \in \reg] = \E[Y \mid X \in \reg]
\end{aligned}
\end{equation*}
as long as the latter expectation exists. Under additional standard regularity/strong overlap assumptions, the sample mean of the transformed outcomes obeys a central limit theorem. The variance of $\tilde{Y}_i$ is minimized when $g_1(\cdot)$ and $g_0(\cdot)$ are equal to $\E[Y'(1) \mid X = x]$ and $\E[Y'(0) \mid X = x]$ respectively.

When the propensity $e(\cdot)$ is known and the functions $g_1(\cdot)$ and $g_0(\cdot)$ are fixed \textit{a priori}, then we may simply run chiseling on the dataset $(\tilde{X}_i, Y_i)_{i=1}^n$ where $\tilde{X}_i := (X_i, W_i, Y'_i)$, making sure to only chisel within the $X$-dimension so that regions are subsets of $\mathcal{X}$. We include $W_i$ and $Y'_i$ as part of $\tilde{X}_i$ so that the analyst at least has access to the original pre-transformed data. The theory we have established up to now for general outcomes fully applies to this procedure as long as the transformed outcome obeys the asymptotic conditions outlined in Section~\ref{section:general-test-design} or the more general ones described in Appendix~\ref{appendix:proofs-general-validity}.

We move beyond the scope of the theory established up to now when $g_1(\cdot)$ and $g_0(\cdot)$, and possibly $e(\cdot)$, are estimated from the data. We propose, as is now common, to estimate them via cross-fitting. Let there be $\Kcv \geq 2$ folds. Let $\mathcal{I}_1,...,\mathcal{I}_{\Kcv}$ be random, approximately even partitions of $[n]$. For $w = 0,1$ and $j = 1,...,\Kcv$, let $\hat{g}_w^{(j)}(\cdot)$ be an estimate of $\E[Y(w) \mid X = x]$ and $\hat{e}^{(j)}(\cdot)$ be an estimate of $\Prb(W = 1 \mid X = x)$ using the data in $(X_i, W_i, Y'_i)_{i \not\in \mathcal{I}_j}$. Define
\begin{equation*}
\begin{aligned}
\hat{Y}_{i,1} &:= \sum_{j=1}^{\Kcv} \paren*{ \hat{g}_1^{(j)}(X_i) + \frac{W_i (Y'_i - \hat{g}_1^{(j)}(X_i))}{\hat{e}^{(j)}(X_i)}} \indic\set{i \in \mathcal{I}_j},\\
\hat{Y}_{i,0} &:= \sum_{j=1}^{\Kcv} \paren*{ \hat{g}_0^{(j)}(X_i) + \frac{(1 - W_i) (Y'_i - \hat{g}_0^{(j)}(X_i))}{1 - \hat{e}^{(j)}(X_i)}} \indic\set{i \in \mathcal{I}_j},
\end{aligned}
\end{equation*}
and $\hat{Y}_i := \hat{Y}_{i,1} - \hat{Y}_{i,0}$. The \emph{AIPW variant} of our procedure runs Algorithm~\ref{alg:chiseling-testing} on $(X_i, W_i, Y'_i)_{i=1}^n$ (only shrinking in the $X$ dimension) while using $(\hat{Y}_i)_{i=1}^n$ to calculate the tests. In particular, no subset of $(\hat{Y}_i)_{i=1}^n$ is ever revealed to the analyst.

We will establish theory for the case when the propensity $e(\cdot)$ is known and used to construct $\hat{Y}_{i}$, i.e. $\hat{e}^{(j)}(\cdot) = e(\cdot)$ for $j=1,...,\Kcv$. This is the case in randomized experiments, where the role of the AIPW adjustment is to increase efficiency by eliminating nuisance variation via $\hat{g}_w^{(j)}(\cdot)$. In this case, we only need the following assumption on the learned nuisance functions. We introduce indexing by $n$ to accomodate the triangular array asymptotic setup we described in Appendix~\ref{appendix:formalizing-asymptotic-conditions}, but will subsequently drop it for clarity.

\begin{assumption}[Stable nuisance estimation]
\label{assumption:stable-nuisance}
There exist functions $g_{1,n}(\cdot)$ and $g_{0,n}(\cdot)$ such that
\begin{equation*}
\begin{aligned}
\Lstable^{(n)} := \sup_{x \in \mathrm{supp}(X^{(n)})} \abs{\hat{g}_{w,n}^{(j)}(x) - g_{w,n}(x)} \stackrel{p}{\to} 0
\end{aligned}
\end{equation*}
for all $w = 0,1$ and $j = 1,...,\Kcv$, where $\mathrm{supp}(X^{(n)})$ denotes the support of the random variable $X^{(n)}$.
\end{assumption}

In particular, Assumption~\ref{assumption:stable-nuisance} does not require $g_w(\cdot)$ to be the actual CATE function $\E[Y(w) \mid X = x]$, though the closer it is the greater the efficiency gain. It only requires that the nuisance estimators stabilize in the sense of being coupled to some deterministic sequence, and is thus extremely mild and satisfied for reasonable low-dimensional models and well-regularized estimation procedures.

When the propensities $e(\cdot)$ must also be estimated via cross-fitting, we will at minimum need to additionally impose the usual rate conditions from semiparametric efficiency theory on the nuisance estimators. One version of these is to require that $\hat{e}^{(j)}(\cdot)$ is $\sup$ norm consistent for $e(x)$ and that
\begin{equation*}
\begin{aligned}
\E\bkt*{(\hat{g}_w^{(j)}(X) - g_w(X))^2} \E\bkt*{(\hat{e}^{(j)}(X) - e(X))^2} = o\paren*{\frac{1}{n}} \quad \text{ for } w\in\set{0,1}, j\in\set{1,...,\Kcv}.
\end{aligned}
\end{equation*}
See Chapter 3.1 on ``Double machine learning" in \textcite{wager2024causal} for a discussion of these rate conditions and exemplar calculations utilizing them. For our purposes, some additional mild regularity conditions may be required. We leave an investigation to future work.

\begin{remark}[Translation invariance]
In practice, we recommend at least letting $\hat{g}^{(j)}_1(\cdot)$ and $\hat{g}^{(j)}_0(\cdot)$ be intercept models for each treatment group calibrated via cross-fitting, since the IPW estimator has the unappealing property of lacking translation invariance, while the AIPW with intercept model is asymptotically translation invariant. Of course, greater efficiency gains are possible by leveraging simple, low-dimensional models incorporating explanatory variables.
\end{remark}

\subsubsection{Defining the AIPW-based tests}
\label{appendix:defining-aipw}

Let $(Y_i)_{i=1}^n$ and $(\hat{Y}_i)_{i=1}^n$ be defined as in Appendix~\ref{appendix:aipw-setup}. Throughout, we will suppose that chiseling only shrinks in the $X$ dimension; that is, all regions can be thought of subsets of $\mathcal{X}$. Throughout Appendix~\ref{appendix:proofs-aipw} only, for $s \leq t$ we define
\begin{equation}
\label{eq:aipw-all-statistics}
\begin{aligned}
\hat{\mu}_t &:= \frac{1}{\un_t} \sum_{i : X_i \in \reg_t} \hat{Y}_i,\\
\hsigsq_t &:= \frac{1}{\un_t} \sum_{i : X_i \in \reg_t} (\hat{Y}_i - \hat{\mu}_t)^2\\
\hat{\interR}_{s,t} &:= \frac{1}{\sqrt{\hsigsq_s} \cdot \sqrt{\un_s}} \sum_{i : X_i \in \reg_s \setminus \reg_t} \hat{Y}_i\\
\hat{\interv}_{s,t} &:= \sqrt{\frac{\hsigsq_t}{\hsigsq_s}} \sqrt{\frac{\un_t}{\un_s}}.
\end{aligned}
\end{equation}
Note that compared to the definitions in Appendix~\ref{appendix:formal-oracle-empirical}, $\hsigsq_t$, $\hat{\interR}_{s,t}$, and $\hat{\interv}_{s,t}$ not only use the empirical variances but also $\hat{Y}_i$ in place of $Y_i$. The following formalizes the AIPW test.

\begin{definition}{(AIPW test statistics and critical values)}
\label{def:aipw-test-stat-crit-val}
Define $\hat{\ts}_t := \sqrt{\un_t} \cdot \hsigsq_t^{-1/2} \hat{\mu}_t$. Also, recursively define
\begin{equation*}
\begin{aligned}
\hat{\trunc}_t &:= \min_{s < t : \un_t / \un_s \geq \minprop} \set*{ \frac{\hat{\critval}_s - \hat{\interR}_{s,t}}{\hat{\interv}_{s,t}} } \quad \text{ and } \quad \hat{\critval}_t &:= \max\set{ 0, \Phi^{-1}(1 - \allocalpha_t; \hat{\trunc}_t) }
\end{aligned}
\end{equation*}
where the minimum is $\infty$ if the set is empty (i.e. when $t = 0$). The empirical tests are defined as $\hat{\gentest}_t := \indic\set{\hat{\ts}_t > \hat{\critval}_t}$.
\end{definition}

\subsubsection{Guarantees and proof outline}

Recall the region moment functions from Equation~\eqref{eq:region-moment-functions}, which we will also use here. We have aligned our notation so that Equation~\eqref{eq:region-moment-functions} should be imported exactly as is: for instance, $\Mom{k}(\reg) = \E[\abs{Y}^k \mid X \in \reg] \neq \E[\abs{Y'}^k \mid X \in \reg]$, which is an important distinction. The former refers to the moments of the oracle transformed outcome (defined in Appendix~\ref{appendix:aipw-setup}), while the latter refers to moments of the raw outcome that the analyst sees. We can establish the following.

\begin{theorem}[Validity of AIPW test]
\label{theorem:aipw-validity}
Suppose that Algorithm~\ref{alg:chiseling-testing} is run on the dataset $(X_{i,n}, W_{i,n}, Y'_{i,n})_{i=1}^n$ while only shrinking in the $X$ dimension so that the $n$th region is a subset of $\mathcal{X}^{(n)}$, the ambient space of $X^{(n)}$. Suppose that the chiseled regions satisfy Assumption~\ref{assumption:gen-dgp-regularity} and that we additionally satisfy Assumption~\ref{assumption:stable-nuisance}, Constraint~\ref{constraint:tail-error-restriction}, and Condition~\ref{condition:diverging-sample-size}. Let $\reg_{\tsel}^{(n)}$ be the region reported by Algorithm~\ref{alg:chiseling-testing} when the sample size is $n$, using the tests $\hat{\gentest}_t^{(n)}$ given in Definition~\ref{def:aipw-test-stat-crit-val}. Lastly, suppose that we use the true propensities, i.e. $e^{(j)}_n(x) = e_n(x) = \Prb(W^{(n)} = 1 \mid X^{(n)} = x)$ for all $j$, and that the propensities satisfy strong overlap, i.e. $\eta \leq e_n(X) \leq 1 - \eta$ almost surely for some $0 < \eta \leq 0.5$. Then
\begin{equation*}
\begin{aligned}
\limsup_{n \to \infty} \Prb\paren*{ \mean(\reg_{\tsel}^{(n)}) \leq 0 } \leq \alpha.
\end{aligned}
\end{equation*}
\end{theorem}

To reiterate: when we read Assumption~\ref{assumption:gen-dgp-regularity} in the above, we should understand it as applying to the moments of the oracle transformed outcomes, not the raw outcomes. We prove Theorem~\ref{theorem:aipw-validity} in the next subsubsection (Appendix~\ref{appendix:aipw-validity-proof}).

\subsubsection{Proof of Theorem~\ref{theorem:aipw-validity}}
\label{appendix:aipw-validity-proof}

First, we note that the AIPW instance of chiseling which uses Definition~\ref{def:aipw-test-stat-crit-val} to calculate the tests can be coupled with an oracle instance of chiseling that uses the dataset $(\tilde{X}_i, Y_i)_{i=1}^n$ where $\tilde{X}_i := (X_i, W_i, Y_i')$ and which uses Definition~\ref{def:oracle-normal-test-stat-crit-val} to calculate the tests. Note that this oracle instance (1) uses $(Y_i)_{i=1}^n$ to calculate the test, as opposed to the AIPW instance which uses $(\hat{Y}_i)_{i=1}^n$, and (2) uses the true region variances $\sigsq_t := \sigma^2(\reg_t)$ to calculate the oracle critical values $\critval_t$. We can specify these two instances of chiseling to produce the same sequence of regions $(\reg_t)_{t=0}^{\tmax}$ because the oracle has as much information as the AIPW instance (as encoded in $\tilde{X}_i$). We assume that these two instances also set the same sequence of target levels $(\allocalpha_t)_{t=0}^{\tmax}$.

Noting that the conditions of this theorem include those necessary to apply Proposition~\ref{prop:oracle-subgroup-mean-validity}, we have that the oracle instance of chiseling is asymptotically valid. We wish to show now that the AIPW instance rejects the same region as the oracle instance with probability going to $1$. In this proof, define
\begin{equation}
\label{eq:aipw-orac-emp-shifted-quant}
\begin{aligned}
R_t &:= \frac{1}{\sqrt{\sigsq_t} \cdot \sqrt{\un_t}} \sum_{i : X_i \in \reg_t} (\hat{Y}_i - Y_i), & \text{(remainder)}\\
d_t &:= \critval_{t} - \sqrt{\frac{\un_{t}}{\sigsq_{t}}} \cdot \mean(\reg_{t}), & \text{(critical value minus normalized mean)}\\
\hat{d}_t &:= \sqrt{\frac{\hsigsq_{t}}{\sigsq_{t}}} \cdot \hat{\critval}_{t} - R_t - \sqrt{\frac{\un_{t}}{\sigsq_{t}}} \cdot \mean(\reg_{t}), & \text{(empirical variant of $d_t$)}\\
Z_t &:= \ts_{t} - \sqrt{\frac{\un_{t}}{\sigsq_{t}}} \cdot \mean(\reg_{t}) & \text{($Z$-stat.)}
\end{aligned}
\end{equation}
We will need the following result, which is analogous to Lemma~\ref{lemma:uniform-conv-shift-crit-val}.

\begin{lemma}[Uniform convergence of AIPW shifted critical values]
\label{lemma:aipw-uniform-conv-shift-crit-val}
Under the conditions of the Theorem~\ref{theorem:aipw-validity}, we have
\begin{equation*}
\begin{aligned}
\sup_{t \in [\tmax(n)]_0 : \allocalpha_t^{(n)} > 0} \abs{d_t^{(n)} - \hat{d}_t^{(n)}} \stackrel{p}{\to} 0
\end{aligned}
\end{equation*}
where here we take the convention that the supremum over an empty set is $0$.
\end{lemma}

We prove this in Appendix~\ref{proof:lemma-aipw-uniform-conv-shift-crit-val}. Now letting $\ts_t$ be the oracle test statistic given by Definition~\ref{def:oracle-normal-test-stat-crit-val}, note that $\hat{\ts}_t = \sqrt{\frac{\sigsq_t}{\hsigsq_t}} \cdot \paren*{ \ts_t + R_t }$, and thus
\begin{equation*}
\begin{aligned}
& \indic\set{\ts_t \leq \critval_t} \neq \indic\set{\hat{\ts}_t \leq \hat{\critval}_t}\\
\iff & \min\set*{ \critval_t, \sqrt{\frac{\hsigsq_t}{\sigsq_t}} \cdot \hat{\critval}_t - R_t } < \ts_t \leq \max\set*{ \critval_t, \sqrt{\frac{\hsigsq_t}{\sigsq_t}} \cdot \hat{\critval}_t - R_t }\\
\iff & \min\set{d_t, \hat{d}_t} < Z_t \leq \max\set{d_t, \hat{d}_t}.
\end{aligned}
\end{equation*}
Given Lemma~\ref{lemma:aipw-uniform-conv-shift-crit-val}, the proof that the AIPW instance of chiseling and the oracle instance of chiseling reject the same region with probability tending to $1$ follows the proof of Lemma~\ref{lemma:oracle-empirical-convergence} in Appendix~\ref{appendix:proof-oracle-empirical-conv} exactly. Then the asymptotic validity of the AIPW instance of chiseling follows from copying the steps of the proof of Theorem~\ref{theorem:empirical-subgroup-mean-validity} in Appendix~\ref{appendix:proof-empirical-validity} exactly.

\section{Proofs of supporting lemmas}
\label{appendix:supporting-proofs}

\subsection{Proof of Lemma~\ref{lemma:implied-convergence-properties}}
\label{proof:lemma-implied-convergence-properties}

For readability, we separate this into three separate proofs, one for each condition.

\begin{proof}[Proof of Lemma~\ref{lemma:implied-convergence-properties} ($\implies$ Property~\ref{property:consist-approx})]
Recall that Corollary~\ref{corollary:distribution-subsamples} states that conditional on $\F_t$, the $\un_t$ pairs that fall within the region $\reg_t$ are an i.i.d. sample of $\un_t$ pairs where each pair has the distribution $(X, Y) \mid X \in \reg_t$. Then by the Berry--Esseen theorem, for some absolute constant $\berryessbound$, the distance between the empirical CDF of the standardized sample mean in the region and the standard normal distribution can be bounded by
\begin{equation*}
\begin{aligned}
\sup_{x \in \mathbb{R}} \abs*{ \Prb\paren*{\frac{\sqrt{\un_t}}{\sqrt{\sigsq_t}} \paren*{\meanest_t - \mean(\reg_t)} \leq x \bigcond \F_t} - \Phi(x) } \leq \frac{\berryessbound \cdot \CMom{3}(\reg_t)}{\CStd{1}^3(\reg_t) \cdot \sqrt{\un_t}} \quad \text{ a.s.}
\end{aligned}
\end{equation*}
Then the supremum of the left-hand side over $t \in [\tmax]_0$ such that $\allocalpha_t > 0$ is almost surely upper bounded by
\begin{equation*}
\begin{aligned}
\sup_{t \in [\tmax]_0 : \allocalpha_t > 0} \frac{\berryessbound \cdot \CMom{3}(\reg_t)}{\CStd{1}^3(\reg_t) \cdot \sqrt{\un_t}} \leq \berryessbound \cdot \nmin^{-1/2} \cdot \paren*{\sup_{t \in [\tmax]_0 : \allocalpha_t > 0} \CMom{3}(\reg_t)} \paren*{\sup_{t \in [\tmax]_0 : \allocalpha_t > 0} \CStd{1}^{-3}(\reg_t)}
\end{aligned}
\end{equation*}
where we have used the fact that $\sup_{t \in [\tmax]_0 : \allocalpha_t > 0} \un_t^{-1/2} = \nmin^{-1/2}$ by definition. Since $\nmin \stackrel{p}{\to} \infty$ by Condition~\ref{condition:diverging-sample-size}, it suffices that the last two supremum terms above are bounded in probability. The second supremum is directly bounded in probability by Assumption~\ref{assumption:gen-dgp-regularity}. For the first supremum, we note that for all fixed $\reg \subseteq \mathcal{X}$,
\begin{equation*}
\begin{aligned}
\CMom{3}(\reg) = \E[\abs{Y - \mean(\reg)}^3 \mid X \in \reg] \leq 8 \cdot \E[\abs{Y}^3 \mid X \in \reg] = 8 \cdot \Mom{3}(\reg)
\end{aligned}
\end{equation*}
by a standard bound on centered moments by raw moments (Lemma~\ref{lemma:raw-bound-center-moment}) applied to the random variable with distribution $Y \mid \set{X \in \reg}$. Then
\begin{equation*}
\begin{aligned}
\sup_{t \in [\tmax]_0 : \allocalpha_t > 0} \CMom{3}(\reg_t) \leq 8 \cdot \sup_{t \in [\tmax]_0 : \allocalpha_t > 0} \Mom{3}(\reg_t)
\end{aligned}
\end{equation*}
which is bounded in probability by Assumption~\ref{assumption:gen-dgp-regularity} (note that boundedness in probability of the fourth absolute region conditional moment automatically translates to boundedness in probability for lower absolute region conditional moments; we may see this formally by applying standard $L^p$ norm inequalities). Hence, the result follows.
\end{proof}

\begin{proof}[Proof of Lemma~\ref{lemma:implied-convergence-properties} ($\implies$ Property~\ref{property:var-ratio-bound})]

Note that by bounding the supremum of products by the product of supremums, we can almost surely bound $\vratbound$ by the product of the following quantities:
\begin{equation*}
\begin{aligned}
S_1 &:= \sup_{s, t \in [\tmax]_0 : \allocalpha_s, \allocalpha_t > 0} \frac{\un_s}{\un_t} \cdot \indic\set{\minprop \leq \un_s / \un_t \leq \minprop^{-1}},\\
S_2 &:= \sup_{s, t \in [\tmax]_0 : \allocalpha_s, \allocalpha_t > 0} \sigsq_s = \sup_{t \in [\tmax]_0 : \allocalpha_t > 0} \sigsq_t,\\
S_3 &:= \sup_{s, t \in [\tmax]_0 : \allocalpha_s, \allocalpha_t > 0} \sigsq_t^{-1} = \sup_{t \in [\tmax]_0 : \allocalpha_t > 0} \sigsq_t^{-1}.
\end{aligned}
\end{equation*}
Note that $S_1 \leq \minprop^{-1}$ by definition and $S_3$ is bounded in probability by Assumption~\ref{assumption:gen-dgp-regularity}. As for $S_2$, we note that $\sigsq_t \leq \Mom{2}(\reg_t)$ almost surely. Then by Assumption~\ref{assumption:gen-dgp-regularity}, which requires that $\sup_{t \in [\tmax]_0 : \allocalpha_t > 0} \Mom{4}(\reg_t)$ is bounded in probability, $S_2$ is also bounded in probability. Hence, the product and thus $\vratbound$ are bounded in probability.
\end{proof}

\begin{proof}[Proof of Lemma~\ref{lemma:implied-convergence-properties} ($\implies$ Property~\ref{property:consist-var})]
First note that
\begin{equation*}
\begin{aligned}
\sup_{t \in [\tmax]_0 : \allocalpha_t > 0} \abs*{ \frac{\hsigsq_t}{\sigsq_t} - 1 } = \sup_{t \in [\tmax]_0 : \allocalpha_t > 0} \abs*{ \frac{\hsigsq_t - \sigsq_t}{\sigsq_t}} \leq \sup_{t \in [\tmax]_0 : \allocalpha_t > 0} \sigsq_t^{-1} \cdot \sup_{t \in [\tmax]_0 : \allocalpha_t > 0} \abs{\hsigsq_t - \sigsq_t}.
\end{aligned}
\end{equation*}
The first supremum on the right-hand side is bounded in probability by Assumption~\ref{assumption:gen-dgp-regularity}, so it suffices to show that the second supremum converges to $0$ in probability. Write
\begin{equation*}
\begin{aligned}
\momtwoest_t = \frac{1}{\un_t} \sum_{i=1}^n Y_i^2 \indic\set{X_i \in \reg_t}
\end{aligned}
\end{equation*}
and note that $\hsigsq_t = \momtwoest_t - \meanest_t^2$. Also note that $\sigsq_t = \Mom{2}(\reg_t) - \mean(\reg_t)^2$. Then
\begin{equation}
\label{eq:implication-var-split-sup}
\begin{aligned}
\sup_{t \in [\tmax]_0 : \allocalpha_t > 0} \abs{\hsigsq_t - \sigsq_t} \leq \sup_{t \in [\tmax]_0 : \allocalpha_t > 0} \abs{\momtwoest_t - \Mom{2}(\reg_t)} + \sup_{t \in [\tmax]_0 : \allocalpha_t > 0} \abs{\meanest_t^2 - \mean(\reg_t)^2}.
\end{aligned}
\end{equation}
We deal with each term separately. Let $\delta > 0$ and focus on the first term. Conditional on $\F_t$, Corollary~\ref{corollary:distribution-subsamples} tells us that $\momtwoest_t$ is just the simple average of $\un_t$ i.i.d. points, each of which has mean $\Mom{2}(\reg_t)$ and variance $\CStd{2}^2(\reg_t)$. Hence $\momtwoest_t$ is conditionally unbiased for $\Mom{2}(\reg_t)$ and has variance $\CStd{2}^2(\reg_t) / \un_t$, so that Chebyshev's inequality yields
\begin{equation*}
\begin{aligned}
\Prb\paren*{\abs{\momtwoest_t - \Mom{2}(\reg_t)} > \delta \mid \F_t} \leq \min\set*{1,  \frac{\CStd{2}^2(\reg_t)}{\un_t \cdot \delta^2} } \leq \min\set*{1,  \frac{\Mom{4}(\reg_t)}{\un_t \cdot \delta^2} }
\end{aligned}
\end{equation*}
and thus
\begin{equation*}
\begin{aligned}
\Prb\paren*{\abs{\momtwoest_t - \Mom{2}(\reg_t)} > \delta \mid \F_t} \cdot \indic\set{\allocalpha_t > 0} \leq \min\set*{1,  \frac{\Mom{4}(\reg_t)}{\nmin \cdot \delta^2} } \cdot \indic\set{\allocalpha_t > 0}.
\end{aligned}
\end{equation*}
Then
\begin{equation*}
\begin{aligned}
\Prb\paren*{ \sup_{t \in [\tmax]_0 : \allocalpha_t > 0} \abs{\momtwoest_t - \Mom{2}(\reg_t)} > \delta} \leq & \sum_{t=0}^{\tmax} \Prb\paren*{\abs{\momtwoest_t - \Mom{2}(\reg_t)} > \delta \text{ and } \allocalpha_t > 0}\\
= &  \sum_{t=0}^{\tmax} \E\bkt*{ \Prb\paren*{\abs{\momtwoest_t - \Mom{2}(\reg_t)} > \delta \mid \F_t } \cdot \indic\set{\allocalpha_t > 0}}\\
\leq & \E\bkt*{ \min\set*{1, \sup_{t \in [\tmax]_0 : \allocalpha_t > 0} \frac{\Mom{4}(\reg_t)}{\nmin \cdot \delta^2} } \cdot \sum_{t=0}^{\tmax} \indic\set{\allocalpha_t > 0} }\\
\leq & \maxreg \cdot \E\bkt*{ \min\set*{1, \sup_{t \in [\tmax]_0 : \allocalpha_t > 0} \frac{\Mom{4}(\reg_t)}{\nmin \cdot \delta^2} } }\\
\to & 0
\end{aligned}
\end{equation*}
where to go to the second to last line we use Constraint~\ref{constraint:tail-error-restriction} to bound the sum of the indicators $\indic\set{\allocalpha_t > 0}$ by $\maxreg$. To go the last line we use the fact that $\sup_{t \in [\tmax]_0 : \allocalpha_t > 0} \Mom{4}(\reg_t)$ is bounded in probability by Assumption~\ref{assumption:gen-dgp-regularity} and that $\nmin \stackrel{p}{\to} \infty$ by Condition~\ref{condition:diverging-sample-size}, so that the minimum term converges to $0$ in probability and hence in $L^1$ because it is bounded by $1$.

Now we consider the second supremum in Equation~\eqref{eq:implication-var-split-sup}. First, we control the convergence of the mean estimate. Again, conditional on $\F_t$, Corollary~\ref{corollary:distribution-subsamples} tells us that $\meanest_t$ is just the simple average of $\un_t$ i.i.d. points, each of which has mean $\mean(\reg_t)$ and variance $\CStd{1}^2(\reg_t)$, so that $\meanest_t$ is conditionally unbiased for $\mean(\reg_t)$ and has variance $\CStd{1}^2(\reg_t) / \un_t$. Then we may carry out the same steps as before to conclude that
\begin{equation*}
\begin{aligned}
\sup_{t \in [\tmax]_0 : \allocalpha_t > 0} \abs{\meanest_t - \mean(\reg_t)} \stackrel{p}{\to} 0.
\end{aligned}
\end{equation*}
Now since $\abs{\meanest_t^2 - \mean(\reg_t)^2} = \abs{\meanest_t + \mean(\reg_t)} \abs{\meanest_t - \mean(\reg_t)}$, we have
\begin{equation*}
\begin{aligned}
\sup_{t \in [\tmax]_0 : \allocalpha_t > 0} \abs{\meanest_t^2 - \mean(\reg_t)^2} \leq \sup_{t \in [\tmax]_0 : \allocalpha_t > 0} \abs{\meanest_t + \mean(\reg_t)} \cdot \sup_{t \in [\tmax]_0 : \allocalpha_t > 0} \abs{\meanest_t - \mean(\reg_t)}.
\end{aligned}
\end{equation*}
The second supremum converges to $0$ in probability, so it suffices to show that the first supremum is bounded in probability. We can see this from
\begin{equation*}
\begin{aligned}
\sup_{t \in [\tmax]_0 : \allocalpha_t > 0} \abs{\meanest_t + \mean(\reg_t)} \leq \sup_{t \in [\tmax]_0 : \allocalpha_t > 0} \abs{\meanest_t - \mean(\reg_t)} + \sup_{t \in [\tmax]_0 : \allocalpha_t > 0}  2\abs{\mean(\reg_t)}.
\end{aligned}
\end{equation*}
The first term converges to $0$ in probability. Since $\abs{\mean(\reg_t)} \leq \Mom{1}(\reg)$, Assumption~\ref{assumption:gen-dgp-regularity} implies that the second term is bounded in probability. Thus, the sum is bounded in probability, and from this, as discussed, the second supremum in Equation~\eqref{eq:implication-var-split-sup} converges to $0$ in probability. From this, the main result follows.
\end{proof}

\subsection{Proof of Lemma~\ref{lemma:uniform-conv-shift-crit-val}}
\label{proof:lemma-uniform-conv-shift-crit-val}

Before we proceed to the proof of Lemma~\ref{lemma:uniform-conv-shift-crit-val}, we first establish that a few intermediate quantities converge. Let us introduce some notation to track the indices of stages where the analyst allocates non-zero error to the test. Since by Constraint~\ref{constraint:chisel-cons} at most $\maxreg$ stages result in non-trivial tests, we can define $t_j$ to be the stage number of the $j$th non-trivial test for $j=1,...,\maxreg$. Formally,
\begin{equation*}
\begin{aligned}
t_1 &:= \min\set{ t \in [\tmax]_0 : \allocalpha_t > 0},\\
t_j &:= \min\set{ t \in [\tmax]_0 : t > t_{j-1}, \allocalpha_t > 0}.
\end{aligned}
\end{equation*}
If the set $\set*{ t \in [\tmax]_0 : t > t_{j-1}, \allocalpha_t > 0}$ is ever empty (i.e. there are fewer than $\maxreg$ non-trivial tests), then we will define $t_j = t_{j-1}$. If $\allocalpha_t = 0$ for all $t$, then we will define $t_1 = ... = t_{\maxreg} = 0$, though the choice is not important. These quantities implicitly depend on $n$ and should formally be understood as $t_j^{(n)}$; we will suppress redundant indexing by $n$, for instance using the shorthand $\interR_{t_s, t_j}^{(n)}$ while understanding that both $t_s$ and $t_j$ also depend on $n$ in this expression.

Recall that ultimately in Lemma~\ref{lemma:oracle-empirical-convergence} where this is applied, we assume that both the oracle and empirical versions of Algorithm~\ref{alg:chiseling-testing} are run on the same dataset and make the same choices, so that $\allocalpha_0,...,\allocalpha_n$ are identical for both versions, and thus $t_1,...,t_{\maxreg}$ can be referred to unambiguously for both. The next lemma tells us that it suffices to restrict our attention to these indices when calculating the truncation levels. Recall the definitions of $\trunc_t$ and $\hat{\trunc}_t$ from Definitions~\ref{def:oracle-normal-test-stat-crit-val} and~\ref{def:empirical-normal-test-stat-crit-val}.

\begin{lemma}[Index reduction for truncation]
\label{lemma:index-reduction-truncation}
For $j = 1,...,\maxreg$,
\begin{equation*}
\begin{aligned}
\trunc_{t_j} &= \min_{s < j : \un_{t_j} / \un_{t_s} \geq \minprop} \set*{\frac{\critval_{t_s} - \interR_{t_s, t_j}}{\interv_{t_s, t_j}}},\\
\hat{\trunc}_{t_j} &= \min_{s < j : \un_{t_j} / \un_{t_s} \geq \minprop} \set*{\frac{\hat{\critval}_{t_s} - \hat{\interR}_{t_s, t_j}}{\hat{\interv}_{t_s, t_j}}}.
\end{aligned}
\end{equation*}
\end{lemma}

\begin{proof}[Proof of Lemma~\ref{lemma:index-reduction-truncation}]
We will focus on the proof of the equivalence for the oracle truncation levels. The proof for the empirical truncation levels is algebraically equivalent. If $\allocalpha_t = 0$ for all $t$, then $t_1 = ... = t_{\maxreg} = 0$, and the result follows by definition. Thus, suppose $\allocalpha_t > 0$ for at least one $t$. Note that for any $s < t$, if $\allocalpha_s = 0$ then
\begin{equation*}
\begin{aligned}
\frac{\critval_s - \interR_{s,t}}{\interv_{s,t}} &\geq \frac{\trunc_s - \interR_{s,t}}{\interv_{s,t}}\\
&= \min_{u < s : \un_s / \un_u \geq \minprop} \set*{ \paren*{\frac{\critval_u - \interR_{u,s}}{\interv_{u,s}} - \interR_{s,t}} \frac{1}{\interv_{s,t}} }\\
&= \min_{u < s : \un_s / \un_u \geq \minprop} \set*{ \frac{\critval_u - (\interR_{u,s} + \interv_{u,s} \interR_{s,t})}{\interv_{u,s} \interv_{s,t}} }\\
&= \min_{u < s : \un_s / \un_u \geq \minprop} \set*{ \frac{\critval_u - \interR_{u,t}}{\interv_{u,t}} }
\end{aligned}
\end{equation*}
where the last line follows by simple algebraic manipulations of the definitions. Now recall the definition
\begin{equation*}
\begin{aligned}
\trunc_t = \min_{s < t : \un_t / \un_s \geq \minprop} \set*{ \frac{\critval_s - \interR_{s,t}}{\interv_{s,t}} }.
\end{aligned}
\end{equation*}
We have just shown that for any index $s$ where $\allocalpha_s = 0$, the term $(\critval_s - \interR_{s,t}) / \interv_{s,t}$ must exceed the minimum of the first $s - 1$ terms whose indices $u$ satisfy $\un_s / \un_u \geq \minprop$. Hence, the overall minimum is unchanged if we remove such indices from the minimization. That is,
\begin{equation*}
\begin{aligned}
\trunc_{t} = \min_{\substack{s < t : \un_t / \un_s \geq \minprop \\ \allocalpha_s > 0}} \set*{ \frac{\critval_s - \interR_{s,t}}{\interv_{s,t}} }.
\end{aligned}
\end{equation*}
But this exactly coincides with the definition of $\trunc_{t_j}$ when $t = t_j$ (as long as there exists at least one $\allocalpha_s > 0$ for $s < t$).
\end{proof}

Next, we establish a number of intermediate convergence results. We say that a sequence of positive random variables is \textit{positively log-bounded in probability} if both the sequence and the sequence of reciprocals are bounded in probability. We will also make use of a notion we call $\Phi$-convergence. We elaborate on the properties of $\Phi$-convergence in Appendix~\ref{appendix:conv-bounded-metrics}. Here, we simply note that for two random sequences $G_n$ and $H_n$, we say that \textit{$G_n$ converges to $H_n$ under the $\Phi$-metric} if
\begin{equation*}
\begin{aligned}
\abs{\Phi(G_n) - \Phi(H_n)} \stackrel{p}{\to} 0.
\end{aligned}
\end{equation*}
In this case, we will write $G_n \fconv{\Phi} H_n$. Throughout, we will make frequent use of Lemma~\ref{lemma:concrete-f-composition-rules}, which describes various basic operations that preserve $\Phi$-convergence. We also recall here that by Lemma~\ref{lemma:implied-convergence-properties}, Assumption~\ref{assumption:gen-dgp-regularity}, Constraint~\ref{constraint:tail-error-restriction}, and Condition~\ref{condition:diverging-sample-size} imply Properties~\ref{property:consist-approx},~\ref{property:var-ratio-bound}, and~\ref{property:consist-var}.

\begin{lemma}[Convergence of intermediate quantities]
\label{lemma:conv-intermediate}
Let $\interR_{s,t}$ and $\interv_{s,t}$ be defined as in Equation~\eqref{eq:oracle-intermediate-statistics} and $\hat{\interR}_{s,t}$ and $\hat{\interv}_{s,t}$ be defined analogously but with $\hsigsq_t$ used in place of $\sigsq_t$ wherever it appears. Let $0 \leq s \leq j \leq \maxreg$ be fixed. Under Assumption~\ref{assumption:gen-dgp-regularity}, Constraint~\ref{constraint:tail-error-restriction}, and Condition~\ref{condition:diverging-sample-size},
\begin{equation*}
\begin{aligned}
\hat{\interR}_{t_s, t_j}^{(n)} &\fconv{\Phi} \interR_{t_s, t_j}^{(n)}.
\end{aligned}
\end{equation*}
Furthermore, there are positively log-bounded in probability random sequences $\hat{w}_{s,j}^{(n)}$ and $w_{s,j}^{(n)}$ such that $(\hat{w}_{s,j}^{(n)})^{-1} \fconv{\Phi} (w_{s, j}^{(n)})^{-1}$ and
\begin{equation*}
\begin{aligned}
\hat{\interv}_{t_s, t_j}^{(n)} &= \hat{w}_{s,j}^{(n)} \cdot \indic\set{\un_{t_j}^{(n)} / \un_{t_s}^{(n)} \geq \minprop} \quad \text{ and } \quad \interv_{t_s, t_j}^{(n)} &= w_{s,j}^{(n)} \cdot \indic\set{\un_{t_j}^{(n)} / \un_{t_s}^{(n)} \geq \minprop}.
\end{aligned}
\end{equation*}
\end{lemma}

\begin{proof}[Proof of Lemma~\ref{lemma:conv-intermediate}]
We proceed in a few steps.
\begin{itemize}
\item First note that since $1 / x$ is uniformly continuous in a neighborhood of $x = 1$, then for all $\delta > 0$, there exists an $s > 0$ such that $x \in [1 - s, 1 + s]$ implies $|1/x - 1| < \delta$. Then
\begin{equation*}
\begin{aligned}
\Prb\paren*{\sup_{t \in [\tmax]_0 : \allocalpha_t > 0} \abs*{ \frac{\sigsq_t}{\hsigsq_t} - 1 } > \delta} &\leq \Prb\paren*{\exists t \in [\tmax]_0 \text{ s.t. } \allocalpha_t > 0 \text{ and } \frac{\hsigsq_t}{\sigsq_t} \not\in [1 - s, 1 + s] }\\
&= \Prb\paren*{\sup_{t \in [\tmax] : \allocalpha_t > 0} \abs*{ \frac{\hsigsq_t}{\sigsq_t} - 1 } > s}\\
&\stackrel{p}{\to} 0
\end{aligned}
\end{equation*}
by Property~\ref{property:consist-var}. Then since $\abs{\sqrt{x} - 1} \leq \sqrt{\abs{x - 1}}$ note that for all $j = 1,...,\maxreg$,
\begin{equation*}
\begin{aligned}
\abs*{ \Phi\paren*{\sqrt{\frac{\hsigsq_{t_j}}{\sigsq_{t_j}}}} - \Phi(1) } \leq \abs*{\sqrt{\frac{\hsigsq_{t_j}}{\sigsq_{t_j}}} - 1} \leq \sqrt{\abs*{\frac{\hsigsq_{t_j}}{\sigsq_{t_j}} - 1}} \leq \sqrt{\sup_{t \in [\tmax] : \allocalpha_t > 0} \abs*{ \frac{\hsigsq_t}{\sigsq_t} - 1 }} \stackrel{p}{\to} 0
\end{aligned}
\end{equation*}
where we use the fact that $\Phi$ is $1$-Lipschitz. Hence, $\sqrt{\frac{\hsigsq_{t_j}}{\sigsq_{t_j}}} \fconv{\Phi} 1$ for all $j$. An analogous argument shows that $\sqrt{\frac{\sigsq_{t_j}}{\hsigsq_{t_j}}} \fconv{\Phi} 1$ for all $j$.
\item Note that for all $s < j$, we have that $\hat{\interR}_{t_s, t_j} \fconv{\Phi} \hat{\interR}_{t_s, t_j}$ trivially, and that $\sqrt{\frac{\hsigsq_{t_s}}{\sigsq_{t_s}}} \fconv{\Phi} 1$ as we showed above. Note also that the sequence consisting of all $1$'s is positively log-bounded in probability. Hence, using the multiplication rule of Lemma~\ref{lemma:concrete-f-composition-rules} yields that $\sqrt{\frac{\hsigsq_{t_s}}{\sigsq_{t_s}}} \cdot \hat{\interR}_{t_s, t_j} \fconv{\Phi} 1 \cdot \hat{\interR}_{t_s, t_j}$. Since $\sqrt{\frac{\hsigsq_{t_s}}{\sigsq_{t_s}}} \cdot \hat{\interR}_{t_s, t_j} = \interR_{t_s, t_j}$, this is equivalent to $\interR_{t_s, t_j} \fconv{\Phi} \hat{\interR}_{t_s, t_j}$.

\item Note that for all $s < j$, we have $\sqrt{\frac{\sigsq_{t_j}}{\hsigsq_{t_j}}} \cdot \sqrt{\frac{\hsigsq_{t_s}}{\sigsq_{t_s}}} \cdot \hat{\interv}_{t_s, t_j}^{-1} = \interv_{t_s, t_j}^{-1}$. Same as with the previous bullet, two applications of the multiplication rule of Lemma~\ref{lemma:concrete-f-composition-rules} yield that $\interv_{t_s, t_j}^{-1} \fconv{\Phi} \hat{\interv}_{t_s, t_j}^{-1}$.

\item In the below only, for a random variable $A$ and event $E$, we will appropriate the notation $A \otimes \indic\set{E} = A \indic\set{E} + (1 - \indic\set{E})$. Note that $(A \otimes \indic\set{E}) \times  \indic\set{E} = A \indic\set{E}$. Furthermore, if $A_n \fconv{\Phi} B_n$, then $A_n \otimes \indic\set{E_n} \fconv{\Phi} B_n \otimes \indic\set{E_n}$; we can see this by first applying Lemma~\ref{lemma:indic-f-conv-composition} to conclude that $A_n \indic\set{E_n} \fconv{\Phi} B_n \indic\set{E_n}$. Then since $1 - \indic\set{E_n} \fconv{\Phi} 1 - \indic\set{E_n}$ trivially and is bounded in probability, the addition rule of Lemma~\ref{lemma:concrete-f-composition-rules} shows that $A_n \otimes \indic\set{E_n} \fconv{\Phi} B_n \otimes \indic\set{E_n}$.

\item Define
\begin{equation*}
\begin{aligned}
\hat{w}_{s,j} = \hat{\interv}_{t_s, t_j} \otimes \indic\set{\un_{t_j} / \un_{t_s} \geq \minprop} \quad \text{ and } \quad w_{s,j} = \interv_{t_s, t_j} \otimes \indic\set{\un_{t_j} / \un_{t_s} \geq \minprop}.
\end{aligned}
\end{equation*}
Then by the previous two bullets, we have $\hat{w}_{s,j}^{-1} \fconv{\Phi} w_{s,j}^{-1}$.

\item Lastly, recalling the definition of $\vratbound$ from Property~\ref{property:var-ratio-bound}, we have
\begin{equation*}
\begin{aligned}
\interv_{t_s, t_j} \cdot \indic\set{\un_{t_j} / \un_{t_s} \geq \minprop} &\leq \max\set{1, \vratbound^{1/2}},\\
\interv_{t_s, t_j}^{-1} \cdot \indic\set{\un_{t_j} / \un_{t_s} \geq \minprop} &\leq \max\set{1, \vratbound^{1/2}}.
\end{aligned}
\end{equation*}
This follows by the definition of $\vratbound$ when $\allocalpha_t > 0$ for at least one $t$ (otherwise, when $\allocalpha_t = 0$ for all $t$, then $\interv_{t_s, t_j} = 1$ while the right-hand sides are at least $1$). Then both $w_{s,j}$ and $w_{s,j}^{-1}$ are bounded by $\max\set{1, \vratbound^{1/2}}$ which is bounded in probability by Property~\ref{property:var-ratio-bound}. Thus $w_{s,j}$ is positively log-bounded in probability. By the transitivity of positively log-boundedness in probability (Lemma~\ref{lemma:b-bounded-equivalence}), we also have that $\hat{w}_{s,j}$ is positively log-bounded in probability.
\end{itemize}
\end{proof}

Now we may use these results to establish the uniform convergence of the critical values.

\begin{lemma}[Uniform convergence of critical values]
\label{lemma:uniform-conv-crit-val}
Under Assumption~\ref{assumption:gen-dgp-regularity}, Constraint~\ref{constraint:tail-error-restriction}, and Condition~\ref{condition:diverging-sample-size}, we have
\begin{equation*}
\begin{aligned}
\sup_{t \in [\tmax(n)]_0 : \allocalpha_t^{(n)} > 0} \abs{\hat{\critval}_t^{(n)} - \critval_t^{(n)}} \stackrel{p}{\to} 0
\end{aligned}
\end{equation*}
where here we take the convention that the supremum over an empty set is $0$.
\end{lemma}

\begin{proof}[Proof of Lemma~\ref{lemma:uniform-conv-crit-val}]
First define the indicator $H := \indic\set{ \exists t \text{ s.t. } \allocalpha_t > 0 }$, and define $\critval_t' := \critval_t \cdot H$ and $\hat{\critval}_t' := \hat{\critval}_t \cdot H$. Here we take the convention that $\infty \cdot 0 = 0$. Then
\begin{equation*}
\begin{aligned}
\sup_{t \in [\tmax]_0 : \allocalpha_t > 0} \abs{ \hat{\critval}_{t} - \critval_{t} } = \sup_{j=1,...,\maxreg} \abs{ \hat{\critval}_{t_j}' - \critval_{t_j}' }
\end{aligned}
\end{equation*}
since if $H = 1$ then the supremum on the left-hand side is precisely over the indices $t_1,...,t_{\maxreg}$ (with possibly some redundancy), and otherwise both sides of the above are $0$.

Next note that for $j=1,...,\maxreg$ and for all $y$,
\begin{equation*}
\begin{aligned}
\Phi^{-1}(1 - \allocalpha_{t_j}; y) \cdot H \leq \Phi^{-1}(1 - \allocalpha_{t_j}) \cdot H \leq \Phi^{-1}(1 - \alphamin) < \infty.
\end{aligned}
\end{equation*}
The second inequality follows since if $H = 1$, then $\allocalpha_{t_j} > 0$ and hence $\allocalpha_{t_j} \geq \alphamin$ by Constraint~\ref{constraint:tail-error-restriction}. Hence $\critval_{t_j}', \hat{\critval}_{t_j}' \in [0, \Phi^{-1}(1 - \alphamin)]$ a.s., and are hence also bounded in probability.

Next, we will show by induction that $\hat{\critval}_{t_j}' \fconv{\Phi} \critval_{t_j}'$ for $j=1,...,\maxreg$. The base case of $j = 1$ follows from the fact that $\hat{\trunc}_{t_1} = \trunc_{t_1} = \infty$ by Lemma~\ref{lemma:index-reduction-truncation}, and so $\hat{\critval}_{t_1} = \critval_{t_1}$, which also implies that $\hat{\critval}_{t_1}' = \critval_{t_1}'$. Now suppose $\hat{\critval}_{t_s}' = \critval_{t_s}'$ for all $s < j$. We will show that $\hat{\critval}_{t_j}' \fconv{\Phi} \critval_{t_j}'$.

Let $w_{s,j}$ and $\hat{w}_{s,j}$ be defined as in Lemma~\ref{lemma:conv-intermediate}. Since $\critval_{t_s}'$ is bounded in probability, the subtraction rule of Lemma~\ref{lemma:concrete-f-composition-rules}, the inductive hypothesis, and Lemma~\ref{lemma:conv-intermediate} yield that $\hat{\critval}_{t_s}' - \hat{\interR}_{t_s, t_j} \fconv{\Phi} \critval_{t_s}' - \interR_{t_s, t_j}$ for every $s < j$. Then this fact, the multiplication rule of Lemma~\ref{lemma:concrete-f-composition-rules}, and Lemma~\ref{lemma:conv-intermediate} yield that $\frac{\hat{\critval}_{t_s}' - \hat{\interR}_{t_s, t_j}}{\hat{w}_{s, j}} \fconv{\Phi} \frac{\critval_{t_s}' - \interR_{t_s, t_j}}{w_{s, j}}$ for each $s < j$. Then by repeated application of the minimum rule of Lemma~\ref{lemma:concrete-f-composition-rules},
\begin{equation*}
\begin{aligned}
\hat{m}_j := \min_{s < j : \un_{t_j} / \un_{t_s} \geq \minprop} \set*{\frac{\hat{\critval}_{t_s}' - \hat{\interR}_{t_s, t_j}}{\hat{w}_{s, j}}} \fconv{\Phi} \min_{s < j : \un_{t_j} / \un_{t_s} \geq \minprop} \set*{\frac{\critval_{t_s}' - \interR_{t_s, t_j}}{w_{s, j}}} =: m_j.
\end{aligned}
\end{equation*}
Also note that in the above, we may replace $w_{s,j}$ with $\interv_{t_s, t_j}$ and $\hat{w}_{s,j}$ with $\hat{\interv}_{t_s, t_j}$, since these quantities agree whenever $\un_{t_j} / \un_{t_s} \geq \minprop$.

Now letting $\mathrm{Clip}_{a, b}(x) = \max\{ a, \min\{ x, b \} \}$, using the clipped truncated normal quantile rule of Lemma~\ref{lemma:concrete-f-composition-rules} and the fact that $\allocalpha_{t_j} \fconv{\Phi} \allocalpha_{t_j}$ trivially, we have
\begin{equation*}
\begin{aligned}
\mathrm{Clip}_{0,\infty}(\Phi^{-1}(\mathrm{Clip}_{0,b}(1 - \allocalpha_{t_j})) \cdot \Phi(\hat{m}_{j})) \fconv{\Phi} \mathrm{Clip}_{0,\infty}(\Phi^{-1}(\mathrm{Clip}_{0,b}(1 - \allocalpha_{t_j})) \cdot \Phi(m_{j}))
\end{aligned}
\end{equation*}
where $b = 1 - \alphamin$. Observe that
\begin{equation*}
\begin{aligned}
\hat{\critval}_{t_j}' &= \max\set{0, \Phi^{-1}(1 - \allocalpha_{t_j}; \hat{\trunc}_{t_j})} \cdot H &= \mathrm{Clip}_{0,\infty}(\Phi^{-1}(\mathrm{Clip}_{0,b}(1 - \allocalpha_{t_j})) \cdot \Phi(\hat{m}_{j})) \cdot H.
\end{aligned}
\end{equation*}
The first equality follows by definition. For the second equality, note that if $H = 0$, then both sides of the above are $0$. If $H = 1$, then $\allocalpha_{t_j} > 0$ and hence $\allocalpha_{t_j} \geq \alphamin$ almost surely, and so $1 - \allocalpha_{t_j} = \mathrm{Clip}_{0,b}(1 - \allocalpha_{t_j})$. Also, if $H = 1$, then $\hat{\critval}_{t_s}' = \hat{\critval}_{t_s}$ for all $s < j$, and hence $\hat{m}_j = \hat{\trunc}_{t_j}$ by definition of the former and Lemma~\ref{lemma:index-reduction-truncation}. So the second equality follows after making the appropriate substitutions. By the same reasoning,
\begin{equation*}
\begin{aligned}
\critval_{t_j}' &= \max\set{0, \Phi^{-1}(1 - \allocalpha_{t_j}; \trunc_{t_j})} \cdot H &= \mathrm{Clip}_{0,\infty}(\Phi^{-1}(\mathrm{Clip}_{0,b}(1 - \allocalpha_{t_j})) \cdot \Phi(m_{j})) \cdot H.
\end{aligned}
\end{equation*}
Finally, Lemma~\ref{lemma:indic-f-conv-composition} implies that $\hat{\critval}_{t_j}' \fconv{\Phi} \critval_{t_j}'$, which completes the proof of the inductive step.

Hence, we have shown that $\hat{\critval}_{t_j}' \fconv{\Phi} \critval_{t_j}'$ for $j=1,...,\maxreg$. Recalling that $\hat{\critval}_{t_j}'$ and $\critval_{t_j}'$ both lie in the bounded interval $[0, \Phi^{-1}(1 - \alphamin)]$ a.s., we may use the fact that bounded $\Phi$-convergent sequences converge in the ordinary sense (Lemma~\ref{lemma:f-conv-ordinary-bounded}), so that $\abs{\hat{\critval}_{t_j}' - \critval_{t_j}' } \stackrel{p}{\to} 0$. Finally, we have
\begin{equation*}
\begin{aligned}
\sup_{t \in [\tmax]_0 : \allocalpha_t > 0} \abs{\hat{\critval}_t - \critval_t} = \sup_{j=1,...,\maxreg} \abs{\hat{\critval}_{t_j}' - \critval_{t_j}'} \stackrel{p}{\to} 0
\end{aligned}
\end{equation*}
where the final convergence follows from the continuous mapping theorem applied to the function $g(x_1,...,x_{\maxreg}) = \sup_{j=1,...,\maxreg} x_j$.
\end{proof}

We are finally ready to prove Lemma~\ref{lemma:uniform-conv-shift-crit-val}.

\begin{proof}[Proof of Lemma~\ref{lemma:uniform-conv-shift-crit-val}]
Recall, as in the proof of Lemma~\ref{lemma:uniform-conv-crit-val}, that by Constraint~\ref{constraint:tail-error-restriction}, whenever $\allocalpha_t > 0$, we also have $\allocalpha_t \geq \alphamin$. So for $j=1,...,\maxreg$,
\begin{equation*}
\begin{aligned}
\Phi^{-1}(1 - \allocalpha_{t_j}; y) \leq \Phi^{-1}(1 - \allocalpha_{t_j}) \leq \Phi^{-1}(1 - \alphamin) < \infty.
\end{aligned}
\end{equation*}
Hence $\critval_{t_j}, \hat{\critval}_{t_j} \in [0, \Phi^{-1}(1 - \alphamin)]$ a.s. Then
\begin{equation}
\label{eq:d-conv-expr}
\begin{aligned}
& \sup_{t \in [\tmax]_0 : \allocalpha_t > 0} \abs{d_t - \hat{d}_t}\\
= & \sup_{t \in [\tmax]_0 : \allocalpha_t > 0} \abs*{\critval_t - \sqrt{\frac{\hsigsq_t}{\sigsq_t}} \cdot \hat{\critval}_t}\\
\leq & \sup_{t \in [\tmax]_0 : \allocalpha_t > 0} \abs*{\critval_t - \sqrt{\frac{\hsigsq_t}{\sigsq_t}} \cdot \critval_t} + \abs*{\sqrt{\frac{\hsigsq_t}{\sigsq_t}} \cdot \critval_t - \sqrt{\frac{\hsigsq_t}{\sigsq_t}} \cdot \hat{\critval}_t}\\
\leq & \Phi^{-1}(1 - \alphamin) \sup_{t \in [\tmax]_0 : \allocalpha_t > 0} \abs*{\sqrt{\frac{\hsigsq_t}{\sigsq_t}} - 1} + \sup_{t \in [\tmax]_0 : \allocalpha_t > 0} \abs*{\sqrt{\frac{\hsigsq_t}{\sigsq_t}}} \sup_{t \in [\tmax]_0 : \allocalpha_t > 0} \abs{\critval_t - \hat{\critval}_t}.
\end{aligned}
\end{equation}
Then since $\abs{\sqrt{x} - 1} \leq \sqrt{\abs{x - 1}}$,
\begin{equation*}
\begin{aligned}
\sup_{t \in [\tmax]_0 : \allocalpha_t > 0} \abs*{\sqrt{\frac{\hsigsq_t}{\sigsq_t}} - 1} &\leq \sup_{t \in [\tmax]_0 : \allocalpha_t > 0} \sqrt{\abs*{\frac{\hsigsq_t}{\sigsq_t} - 1}} = \sqrt{\sup_{t \in [\tmax]_0 : \allocalpha_t > 0} \abs*{\frac{\hsigsq_t}{\sigsq_t} - 1}} \stackrel{p}{\to} 0.
\end{aligned}
\end{equation*}
by the continuous mapping theorem and Property~\ref{property:consist-var}. Next, note that
\begin{equation*}
\begin{aligned}
\sup_{t \in [\tmax]_0 : \allocalpha_t > 0} \abs*{\sqrt{\frac{\hsigsq_t}{\sigsq_t}}} \leq 1 + \sup_{t \in [\tmax]_0 : \allocalpha_t > 0} \abs*{\sqrt{\frac{\hsigsq_t}{\sigsq_t}} - 1} \stackrel{p}{\to} 1.
\end{aligned}
\end{equation*}
Lastly, by Lemma~\ref{lemma:uniform-conv-crit-val} we have $\sup_{t \in [\tmax]_0 : \allocalpha_t > 0} \abs{\critval_t - \hat{\critval}_t} \stackrel{p}{\to} 0$. Then applying these results to the last line in Equation~\eqref{eq:d-conv-expr} along with Slutsky's theorem concludes the lemma.
\end{proof}

\subsection{Proof of Lemma~\ref{lemma:aipw-uniform-conv-shift-crit-val}}
\label{proof:lemma-aipw-uniform-conv-shift-crit-val}

The proof ideas here are very similar to the proof of Lemma~\ref{lemma:uniform-conv-shift-crit-val} in Appendix~\ref{proof:lemma-uniform-conv-shift-crit-val}. In fact, we borrow the definition of $t_j$ for $j=1,...,\maxreg$ at the beginning of Appendix~\ref{proof:lemma-uniform-conv-shift-crit-val}, and we may utilize Lemma~\ref{lemma:index-reduction-truncation} here also since it is an algebraic fact. Furthermore, we will also require the notion of \emph{positively log-bounded in probability} and $\Phi$-convergence. To avoid repeating too much, we refer the reader to the part of Appendix~\ref{proof:lemma-uniform-conv-shift-crit-val} preceding the statement of Lemma~\ref{lemma:conv-intermediate}. Fundamentally, the proof framework is the same, except that there are more intermediate quantities to address.

\begin{lemma}[Convergence of AIPW remainders]
\label{lemma:conv-aipw-remainders}
Recall the definition of $R_t$ from Equation~\eqref{eq:aipw-orac-emp-shifted-quant}, and further define
\begin{equation*}
\begin{aligned}
J_t := \frac{1}{\un_t} \sum_{i : X_i \in \reg_t} (\hat{Y}_i^2 - Y_i^2).
\end{aligned}
\end{equation*}
Then under the conditions of Theorem~\ref{theorem:aipw-validity}, both $\sup_{t \in [\tmax]_0 : \allocalpha_t > 0} \abs{R_t}$ and $\sup_{t \in [\tmax]_0 : \allocalpha_t > 0} \abs{J_t}$ converge to $0$ in probability. Furthermore, $\sup_{t \in [\tmax]_0 : \allocalpha_t > 0} \abs{\hat{\mu}_t - \meanest_t} \stackrel{p}{\to} 0$.
\end{lemma}

\begin{proof}[Proof of Lemma~\ref{lemma:conv-aipw-remainders}]
We first define
\begin{equation*}
\begin{aligned}
\tilde{R}_t := \frac{1}{\sqrt{\un_t}} \sum_{i : X_i \in \reg_t} (\hat{Y}_i - Y_i) = \frac{1}{\sqrt{\un_t}} \sum_{j=1}^{\Kcv} \sum_{i : X_i \in \reg_t} (\hat{Y}_{i,1} - Y_{i,1} - \hat{Y}_{i,0} + Y_{i,0}) \indic\set{i \in \mathcal{I}_j}
\end{aligned}
\end{equation*}
and note that since $\abs{R_t} = \sigsq_t^{-1/2} \times \abs{\tilde{R}_t}$,
\begin{equation*}
\begin{aligned}
\sup_{t : \allocalpha_t > 0} \abs{R_t} \leq \sqrt{\sup_{t : \allocalpha_t > 0} \sigsq_t^{-1}} \times \sup_{t : \allocalpha_t > 0} \abs{\tilde{R}_t}.
\end{aligned}
\end{equation*}
The first supremum on the right-hand side is bounded in probability by Assumption~\ref{assumption:gen-dgp-regularity}, so it suffices to show that the second supremum on the right-hand side converges to $0$. In fact, by the triangle inequality, it suffices to show that
\begin{equation*}
\begin{aligned}
\sup_{t : \allocalpha_t > 0} \abs{\tilde{R}_t^{(j,w)}} \stackrel{p}{\to} 0 \quad \text{ where } \quad \tilde{R}_t^{(j,w)} := \frac{1}{\sqrt{\un_t}} \sum_{i \in \mathcal{I}_j : X_i \in \reg_t} (\hat{Y}_{i,w} - Y_{i,w}).
\end{aligned}
\end{equation*}
Without loss of generality, we will let $j = 1$ and $w = 1$. The proof when $w = 0$ is essentially identical. First note that if $i \in \mathcal{I}_1$ then we can expand $\hat{Y}_{i,1} - Y_{i,1}$ as
\begin{equation*}
\begin{aligned}
\hat{g}_1^{(1)}(X_i) + \frac{W_i (Y'_i - \hat{g}_1^{(1)}(X_i)}{e(X_i)} - g_1(X_i) - \frac{W_i (Y'_i - g(X_i)}{e(X_i)} = (\hat{g}_1^{(1)}(X_i) - g_1(X_i))\paren*{1 - \frac{W_i}{e(X_i)}}.
\end{aligned}
\end{equation*}
Define $\mathcal{G}_t := \sigma(\F_t, \mathcal{I}_1, (X_i, W_i, Y_i')_{i \not\in \mathcal{I}_1}, (X_i)_{i=1}^n)$. Conditionally on $\mathcal{G}_t$, we have that $\hat{g}_1^{(1)}(X_i) - g_1(X_i)$ is fixed for all $i \in \mathcal{I}_1$. Furthermore, we still have $\Prb(W_i = 1 \mid \mathcal{G}) = e(X_i)$ for all $i \in \mathcal{I}_1$ such that $X_i \in \reg_t$ and that $(W_i)_{i \in \mathcal{I}_1 : X_i \in \reg_t}$ are independent. This is because conditioning on $\F_t$ and $\mathcal{I}_1$ tells us that $(X_i, W_i, Y_i')_{i \in \mathcal{I}_1 : X_i \in \reg_t}$ is an i.i.d. sample from $(X, W, Y') \mid X \in \reg_t$ by Corollary~\ref{corollary:distribution-subsamples} and the fact that the folds are generated independently. Further conditioning on the data in $\mathcal{I}_1^{\comp}$ does not affect the distribution of $(X_i, W_i, Y_i')_{i \in \mathcal{I}_1 : X_i \in \reg_t}$ since $\mathcal{I}_1$ and $\mathcal{I}_1^{\comp}$ are disjoint, and further conditioning on $\paren{X_i}_{i=1}^n$ fixes the values of $X_i$ but does not change the conditional distribution of $W_i \mid X_i$ for $i \in \mathcal{I}_1$ such that $X_i \in \reg_t$. Then $1 - W_i / e(X_i)$ has mean $0$ conditionally on $\mathcal{G}_t$ and
\begin{equation*}
\begin{aligned}
\Ec*{ \tilde{R}_{t}^{(1,1)} }{\mathcal{G}_t} = 0
\end{aligned}
\end{equation*}
and
\begin{equation*}
\begin{aligned}
\beta_t := \Var(\tilde{R}_t^{(1,1)} \mid \mathcal{G}_t) &= \frac{1}{\un_t} \sum_{i \in \mathcal{I}_1 : X_i \in \reg_t} \Var(\hat{Y}_{i,w} - Y_{i,w} \mid \mathcal{G}_t)\\
&= \frac{1}{\un_t} \sum_{i \in \mathcal{I}_1 : X_i \in \reg_t} (\hat{g}_1^{(1)}(X_i) - g_1(X_i))^2 \cdot \Ec*{ \paren*{1 - \frac{W_i}{e(X_i)}}^2 }{\mathcal{G}_t}\\
&= \frac{1}{\un_t} \sum_{i \in \mathcal{I}_1 : X_i \in \reg_t} (\hat{g}_1^{(1)}(X_i) - g_1(X_i))^2 \cdot \frac{1 - e(X_i)}{e(X_i)}\\
&\leq \frac{\abs{\set{i \in \mathcal{I}_1 : X_i \in \reg_t}}}{\un_t} \cdot \Lstable^2 \cdot \frac{1 - \eta}{\eta}\\
&\leq \Lstable^2 \cdot \frac{1 - \eta}{\eta}
\end{aligned}
\end{equation*}
where we recall the definition of $\Lstable$ from Assumption~\ref{assumption:stable-nuisance}. This is true for all $t=0,...,\tmax$ as long as $\allocalpha_t > 0$, so $\beta^+ := \sup_{t \in [\tmax]_0 : \allocalpha_t > 0} \beta_t \leq \Lstable^2 \cdot \frac{1 - \eta}{\eta}$. Since the latter converges to $0$ in probability, for any $\delta, \epsilon > 0$ let $b = (\epsilon \delta^2) / (2 \maxreg)$ and choose $N$ so that $\Prb(\beta^+ > b) \leq \epsilon / 2$ for all $n \geq N$. Define the event $E := \set{\beta^+ \leq b}$. Then
\begin{equation*}
\begin{aligned}
\Prb\paren*{\sup_{t : \allocalpha_t > 0} \abs{\tilde{R}_t^{(1,1)}} > \delta, E} &\leq \sum_{t=0}^{\tmax} \Prb(\abs{\tilde{R}_t^{(1,1)}} > \delta, \allocalpha_t > 0, E)\\
&= \sum_{t=0}^{\tmax} \E\bkt{ \indic\set{\allocalpha_t > 0} \Prb(\abs{\tilde{R}_t^{(1,1)}} > \delta, E \mid \mathcal{G}_t) }\\
&\leq \sum_{t=0}^{\tmax} \E\bkt{ \indic\set{\allocalpha_t > 0} \Prb(\abs{\tilde{R}_t^{(1,1)}} > \delta, \beta_t \leq b \mid \mathcal{G}_t) }\\
&\leq \sum_{t=0}^{\tmax} \E\bkt*{ \indic\set{\allocalpha_t > 0} \cdot \frac{b}{\delta^2} }\\
&\leq \frac{b}{\delta^2} \cdot \maxreg\\
&= \epsilon / 2.
\end{aligned}
\end{equation*}
To go to the fourth line we have noted that $\Prb(\abs{\tilde{R}_t^{(1,1)}} > \delta, \beta_t \leq b \mid \mathcal{G}_t) = \Prb(\abs{\tilde{R}_t^{(1,1)}} > \delta \mid \mathcal{G}_t) \indic\set{\beta_t \leq b}$ since $\beta_t$ is $\mathcal{G}_t$-measurable and then applied the conditional Chebyshev inequality. Then
\begin{equation*}
\begin{aligned}
\Prb\paren*{\sup_{t : \allocalpha_t > 0} \abs{\tilde{R}_t^{(1,1)}} > \delta} \leq \Prb\paren*{\sup_{t : \allocalpha_t > 0} \abs{\tilde{R}_t^{(1,1)}} > \delta, E} + \Prb(E^{\comp}) \leq \epsilon / 2 + \epsilon / 2 = \epsilon
\end{aligned}
\end{equation*}
for all $n \geq N$ as desired. Then again, repeating the above across all $j$ and $w$ and bounding using the triangle inequality, we obtain that $\sup_{t : \allocalpha_t > 0} \abs{\tilde{R}_t} \stackrel{p}{\to} 0$. Note that as an immediate consequence we obtain
\begin{equation*}
\begin{aligned}
\sup_{t : \allocalpha_t > 0} \abs{\hat{\mu}_t - \meanest_t} \leq \sup_{t : \allocalpha_t > 0} \abs{\tilde{R}_t} \stackrel{p}{\to} 0.
\end{aligned}
\end{equation*}

Lastly, we turn our attention to $\sup_{t : \allocalpha_t > 0} \abs{J_t}$. Note that by Cauchy-Schwarz we have
\begin{equation*}
\begin{aligned}
\abs{J_t} = \frac{1}{\un_t} \abs*{\sum_{i : X_i \in \reg_t} (\hat{Y}_i + Y_i) (\hat{Y}_i - Y_i)} \leq \sqrt{\frac{1}{\un_t} \sum_{i : X_i \in \reg_t} (\hat{Y}_i + Y_i)^2} \sqrt{\frac{1}{\un_t} \sum_{i : X_i \in \reg_t} (\hat{Y}_i - Y_i)^2}
\end{aligned}
\end{equation*}
and thus
\begin{equation*}
\begin{aligned}
\sup_{t : \allocalpha_t > 0} \abs{J_t} \leq \sqrt{ \sup_{t : \allocalpha_t > 0} \frac{1}{\un_t} \sum_{i : X_i \in \reg_t} (\hat{Y}_i + Y_i)^2} \sqrt{\sup_{t : \allocalpha_t > 0} \frac{1}{\un_t} \sum_{i : X_i \in \reg_t} (\hat{Y}_i - Y_i)^2}.
\end{aligned}
\end{equation*}
We can write $\hat{Y}_i + Y_i = 2Y_i + (\hat{Y}_i - Y_i)$ and use the triangle inequality for norms to obtain
\begin{equation*}
\begin{aligned}
\sqrt{ \sup_{t : \allocalpha_t > 0} \frac{1}{\un_t} \sum_{i : X_i \in \reg_t} (\hat{Y}_i + Y_i)^2} \leq 2 \sqrt{ \sup_{t : \allocalpha_t > 0} \frac{1}{\un_t} \sum_{i : X_i \in \reg_t} Y_i^2} + \sqrt{ \sup_{t : \allocalpha_t > 0} \frac{1}{\un_t} \sum_{i : X_i \in \reg_t} (\hat{Y}_i - Y_i)^2}.
\end{aligned}
\end{equation*}
It is straightforward to show that the first supremum term on the right-hand side is bounded in probability. We can show this by again expanding $Y_i = \mean(\reg_t) + (Y_i - \mean(\reg_t))$ and applying the triangle inequality for norms, then using the probability bounds on the supremum over moments granted by Assumption~\ref{assumption:gen-dgp-regularity}. The second supremum term is repeated from the previous display, and thus to show that $\sup_{t : \allocalpha_t > 0} \abs{J_t} \stackrel{p}{\to}$ it suffices to show that it converges to $0$ in probability as then $\sup_{t : \allocalpha_t > 0} \abs{J_t}$ is bounded by the product of a term that is bounded in probability and a term that converges to $0$ in probability. We focus on this term now. We expand $(\hat{Y}_i - Y_i)^2 = (\hat{Y}_{i,1} - Y_{i,1})^2 + 2(\hat{Y}_{i,1} - Y_{i,1})(\hat{Y}_{i,0} - Y_{i,0}) + (\hat{Y}_{i,0} - Y_{i,0})^2$ to bound the supremum by the sum of the following three terms that we now characterize.
\begin{equation*}
\begin{aligned}
\sup_{t : \allocalpha_t > 0} \frac{1}{\un_t} \sum_{i : X_i \in \reg_t} (\hat{Y}_{i,1} - Y_{i,1})^2 &\leq \Lstable^2 \cdot \sup_{t : \allocalpha_t > 0} \frac{1}{\un_t} \sum_{i : X_i \in \reg_t} \paren*{1 - \frac{W_i}{e(X_i)}}^2\\
&\leq \Lstable^2 \cdot \sup_{t : \allocalpha_t > 0} \frac{1}{\un_t} \sum_{i : X_i \in \reg_t} \paren*{ \frac{1 - \eta}{\eta} }^2\\
&\leq \Lstable^2 \cdot \paren*{\frac{1 - \eta}{\eta}}^2\\
&\stackrel{p}{\to} 0.
\end{aligned}
\end{equation*}
A similar calculation shows that
\begin{equation*}
\begin{aligned}
\sup_{t : \allocalpha_t > 0} \frac{1}{\un_t} \sum_{i : X_i \in \reg_t} (\hat{Y}_{i,0} - Y_{i,0})^2 \leq \Lstable^2 \cdot \paren*{\frac{1 - \eta}{\eta}}^2 \stackrel{p}{\to} 0.
\end{aligned}
\end{equation*}
Lastly, note that by Cauchy-Schwarz,
\begin{equation*}
\begin{aligned}
& \sup_{t : \allocalpha_t > 0} \frac{2}{\un_t} \abs*{\sum_{i : X_i \in \reg_t} (\hat{Y}_{i,1} - Y_{i,1})(\hat{Y}_{i,0} - Y_{i,0})}\\
\leq & 2 \sup_{t : \allocalpha_t > 0} \sqrt{ \frac{1}{\un_t} \sum_{i : X_i \in \reg_t} (\hat{Y}_{i,1} - Y_{i,1})^2} \sqrt{ \frac{1}{\un_t} \sum_{i : X_i \in \reg_t} (\hat{Y}_{i,0} - Y_{i,0})^2}\\
\leq & 2 \sqrt{ \sup_{t : \allocalpha_t > 0} \frac{1}{\un_t} \sum_{i : X_i \in \reg_t} (\hat{Y}_{i,1} - Y_{i,1})^2} \sqrt{ \sup_{t : \allocalpha_t > 0} \frac{1}{\un_t} \sum_{i : X_i \in \reg_t} (\hat{Y}_{i,0} - Y_{i,0})^2}\\
\stackrel{p}{\to}&  0
\end{aligned}
\end{equation*}
by the bounds we just showed. Combining these bounds yields $\sup_{t : \allocalpha_t > 0} \abs{J_t} \stackrel{p}{\to} 0$.
\end{proof}

\begin{lemma}[Convergence of AIPW variances]
\label{lemma:conv-aipw-variances}
Recall that $\sigsq_t := \sigma^2(\reg_t)$ and $\hsigsq_t$ is defined as in Equation~\eqref{eq:aipw-all-statistics}. Then under the conditions of Theorem~\ref{theorem:aipw-validity},
\begin{equation*}
\begin{aligned}
\sup_{t \in [\tmax]_0 : \allocalpha_t > 0} \abs*{ \frac{\hsigsq_t}{\sigsq_t} - 1 } \stackrel{p}{\to} 0.
\end{aligned}
\end{equation*}
\end{lemma}

\begin{proof}[Proof of Lemma~\ref{lemma:conv-aipw-variances}]
First note that as in the proof of Property~\ref{property:consist-var} in Appendix~\ref{proof:lemma-implied-convergence-properties}, it suffices to show that
\begin{equation*}
\begin{aligned}
\sup_{t \in [\tmax]_0 : \allocalpha_t > 0} \abs{\hsigsq_t - \sigsq_t} \stackrel{p}{\to} 0.
\end{aligned}
\end{equation*}
Letting
\begin{equation*}
\begin{aligned}
\tilde{\sigsq}_t := \frac{1}{\un_t} \sum_{i : X_i \in \reg_t} (Y_i - \meanest_t)^2
\end{aligned}
\end{equation*}
note that
\begin{equation*}
\begin{aligned}
\sup_{t \in [\tmax]_0 : \allocalpha_t > 0} \abs{\hsigsq_t - \sigsq_t} \leq \sup_{t \in [\tmax]_0 : \allocalpha_t > 0} \abs{\hsigsq_t - \tilde{\sigsq}_t} + \sup_{t \in [\tmax]_0 : \allocalpha_t > 0} \abs{\tilde{\sigsq}_t - \sigsq_t}
\end{aligned}
\end{equation*}
The second supremum term on the right-hand side is exactly that which is addressed in the proof of Property~\ref{property:consist-var} in Appendix~\ref{proof:lemma-implied-convergence-properties}, where we show that it converges to $0$ in probability. Thus, we focus on the first term. Note that we may write
\begin{equation*}
\begin{aligned}
\hsigsq_t = \paren*{\frac{1}{\un_t} \sum_{i : X_i \in \reg_t} \hat{Y}_i^2} - \hat{\mu}_t^2 \quad \text{ and } \quad \tilde{\sigsq}_t = \paren*{\frac{1}{\un_t} \sum_{i : X_i \in \reg_t} Y_i^2} - \meanest_t^2.
\end{aligned}
\end{equation*}
Then we may bound
\begin{equation*}
\begin{aligned}
\sup_{t : \allocalpha_t > 0} \abs{\hsigsq_t - \tilde{\sigsq_t}} \leq \sup_{t : \allocalpha_t > 0} \abs*{ \frac{1}{\un_t} \sum_{i : X_i \in \reg_t} (\hat{Y}_i^2 - Y_i^2) } + \sup_{t : \allocalpha_t > 0} \abs{ \hat{\mu}_t^2 - \meanest_t^2 }.
\end{aligned}
\end{equation*}
The first supremum on the right-hand side converges to $0$ in probability by Lemma~\ref{lemma:conv-aipw-remainders}. For the second term, we may write
\begin{equation*}
\begin{aligned}
\sup_{t : \allocalpha_t > 0} \abs{ \hat{\mu}_t^2 - \meanest_t^2 } \leq \sup_{t : \allocalpha_t > 0} \abs{ \hat{\mu}_t + \meanest_t } \times \sup_{t : \allocalpha_t > 0} \abs{ \hat{\mu}_t - \meanest_t }.
\end{aligned}
\end{equation*}
The first supremum is bounded in probability since it can be coupled to $2 \sup_{t : \allocalpha_t > 0} \abs{\mean(\reg_t)} \leq 2 \sup_{t : \allocalpha_t > 0} \Mom{1}(\reg_t)$ which is bounded in probability by Assumption~\ref{assumption:gen-dgp-regularity}. The second converges to $0$ by Lemma~\ref{lemma:conv-aipw-remainders}.
\end{proof}

We now state and prove the analogue of Lemma~\ref{lemma:conv-intermediate} adapted for our current purposes.

\begin{lemma}[Convergence of AIPW intermediate quantities]
\label{lemma:aipw-conv-intermediate}
Let $\interR_{s,t}$ and $\interv_{s,t}$ be defined as in Equation~\eqref{eq:oracle-intermediate-statistics} and $\hat{\interR}_{s,t}$ and $\hat{\interv}_{s,t}$ be defined as in Equation~\eqref{eq:aipw-all-statistics}. Let $0 \leq s \leq j \leq \maxreg$ be fixed. Under the conditions of Theorem~\ref{theorem:aipw-validity},
\begin{equation*}
\begin{aligned}
\hat{\interR}_{t_s, t_j}^{(n)} &\fconv{\Phi} \interR_{t_s, t_j}^{(n)}.
\end{aligned}
\end{equation*}
Furthermore, there are positively log-bounded in probability random sequences $\hat{w}_{s,j}^{(n)}$ and $w_{s,j}^{(n)}$ such that $(\hat{w}_{s,j}^{(n)})^{-1} \fconv{\Phi} (w_{s, j}^{(n)})^{-1}$ and
\begin{equation*}
\begin{aligned}
\hat{\interv}_{t_s, t_j}^{(n)} &= \hat{w}_{s,j}^{(n)} \cdot \indic\set{\un_{t_j}^{(n)} / \un_{t_s}^{(n)} \geq \minprop} \quad \text{ and } \quad \interv_{t_s, t_j}^{(n)} &= w_{s,j}^{(n)} \cdot \indic\set{\un_{t_j}^{(n)} / \un_{t_s}^{(n)} \geq \minprop}.
\end{aligned}
\end{equation*}
\end{lemma}

\begin{proof}[Proof of Lemma~\ref{lemma:aipw-conv-intermediate}]
First note that the proofs of all of the properties of $\hsigsq_{t_s}$, $\sigsq_{t_s}$, their ratios, $\hat{\interv}_{t_s, t_j}$, and $\interv_{t_s, t_j}$ can be exactly lifted from the proof of Lemma~\ref{lemma:conv-intermediate}. Also, the construction of $\hat{w}_{s,j}^{-1}$ and $w_{s,j}^{-1}$ can be exactly copied from the proof of Lemma~\ref{lemma:conv-intermediate}, and the proof of their properties applied. Thus, we focus on the first part of the theorem involving $\hat{\interR}_{t_s, t_j}$ and $\interR_{t_s, t_j}$. Define
\begin{equation*}
\begin{aligned}
A_{s,t} := \frac{1}{\sqrt{\un_s}} \sum_{i : X_i \in \reg_{s} \setminus \reg_{t}} (\hat{Y}_i - Y_i)
\end{aligned}
\end{equation*}
noting that $A_{s,t}$ depends on $n$ but we have suppressed dependence on $n$. We have by the multiplication rule of Lemma~\ref{lemma:concrete-f-composition-rules} that
\begin{equation*}
\begin{aligned}
\sqrt{\frac{\hsigsq_{t_s}}{\sigsq_{t_s}}} \cdot \hat{\interR}_{t_s, t_j} \fconv{\Phi} \hat{\interR}_{t_s, t_j}
\end{aligned}
\end{equation*}
since $\sqrt{\frac{\hsigsq_{t_s}}{\sigsq_{t_s}}} \fconv{\Phi} 1$ (as shown in the proof of Lemma~\ref{lemma:conv-intermediate}) which is log bounded in probability. In a moment we will show that $\sigsq_{t_s}^{-1/2} A_{t_s,t_j} \stackrel{p}{\to} 0$. Then we argue that this suffices, since
\begin{equation*}
\begin{aligned}
\abs*{\Phi(\interR_{t_s, t_j}) - \Phi(\hat{\interR}_{t_s, t_j})} = & \abs*{\Phi\paren*{\sqrt{\frac{\hsigsq_{t_s}}{\sigsq_{t_s}}} \cdot \hat{\interR}_{t_s, t_j} - \frac{1}{\sqrt{\sigsq_{t_s}}} \cdot A_{t_s,t_j}} - \Phi(\hat{\interR}_{t_s, t_j})}\\
\leq & \abs*{\Phi\paren*{\sqrt{\frac{\hsigsq_{t_s}}{\sigsq_{t_s}}} \cdot \hat{\interR}_{t_s, t_j}} - \Phi(\hat{\interR}_{t_s, t_j})} + \abs{\sigsq_{t_s}^{-1/2} A_{t_s,t_j}}
\end{aligned}
\end{equation*}
by the fact the $\Phi$ is $1$-Lipschitz. Note that the first absolute value term in the last expression converges to $0$ by the definition of $\Phi$-convergence, and the second term converges in probability to $0$ as we will show. Then, $\interR_{t_s, t_j} \fconv{\Phi} \hat{\interR}_{t_s, t_j}$, as desired.

Note that
\begin{equation*}
\begin{aligned}
\abs{\sigsq_{t_s}^{-1/2} A_{t_s,t_j}} \leq \sqrt{\sup_{t : \allocalpha_t > 0} \sigsq_t^{-1}} \cdot A_{t_s,t_j}
\end{aligned}
\end{equation*}
and the supremum is bounded in probability by Assumption~\ref{assumption:gen-dgp-regularity} (this is true even when $\allocalpha_t = 0$ for all $t$ since then $A_{t_s,t_j} = 0$ by our convention). So it suffices to show that $A_{t_s,t_j} \stackrel{p}{\to} 0$. Let us break this term down even further:
\begin{equation*}
\begin{aligned}
\tilde{A}_{s,t}^{(k,w)} := \frac{1}{\sqrt{\un_s}} \sum_{i \in \mathcal{I}_k : X_i \in \reg_s \setminus \reg_t} (\hat{Y}_{i,w} - Y_{i,w})
\end{aligned}
\end{equation*}
and note that $A_{s,t}$ is a fixed linear combination of the above over $w = 0,1$ and $k=1,...,\Kcv$. So it suffices to show that $\tilde{A}_{t_s,t_j}^{(k,w)} \stackrel{p}{\to} 0$ for all such $w$ and $k$. Without loss of generality, we will consider $w = 1$ and $k = 1$ and drop the superscript on $(k,w) = (1,1)$ for notational clarity.

Next, we slightly redefine $t_j$ for $j=1,...,\maxreg$ and introduce some conventions to make them more convenient to work with. Let
\begin{equation*}
\begin{aligned}
t_1 &:= \min\set{ t \in [\tmax]_0 : \allocalpha_t > 0},\\
t_j &:= \min\set{ t \in [\tmax]_0 : t > t_{j-1}, \allocalpha_t > 0},
\end{aligned}
\end{equation*}
but where now we take the convention that the minimum of an empty set is $\tmax + 1$. Then take the convention that $\tilde{A}_{s,t} = 0$ if either $s$ or $t$ is equal to $\tmax + 1$. This does not modify the value of $\tilde{A}_{t_s,t_j}$ from the definition we had before, since $t_s$ or $t_j$ equals $\tmax + 1$ if and only if $t_s = t_j$ which implies that $\tilde{A}_{t_s,t_j} = 0$, as we can check by cases. This definition ensures that for all $j=1,...,\maxreg$, $t_j$ is a stopping time with respect to $(\F_t)_{t=0}^{\tmax}$.

Define $\mathcal{G}_t := \sigma(\F_t, \mathcal{I}_1, (X_i, W_i, Y_i')_{i \not\in \mathcal{I}_1}, (X_i)_{i=1}^n)$ for $t=0,...,\tmax + 1$ where we let $\F_{\tmax + 1} := \F_{\tmax}$. Subsequently, we will work conditionally on $\mathcal{G}_{t_s}$, so treat all $\mathcal{G}_{t_s}$-measurable quantities as fixed. In particular, we may treat $t_s$ as a fixed value. Let us slightly simplify notation again and define the following process $r=t_s,...,\tmax$:
\begin{equation*}
\begin{aligned}
M_r := \tilde{A}_{t_s, r}
\end{aligned}
\end{equation*}
and $M_{\tmax + 1} := M_{\tmax}$. First note that $\abs{\tilde{A}_{t_s, t_j}} \leq \abs{M_{t_j}}$ almost surely (with inequality only possible if $t_j = \tmax + 1$), so it suffices to show that the latter converges to $0$ in probability. Our idea is to show that $\E[M_{t_j} \mid \mathcal{G}_{t_s}] = 0$ and $\Var(M_{t_j} \mid \mathcal{G}_{t_s}) \stackrel{p}{\to} 0$. Then the result follows by Lemma~\ref{lemma:conditional-p0-implies-p0}. We first show that conditionally on $\mathcal{G}_{t_s}$, $(M_r)_{r=t_s}^{\tmax + 1}$ is a $0$-mean martingale with respect to $(\mathcal{G}_r)_{r=t_s}^{\tmax + 1}$. First observe that $\tilde{A}_{t_s, r}$ is $\mathcal{G}_r$-measurable since the terms which compose it are $\mathcal{G}_r$-measurable, so $M_r$ is $\mathcal{G}_r$-measurable.

Next, we show that $\E[M_{r+1} - M_r \mid \mathcal{G}_r] = 0$. When $r = \tmax$ this follows by construction, so let $r < \tmax$. Observe, importantly, that the scoring function and cap used to define $\reg_{r+1}$ is $\F_r$-measurable, and since we get to see all of $(X_i)_{i=1}^n$ in $\mathcal{G}_r$ we in fact have that $\reg_{r+1}$ is $\mathcal{G}_r$-measurable and also we can calculate how many points $\un_{r+1}$ fall within it. Then
\begin{equation*}
\begin{aligned}
M_{r + 1} - M_r &= \frac{1}{\sqrt{\un_{t_s}}} \sum_{i \in \mathcal{I}_1 : X_i \in \reg_r \setminus \reg_{r + 1}} (Y_{i,w} - \hat{Y}_{i,w})\\
&= \frac{1}{\sqrt{\un_{t_s}}} \sum_{i \in \mathcal{I}_1 : X_i \in \reg_r \setminus \reg_{r + 1}} (g_1(X_i) - \hat{g}_1^{(1)}(X_i))\paren*{ 1 - \frac{W_i}{e(X_i)} }.
\end{aligned}
\end{equation*}
Note that $\un_{t_s}$, $\reg_r \setminus \reg_{r + 1}$, $(X_i)_{i=1}^n$, and $\hat{g}_1^{(1)}(\cdot)$ are $\mathcal{G}_r$-measurable. Then the only random quantities, conditional on $\mathcal{G}_r$, in the above expression, are the $W_i$. But as in the proof of Lemma~\ref{lemma:conv-aipw-remainders}, conditioning on $\mathcal{G}_r$ does not affect the conditional distribution of $W_i \mid X_i$, and thus $1 - W_i / e(X_i)$ have mean $0$ conditional on $\mathcal{G}_r$ and $\E[M_{r + 1} - M_r \mid \mathcal{G}_r] = 0$. Now $t_j$ is a bounded stopping time conditionally on $\mathcal{G}_{t_s}$, so we may apply the optional stopping theorem to conclude that $\E[M_{t_j} \mid \mathcal{G}_{t_s}] = 0$.

Now we consider the variance, which we decompose in terms of martingale differences as
\begin{equation*}
\begin{aligned}
\Var(M_{t_j} \mid \mathcal{G}_{t_s}) &= \Ec*{ \sum_{r=t_s + 1}^{t_j} \Var(M_{r + 1} - M_r \mid \mathcal{G}_r )}{\mathcal{G}_{t_s}}.
\end{aligned}
\end{equation*}
Then again recalling that conditionally on $\mathcal{G}_r$ the only random terms in the difference $M_{r + 1} - M_r$ are the treatment indicators $W_i$, which are also independent,
\begin{equation*}
\begin{aligned}
\Var(M_{r + 1} - M_r \mid \mathcal{G}_r) &= \frac{1}{\un_{t_s}} \sum_{i \in \mathcal{I}_1 : X_i \in \reg_r \setminus \reg_{r + 1}} (g_1(X_i) - \hat{g}_1^{(1)}(X_i))^2 \Varc*{1 - \frac{W_i}{e(X_i)}}{\mathcal{G}_r}\\
&\leq \frac{1}{\un_{t_s}} \cdot \Lstable^2 \sum_{i \in \mathcal{I}_1 : X_i \in \reg_r \setminus \reg_{r + 1}} \Ec*{ \paren*{1 - \frac{W_i}{e(X_i)}}^2 }{\mathcal{G}_r}\\
&= \frac{1}{\un_{t_s}} \cdot \Lstable^2 \sum_{i \in \mathcal{I}_1 : X_i \in \reg_r \setminus \reg_{r + 1}} \frac{1 - e(X_i)}{e(X_i)}\\
&\leq \frac{\abs{\set{i \in \mathcal{I}_1 : X_i \in \reg_r \setminus \reg_{r + 1}}}}{\un_{t_s}} \cdot \Lstable^2 \cdot \frac{1 - \eta}{\eta}.
\end{aligned}
\end{equation*}
Therefore,
\begin{equation*}
\begin{aligned}
& \sum_{r=t_s + 1}^{t_j} \frac{\abs{\set{i \in \mathcal{I}_1 : X_i \in \reg_r \setminus \reg_{r + 1}}}}{\un_{t_s}} \cdot \Lstable^2 \cdot \frac{1 - \eta}{\eta}\\
=& \frac{\Lstable^2}{\un_{t_s}} \frac{1 - \eta}{\eta} \sum_{r=t_s + 1}^{t_j} \abs{\set{i \in \mathcal{I}_1 : X_i \in \reg_r \setminus \reg_{r + 1}}}\\
\leq& \frac{\Lstable^2}{\un_{t_s}} \frac{1 - \eta}{\eta} \cdot \un_{t_s}\\
=& \Lstable^2 \cdot \frac{1 - \eta}{\eta}.
\end{aligned}
\end{equation*}
We have bounded the sum in the second line by $\un_{t_s}$ by noticing that the sum is exactly adding up the number of points in-between $\reg_{t_s + 1}$ and $\reg_{t_s + 2}$, between $\reg_{t_s + 2}$ and $\reg_{t_s + 3}$, and so on. This can be no greater than $\un_{t_s}$, the number of points in $\reg_{t_s}$, which contains all of these regions. Then as $\Lstable^2 (1 - \eta) / \eta$ is $\mathcal{G}_{t_s}$-measurable, we have
\begin{equation*}
\begin{aligned}
\Var(M_{t_j} \mid \mathcal{G}_{t_s}) \leq \Lstable^2 \cdot \frac{1 - \eta}{\eta} \stackrel{p}{\to} 0
\end{aligned}
\end{equation*}
by Assumption~\ref{assumption:stable-nuisance}, as desired.
\end{proof}

With these facts, we are prepared to prove Lemma~\ref{lemma:aipw-uniform-conv-shift-crit-val}.

\begin{proof}[Proof of Lemma~\ref{lemma:aipw-uniform-conv-shift-crit-val}]
Note that as we have established Lemma~\ref{lemma:aipw-conv-intermediate}, we may exactly imitate the proof of Lemma~\ref{lemma:uniform-conv-crit-val} to obtain that
\begin{equation*}
\begin{aligned}
\sup_{t \in [\tmax]_0 : \allocalpha_t > 0} \abs{\hat{\critval}_t - \critval_t} \stackrel{p}{\to} 0.
\end{aligned}
\end{equation*}
Now we may write
\begin{equation*}
\begin{aligned}
\sup_{t \in [\tmax]_0 : \allocalpha_t > 0} \abs{d_t - \hat{d}_t} &= \sup_{t \in [\tmax]_0 : \allocalpha_t > 0} \abs*{ \critval_t - \sqrt{\frac{\hsigsq_t}{\sigsq_t}} \cdot \critval_t - R_t }\\
&\leq \sup_{t \in [\tmax]_0 : \allocalpha_t > 0} \abs*{ \critval_t - \sqrt{\frac{\hsigsq_t}{\sigsq_t}} \cdot \critval_t } + \sup_{t \in [\tmax]_0 : \allocalpha_t > 0} \abs{R_t}.
\end{aligned}
\end{equation*}
That the first supremum term in the final expression converges to $0$ follows the proof of Lemma~\ref{lemma:uniform-conv-shift-crit-val} identically, where we note that this exact term appears in that proof. That the second supremum term converges to $0$ follows from Lemma~\ref{lemma:conv-aipw-remainders}.
\end{proof}

\subsection{Distributional properties of chiseled regions}
\label{appendix:chisel-mass-var-props}

We work out some properties regarding the masses and conditional variances of chiseled regions. These are helpful for connecting some of the primitives described in Appendix~\ref{appendix:primitives} to the general asymptotic conditions described in Appendix~\ref{appendix:formalizing-asymptotic-conditions}. Some similar results can be found in \textcite{Fraser1951}, though our proofs are self-contained.

In the following discussion we will allow the regions and filtrations to be defined for all positive integers $t$ as in Equation~\eqref{eq:formalized-chiseling}. That is, we are working with $(\reg_t)_{t=0}^{\infty}$ and $(\F_t)_{t=0}^{\infty}$. Recall, also, that $\un_t := \abs{\set{i : X_i \in \reg_t}}$. For simplicity, we assume that Algorithm~\ref{alg:chiseling} chooses $\threshlim < \infty$ at most a finite number $N$ of times, where $N$ can be arbitrary, but is fixed. This way, $\reg_t = \emptyset$ for all $t > n + N$.

\begin{lemma}[Joint distribution of region masses]
\label{lemma:joint-dist-mass}
Suppose that $\un_t - \un_{t + 1} = 1$ for all $t=0,...,n-1$ almost surely. Then
\begin{equation*}
\begin{aligned}
(\Vol(\reg_1),...,\Vol(\reg_n)) \stackrel{d}{=} (U_{(n)},...,U_{(1)})
\end{aligned}
\end{equation*}
where $(U_{(1)},...,U_{(n)})$ are the joint order statistics of $n$ independent standard uniforms.
\end{lemma}

\begin{proof}[Proof of Lemma~\ref{lemma:joint-dist-mass}]
Let $U_{m, n}$ be a random variable representing the $m$th order statistic of $n$ uniform random variables. We will inductively characterize the distribution of $\Vol(\reg_t)$ in increasing order of $t$. In particular, we will show that for $t \geq 0$, $\Vol(\reg_{t + 1}) \stackrel{d}{=} \Vol(\reg_t) \times U_{n - t, n - t}$ conditionally on $\F_t$. Since $\Vol(\reg_0) = 1$, we can thus sample $(\Vol(\reg_1),...,\Vol(\reg_n))$ by independently generating $U_{n,n},...,U_{1,1}$ and setting $\Vol(\reg_t) = U_{n,n} \times ... \times U_{n-t, n-t}$ for each $t$. Lemma~\ref{lemma:unif-ord-stat-rep} then directly shows that this is equivalent to sampling the order statistics of $n$ independent uniforms, thus yielding the conclusion.

Let $t$ be fixed, and let $\score(\cdot)$ and $\threshlim$ be what Algorithm~\ref{alg:chiseling} chooses based on $\F_t$. First define $\pthresh := \threshlim \wedge \min_{i : X_i \in \reg_t} \score(X_i)$. Note that since we assume exactly one point is revealed with every application of chiseling, we must have that $\pthresh = \min_{i : X_i \in \reg_t} \score(X_i)$ almost surely, and also that $\score(X_i)$ has no ties almost surely conditionally on $\F_t$. Letting $F(\cdot)$ denote the (random, $\F_t$-measurable) CDF of $\score(X) \mid X \in \reg_t$, the latter condition implies that $F(Z)$ is uniformly distributed when $Z$ is sampled from $X \mid X \in \reg_t$. Now we implicitly condition on $\F_t$ and note that
\begin{equation*}
\begin{aligned}
\frac{\Vol(\reg_{t+1})}{\Vol(\reg_t)} &= \frac{\Prb(X \in \reg_{t+1})}{\Prb(X \in \reg_t)}\\
&= \Prb(X \in \reg_{t+1} \mid X \in \reg_t)\\
&= \Prb(Z \in \reg_{t+1})\\
&= \Prb(\score(Z) > \pthresh).
\end{aligned}
\end{equation*}
To be clear: $\reg_{t+1}$ and $\pthresh$ are still random conditionally on $\F_t$, and hence we read $\Prb(X \in \reg_{t+1})$ as the random variable that arises from plugging the random variable $\reg_{t+1}$ into the fixed function that evaluates $\Prb(X \in \reg)$, and so on. The first equality above is thus by definition and the second equality uses the fact that $\reg_{t+1} \subseteq \reg_t$ almost surely. The third equality is simply renaming; the fourth equality follows because $Z \in \reg_t$ almost surely and hence belongs to $\reg_{t+1}$ if and only if it additionally satisfies $\score(Z) > \pthresh$.

We have thus observed that the relative mass $\Vol(\reg_{t + 1}) / \Vol(\reg_t)$ is precisely the proportion of $\score(Z)$'s mass that falls above $\pthresh$; that is, $1 - F(\pthresh)$. But because $F(\cdot)$ is non-decreasing,
\begin{equation*}
\begin{aligned}
1 - F(\pthresh) = 1 - \min_{i : X_i \in \reg_t} F(\score(X_i)) \stackrel{d}{=} 1 - \min_{j=1,...,n-t} U_j \stackrel{d}{=} \max_{j=1,...,n-t} U_j
\end{aligned}
\end{equation*}
where $U_j$ are i.i.d. uniform random variables. Above, we have applied Corollary~\ref{corollary:distribution-subsamples} to conclude that $(\score(X_i))_{i : X_i \in \reg_t}$ are i.i.d. from the distribution $f(X) \mid X \in \reg_t$ and hence $(F(\score(X_i)))_{i : X_i \in \reg_t}$ are i.i.d. uniforms. This shows us that conditionally on $\F_t$, $\Vol(\reg_{t + 1}) / \Vol(\reg_t)$ is distributed as $U_{n - t, n - t}$. Thus we have that $\Vol(\reg_{t + 1}) = \Vol(\reg_t) \times (\Vol(\reg_{t + 1}) / \Vol(\reg_t)) \stackrel{d}{=} \Vol(\reg_t) \times U_{n-t, n-t}$ conditionally on $\F_t$ as desired.
\end{proof}

\begin{remark}[Generalized CDF]
The proof of Lemma~\ref{lemma:joint-dist-mass} works in general for scores taking values in $\mathbb{R}^2$ with lexicographic ordering (Appendix~\ref{appendix:total-order}). We mention this because we will leverage a construction based on lexicographic order in the subsequent proofs which reference this lemma. Note the following generalization of a CDF $F$: if $A$ is a random variable in $\mathbb{R}^2$ and $a \in \mathbb{R}^2$, define $F(a) := \Prb(A \leq_{\mathrm{lo}} a)$ where $\leq_{\mathrm{lo}}$ is lexicographic ordering. $F$ retains the useful properties that (1) if $a \leq_{\mathrm{lo}} b$, then $F(a) \leq F(b)$, and (2) $F(A)$ has the standard uniform distribution as long as $A$ is non-atomic, i.e. $\Prb(A = a) = 0$ for all $a$. Then $\leq_{\mathrm{lo}}$ may be used in place of the usual $\leq$ in the proof of Lemma~\ref{lemma:joint-dist-mass} without modification.
\end{remark}

Lemma~\ref{lemma:joint-dist-mass} characterizes the masses of the chiseled regions when they reveal one point at a time. We will soon prove a lemma that characterizes the distribution of masses more generally. But first we need a construction that allows us to ``consolidate" a sequence of chiseled regions so that there is at least one point in-between every contiguous pair of non-empty regions, and one that allows us to ``split" a sequence so that there is at most one point in-between every contiguous pair of regions. It is helpful to first offer an intuitive picture. One way to view chiseling is as a continuous process of shrinking the covariate space which reacts to data points as they are excluded from the region. The sequence of regions $(\reg_t)_{t=0}^{\infty}$ are simply discrete ``snapshots" of this continuous process, where the choice of whether to take a snapshot at any particular point depends only on the data points that have been excluded from the current region. Intuitively, then, we may construct instantiations of chiseling that correspond to fundamentally the same shrinking process but which take different snapshots---perhaps recording more intermediate stages of the shrinking process, or skipping over some previously recorded regions. The following lemma simply states that we may choose to not snapshot regions that have not revealed any new data points since the last snapshotted region. The lemma which follows states that we may snapshot more often so that each snapshot captures a change in at most one data point of information (in an augmented covariate space).

In the below, by an ``instantiation of chiseling," we mean a way of choosing $\score(\cdot)$ and $\threshlim$ in Algorithm~\ref{alg:chiseling} from the data. This way of choosing is itself a function from data, outputting the pair $(\score(\cdot), \threshlim)$.

\begin{lemma}[Consolidating regions]
\label{lemma:consolidation}
Let $(\reg_t)_{t=0}^{\infty}$ be the sequence of regions produced by a particular instantiation of chiseling. Also, for some fixed $N < \infty$, suppose that in this instantiation, Algorithm~\ref{alg:chiseling} sets $\threshlim < \infty$ at most $N$ times almost surely. Then there exists an instantiation of chiseling that produces a subsequence of regions $(\tilde{\reg}_t)_{t=0}^{\infty} \subseteq (\reg_t)_{t=0}^{\infty}$ with the property that for all $t \geq 0$, either $\set{i : X_i \in \tilde{\reg}_{t-1} \setminus \tilde{\reg}_t}$ is nonempty or $\tilde{\reg}_t = \emptyset$ almost surely. Furthermore, we have that almost surely, for all $t$ such that $\set{i : X_i \in \reg_{t-1} \setminus \reg_t}$ is nonempty, $\reg_t \in (\tilde{\reg}_t)_{t=0}^{\infty}$.
\end{lemma}

\begin{proof}[Proof of Lemma~\ref{lemma:consolidation}]
Let us index the scoring functions as $\score_t(\cdot)$ and caps as $\threshlim_t$; that is, the first application of chiseling will use $\score_1(\cdot)$, and $\threshlim_1$ in Algorithm~\ref{alg:chiseling}, and so on. Suppose that $r$ is the smallest index such that $\threshlim_r < \infty$; note that $r$ is random, a stopping time, and that we may have $r = \infty$. We describe a modified instantiation of chiseling that uses the scoring functions $\tilde{\score}_t(\cdot)$ and caps $\tilde{\threshlim}_t$, producing the regions $(\tilde{\reg}_t)_{t=0}^{\infty}$. For all $t < r$, we let $\tilde{\score}_t(\cdot) = \score_t(\cdot)$ and $\tilde{\threshlim}_t = \threshlim_t$. If $r = \infty$, then this corresponds to no modification. Otherwise if $r < \infty$, let $\score'(\cdot)$ and $\threshlim'$ be the values that $\score_{r+1}(\cdot)$ and $\threshlim_{r+1}$ \emph{would} take in the event that $\set{i : X_i \in \reg_{r-1} \setminus \reg_r}$ is empty. Note that both $\score'(\cdot)$ and $\threshlim'$ can be determined without actually having to observe whether $\set{i : X_i \in \reg_{r-1} \setminus \reg_r}$ is empty, i.e. they are $\F_r$-measurable (i.e. the analyst can ``think ahead"). Then we let
\begin{equation*}
\begin{aligned}
\tilde{\score}_r(x) = \begin{cases}
\arctan(\score_{r}(x)) & \text{ if } \score_{r}(x) \leq \threshlim_{r},\\
\pi + \arctan(\score'(x)) & \text{ if } \score_{r}(x) > \threshlim_{r},\\
\end{cases}
\quad \text{ and } \quad
\tilde{\threshlim}_r = \begin{cases}
\infty & \text{ if } \threshlim' = \infty,\\
\pi + \arctan(\threshlim') & \text{ otherwise}.
\end{cases}
\end{aligned}
\end{equation*}
Straightforward, though tedious, algebra and case-checking verifies that $\tilde{\reg}_r = \reg_r$ if $\set{i : X_i \in \reg_{r-1} \setminus \reg_r}$ is nonempty and $\tilde{\reg}_r = \reg_{r+1}$ otherwise. Intuitively, all the above is doing is shrinking according to $\score_r(\cdot)$ as long as we are below the cap $\threshlim_r$, and switching to $\score'(\cdot)$ thereafter, stopping at $\threshlim'$. The role of $\arctan(\cdot)$ and the shifting by $\pi$ is simply to separate the two scores into non-overlapping ranges so that we can shrink according to one scoring function followed by the other.

Finally, for all $t > r$ we define $\tilde{\score}_t(\cdot) = \score_t(\cdot)$ and $\tilde{\threshlim}_t = \threshlim_t$ if $\set{i : X_i \in \reg_{r-1} \setminus \reg_r}$ is nonempty, and $\tilde{\score}_{t+1}(\cdot) = \score_{t+1}(\cdot)$ and $\tilde{\threshlim}_{t+1} = \threshlim_{t+1}$ otherwise. Note that this last construction only makes sense if $\tilde{\score}_t(\cdot)$ and $\tilde{\threshlim}_t(\cdot)$ can be calculated from the revealed data points in the modified instantiation. But this is indeed the case, since by the above we have access to all the data points outside of $\reg_r$ when $\set{i : X_i \in \reg_{r-1} \setminus \reg_r}$ is nonempty, and all the data points outside of $\reg_{r+1}$ otherwise. Thus, the information necessary to determine $\score_{r+1}(\cdot)$ in the former case and $\score_{r+2}(\cdot)$ in the latter case is available. It follows that $\tilde{\reg}_{r+1} = \reg_{r+1}$ in the former case and $\tilde{\reg}_{r+1} = \reg_{r+2}$ in the latter case. Then applying this argument inductively, for all $t > r$ we have that $\tilde{\score}_t(\cdot)$ and $\tilde{\threshlim}_t$ can be calculated from the revealed information (i.e. this is a valid instantiation of chiseling), and that $\tilde{\reg}_t = \reg_t$ when $\set{i : X_i \in \reg_{r-1} \setminus \reg_r}$ is nonempty and $\tilde{\reg}_{t} = \reg_{t+1}$ otherwise.

Thus, we have described a valid instantiation of chiseling that produces regions $(\tilde{\reg}_t)_{t=0}^{\infty} \subseteq (\reg_t)_{t=0}^{\infty}$. Furthermore, the modified instantiation sets $\tilde{\threshlim} < \infty$ in Algorithm~\ref{alg:chiseling} at most $N - 1$ times almost surely, because the modified instantiation almost surely sets one fewer finite cap than the original instantiation (in particular, cutting out the finite cap at stage $r$) as long as the latter sets at least one finite cap. Lastly, it preserves all regions $\reg_t$ where $\set{i : X_i \in \reg_{t-1} \setminus \reg_t}$ is nonempty by construction, since the only possibility is that it removes $\reg_r$ if $\reg_r$ does not satisfy this property.

By again modifying the modified instantiation using the same construction as above, we can obtain an instantiation that sets at most $N - 2$ finite caps, and so on until we obtain an instantiation that sets zero finite caps. Each modified instantiation produces regions that are subsequences of the regions outputted by the instantiation it modifies, and never does it remove a region $\reg_t$ with the property that $\set{i : X_i \in \reg_{t-1} \setminus \reg_t}$ is nonempty. Since in the final modification all caps are infinite, we must have that each application of chiseling reveals at least one point as long as there are points left. That is, if $(\tilde{\reg}_t)_{t=0}^{\infty} \subseteq (\reg_t)_{t=0}^{\infty}$ denotes the regions produced by the final modified instantiation, then $\set{i : X_i \in \tilde{\reg}_{t-1} \setminus \tilde{\reg}_t}$ is nonempty as long as $\reg_t \neq \emptyset$, as desired.
\end{proof}

\begin{lemma}[Splitting regions]
\label{lemma:splitting-regions}
Let $(\reg_t)_{t=0}^{\infty}$ be the sequence of regions produced by a particular instantiation of chiseling. Also, for some fixed $N < \infty$, suppose that in this instantiation, Algorithm~\ref{alg:chiseling} sets $\threshlim < \infty$ at most $N$ times almost surely. Let $U_1,...,U_n$ be i.i.d. standard uniforms, and define $\tilde{X}_i := (X_i, U_i)$ for $i=1,...,n$. Then there exists an instantiation of chiseling applied to $(\tilde{X}_i, Y_i)_{i=1}^n$ that produces a supersequence of augmented regions $(\tilde{\reg}_t)_{t=0}^{\infty} \supseteq (\reg_t \times [0,1])_{t=0}^{\infty}$ with the property that $\set{i : \tilde{X}_i \in \tilde{\reg}_{t-1} \setminus \tilde{\reg_t}}$ has at most one element almost surely for all $t \geq 0$.
\end{lemma}

\begin{proof}[Proof of Lemma~\ref{lemma:splitting-regions}]
In this proof we use the tiebreaking construction described in Appendix~\ref{appendix:tiebreaking}. Thus, orderings here should be understood lexicographically. We constructively describe an instantiation of chiseling and the chiseled regions $\tilde{\reg}_t$. First, we note that $\tilde{\reg}_0 = \mathcal{X} \times [0,1] = \reg_0 \times [0,1]$ by construction. We describe a sequence of chiseled regions $\tilde{\reg}_1,...,\tilde{\reg}_k$ of variable length $k$ such that $\tilde{\reg}_k$ coincides with $\reg_1 \times [0,1]$ and $\tilde{\reg}_{k-1} \setminus \tilde{\reg}_k$ contains at most one data point. As in the proof of Lemma~\ref{lemma:consolidation}, let us index the scoring functions in the original instantiation as $\score_t(\cdot)$ and caps as $\threshlim_t$. In the modified instantiation, run Algorithm~\ref{alg:chiseling} with
\begin{equation*}
\begin{aligned}
\tilde{\score}(x, u) = (\score_1(x), u) \quad \text{ and } \quad \tilde{\threshlim} = (\threshlim_1, 1)
\end{aligned}
\end{equation*}
once, and let the outputted region be $\tilde{\reg}_1$. If $\set{i : \tilde{X_i} \in \tilde{\reg}_0 \setminus \tilde{\reg}_1}$ is empty, then $\tilde{\reg}_1 = \reg_1 \times [0,1]$. Otherwise, let $\pthresh = \min_{i : (X_i, U_i) \in \tilde{\reg}_0 \setminus \tilde{\reg}_1} \score_1(X_i)$ and repeatedly run Algorithm~\ref{alg:chiseling} with the same $\tilde{\score}(x, u)$ as above but with $\tilde{\threshlim} = (\pthresh, 1)$ until no new points are revealed, which must happen eventually since there are finitely many data points. That is, let $\tilde{\reg}_2,...,\tilde{\reg}_k$ be obtained by chiseling with the above scoring function and cap and such that $\set{i : \tilde{X_i} \in \tilde{\reg}_{t-1} \setminus \tilde{\reg}_t}$ has exactly one element for $t = 3,...,k-1$ (since by using the auxiliary dimension we encounter no ties almost surely) and $\set{i : \tilde{X_i} \in \tilde{\reg}_{k-1} \setminus \tilde{\reg}_k}$ is empty. Then
\begin{equation*}
\begin{aligned}
\tilde{\reg}_k &= \tilde{\reg}_0 \cap \set{(x, u) \in \mathcal{X} \times [0,1] : (\score_1(x), u) > (\pthresh, 1) }\\
&= \reg_0 \cap \set{(x, u) \in \mathcal{X} \times [0,1] : \score_1(x) > \pthresh}\\
&= \reg_1 \times [0,1].
\end{aligned}
\end{equation*}
The first equality follows since $\tilde{\reg}_1,...,\tilde{\reg}_k$ are chiseled using the same $\tilde{\score}(x, u)$ and thus it suffices to consider just the most stringent threshold for $\tilde{\score}(x, u)$, which must be $(\pthresh, 1)$ since that is the cap and we chisel until no new points are revealed. Now since we have revealed the same data points that would have been revealed in the original instantiation after shrinking to $\reg_1$, we can deduce $\score_2(\cdot)$ and $\threshlim_2$ from this information in the modified instantiation. Then we may repeat this construction to obtain a further sequence of regions $\tilde{\reg}_{k+1},...,\tilde{\reg}_{k + l}$ such that $\tilde{\reg}_{k + l} = \reg_2 \times [0,1]$ almost surely. Recursively repeating this construction results in a supersequence $(\tilde{\reg}_t)_{t=0}^{\infty}$ of $(\reg_t \times [0,1])_{t=0}^{\infty}$. Also, we have already seen that since we use the auxiliary dimension to tiebreak that we must reveal at most one element with every application of Algorithm~\ref{alg:chiseling}, i.e. $\set{i : \tilde{X}_i \in \tilde{\reg}_{t-1} \setminus \tilde{\reg}_t}$ has at most one element as desired.
\end{proof}

Now we state a general characterization of the masses of chiseled regions.

\begin{lemma}[Distribution of smallest mass with sufficient samples]
\label{lemma:dist-smallest-mass}
Let $1 < m \leq n$ be fixed. Define
\begin{equation*}
\begin{aligned}
\volbound^{-1} := \inf_{t \geq 0 : \un_t \geq m} \Vol(\reg_t).
\end{aligned}
\end{equation*}
Let $U_{m, n}$ be a random variable representing the $m$th order statistic of $n$ uniform random variables. Then $\volbound^{-1}$ stochastically dominates $U_{m,n}$.
\end{lemma}

\begin{proof}[Proof of Lemma~\ref{lemma:dist-smallest-mass}]
We will construct a dataset $(\tilde{X}_i, Y_i)_{i=1}^n$ coupled to the same probability space as $(X_i, Y_i)_{i=1}^n$ and a sequence of chiseled regions $(\tilde{\reg}_t)_{t=0}^{\infty}$ obtained from the coupled dataset with certain properties. First, define $\tilde{\un}_t := \abs{\set{i : \tilde{X}_i \in \tilde{\reg}_t}}$ and $\tilde{\Vol}(\tilde{\reg}) := \Prb(\tilde{X} \in \tilde{\reg})$. We will show that
\begin{equation}
\label{eq:vol-coupling}
\begin{aligned}
\volbound^{-1} = \inf_{t \geq 0 : \un_t \geq m} \Vol(\reg_t) \geq \inf_{t \geq 0 : \tilde{\un}_t \geq m - 1} \tilde{\Vol}(\tilde{\reg}_t) \quad \text{ a.s.}
\end{aligned}
\end{equation}
and that $\tilde{\un}_t - \tilde{\un}_{t + 1} = 1$ for all $t = 0,...,n-1$ almost surely. Then as $\tilde{\un}_t = n - t$, the minimum is attained when $m - 1 = n - t$, or equivalently when $t = n - m + 1$. Thus, the right-hand side further reduces to $\tilde{\Vol}(\tilde{\reg}_{n - m + 1})$, and since this term arises from an instantiation of chiseling, by Lemma~\ref{lemma:joint-dist-mass} it is distributed as $U_{m, n}$, and this will complete the proof.

Let $(\tilde{X}_i, Y_i)_{i=1}^n$ be as in the statement of Lemma~\ref{lemma:splitting-regions}, and let $(\reg_t')_{t=0}^{\infty}$ be the supersequence of regions given by the lemma. Then let $(\tilde{\reg}_t)_{t=0}^{\infty}$ be the subsequence of $(\reg_t')_{t=0}^{\infty}$ obtained by consolidating the latter using Lemma~\ref{lemma:consolidation}. Since $(\reg_t')_{t=0}^{\infty}$ is such that each successive region reveals at most one data point, and $(\tilde{\reg}_t)_{t=0}^{\infty}$ is precisely the subsequence that skips over regions that do not reveal any new data points, we have that $\set{i : \tilde{X}_i \in \tilde{\reg}_{t-1} \setminus \tilde{\reg}_t}$ has exactly one element for all $t = 1,...,n$. That is, $\tilde{\un}_{t-1} - \tilde{\un}_t = 1$ for all $t = 1,...,n$. Furthermore, we note that for every $t$, there is a $\tilde{\reg}_{s(t)}$ such that $\reg_t \times [0,1] \supseteq \tilde{\reg}_{s(t)}$ and $\un_t - \tilde{\un}_{s(t)} \leq 1$. That is, for every region $\reg_t$ there is a region $\tilde{\reg}_{s(t)}$ contained in $\reg_t \times [0,1]$ that excludes at most one additional point. To see why, note that $\reg_t \times [0,1] = \reg_u'$ for some $u$. Let $s$ be the smallest integer such that $\tilde{\reg}_s \subseteq \reg_u'$. Then $\un_t - \tilde{\un}_s \leq 1$, since splitting creates a sequence that reveals at most one data point at each stage, consolidation only excludes regions that revealed no data points since the last stage, and $(\tilde{\reg}_t)_{t=0}^{\infty}$ is obtained by consolidating $(\reg_t')_{t=0}^{\infty}$, which is obtained by splitting.

Finally, we must show the domination in Equation~\eqref{eq:vol-coupling}. But this simply follows because if $\un_t \geq m$, then $\tilde{\un}_{s(t)} \geq m - 1$ and $\Vol(\reg_t) = \tilde{\Vol}(\reg_t \times [0,1]) \geq \tilde{\Vol}(\tilde{\reg}_{s(t)})$, and hence for every element in the left-hand side infimum there is a smaller element in the right-hand side infimum.
\end{proof}

This result allows us to easily establish the following corollary.

\begin{corollary}[$L^1$-bound on inverse volume]
\label{corollary:l1-bound-inverse-volume}
Let $\minprop \in (0,1)$ be fixed, and let
\begin{equation*}
\begin{aligned}
\volbound := \sup_{t \geq 0 : \un_t \geq \minprop \cdot n} \Vol(\reg_t)^{-1}.
\end{aligned}
\end{equation*}
Then as long as $\minprop \cdot n > 1$,
\begin{equation*}
\begin{aligned}
\E[\volbound] \leq \frac{1}{\minprop - n^{-1}}.
\end{aligned}
\end{equation*}
In particular, $\E[\volbound] \leq 2 / \minprop$ for $n > 2 / \minprop$.
\end{corollary}

\begin{proof}[Proof of Corollary~\ref{corollary:l1-bound-inverse-volume}]
Letting $m = \ceil{\minprop \cdot n}$, Lemma~\ref{lemma:dist-smallest-mass} tells us that $\volbound$ is stochastically dominated by $U_{m, n}^{-1}$. We note that $U_{m,n} \sim \text{Beta}(m, n - m + 1)$, which has the same distribution as $Z_1 / (Z_1 + Z_2)$ where $Z_1 \sim \text{Gamma}(m, 1)$ and $Z_2 \sim \text{Gamma}(n - m + 1, 1)$ are independent. Thus, $U_{m,n}^{-1}$ has the same distribution as $(Z_1 + Z_2) / Z_1 = 1 + Z_2 / Z_1 = 1 + B'$ where $B' \sim \text{BetaPrime}(n - m  + 1, m)$. The mean of the beta prime distribution is a standard fact and in this case given by $\frac{n - m + 1}{m - 1}$ (note that $m > 1$), so
\begin{equation*}
\begin{aligned}
\E[\volbound] \leq 1 + \frac{n - m + 1}{m - 1} = \frac{n}{m - 1} \leq \frac{n}{\minprop \cdot n - 1} = \frac{1}{\minprop - n^{-1}}.
\end{aligned}
\end{equation*}
The right-hand side is bounded by $2/\minprop$ for $n > 2 / \minprop$.
\end{proof}

The bound on the inverse volume helps us establish another useful bound on the smallest conditional variance of any tested region in the pointwise setting. The following lemma reintroduces dependence on $n$, since we will show that a certain random sequence is bounded in probability. As in Appendix~\ref{appendix:proofs-general-validity}, formal statements will include $n$ superscripts where appropriate but otherwise we will suppress dependence on $n$ for readability.

\begin{lemma}[Probability bound on inverse of smallest variance]
\label{lemma:bound-inverse-variance}
Suppose that we work under the pointwise asymptotics defined in Section~\ref{section:general-test-design}. Let $\minprop \in (0,1)$ be fixed and define
\begin{equation*}
\begin{aligned}
\minselstd^{(n)} := \paren*{ \inf_{t \geq 0 : \un_t^{(n)} \geq \minprop \cdot n} \CStd{1}(\reg_t^{(n)}) }^{-1}.
\end{aligned}
\end{equation*}
Then under Assumption~\ref{assumption:dgp-regularity}, $\minselstd^{(n)}$ is bounded in probability.
\end{lemma}

\begin{proof}[Proof of Lemma~\ref{lemma:bound-inverse-variance}]
Let us define
\begin{equation*}
\begin{aligned}
B(q) := \inf_{\reg \subseteq \mathcal{X} : \Vol(\reg) \geq q} \CStd{1}(\reg).
\end{aligned}
\end{equation*}
Note that Assumption~\ref{assumption:dgp-regularity} implies that $B(q)^2 > 0$ for all $0 < q \leq 1$ (Lemma~\ref{lemma:positive-region-cond-var}). Hence, $B(q) > 0$ for all $0 < q \leq 1$.

Let $\epsilon > 0$. Recall the definition of $\volbound$ from Corollary~\ref{corollary:l1-bound-inverse-volume} and note that
\begin{equation*}
\begin{aligned}
\volbound \leq q^{-1} &\implies \inf_{t \geq 0 : \un_t \geq \minprop \cdot n} \Vol(\reg_t) \geq q\\
&\implies \inf_{t \geq 0 : \un_t \geq \minprop \cdot n} \CStd{1}(\reg_t) \geq B(q)\\
&\implies \minselstd \leq B(q)^{-1}.
\end{aligned}
\end{equation*}
Since $\volbound$ is $L^1$-bounded by Corollary~\ref{corollary:l1-bound-inverse-volume} (where we have now made $\volbound$ dependent upon $n$), it is also bounded in probability. So we may choose $q > 0$ such that $\Prb(\volbound > q^{-1}) < \epsilon$ for sufficiently large $n$. Then
\begin{equation*}
\begin{aligned}
\Prb(\minselstd > B(q)^{-1}) \leq \Prb(\volbound > q^{-1}) < \epsilon
\end{aligned}
\end{equation*}
for sufficiently large $n$, and hence $\minselstd$ is bounded in probability.
\end{proof}

\section{Proofs of technical lemmas}
\label{appendix:technical-proofs}

The proofs in Appendix~\ref{appendix:technical-proofs} are self-contained. It will help to consider the notational choices used here as independent of those made in the rest of the paper.

\subsection{Measure theory refreshers}

\begin{lemma}[Nonzero conditional probability if event happens]
\label{lemma:nonzero-cond-prob-bound}
Let $A$ be an event and let $\mathcal{F}$ be a $\sigma$-algebra. Then
\begin{equation*}
\begin{aligned}
\E[\indic\set{A} \mid \mathcal{F}] > 0 \text{ a.s. conditional on } A.
\end{aligned}
\end{equation*}
\end{lemma}

\begin{proof}[Proof of Lemma~\ref{lemma:nonzero-cond-prob-bound}]
Let $Z := \indic\set{\E[\indic\set{A} \mid \mathcal{F}] = 0}$. Note that $Z$ is $\F$-measurable. Then
\begin{equation*}
\begin{aligned}
0 = \E[0] = \E[\E[\indic\set{A} \mid \mathcal{F}] \cdot Z] = \E[\E[\indic\set{A} \cdot Z \mid \mathcal{F}]] = \E[\indic\set{A} \cdot Z].
\end{aligned}
\end{equation*}
Then $\indic\set{A} \cdot Z = 0$ almost surely, since it is nonnegative with expectation $0$. Then for almost all $\omega \in A$, this implies that
\begin{equation*}
\begin{aligned}
0 = \indic\set{A}(\omega) \cdot Z(\omega) = Z(\omega) \implies \E[\indic\set{A} \mid \mathcal{F}](\omega) > 0
\end{aligned}
\end{equation*}
since $\indic\set{A}(\omega) = 1$.
\end{proof}

\begin{lemma}[Measure-theoretic conditional expectation]
\label{lemma:measure-theory-conditional-prob}
Let $X$ be a random variable with finite first moment, $A$ be an event, and $\mathcal{F}$ be a $\sigma$-algebra. Then
\begin{equation*}
\begin{aligned}
\E[X \mid \mathcal{F}, A] = \E[X \mid \mathcal{F}, \indic\set{A}] \quad \text{ a.s. conditional on } A.
\end{aligned}
\end{equation*}
\end{lemma}

\begin{proof}[Proof of Lemma~\ref{lemma:measure-theory-conditional-prob}]
Recall that we have defined
\begin{equation*}
\begin{aligned}
\E[X \mid \mathcal{F}, A] = \frac{\E[X \indic\set{A} \mid \mathcal{F}]}{\E[\indic\set{A} \mid \mathcal{F}]}
\end{aligned}
\end{equation*}
where originally we have taken the convention that $0/0$ is undefined. Note that since $\E[\indic\set{A} \mid \mathcal{F}] > 0$ almost surely conditional on $A$ (Lemma~\ref{lemma:nonzero-cond-prob-bound}), then for the purposes of this proof, we may as well take the convention $0/0 = 0$, since these two conventions result in definitions that agree almost everywhere on $A$. Thus, we subsequently think of $\E[X \mid \mathcal{F}, A]$ as being well-defined almost everywhere. This will be a much more convenient object to work with. Now note that
\begin{equation*}
\begin{aligned}
\frac{\E[X \indic\set{A} \mid \mathcal{F}]}{\E[\indic\set{A} \mid \mathcal{F}]} \cdot \E\bkt*{\indic\set{A} \mid \mathcal{F}} = \E[X \indic\set{A} \mid \mathcal{F}]
\end{aligned}
\end{equation*}
since when $\E[\indic\set{A} \mid \mathcal{F}] = 0$, the left-hand side evaluates to $0 \cdot 0 = 0$ and the right-hand side evaluates to $0$. Otherwise, the denominator cancels ordinarily.

Next, note that $\mathcal{P} := \set{ S_1 \cap S_2 : S_1 \in \mathcal{F}, S_2 \in \set{A, A^{\comp}, \Omega}}$ is a $\pi$-system and $\sigma(\mathcal{P}) = \sigma(\mathcal{F}, \indic\set{A})$. We will check that
\begin{equation}
\label{eq:measure-theory-cond-prob}
\begin{aligned}
\E\bkt*{\frac{\E[X \indic\set{A} \mid \mathcal{F}]}{\E[\indic\set{A} \mid \mathcal{F}]} \cdot \indic\set{A} \cdot \indic\set{S}} = \E[X\indic\set{A} \cdot \indic\set{S}]
\end{aligned}
\end{equation}
for all $S \in \mathcal{P}$. But probability measures are uniquely defined by their values on $\pi$-systems, and thus the above will hold for all $S \in \sigma(\mathcal{F}, \indic\set{A})$. Hence,
\begin{equation*}
\begin{aligned}
\E[X \mid \mathcal{F}, \indic\set{A}] \cdot \indic\set{A} = \E[X \indic\set{A} \mid \mathcal{F}, \indic\set{A}] = \frac{\E[X \indic\set{A} \mid \mathcal{F}]}{\E[\indic\set{A} \mid \mathcal{F}]} \cdot \indic\set{A} \quad \text{ a.s.}
\end{aligned}
\end{equation*}
by the measure-theoretic definition and uniqueness of conditional expectation. But this is equivalent to the statement in the lemma. Now we check Equation~\eqref{eq:measure-theory-cond-prob}. Below, $S_1 \in \mathcal{F}$.
\begin{enumerate}
\item \textbf{Case 1:} $S = S_1 \cap A$. Then
\begin{equation*}
\begin{aligned}
\E\bkt*{\frac{\E[X \indic\set{A} \mid \mathcal{F}]}{\E[\indic\set{A} \mid \mathcal{F}]} \cdot \indic\set{A} \cdot \indic\set{S}} &= \E\bkt*{\frac{\E[X \indic\set{A} \mid \mathcal{F}]}{\E[\indic\set{A} \mid \mathcal{F}]} \cdot \indic\set{A} \cdot \indic\set{S_1} \cdot \indic\set{A}}\\
&= \E\bkt*{\E\bkt*{\frac{\E[X \indic\set{A} \mid \mathcal{F}]}{\E[\indic\set{A} \mid \mathcal{F}]} \cdot \indic\set{A} \cdot \indic\set{S_1} \bigcond \mathcal{F}}}\\
&= \E\bkt*{\frac{\E[X \indic\set{A} \mid \mathcal{F}]}{\E[\indic\set{A} \mid \mathcal{F}]} \cdot \indic\set{S_1} \cdot \E\bkt*{\indic\set{A} \mid \mathcal{F}}}\\
&= \E\bkt*{\E[X \indic\set{A} \mid \mathcal{F}] \cdot \indic\set{S_1} }\\
&= \E\bkt*{\E[X \indic\set{A} \cdot \indic\set{S_1} \mid \mathcal{F}] }\\
&= \E\bkt{X \indic\set{A} \cdot \indic\set{S_1}}\\
&= \E\bkt{X \indic\set{A} \cdot \indic\set{S}}
\end{aligned}
\end{equation*}
where to go to the last line we use the fact that $\indic\set{A} \cdot \indic\set{S_1} = \indic\set{A} \cdot \indic\set{S}$.
\item \textbf{Case 2:} $S = S_1 \cap A^{\comp}$. Then 
\begin{equation*}
\begin{aligned}
\E\bkt*{\frac{\E[X \indic\set{A} \mid \mathcal{F}]}{\E[\indic\set{A} \mid \mathcal{F}]} \cdot \indic\set{A} \cdot \indic\set{S}} = \E\bkt*{\frac{\E[X \indic\set{A} \mid \mathcal{F}]}{\E[\indic\set{A} \mid \mathcal{F}]} \cdot \indic\set{A} \cdot \indic\set{S_1} \cdot \indic\set{A^{\comp}}} = 0
\end{aligned}
\end{equation*}
and $\E[X\indic\set{A} \cdot \indic\set{S}] = \E[X\indic\set{A} \cdot \indic\set{S_1} \cdot \indic\set{A^{\comp}}] = 0$.
\item \textbf{Case 3:} $S = S_1 \cap \Omega = S_1$. Then we may take the exact same calculation and reasoning as in \textbf{Case 1}.
\end{enumerate}
\end{proof}

\begin{lemma}[Conditional equivalence at locally invertible points]
\label{lemma:equivalence-local-invertible}
Let $X$ and $Z$ be random variables where $Z$ is discrete. Let $h$ be a measurable function and define $A := \set{\omega \in \Omega : h^{-1}(h(Z(\omega))) \text{ is a singleton}}$. Then
\begin{equation*}
\begin{aligned}
\E[X \mid Z] = \E[X \mid h(Z)] \quad \text{ a.s. conditional on } A.
\end{aligned}
\end{equation*}
\end{lemma}

\begin{proof}[Proof of Lemma~\ref{lemma:equivalence-local-invertible}]
This follows directly from
\begin{equation*}
\begin{aligned}
\E[X \mid h(Z)](\omega) = \E[X \mid h(Z) = h(Z(\omega))] = \E[X \mid Z = Z(\omega)] = \E[X \mid Z](\omega)
\end{aligned}
\end{equation*}
because $h(Z) = h(Z(\omega)) \iff Z = Z(\omega)$ for $\omega \in \set{\omega \in \Omega : h^{-1}(h(Z(\omega))) \text{ is a singleton}}$.
\end{proof}

\begin{lemma}[Stopped conditional expectation evaluated at a point]
\label{lemma:stopped-sigma-algebra}
Let $Z$ be a bounded random variable and $(\mathcal{F}_t)_{t=0}^{\infty}$ be a filtration. Let $\nu$ be an almost surely finite stopping time with respect to $(\mathcal{F}_t)_{t=0}^{\infty}$ and define the stopped $\sigma$-algebra $\mathcal{F}_{\nu} := \set{A : \set{\nu \leq t} \cap A \in \F_t \text{ for all } t}$. Then
\begin{equation*}
\begin{aligned}
\E[Z \mid \mathcal{F}_{\nu}] = \sum_{t=0}^{\infty} \E[Z \mid \mathcal{F}_t] \indic\set{\nu = t} \quad \text{ a.s.}
\end{aligned}
\end{equation*}
\end{lemma}

\begin{proof}[Proof of Lemma~\ref{lemma:stopped-sigma-algebra}]
Write the sum on the right-hand side as $M$. We first show that $M$ is $\mathcal{F}_{\nu}$-measurable. Let $B \subseteq \mathbb{R}$ be Borel measurable. We will show that $\set{M \in B} \in \mathcal{F}_{\nu}$, or equivalently that
\begin{equation*}
\begin{aligned}
\set{\nu \leq t} \cap \set{M \in B} \in \mathcal{F}_t \quad \text{ for all } t.
\end{aligned}
\end{equation*}
But this follows since conditional on $\set{\nu \leq t}$, we have $M = \sum_{s=0}^t \E[Z \mid \mathcal{F}_s] \indic\set{\nu = s}$. Each term in the sum is an $\mathcal{F}_t$-measurable quantity, so conditional on $\set{\nu \leq t}$, $M$ is $\mathcal{F}_t$-measurable.

Now let $A \in \mathcal{F}_{\nu}$. Note that $\set{\nu \leq t} \cap A$ and $\set{\nu \leq t - 1} \cap A$ are both $\mathcal{F}_t$-measurable by definition (the latter since $\mathcal{F}_{t-1} \subseteq \mathcal{F}_t$). Taking the difference implies that $\set{\nu = t} \cap A$ is $\F_t$-measurable. Then
\begin{equation*}
\begin{aligned}
\E[Z \indic\set{A}] &= \sum_{t=0}^{\infty} \E\bkt*{ Z \indic\set{A}  \indic\set{\nu = t}}\\
&= \sum_{t=0}^{\infty} \E\bkt*{ \E[Z \indic\set{A}  \indic\set{\nu = t} \mid \mathcal{F}_t] }\\
&= \sum_{t=0}^{\infty} \E\bkt*{ \indic\set{A} \indic\set{\nu = t} \E[Z \mid \mathcal{F}_t] }\\
&= \E\bkt*{ \indic\set{A} \sum_{t=0}^{\infty} \indic\set{\nu = t} \E[Z \mid \mathcal{F}_t] }.
\end{aligned}
\end{equation*}
The first equality follows since $\sum_{t=0}^{\infty} \indic\set{\nu = t} = 1$ as $\nu$ is almost surely finite. The second equality follows from the tower rule, the third equality follows since $\indic\set{A} \indic\set{\nu = t} = \indic\set{A \cap \set{\nu = t}}$ which is $\F_t$-measurable, so the product can be pulled out of the expectation. The last equality simply pushes the sum back inside the expectation. The result then follows by the uniqueness of conditional expectation.
\end{proof}

\subsection{Proof of Lemma~\ref{lemma:distribution-min-cond}}
\label{appendix:proof-distribution-min-cond}

\begin{proof}
Let $\Omega$ denote the sample space of the base probability measure. Let $\omega \in \Omega$ be fixed. Then
\begin{equation*}
\begin{aligned}
\pthresh = \pthresh(\omega) \text{ and } \I = \I(\omega) \iff S_i = \pthresh(\omega) \text{ for all } i \not\in \I(\omega) \text{ and } S_i > \pthresh(\omega) \text{ for all } i \in \I(\omega).
\end{aligned}
\end{equation*}
Hence, for any measurable $B \in \mathbb{R}^{\abs{\I(\omega)}}$,
\begin{equation*}
\begin{aligned}
& \Prb((S_i)_{i \in \I} \in B \mid \pthresh, \I)(\omega)\\
=& \Prb((S_i)_{i \in \I(\omega)} \in B \mid \pthresh = \pthresh(\omega), \I = \I(\omega))\\
=& \Prb((S_i)_{i \in \I(\omega)} \in B \mid S_i = \pthresh(\omega) \text{ for all } i \not\in \I(\omega), S_i > \pthresh(\omega) \text{ for all } i \in \I(\omega))\\
=& \Prb((S_i)_{i \in \I(\omega)} \in B \mid S_i > \pthresh(\omega) \text{ for all } i \in \I(\omega)).
\end{aligned}
\end{equation*}
The last line follows because $(S_i)_{i \in \I(\omega)}$ is independent of $(S_i)_{i \not\in \I(\omega)}$; note here that $\I(\omega)$ is a fixed set. Also, the last line exactly describes the measure of $\abs{\I(\omega)}$ draws from the distribution of $S \mid S > \pthresh(\omega)$.
\end{proof}

\subsection{Convergences, inequalities, and distributional facts}

\begin{lemma}[Discrete conditional stochastic dominance]
\label{lemma:discrete-conditional-stochastic-dominance}
Let $X, Y$ be discrete random variables on a common support $[n] := \set{ 0,...,n }$. Denote the PMF and CDF of $X$ by $f$ and $F$, and the PMF and CDF of $Y$ by $g$ and $G$, respectively. Then
\begin{equation*}
\begin{aligned}
\frac{F(c)}{F(\gamma)} \leq \frac{G(c)}{G(\gamma)} \quad \text{ for all } c \leq \gamma \in [n] \iff \frac{f(k)}{F(k)} \geq \frac{g(k)}{G(k)} \quad \text{ for all } k \in [n].
\end{aligned}
\end{equation*}
\end{lemma}

\begin{proof}[Proof of Lemma~\ref{lemma:discrete-conditional-stochastic-dominance}]
Since $\frac{F(c)}{F(\gamma)} \leq \frac{G(c)}{G(\gamma)} \iff \frac{F(c)}{G(c)} \leq \frac{F(\gamma)}{G(\gamma)}$, the first statement is equivalent to the claim that $F(k) / G(k)$ is increasing in $k$. But this holds if and only if for all $k \in [n]$,
\begin{equation*}
\begin{aligned}
\frac{F(k)}{G(k)} - \frac{F(k-1)}{G(k-1)} \geq 0 &\iff \frac{F(k)}{G(k)} \geq \frac{F(k-1)}{G(k-1)}\\
&\iff \frac{G(k-1)}{G(k)} \geq \frac{F(k-1)}{F(k)}\\
&\iff 1 - \frac{G(k-1)}{G(k)} \leq 1 - \frac{F(k-1)}{F(k)}\\
&\iff \frac{G(k) - G(k-1)}{G(k)} \leq \frac{F(k) - F(k-1)}{F(k)}\\
&\iff \frac{g(k)}{G(k)} \leq \frac{f(k)}{F(k)}.
\end{aligned}
\end{equation*}
In the above, we have defined $0/0 = 0$ wherever it may appear. Note that this does not affect the relations above since $F(k) / G(k) = 0/0$ immediately implies that $F(k-1) / G(k-1) = f(k) / F(k) = g(k) / G(k) = 0/0$. Other divide by $0$ issues can be ruled out since taking $\gamma = n$, the first condition implies $F(c) \leq G(c)$ for all $c \in [n]$.
\end{proof}

\begin{lemma}[Truncated binomial stochastic dominance]
\label{lemma:truncated-binomial-stochastic-dominance}
Let $Z \sim \text{Binom}(n, p_1)$ and $Z' \sim \text{Binom}(n, p_2)$ with $p_1 \leq p_2$. Then for all $c, \gamma$, we have
\begin{equation*}
\begin{aligned}
\Prb(Z \leq c \mid Z \leq \gamma) \geq \Prb(Z' \leq c \mid Z' \leq \gamma).
\end{aligned}
\end{equation*}
The inequality remains true if we replace $c$ by a random variable $C$ that is independent of $Z$ and $Z'$.
\end{lemma}

\begin{proof}[Proof of Lemma~\ref{lemma:truncated-binomial-stochastic-dominance}]
First consider non-random $c$. Both sides of the inequality equal $1$ when $c > \gamma$, so suppose that $c \leq \gamma$. To make use of Lemma~\ref{lemma:discrete-conditional-stochastic-dominance}, we must verify that for all $k$,
\begin{equation*}
\begin{aligned}
\frac{\binom{n}{k} p_1^k (1 - p_1)^{n-k}}{\sum_{j=0}^k \binom{n}{j} p_1^j (1 - p_1)^{n-j}} \leq \frac{\binom{n}{k} p_2^k (1 - p_2)^{n-k}}{\sum_{j=0}^k \binom{n}{j} p_2^j (1 - p_2)^{n-j}}.
\end{aligned}
\end{equation*}
It suffices to show that for any fixed $k$, the left-hand side is non-decreasing in $p_1$. Writing $p := p_1$, we can rewrite the left-hand side as
\begin{equation*}
\begin{aligned}
\frac{\binom{n}{k} p^k (1 - p)^{n-k}}{\sum_{j=0}^k \binom{n}{j} p^j (1 - p)^{n-j}} = \paren*{ \sum_{j=0}^k a_j p^{j-k} (1 - p)^{n - j - (n - k)} }^{-1} = \paren*{ \sum_{j=0}^k a_j \paren*{ \frac{p}{1-p} }^{j - k} }^{-1}
\end{aligned}
\end{equation*}
where $a_j = \binom{n}{j} / \binom{n}{k}$. Since $p \to \frac{p}{1 - p}$ is increasing, $j - k \leq 0$, and $a_j \geq 0$, we have that $p \to a_j \paren*{\frac{p}{1 - p}}^{j - k}$ is non-increasing. The sum of non-increasing functions is non-increasing, so the sum inside the parentheses is non-increasing and positive, and therefore its inverse is non-decreasing, as desired. Hence, Lemma~\ref{lemma:discrete-conditional-stochastic-dominance} implies that
\begin{equation*}
\begin{aligned}
\frac{\Prb(Z' \leq c)}{\Prb(Z' \leq \gamma)} \leq \frac{\Prb(Z \leq c)}{\Prb(Z \leq \gamma)} \iff \Prb(Z' \leq c \mid Z' \leq \gamma) \leq \Prb(Z \leq c \mid Z \leq \gamma)
\end{aligned}
\end{equation*}
for $c \leq \gamma$. Now consider random $C$ and let $\gamma$ be fixed. Define $h(c) = \Prb(Z \leq c \mid Z \leq \gamma)$ and $h'(c) = \Prb(Z' \leq c \mid Z' \leq \gamma)$. Since $C$ is independent of $Z$ and $Z'$, we have that
\begin{equation*}
\begin{aligned}
\Prb(Z \leq C \mid Z \leq \gamma) &= \E[h(C)],\\
\Prb(Z' \leq C \mid Z' \leq \gamma) &= \E[h'(C)].
\end{aligned}
\end{equation*}
But $h$ dominates $h'$ pointwise, so $h(C) \geq h'(C)$ almost surely, and $\E[h(C)] \geq \E[h'(C)]$.
\end{proof}

\begin{lemma}[Continuous conditional stochastic dominance]
\label{lemma:continuous-conditional-stochastic-dominance}
Let $X, Y$ be continuous random variables supported on all of $\mathbb{R}$. Denote the PMF and CDF of $X$ by $f$ and $F$, and the PMF and CDF of $Y$ by $g$ and $G$, respectively. Then
\begin{equation*}
\begin{aligned}
\frac{F(c)}{F(\gamma)} \leq \frac{G(c)}{G(\gamma)} \quad \text{ for all } c \leq \gamma \in \mathbb{R} \iff \frac{f(x)}{F(x)} \geq \frac{g(x)}{G(x)} \quad \text{ for all } x \in \mathbb{R}.
\end{aligned}
\end{equation*}
\end{lemma}

\begin{proof}[Proof of Lemma~\ref{lemma:continuous-conditional-stochastic-dominance}]
This is a standard result, but we replicate the proof here for completeness. Since $\frac{F(c)}{F(\gamma)} \leq \frac{G(c)}{G(\gamma)} \iff \frac{F(c)}{G(c)} \leq \frac{F(\gamma)}{G(\gamma)}$, the first statement is equivalent to the claim that $F(x) / G(x)$ is increasing in $x$, or equivalently that $\log F(x) - \log G(x)$ is increasing in $x$. But
\begin{equation*}
\begin{aligned}
\frac{d}{dx} \left(\log F(x) - \log G(x)\right) = \frac{f(x)}{F(x)} - \frac{g(x)}{G(x)}
\end{aligned}
\end{equation*}
which is greater than or equal to $0$ for all $x$ if and only if $\frac{f(x)}{F(x)} \geq \frac{g(x)}{G(x)}$ for all $x$.
\end{proof}

\begin{lemma}[Truncated normal stochastic dominance]
\label{lemma:truncated-normal-stochastic-dominance}
Let $Z \sim \mathcal{N}(\mu_1, 1)$ and $Z' \sim \mathcal{N}(\mu_2, 1)$ with $\mu_1 \leq \mu_2$. Then for all $c, \gamma$, we have
\begin{equation*}
\begin{aligned}
\Prb(Z \leq c \mid Z \leq \gamma) \geq \Prb(Z' \leq c \mid Z' \leq \gamma).
\end{aligned}
\end{equation*}
Letting $\Phi(c; \gamma) = \Prb(Z \leq c \mid Z \leq \gamma)$, this is equivalently stated as $\Phi(c + s; \gamma + s) \geq \Phi(c; \gamma)$ for all $c, \gamma \in \mathbb{R}$ and $s \geq 0$.
\end{lemma}

\begin{proof}[Proof of Lemma~\ref{lemma:truncated-normal-stochastic-dominance}]
Both sides of the inequality equal $1$ if $c > \gamma$, so suppose that $c \leq \gamma$. To make use of Lemma~\ref{lemma:continuous-conditional-stochastic-dominance}, we must show that for all $x$,
\begin{equation*}
\begin{aligned}
\frac{\phi(x - \mu_1)}{\Phi(x - \mu_1)} \leq \frac{\phi(x - \mu_2)}{\Phi(x - \mu_2)}.
\end{aligned}
\end{equation*}
Since $x - \mu_1 \geq x - \mu_2$, it suffices to show that $\phi(y) / \Phi(y)$ is decreasing in $y$. A well-known fact is that $\phi(y) / \Phi(y)$, known as the inverse Mill's ratio, is decreasing in $y$. Hence it follows from Lemma~\ref{lemma:continuous-conditional-stochastic-dominance} that
\begin{equation*}
\begin{aligned}
\frac{\Prb(Z' \leq c)}{\Prb(Z' \leq \gamma)} \leq \frac{\Prb(Z \leq c)}{\Prb(Z \leq \gamma)} \iff \Prb(Z' \leq c \mid Z' \leq \gamma) \leq \Prb(Z \leq c \mid Z \leq \gamma)
\end{aligned}
\end{equation*}
for $c \leq \gamma$.
\end{proof}

\begin{lemma}[Convergence of Kolmogorov distance of unconditional normal implies convergence of truncated variant]
\label{lemma:trunc-cdf-conv}
Let $Z_t^{(n)}$ be a doubly-indexed sequence of random variables where $t$ takes values in a non-random set $S_n$. Similarly, let $\mathcal{F}_t^{(n)}$ be a doubly-indexed sequence of $\sigma$-algebras. For each $t \in S_n$, define

\begin{equation*}
\begin{aligned}
F_t^{(n)}(x) &:= \Prb\paren{Z_t^{(n)} \leq x \mid \mathcal{F}_t^{(n)}},\\
F_t^{(n)}(x; y) &:= \Prb\paren{Z_t^{(n)} \leq x \mid \mathcal{F}_t^{(n)}, Z_t^{(n)} \leq y}.
\end{aligned}
\end{equation*}
Write $\Phi$ for the standard normal CDF and $\Phi(x; y) = \Prb(Z \leq x \mid Z \leq y)$ where $Z$ is standard normal. Let $A_n$ be a random subset of $S_n$ for each $n$. Then
\begin{equation*}
\begin{aligned}
\sup_{t \in A_n} \sup_{x \in \mathbb{R}} \abs{F_t^{(n)}(x) - \Phi(x)} \stackrel{p}{\to} 0 \implies \sup_{t \in A_n} \sup_{x \geq 0, y \in \mathbb{R}} \abs{F_t^{(n)}(x; y) - \Phi(x; y)} \stackrel{p}{\to} 0.
\end{aligned}
\end{equation*}
\end{lemma}

\begin{proof}[Proof of Lemma~\ref{lemma:trunc-cdf-conv}]
First note that
\begin{equation*}
\begin{aligned}
\sup_{t \in A_n} \sup_{x \geq 0, y \in \mathbb{R}} \abs{F_t^{(n)}(x; y) - \Phi(x; y)} = \sup_{t \in A_n} \sup_{0 \leq x \leq y} \abs{F_t^{(n)}(x; y) - \Phi(x; y)}
\end{aligned}
\end{equation*}
since when $y < x$, both $F_t^{(n)}(x; y)$ and $\Phi(x; y)$ evaluate to $1$, and hence their difference is $0$. We would like to rewrite $F_t^{(n)}(x; y)$ as the ratio $F_t^{(n)}(x) / F_t^{(n)}(y)$, but this is only well-defined when the denominator is nonzero. Thus, let us first show that the denominator is nonzero with probability tending to $1$:
\begin{equation*}
\begin{aligned}
\inf_{t \in A_n} \inf_{y \geq 0} F_t^{(n)}(y) & \geq \inf_{t \in A_n} \inf_{y \geq 0} \Phi(y) - \abs{F_t^{(n)}(y) - \Phi(y)}\\
& \geq \inf_{t \in A_n} \inf_{y \geq 0} \Phi(y) + \inf_{t \in A_n} \inf_{y \geq 0} \paren{ - \abs{F_t^{(n)}(y) - \Phi(y)} }\\
& \geq 0.5 - \sup_{t \in A_n} \sup_{y \geq 0} \abs{F_t^{(n)}(y) - \Phi(y)}\\
& \stackrel{p}{\to} 0.5.
\end{aligned}
\end{equation*}
Let $\delta, \epsilon > 0$ and choose $0 < s_1 < 0.5$ such that $\Prb(\inf_{t \in A_n} \inf_{y \geq 0} F_t^{(n)}(y) < 0.5 - s_1) < \epsilon / 2$ for all sufficiently large $n$. Also, for notational convenience define the event $E_n := \set{ \inf_{t \in A_n} \inf_{y \geq 0} F_t^{(n)}(y) < 0.5 - s_1 }$.

Next, we observe that $h(x, y) = x / y$ is uniformly continuous on $[0.5 - s_1, 1]^2$. This follows since it is a continuous function on a compact set. Hence, letting $\norm{\cdot}_2$ denote the standard Euclidean norm, there exists some $s_2 > 0$ such that $\norm{(x_1, y_1) - (x_2, y_2)}_2 < s_2 \implies \abs{h(x_1, y_1) - h(x_2, y_2)} < \delta$ for all $(x_1, y_1), (x_2, y_2) \in [0.5 - s_1, 1]^2$. Then for all sufficiently large $n$,
\begin{equation*}
\begin{aligned}
& \Prb\paren*{ \sup_{t \in A_n} \sup_{0 \leq x \leq y} \abs{F_t^{(n)}(x; y) - \Phi(x; y)} > \delta }\\
\leq & \Prb\paren*{ \sup_{t \in A_n} \sup_{0 \leq x \leq y} \abs{F_t^{(n)}(x; y) - \Phi(x; y)} > \delta , E^{\comp}_n} + \Prb(E_n)\\
= & \Prb\paren*{ \sup_{t \in A_n} \sup_{0 \leq x \leq y} \abs*{\frac{F_t^{(n)}(x)}{F_t^{(n)}(y)} - \frac{\Phi(x)}{\Phi(y)}} > \delta , E^{\comp}_n} + \Prb(E_n)\\
= & \Prb\paren*{ \sup_{t \in A_n} \sup_{0 \leq x \leq y} \abs{h(F_t^{(n)}(x), F_t^{(n)}(y)) - h(\Phi(x), \Phi(y))} > \delta , E^{\comp}_n} + \Prb(E_n)\\
\leq & \Prb\paren*{ \sup_{t \in A_n} \sup_{0 \leq x \leq y} \norm{(F_t^{(n)}(x), F_t^{(n)}(y)) - (\Phi(x), \Phi(y))}_2 \geq s_2, E^{\comp}_n} + \Prb(E_n)\\
\leq & \Prb\paren*{ \sup_{t \in A_n} \sup_{0 \leq x \leq y} \norm{(F_t^{(n)}(x), F_t^{(n)}(y)) - (\Phi(x), \Phi(y))}_2 \geq s_2} + \epsilon / 2\\
\stackrel{p}{\to} & \epsilon / 2.
\end{aligned}
\end{equation*}
In particular, the probability is bounded below $\epsilon$ for all sufficiently large $n$, as desired. The third line follows from the fact that $x \leq y$ and the denominator term $F_t^{(n)}(y)$ is positive for all $t \in A_n$ and $y \geq 0$ on the event $E^{\comp}_n$. The fifth line follows from the uniform continuity, where we additionally note that $F_t^{(n)}(x), F_t^{(n)}(y), \Phi(x), \Phi(y)$ are in $[0.5 - s_1, 1]$ for all $t \in A_n$ and $0 \leq x \leq y$ on the event $E^{\comp}_n$. The last line follows from the fact that $\sqrt{a^2 + b^2} \leq \abs{a} + \abs{b}$ and hence
\begin{equation*}
\begin{aligned}
& \sup_{t \in A_n} \sup_{0 \leq x \leq y} \norm{(F_t^{(n)}(x), F_t^{(n)}(y)) - (\Phi(x), \Phi(y))}_2\\
\leq & \sup_{t \in A_n} \sup_{0 \leq x \leq y} \abs{F_t^{(n)}(x) - \Phi(x)} + \abs{F_t^{(n)}(y) - \Phi(y)}\\
\leq & \sup_{t \in A_n} \sup_{x \in \mathbb{R}} \abs{F_t^{(n)}(x) - \Phi(x)} + \sup_{t \in A_n} \sup_{y \in \mathbb{R}} \abs{F_t^{(n)}(y) - \Phi(y)}\\
\stackrel{p}{\to} & 0
\end{aligned}
\end{equation*}
by assumption.
\end{proof}

\begin{lemma}[Representation for uniform order statistics]
\label{lemma:unif-ord-stat-rep}
Let $U_{m,n}$ be the $m$th order statistic of $n$ uniforms. Suppose that $U_{1,1},U_{2,2},...,U_{n,n}$ are independent. For $i=1,...,n$ define
\begin{equation*}
\begin{aligned}
Z_i = U_{n, n} \times ... \times U_{n + 1 - i, n + 1 - i}.
\end{aligned}
\end{equation*}
Then $(Z_n,...,Z_1)$ has the same distribution as $(U_{(1)},...,U_{(n)})$, the joint distribution of the order statistics of $n$ standard uniforms.
\end{lemma}

\begin{proof}[Proof of Lemma~\ref{lemma:unif-ord-stat-rep}]
We proceed by induction. First, $Z_1 \stackrel{d}{=} U_{(n)}$ by definition. Now suppose
\begin{equation*}
\begin{aligned}
(Z_1,...,Z_i) \stackrel{d}{=} (U_{(n)},...,U_{(n + 1 - i)}).
\end{aligned}
\end{equation*}
If $(U_1,...,U_n)$ are standard uniform and we condition on the top $i$ values, the remaining $n - i$ coordinates are independently distributed from $\text{Unif}[0, U_{(n + 1 - i)}]$, where we note that $U_{(n + 1 - i)}$ is the smallest of the top $i$ values. Then the maximum among the remaining $n - i$ coordinates has the distribution $U_{(n + 1 - i)} \times U_{n - i, n - i}$. In other words, the distribution of $U_{(n - i)}$ conditional on $(U_{(n)},...,U_{(n + 1 - i)})$ is $U_{(n + 1 - i)} \times U_{n - i, n - i}$. Similarly, conditional on $(Z_1,...,Z_i)$, the distribution of $Z_{i + 1}$ is $Z_i \times U_{n - i, n - i}$. Thus, applying the inductive hypothesis we have that $(Z_1,...,Z_{i + 1}) \stackrel{d}{=} (U_{(n)},...,U_{(n - i)})$.
\end{proof}

\begin{lemma}[Positive region conditional variance]
\label{lemma:positive-region-cond-var}
Suppose $X \in \mathcal{X}$ and $Y \in \mathbb{R}$ are such that $\Var(Y \mid X) > 0$ almost surely. Assume that all $R \subseteq \mathcal{X}$ referred to are measurable. For $R \subseteq \mathcal{X}$, define $\Vol(R) := \Prb(X \in R)$. Then
\begin{equation*}
\begin{aligned}
\inf_{R \subseteq \mathcal{X} : \Vol(R) \geq q} \Var(Y \mid X \in R) > 0
\end{aligned}
\end{equation*}
for all $0 < q \leq 1$.
\end{lemma}

\begin{proof}[Proof of Lemma~\ref{lemma:positive-region-cond-var}]
First observe that $\Var(Y \mid X \in R) \geq \E[\Var(Y \mid X) \mid X \in R]$ by the law of total variance. Let us simply denote $f(X) := \Var(Y \mid X)$ for notational convenience. It suffices to show that
\begin{equation*}
\begin{aligned}
\inf_{R \subseteq \mathcal{X} : \Vol(R) \geq q} \E[f(X) \mid X \in R] > 0.
\end{aligned}
\end{equation*}
Suppose for contradiction that the infimum is $0$. Then there exists a sequence $(R_n)_{n=1}^{\infty}$ such that $\E[f(X) \mid X \in R_n] \leq n^{-2}$ and $\Vol(R_n) \geq q$ for all $n$. Then
\begin{equation*}
\begin{aligned}
& \Prb( f(X) > n^{-1})\\
= & \Prb( f(X) > n^{-1} \mid X \in R_n ) \Prb( X \in R_n ) + \Prb( f(X) > n^{-1} \mid X \not\in R_n ) \Prb( X \not\in R_n )\\
\leq & n \cdot \E[f(X) \mid X \in R_n] \Prb(X \in R_n) + \Prb(X \not\in R_n)\\
\leq & n \cdot \frac{1}{n^2} \cdot \Prb(X \in R_n) + \Prb(X \not \in R_n)\\
\leq & \frac{1}{n} \cdot q + (1 - q).
\end{aligned}
\end{equation*}
In going to the third line of the above, we use the fact that $\Prb( f(X) > n^{-1} \mid X \in R_n) \leq \Prb( f(X) \geq n^{-1} \mid X \in R_n)$, and then we apply Markov's inequality conditionally on the event $\set{ X \in R_n }$. Next, by the right continuity of CDFs, for all $\epsilon > 0$ there exists an $N$ such that for all $n \geq N$, $\abs{ \Prb( f(X) \leq 0 ) - \Prb( f(X) \leq n^{-1}) } \leq \epsilon$. Then
\begin{equation*}
\begin{aligned}
\Prb( f(X) = 0 ) &= \Prb( f(X) \leq 0 )\\
& \geq \Prb( f(X) \leq n^{-1}) - \epsilon\\
& \geq 1 - \frac{1}{n} \cdot q - (1 - q) - \epsilon\\
& = q \paren*{ 1 - \frac{1}{n} } - \epsilon
\end{aligned}
\end{equation*}
for all $n \geq N$. In particular, choosing $\epsilon = q / 2$, for sufficiently large $n$ the above implies that $\Prb(f(X) = 0)$ is positive, which contradicts our assumption that $\Var(Y \mid X) > 0$ almost surely. Hence, the infimum cannot be $0$ and must be positive.
\end{proof}

\begin{lemma}[Simple centered moment bound using raw moments]
\label{lemma:raw-bound-center-moment}
Let $k$ be a fixed positive integer greater than $1$ and let $Y$ be a random variable with finite $k$th moment. Then
\begin{equation*}
\begin{aligned}
\E[\abs{Y - \E[Y]}^k] \leq 2^k \cdot \E[\abs{Y}^k].
\end{aligned}
\end{equation*}
\end{lemma}

\begin{proof}[Proof of Lemma~\ref{lemma:raw-bound-center-moment}]
It is a well-known fact that for random variables $X_1, X_2$ such that the $k$th moment of each exists, we have
\begin{equation*}
\begin{aligned}
\E[\abs{X_1 + X_2}^k] \leq 2^{k-1} \paren{ \E[\abs{X_1}^k] + \E[\abs{X_2}^k] }.
\end{aligned}
\end{equation*}
See, for instance, the introduction of \textcite{von-bahr-esseen}. Letting $X_1 = Y$ and $X_2 = -\E[Y]$, the above yields
\begin{equation*}
\begin{aligned}
\E[\abs{Y - \E[Y]}^k] &\leq 2^{k-1} \paren{ \E[\abs{Y}^k] + \E[\abs{\E[Y]}^k] }\\
&= 2^{k-1} \paren{ \E[\abs{Y}^k] + \abs{\E[Y]}^k }\\
&\leq 2^{k-1} \paren{ \E[\abs{Y}^k] + \E[\abs{Y}^k] }\\
&= 2^k \cdot \E[\abs{Y}^k]
\end{aligned}
\end{equation*}
where we use the fact that $\abs{\E[Y]}^k \leq \E[\abs{Y}^k]$ by Jensen's inequality.
\end{proof}

\begin{lemma}[Conditional vanishing in probability]
\label{lemma:conditional-p0-implies-p0}
Let $X_n$ be a sequence of random variables and $\mathcal{F}_n$ be a sequence of $\sigma$-algebras. Suppose that $\E[X_n \mid \mathcal{F}_n] = 0$ and $\Var(X_n \mid \mathcal{F}_n) \stackrel{p}{\to} 0$. Then $X_n \stackrel{p}{\to} 0$.    
\end{lemma}

\begin{proof}[Proof of Lemma~\ref{lemma:conditional-p0-implies-p0}]
Define $\sigma^2_n := \Var(X_n \mid \mathcal{F}_n)$, which is a random variable. For any $\delta, \epsilon > 0$, let $b := \epsilon \delta^2 / 2$ and choose $N$ so that $\Prb(\sigma^2_n > b) < \epsilon / 2$ for all $n \geq N$. Then since $\sigma^2_n$ is $\mathcal{F}_n$-measurable,
\begin{equation*}
\begin{aligned}
\Prb(\abs{X_n} > \delta, \sigma^2_n \leq b \mid \mathcal{F}_n) = \indic\set{\sigma^2_n \leq b} \Prb(\abs{X_n} > \delta \mid \mathcal{F}_n) \leq \indic\set{\sigma^2_n \leq b} \cdot \frac{\sigma_n^2}{\delta^2} \leq b / \delta^2 = \epsilon / 2
\end{aligned}
\end{equation*}
by conditional Chebyshev's inequality. Then taking expectations of both sides, $\Prb(\abs{X_n} > \delta, \sigma^2_n \leq b) \leq \epsilon / 2$. Then
\begin{equation*}
\begin{aligned}
\Prb(\abs{X_n} > \delta) &\leq \Prb(\abs{X_n} > \delta, \sigma^2_n \leq b) + \Prb(\sigma^2_n > b)\\
&\leq \epsilon / 2 + \epsilon / 2\\
&= \epsilon
\end{aligned}
\end{equation*}
for all $n \geq N$.
\end{proof}

\subsection{Convergence on bounded metrics}
\label{appendix:conv-bounded-metrics}

A difficulty that we encounter throughout is that certain quantities do not converge in the usual sense, but are close in the sense that they are either close to one another, or they diverge together in the same direction. For our purposes, this weaker notion of convergence suffices. To formalize this notion, we introduce a new metric under which such quantities \textit{do} converge, and establish some useful properties. Throughout this section, $f$ will always refer to a strictly increasing function. Occasionally, we will require $f$ to have additional properties, but we will always specify these situations.

\begin{definition}[$f$-metric]
\label{def:f-metric}
Let $f:\mathbb{R} \to \mathbb{R}$ be a strictly increasing function, and by an abuse of notation let $f(x) = (f(x_1),...,f(x_k))$ for $x \in \mathbb{R}^k$. We define $d_{f}: \mathbb{R}^k \times \mathbb{R}^k \to \mathbb{R}$ via
\begin{equation*}
\begin{aligned}
d_{f}(x, y) = \norm{f(x) - f(y)}_2
\end{aligned}
\end{equation*}
where $\norm{\cdot}_2$ is the standard Euclidean norm. We will refer to $d_{f}$ as the \textit{$f$-metric}.
\end{definition}

\begin{lemma}[Metricity]
\label{lemma:metricity}
$d_{f}$ is a metric.
\end{lemma}

\begin{proof}[Proof of Lemma~\ref{lemma:metricity}]
Symmetry and the fact that $d_f(x,x) = 0$ are immediate. Positivity follows from the fact that $f$ is strictly increasing and so $f(x) \neq f(y)$ for all $x \neq y$. Lastly,
\begin{equation*}
\begin{aligned}
d_f(x, z) &= \norm{f(x) - f(z)}_2\\
&= \norm{f(x) - f(y) + f(y) - f(z)}_2\\
& \leq \norm{f(x) - f(y)}_2 + \norm{f(y) - f(z)}_2\\
&= d_f(x, y) + d_f(x, z)
\end{aligned}
\end{equation*}
from the ordinary triangle inequality on $\norm{\cdot}_2$.
\end{proof}

We extend the notion of convergence in probability to $f$-metrics.

\begin{definition}[$f$-convergence]
\label{def:f-convergence}
Let $X_n$ and $Y_n$ be two sequences of random variables. We will say that \textit{$X_n$ converges to $Y_n$ under the $f$-metric} if $d_f(X_n, Y_n) \stackrel{p}{\to} 0$. In this case, we will write $X_n \fconv{f} Y_n$, with the bidirectional arrow to indicate that the relation is symmetric; that is, $X_n \fconv{f} Y_n \iff Y_n \fconv{f} X_n$.
\end{definition}

We will be able to show that $f$-convergence is preserved under uniformly continuous maps, where uniform continuity is defined with respect to the $f$-metric. However, some maps that we need are not uniformly continuous maps. But as long as the random variables that are inputs to this map ``usually" lie on a uniformly continuous restriction, we will face no issues. To this end, we introduce the following notion, which captures the usual notion of bounded in probability, but also handles the idea of a sequence of random variables being ``bounded away" from some threshold. We also show that this notion of boundedness is preserved between $f$-convergent sequences.

\begin{definition}[$\mathcal{B}$-bounded in probability]
\label{def:b-bounded}
Let $X_n$ be a sequence of random variables, and let $\mathcal{B}$ be a collection of open intervals in $\mathbb{R}$ with the property that, for every $(a,b) \in \mathcal{B}$, there exists $(a', b') \in \mathcal{B}$ such that $a' < a$ and $b' > b$, where we take the convention that $-\infty < -\infty$ and $\infty > \infty$. We say that $X_n$ is \textit{$\mathcal{B}$-bounded in probability} if for every $\epsilon > 0$, there exists an integer $N$ and an interval $I \in \mathcal{B}$ such that $\Prb\paren{X_n \not\in I } < \epsilon$ for all $n > N$. We note three special cases:
\begin{enumerate}
\item When $\mathcal{B} = \set{(-c, c) : c > 0}$, we will simply say that $X_n$ is \textit{bounded in probability.}
\item When $\mathcal{B} = \set{(1/c, c) : c > 1}$, we will simply say that $X_n$ is \textit{positively log-bounded in probability.}
\item When $(-\infty, \infty) \in \mathcal{B}$, then every $X_n$ is trivially $\mathcal{B}$-bounded in probability.
\end{enumerate}
\end{definition}

\begin{lemma}[Transitivity of $\mathcal{B}$-boundedness in probability]
\label{lemma:b-bounded-equivalence}
Let $X_n$ and $Y_n$ be sequences of random variables and suppose $X_n \fconv{f} Y_n$. Let $\mathcal{B}$ be any collection that satisfies the conditions in Definition~\ref{def:b-bounded}. Then $X_n$ is $\mathcal{B}$-bounded in probability if and only if $Y_n$ is $\mathcal{B}$-bounded in probability.
\end{lemma}

\begin{proof}[Proof of Lemma~\ref{lemma:b-bounded-equivalence}]
Without loss of generality suppose $X_n$ is $\mathcal{B}$-bounded in probability. Let $\epsilon > 0$. Choose $I_1 \in \mathcal{B}$ and $N_1$ such that $\Prb(X_n \not\in I_1) < \epsilon / 2$ for all $n > N_1$. Writing $I_1 = (l_1, u_1)$ we have that by assumption there exists $I_2 = (l_2, u_2) \in \mathcal{B}$ with $l_2 < l_1$ and $u_2 > u_1$. We will just discuss the case when $l_1 \neq -\infty$ and $u_1 \neq \infty$; when either of the endpoints is infinite, the proof is very similar.

Let $\Delta = \min\set{ f(u_2) - f(u_1), f(l_1) - f(l_2) }$ and note that $\Delta > 0$ since $f$ is strictly increasing. Also note that
\begin{equation*}
\begin{aligned}
Y_n \not\in I_2 \text{ and } X_n \in I_1 \implies f(Y_n) - f(X_n) > \Delta
\end{aligned}
\end{equation*}
because either $Y_n$ falls to the right of $u_2$, in which case $\abs{f(Y_n) - f(X_n)} > f(u_2) - f(u_1)$, or it falls to the left of $l_2$, in which case $\abs{f(Y_n) - f(X_n)} > f(l_1) - f(l_2)$. Thus, the absolute difference is always at least as great as the minimum of these two quantities. Since $X_n \fconv{f} Y_n$, we can choose $N_2$ such that $\Prb\paren{\abs{f(X_n) - f(Y_n)} > \Delta} < \epsilon / 2$ for all $n > N_2$. Then for all $n > \max\set{ N_1, N_2 }$,
\begin{equation*}
\begin{aligned}
\Prb\paren{Y_n \not\in I_2} &\leq \Prb\paren{ Y_n \not\in I_2 \text{ and } X_n \in I_1} + \Prb\paren{ X_n \not\in I_1}\\
&\leq \Prb\paren{ \abs{f(Y_n) - f(X_n)} > \Delta} + \epsilon / 2\\
&\leq \epsilon / 2 + \epsilon / 2\\
&= \epsilon.
\end{aligned}
\end{equation*}
So $Y_n$ is $\mathcal{B}$-bounded in probability.
\end{proof}

We are ready to state the main composition results. We first establish a lemma regarding the leap from univariate $f$-convergence to bivariate $f$-convergence. We then give conditions for transformations that preserve $f$-convergence.

\begin{lemma}[Bivariate $f$-convergence]
\label{lemma:bivariate-fconv}
Let $A_n, B_n, C_n, D_n$ be sequences of random variables. Suppose that $A_n \fconv{f} B_n$ and $C_n \fconv{f} D_n$. Then
\begin{equation*}
\begin{aligned}
d_{f}((A_n, C_n), (B_n, D_n)) \stackrel{p}{\to} 0.
\end{aligned}
\end{equation*}
\end{lemma}

\begin{proof}[Proof of Lemma~\ref{lemma:bivariate-fconv}]
Since $\sqrt{x^2 + y^2} \leq |x| + |y|$,
\begin{equation*}
\begin{aligned}
\norm{(f(A_n), f(C_n)) - (f(B_n), f(D_n))}_2 \leq |f(A_n) - f(B_n)| + |f(C_n) - f(D_n)| \stackrel{p}{\to} & 0.
\end{aligned}
\end{equation*}
\end{proof}

For two metrics $d_X$ and $d_Y$ on sets $M_X$ and $M_Y$ respectively, we will say that a function $g : M_X \to M_Y$ is $(d_X, d_Y)$-uniformly continuous if it is uniformly continuous with respect to the metric $d_X$ on $M_X$ and the metric $d_Y$ on $M_Y$.

\begin{lemma}[Abstract $f$-convergence composition rule]
\label{lemma:abstract-f-composition-rule}
Let $A_n, B_n, C_n, D_n$ be sequences of random variables. Suppose that $A_n \fconv{f} B_n$ and $C_n \fconv{f} D_n$. Suppose that $B_n$ is $\mathcal{B}_1$-bounded in probability and $D_n$ is $\mathcal{B}_2$-bounded in probability. Let $g: \mathbb{R}^2 \to \mathbb{R}$ be a function, and for each $I_1 \in \mathcal{B}_1, I_2 \in \mathcal{B}_2$ define the restriction $g|_{I_1 \times I_2} : I_1 \times I_2 \to \mathbb{R}$ by $g|_{I_1 \times I_2}(x, y) = g(x, y)$ for all $x \in I_1, y \in I_2$. Suppose that $g|_{I_1 \times I_2}$ is uniformly $(d_f, d_f)$-continuous for every $I_1 \in \mathcal{B}_1, I_2 \in \mathcal{B}_2$. Then $g(A_n, C_n) \fconv{f} g(B_n, D_n)$.
\end{lemma}

\begin{proof}[Proof of Lemma~\ref{lemma:abstract-f-composition-rule}]
Let $\epsilon, \delta > 0$. By Lemma~\ref{lemma:b-bounded-equivalence}, we also have that $A_n$ is $\mathcal{B}_1$-bounded in probability and $C_n$ is $\mathcal{B}_2$-bounded in probability. Choose $I_1 \in \mathcal{B}_1, I_2 \in \mathcal{B}_2$ and $N$ such that
\begin{equation*}
\begin{aligned}
1 - \Prb(\text{In}(A_n)) &:= \Prb(\text{Out}(A_n)) &:= \Prb(A_n \not\in I_1) &< \epsilon / 8,\\
1 - \Prb(\text{In}(B_n)) &:= \Prb(\text{Out}(B_n)) &:= \Prb(B_n \not\in I_1) &< \epsilon / 8,\\
1 - \Prb(\text{In}(C_n)) &:= \Prb(\text{Out}(C_n)) &:= \Prb(C_n \not\in I_2) &< \epsilon / 8,\\
1 - \Prb(\text{In}(D_n)) &:= \Prb(\text{Out}(D_n)) &:= \Prb(D_n \not\in I_2) &< \epsilon / 8
\end{aligned}
\end{equation*}
for all $n > N$. Since $g|_{I_1 \times I_2}$ is uniformly $(d_f, d_f)$-continuous by assumption, choose $s > 0$ so that $d_{f}((x_1, y_1), (x_2, y_2)) < s \implies d_{f}(g|_{I_1 \times I_2}(x_1, y_1), g|_{I_1 \times I_2}(x_2, y_2)) < \delta$ for every $(x_1, y_1), (x_2, y_2) \in I_1 \times I_2$. Then for all $n > N$, we have
\begin{equation*}
\begin{aligned}
& \Prb\paren*{ d_{f}(g(A_n, C_n), g(B_n, D_n)) > \delta}\\
\leq & \Prb\paren*{d_{f}(g(A_n, C_n), g(B_n, D_n)) > \delta, \text{In}(A_n), \text{In}(B_n), \text{In}(C_n), \text{In}(D_n)}\\
& + \Prb(\text{Out}(A_n)) + \Prb(\text{Out}(B_n)) + \Prb(\text{Out}(C_n)) + \Prb(\text{Out}(D_n))\\
\leq & \Prb\paren*{d_{f}(g_{I_1 \times I_2}(A_n, C_n), g_{I_1 \times I_2}(B_n, D_n)) > \delta, \text{In}(A_n), \text{In}(B_n), \text{In}(C_n), \text{In}(D_n)}\\
& + 4(\epsilon / 8)\\
\leq & \Prb\paren*{d_{f}((A_n, C_n), (B_n, D_n)) \geq s} + \epsilon / 2.
\end{aligned}
\end{equation*}
By Lemma~\ref{lemma:bivariate-fconv}, $d_{f}((A_n, C_n), (B_n, D_n)) \stackrel{p}{\to} 0$, so that we can choose $N'$ such that
\begin{equation*}
\begin{aligned}
\Prb\paren*{ d_{f}((A_n, C_n), (B_n, D_n)) \geq s} < \epsilon / 2
\end{aligned}
\end{equation*}
for all $n > N'$. Then $\Prb\paren*{d_{f}(g(A_n, C_n), g(B_n, D_n)) > \delta} < \epsilon$ for all $n > \max\set{ N, N' }$.
\end{proof}

Lemma~\ref{lemma:abstract-f-composition-rule} gives us abstract conditions on maps that preserve $f$-convergence. The following lemmas give us a concrete strategy for establishing $f$-convergence. First we establish a reduction to uniform continuity with respect to the standard Euclidean metric.

\begin{lemma}[Reduction to standard uniform continuity]
\label{lemma:standard-uniform-reduction}
Let $f_1$ and $f_2$ be strictly increasing functions from $\mathbb{R}$ to $\mathbb{R}$, and let $g: S \to \mathbb{R}$ where $S \subseteq \mathbb{R}^k$. Recall the abuse of notation that $f(x) = (f(x_1),...,f(x_k))$ for $x \in \mathbb{R}^k$. Further abusing notation, write $f^{-1}(x) = (f^{-1}(x_1),...,f^{-1}(x_k))$ for $x \in \mathbb{R}^k$. Then $g$ is uniformly $(d_{f_1}, d_{f_2})$-continuous if and only if the function
\begin{equation*}
\begin{aligned}
h &: f_1(S) \to f_2(\mathbb{R}),\\
h(x) &= f_2(g(f_1^{-1}(x)))
\end{aligned}
\end{equation*}
is uniformly continuous with respect to the standard Euclidean metric.
\end{lemma}

\begin{proof}[Proof of Lemma~\ref{lemma:standard-uniform-reduction}]
$g$ is uniformly $(d_{f_1}, d_{f_2})$-continuous if and only if 
\begin{equation*}
\begin{aligned}
\forall \epsilon \kern3pt \exists \delta \text{ s.t. } \norm{f_1(x) - f_1(y)}_2 \geq \delta \text{ or } \norm{f_2(g(x)) - f_2(g(y))}_2 < \epsilon \text{ for all } x, y \in S.
\end{aligned}
\end{equation*}
Rewriting $x' = f_1(x), y' = f_1(y)$ and noting that $f_1$ is invertible, the above is equivalent to
\begin{equation*}
\begin{aligned}
\forall \epsilon \kern3pt \exists \delta \text{ s.t. } \norm{x' - y'}_2 \geq \delta \text{ or } \norm{f_2(g(f_1^{-1}(x'))) - f_2(g(f_1^{-1}(y')))}_2 < \epsilon \text{ for all } x', y' \in f_1(S),
\end{aligned}
\end{equation*}
which is equivalent to $h$ being uniformly continuous on the standard Euclidean metric.
\end{proof}

Next, we recall the definition and some properties of uniformly equivalent metrics.

\begin{definition}[Uniformly equivalent metrics]
\label{def:uniformly-equivalent-metric}
Let $d_1$ and $d_2$ be two metrics defined on a set $M$. Define $\text{Id} : (M, d_1) \to (M, d_2)$ via $\text{Id}(x) = x$. Then we will say that $d_1$ and $d_2$ are uniformly equivalent if $\text{Id}$ and $\text{Id}^{-1}$ are uniformly continuous.
\end{definition}

\begin{lemma}[Uniform equivalence of continuous bounded $f$-metrics]
\label{lemma:bounded-uniform-equivalence}
Let $f_1$ and $f_2$ be continuous, strictly increasing, and bounded functions. Then $d_{f_1}$ and $d_{f_2}$ are uniformly equivalent.
\end{lemma}

\begin{proof}[Proof of Lemma~\ref{lemma:bounded-uniform-equivalence}]
Note that $f_2 \circ \text{Id} \circ f_1^{-1} = f_2 \circ f_1^{-1}$. By Lemma~\ref{lemma:standard-uniform-reduction}, to show that $\text{Id}$ is uniformly continuous, it suffices to show that $f_2 \circ f_1^{-1}$, as a function from $f_1(\mathbb{R})^k$ to $f_2(\mathbb{R})^k$, is uniformly continuous with respect to the standard Euclidean metric. Recall that for $x \in f_1(\mathbb{R})^k$,
\begin{equation*}
\begin{aligned}
f_2(f_1^{-1}(x)) = (f_2(f_1^{-1}(x_1)),...,f_2(f_1^{-1}(x_k))).
\end{aligned}
\end{equation*}
Let $h$ denote $f_2 \circ f_1^{-1}$ as a univariate function from $\mathbb{R}$ to $\mathbb{R}$, i.e. $h(x) = f_2(f_1^{-1}(x))$ for $x \in f_1(\mathbb{R})$. Then $h$ is continuous, bounded (since $f_1$ is bounded), and strictly increasing since both $f_1^{-1}$ and $f_2$ are strictly increasing. A standard fact from real analysis is that continuous, bounded, and strictly increasing functions are uniformly continuous, so $h$ is uniformly continuous. Furthermore, the concatenation of uniformly continuous functions is uniformly continuous, so $f_2 \circ f_1^{-1}$ (as a function from $f_1(\mathbb{R})^k$ to $f_2(\mathbb{R})^k$) is indeed uniformly continuous. An analogous argument shows that $f_1 \circ f_2^{-1}$ is uniformly continuous and hence $\text{Id}^{-1}$ is uniformly continuous.
\end{proof}

\begin{lemma}[Uniform equivalence preserves uniform continuity]
\label{lemma:uniform-equivalence-preserves-uniform}
Let $M_X, M_Y$ be sets and suppose that $d_1^X, d_2^X$ are uniformly equivalent metrics on $M_X$ and $d_1^Y, d_2^Y$ are uniformly equivalent metrics on $M_Y$. Let $g$ be a function from $M_X \to M_Y$. Then $g$ is uniformly $(d_1^X, d_1^Y)$-continuous if and only if $g$ is uniformly $(d_2^X, d_2^Y)$-continuous.
\end{lemma}

\begin{proof}[Proof of Lemma~\ref{lemma:uniform-equivalence-preserves-uniform}]
Suppose $g$ is uniformly $(d_1^X, d_1^Y)$-continuous. Let $\text{Id}_X : (M_X, d_1^X) \to (M_X, d_2^X)$ and $\text{Id}_Y : (M_Y, d_1^Y) \to (M_Y, d_2^Y)$ be defined by $\text{Id}_X(x) = x$ and $\text{Id}_Y(y) = y$. Then note that $\text{Id}_Y \circ g \circ \text{Id}_X^{-1}$ is uniformly $(d_2^X, d_2^Y)$-continuous since $g$, $\text{Id}_X^{-1}$, and $\text{Id}_Y$ are uniformly continuous on their respective metrics by assumption, and the composition of uniformly continuous functions is uniformly continuous. But as a function, $\text{Id}_Y \circ g \circ \text{Id}_X^{-1}$ is just $g$, so $g$ is uniformly $(d_2^X, d_2^Y)$-continuous. The proof of the converse statement is identical.
\end{proof}

With these tools, we can finally establish that certain useful transformations preserve $f$-convergence within the class of continuous, bounded $f$. Note that in particular $\Phi$, the standard normal CDF, belongs to this class.

\begin{lemma}[Concrete $f$-convergence composition rules]
\label{lemma:concrete-f-composition-rules}
Let $f: \mathbb{R} \to \mathbb{R}$ be a continuous, strictly increasing, and bounded function. Let $A_n, B_n, C_n, D_n$ be sequences of random variables and suppose that $A_n \fconv{f} B_n$ and $C_n \fconv{f} D_n$. We have the following composition rules:
\begin{enumerate}
\item \textit{(Addition)} $A_n + C_n \fconv{f} B_n + D_n$ if $B_n$ is bounded in probability.
\item \textit{(Subtraction)} $A_n - C_n \fconv{f} B_n - D_n$ if $B_n$ is bounded in probability.
\item \textit{(Multiplication)} $A_n C_n \fconv{f} B_n D_n$ if $B_n$ is positively log-bounded in probability.
\item \textit{(Minimum)} $\min\set{ A_n, C_n } \fconv{f} \min\set{ B_n, D_n }$.
\item \textit{(Clipped truncated normal quantile)} Define $\mathrm{Clip}_{a, b}(x) = \max\{ a, \min\{ x, b \} \}$. For any fixed $b \in (0, 1)$, let
\begin{equation*}
\begin{aligned}
g(x, y) = \mathrm{Clip}_{0,\infty}\paren*{ \Phi^{-1}(\mathrm{Clip}_{0,b}(x) \cdot \Phi(y)) }.
\end{aligned}
\end{equation*}
Then $g(A_n, C_n) \fconv{f} g(B_n, D_n)$.
\end{enumerate}
\end{lemma}

\begin{proof}[Proof of Lemma~\ref{lemma:concrete-f-composition-rules}]
Define $f(x) = \frac{1}{1 + e^{-x}}$ and note that $f^{-1}(x) = \log\frac{x}{1 - x}$. Since $f$ is continuous, strictly increasing, and bounded, by Lemmas~\ref{lemma:bounded-uniform-equivalence} and~\ref{lemma:uniform-equivalence-preserves-uniform}, it suffices to establish the composition rules for this specific choice of $f$. In particular, we will establish the condition in Lemma~\ref{lemma:standard-uniform-reduction} for this choice of $f$, and then apply Lemma~\ref{lemma:abstract-f-composition-rule}.

\hfill

\noindent \textbf{(Addition)} Let $g(x, y) = x + y$. Note that
\begin{equation*}
\begin{aligned}
& g(f^{-1}(x), f^{-1}(y)) = \log\frac{x}{1 - x} + \log\frac{y}{1 - y}\\
\implies & \exp\paren*{g(f^{-1}(x), f^{-1}(y))} = \frac{x}{1 - x} \frac{y}{1 - y}\\
\implies & h(x, y) := f(g(f^{-1}(x), f^{-1}(y))) = \frac{1}{1 + \paren*{\frac{x}{1 - x} \frac{y}{1 - y}}^{-1}} = \frac{xy}{1 - x - y + 2xy}.
\end{aligned}
\end{equation*}
Consider the restriction of $g$ to $(-c, c) \times \mathbb{R}$ for any $c > 0$. Then since $f(\mathbb{R}) = (0,1)$ and $f((-c, c)) = (a, b)$ for some $0 < a < b < 1$, Lemma~\ref{lemma:standard-uniform-reduction} implies that it suffices to show that $h$ is uniformly continuous on $(a, b) \times (0, 1)$ for all $0 < a < b < 1$. For each fixed $x$, the expression $1 - x - y + 2xy$ is monotonic in $y$, and so the minimum is attained either when $y = 0$ or $y = 1$. Using this, it is straightforward to check that $1 - x - y + 2xy \geq \min\set{ a, 1 - b } > 0$ for all $(x, y) \in [a, b] \times [0, 1]$. So we have that $h$ is continuous, and hence uniformly continuous, on $[a, b] \times [0, 1]$. Thus it is also uniformly continuous on the restriction to $(a, b) \times (0, 1)$.

\hfill

\noindent \textbf{(Subtraction)} Let $g(x, y) = x - y$. A similar manipulation to the previous case yields
\begin{equation*}
\begin{aligned}
h(x, y) := f(g(f^{-1}(x), f^{-1}(y))) = \frac{1}{1 + \paren*{\frac{x}{1 - x} \frac{1 - y}{y}}^{-1}} = \frac{x(1 - y)}{x(1 - y) + (1 - x)y}.
\end{aligned}
\end{equation*}
Again, for $h$ restricted to $(a, b) \times (0, 1)$ for any $0 < a < b < 1$, the denominator is no smaller than $\min\set{ a, 1 - b } > 0$, so $h$ is continuous, and hence uniformly continuous, on $[a, b] \times [0, 1]$. Thus, $h$ is also uniformly continuous on the restriction to $(a, b) \times (0, 1)$.

\hfill

\noindent \textbf{(Multiplication)} Let $g(x, y) = xy$. Note that
\begin{equation*}
\begin{aligned}
& g(f^{-1}(x), f^{-1}(y)) = \log\frac{x}{1 - x} \cdot \log\frac{y}{1 - y}\\
\implies & \exp\paren*{g(f^{-1}(x), f^{-1}(y))} = \left(\frac{y}{1 - y}\right)^{\log \frac{x}{1 - x}}\\
\implies & h(x, y) := f(g(f^{-1}(x), f^{-1}(y))) = \frac{1}{1 + \paren*{\frac{y}{1-y}}^{-\log \frac{x}{1 - x}}}.
\end{aligned}
\end{equation*}
Consider the restriction of $g$ to $(1/c, c) \times \mathbb{R}$ for any $c > 1$. Then $f((1/c, c)) = (a, b)$ for some $0.5 < a < b < 1$. Thus, Lemma~\ref{lemma:standard-uniform-reduction} implies that it suffices to show that $h$ is uniformly continuous on $(a, b) \times (0, 1)$ for all $0.5 < a < b < 1$. Note that
\begin{equation*}
\begin{aligned}
h_2(z, y) := \frac{1}{1 + \paren*{\frac{y}{1 - y}}^{-z}} = \frac{1}{\frac{y^z}{y^z} + \frac{(1 - y)^z}{y^z}} = \frac{y^z}{y^z + (1 - y)^z}
\end{aligned}
\end{equation*}
is continuous for $(z, y) \in I \times [0,1]$ for any compact interval $I$ that excludes $0$, since then the denominator $y^z + (1 - y)^z$ is well-defined and strictly greater than $0$. Furthermore, we have that $h_3(x, y) := \paren*{\log\frac{x}{1 - x}, y}$ is a continuous function for $(x, y) \in [a, b] \times [0,1]$, and moreover the range of $h_3$ is contained in $I \times [0,1]$ for some compact interval $I$ that excludes $0$. Hence, $h_2 \circ h_3$ is a continuous function on $[a, b] \times [0, 1]$, and hence uniformly continuous. But $h = h_2 \circ h_3$ on the restriction to $(a, b) \times (0, 1)$, so it is uniformly continuous on $(a, b) \times (0, 1)$.

\hfill

\noindent \textbf{(Minimum)} Let $g(x, y) = \min\set{ x, y }$. Note that
\begin{equation*}
\begin{aligned}
& g(f^{-1}(x), f^{-1}(y)) = \min\set*{\log\frac{x}{1 - x}, \log\frac{y}{1 - y}}\\
\implies & \exp\paren*{g(f^{-1}(x), f^{-1}(y))} = \min\set*{\frac{x}{1 - x}, \frac{y}{1 - y}}\\
\implies & h(x, y) := f(g(f^{-1}(x), f^{-1}(y))) = \frac{1}{1 + \min\set*{\frac{x}{1 - x}, \frac{y}{1 - y}}^{-1}} = \min\set*{x, y}.
\end{aligned}
\end{equation*}
We see that $h$ is continuous and hence uniformly continuous on $[0,1] \times [0,1]$. Thus, it is uniformly continuous on the restriction to $(0,1) \times (0,1)$.

\hfill

\noindent \textbf{(Clipped truncated normal quantile)} For this last rule, we will establish the conditions in Lemma~\ref{lemma:standard-uniform-reduction} with $f = \Phi$. In this case,
\begin{equation*}
\begin{aligned}
& g(f^{-1}(x), f^{-1}(y)) = \mathrm{Clip}_{0, \infty}\paren*{ \Phi^{-1}\paren*{\mathrm{Clip}_{0,b}(\Phi^{-1}(x)) \cdot y} }\\
\implies & h(x, y) := f(g(f^{-1}(x), f^{-1}(y))) = \mathrm{Clip}_{0.5,1}\paren*{\mathrm{Clip}_{0,b}(\Phi^{-1}(x)) \cdot y}.
\end{aligned}
\end{equation*}
Next note that
\begin{equation*}
\begin{aligned}
\mathrm{Clip}_{0.5,1}\paren*{\mathrm{Clip}_{0,b}(\Phi^{-1}(x)) \cdot y} &= \mathrm{Clip}_{0.5,1}\paren*{\mathrm{Clip}_{0.5,b}(\Phi^{-1}(x)) \cdot y}\\
&= \mathrm{Clip}_{0.5,1}\paren*{\Phi^{-1}(\mathrm{Clip}_{\Phi(0.5),\Phi(b)}(x)) \cdot y}
\end{aligned}
\end{equation*}
where the first equality follows from the fact that if $\Phi^{-1}(x) \leq 0.5$, then the largest value that $\mathrm{Clip}_{0,b}(\Phi^{-1}(x)) \cdot y$ can take over $y \in [0,1]$ is $0.5$. Since this quantity gets clipped to the range $[0.5,1]$ regardless, it makes no difference if we apply an additional clipping that affects values outside of this range prior to clipping to $[0.5, 1]$. Then note that $\mathrm{Clip}_{\Phi(0.5),\Phi(b)}(x)$ is a continuous function on $[0,1]$ whose range is $[\Phi(0.5), \Phi(b)]$, and that $\Phi^{-1}$ is continuous on $[\Phi(0.5), \Phi(b)]$. Hence, $h(x, y)$ is continuous, and thus uniformly continuous, on $[0,1] \times [0, 1]$. In particular, it is uniformly continuous on the restriction to $(0,1) \times (0,1)$.
\end{proof}

We state one more general composition rule with a simpler proof than those above.

\begin{lemma}[Indicator $f$-convergence composition rule]
\label{lemma:indic-f-conv-composition}
Suppose that $X_n \fconv{f} Y_n$ and that $A_n$ is a sequence of binary random variables. Then $X_n \cdot A_n \fconv{f} Y_n \cdot A_n$.
\end{lemma}

\begin{proof}[Proof of Lemma~\ref{lemma:indic-f-conv-composition}]
The result follows from
\begin{equation*}
\begin{aligned}
d_f(X_n \cdot A_n, Y_n \cdot A_n) = d_f(X_n, Y_n) \cdot A_n \leq d_f(X_n, Y_n) \stackrel{p}{\to} 0.
\end{aligned}
\end{equation*}
\end{proof}

Lastly, we state a result that connects $f$-convergence to convergence under the usual Euclidean metric for bounded sequences.

\begin{lemma}[$f$-convergence to ordinary convergence]
\label{lemma:f-conv-ordinary-bounded}
Let $f$ be a continuous, strictly increasing, and bounded function, and suppose $X_n \fconv{f} Y_n$. Furthermore, suppose $X_n$ and $Y_n$ are bounded a.s. by some constant $M$ for all $n$. Then
\begin{equation*}
\begin{aligned}
\abs{X_n - Y_n} \stackrel{p}{\to} 0.
\end{aligned}
\end{equation*}
\end{lemma}

\begin{proof}[Proof of Lemma~\ref{lemma:f-conv-ordinary-bounded}]
Let $\epsilon, \delta > 0$. Let $g$ denote the inverse of $f$, which exists and is continuous. Note that $g$ is uniformly continuous on the compact interval $I := [f(-M), f(M)]$. Thus, choose $s$ so that $|x - y| < s \implies |g(x) - g(y)| < \delta$ for all $x, y \in I$. By assumption, there exists an $N$ such that $\Prb\paren*{\abs{f(X_n) - f(Y_n)} \geq s} < \epsilon$ for all $n > N$. Then
\begin{equation*}
\begin{aligned}
\Prb\paren*{\abs{X_n - Y_n} > \delta} = \Prb\paren*{ \abs{g(f(X_n)) - g(f(Y_n))} > \delta} \leq \Prb\paren*{ \abs{f(X_n) - f(Y_n)} \geq s} < \epsilon.
\end{aligned}
\end{equation*}
\end{proof}

\end{appendices}

\end{document}